\documentclass[a4paper,11pt]{article}
\usepackage{jheppub, appendix, mathrsfs} % for details on the use of the package, please see the JINST-author-manual
\usepackage{lineno}

\usepackage{braket}
\usepackage{subfig}
\usepackage{placeins}
\usepackage[normalem]{ulem}
\title{\boldmath (A)Symmetric Complexity and the Quantum Mpemba Effect}

\author{Cameron Beetar, Jeff Murugan and Hendrik J.R. Van Zyl}
\affiliation{The Laboratory for Quantum Gravity \& Strings,\\
Department of Mathematics and Applied Mathematics,\\
University of Cape Town, Private Bag, Rondebosch, 7701,\\
South Africa}

\emailAdd{btrcam001t@myuct.ac.za}
\emailAdd{jeff.murugan@uct.ac.za}
\emailAdd{hjrvanzyl@gmail.com}

\abstract{The Quantum Mpemba Effect (QME) — the counter-intuitive phenomenon where states further from equilibrium can relax faster than those closer to it — challenges standard expectations of quantum thermalization. In this work, we introduce Krylov complexity as a sensitive diagnostic for the QME. We show that Krylov spread complexity encodes the asymmetry essential to the effect, and we define a new class of projective (a)symmetric complexities that sharpen this connection. Strikingly, the structure of these projective complexities at the initial moment (t=0) already carries predictive power for the onset of Mpemba-like inversions, obviating the need for explicit time evolution. Our results suggest that the geometry of states in Krylov space captures deep information about non-monotonic relaxation and provides a powerful framework for diagnosing and anticipating anomalous thermalization phenomena in quantum systems.}
\begin{document}
\maketitle
\flushbottom

\section{Introduction}
\label{sec:intro}
\noindent
The paradoxical phenomenon where hot liquid can, under certain conditions, freeze faster than cold liquid -- known as the Mpemba effect \cite{Mpemba1969} -- has long confused scientists and the lay public alike. Recently though, interest in its quantum analogue, the Quantum Mpemba Effect (QME) \cite{Ares:2025onj}, has surged, driven by its potential to probe nontrivial relaxation dynamics in complex quantum systems. Unlike classical counterparts, where intuitive notions of thermal gradients drive equilibration, quantum systems carry additional layers of structure, such as conserved quantities and sectorial symmetries, which can intricately modulate the route to equilibrium. Capturing such phenomena requires tools sensitive not only to population imbalances but to coherence structures between symmetry sectors. This paper explores how quantum complexity measures, particularly Entanglement Asymmetry \cite{Ares2023} and Krylov Complexity \cite{Parker:2018yvk, Nandy:2024evd, Rabinovici:2025otw, Baiguera:2025dkc}, serve as diagnostic tools for the QME, revealing hidden layers of dynamical behavior.\\

\medskip
\noindent
The entanglement asymmetry, introduced in \cite{Ares2023}, has emerged as one way to quantify the distance between a reduced density matrix $\rho_A$ of a subsystem $A$ and its symmetry-projected counterpart $\rho_{A,Q}$, defined via the relative entropy
\begin{equation}
   \Delta S = S(\rho_{A,Q}) - S(\rho_A),
   \label{eq:EA}
\end{equation}
where $S(\rho) = -\text{Tr}(\rho \ln \rho)$ is the von Neumann entropy of the system and the symmetry-projected density matrix,
\begin{equation}
   \rho_{A,Q} = \sum_q \Pi_q \rho_A \Pi_q,
\end{equation}
is obtained by projecting $\rho_A$ onto the charge sectors labeled by $q$ associated with a conserved symmetry, with projectors $\Pi_q$. If $\rho_A$ is already block-diagonal in these sectors, $\Delta S$ vanishes. Otherwise, $\Delta S$ measures the asymmetry or coherence across these sectors. Importantly, $\Delta S$ is a divergence, rather than a true distance—it is non-negative, zero only when $\rho_A = \rho_{A,Q}$, and lacks both the symmetric property and the triangle inequality \cite{Nielsen:2012yss}.\\

\noindent
In the context of the Quantum Mpemba Effect \cite{Ares:2025onj}, the entanglement asymmetry serves as a dynamic probe to track how coherences across symmetry sectors evolve over time. Specifically, for two initial states $\rho_1$ and $\rho_2$, we say that a Mpemba-like swap occurs if:

\begin{enumerate}
\item Initially, $\Delta S(\rho_1) < \Delta S(\rho_2)$ at $t = t_0$, but
\item There exists a later time $t_M > t_0$ where $\Delta S(\rho_1) > \Delta S(\rho_2)$.
\end{enumerate}
Note that this inversion of the entanglement asymmetry ordering implies that the plots of $\Delta S(\rho_1(t))$ and $\Delta S(\rho_2(t))$ \textit{cross} at some point in time.  A stronger version of the QME demands that this inversion persists for all $t \geq t_M$. Notably, vanishing entanglement asymmetry is necessary -- but not sufficient -- for thermal equilibrium, making it a subtle yet informative diagnostic for nonequilibrium dynamics. On the other hand, Krylov Complexity \cite{Parker:2018yvk, Nandy:2024evd, Rabinovici:2025otw}, or more specifically ``spread complexity” \cite{Balasubramanian:2022tpr}, quantifies how a state spreads in its Krylov basis under time evolution. Originally applied to operator dynamics, its adaptation to state evolution has proven sensitive to phase distinctions, such as thermalization-localization transitions \cite{RevModPhys.91.021001} and $\mathsf{PT}$-symmetry breaking \cite{Beetar:2023mfn, Bhattacharya:2024hto}. Like entanglement asymmetry, Krylov complexity is non-symmetric and lacks a triangle inequality \cite{Aguilar-Gutierrez:2023nyk}, but it faithfully measures the ``distance” (in complexity space) between a fixed initial state and its time-evolved form. \\

\noindent
The interplay between entanglement asymmetry and Krylov complexity is nuanced. The former is sensitive to coherences between symmetry sectors, while the latter captures the overall spreading -- both within and across sectors. This distinction becomes crucial in systems where dynamics exhibit both diffusive motion within symmetry sectors and hopping between them. Prior investigations \cite{Caputa:2025mii} approached Krylov complexity by summing over sector complexities, accurately capturing {\it intra}-section diffusion but missing some of the nuances of {\it inter}-sector hopping -- a crucial component for detecting Mpemba-like anomalous relaxation. In this work, we propose a complementary approach; while entanglement asymmetry naturally isolates inter-sector hopping dynamics, we develop a complexity formalism designed to appropriately weight intra-section diffusion contributions, aligning the sensitivity of Krylov complexity to the ``asymmetry-hopping" dynamics detected by entanglement asymmetry. This alignment not only sharpens the former's utility in probing QME but also bridges complexity-theoretic perspectives with symmetry-resolved quantum dynamics.\\

\noindent
As a concrete setting, we focus on the Aubry-André (AA) model \cite{Aubry1980}, a one-dimensional crystal with periodically varying onsite energies exhibiting a tunable transition between 
Eigenstate Thermalization Hypothesis \cite{Deutsch:2018ulr} (ETH)-dominated phases and Many-Body Localized (MBL) \cite{Nandkishore:2014kca, Altman:2014fhm} regimes. The Hamiltonian is given by
\begin{equation}
H = -\sum_{i}^{N} \left( \sigma_i^x \sigma_{i+1}^x + \sigma_i^y \sigma_{i+1}^y \right) + V \sum_{i=1}^{N} \sigma_i^z \sigma_{i+1}^z + \sum_{i=1}^N W_i \sigma_i^z,
\label{eq:AAHam}
\end{equation}
where $V$ controls nearest-neighbour interactions and $W_i = W \cos(2\pi \alpha i + \phi)$ represents a quasiperiodic potential with $\alpha = (\sqrt{5}+1)/2$ and random phase $\phi$. The nearest-neighbour interaction strength is set to $V=1/2$ throughout this work. The model hosts a well-characterized ETH/MBL transition around $W_c \approx 3.3$ (see appendix \ref{app:criticalW} and \cite{liu2024}). The global $U(1)$ symmetry corresponding to total magnetization $Q = \sum_i \sigma_i^z$ partitions the Hilbert space into charge sectors, furnishing a natural playground for exploring entanglement asymmetry and Krylov complexity in tandem.\\

\noindent
In this paper, we demonstrate that a suitably refined Krylov complexity can serve as order parameters for the quantum Mpemba effect in the AA model, much like entanglement asymmetry $\Delta S$ \cite{liu2024}. We show that identical (pre-quench) initial states have differing asymmetry profiles due to the quench (tilt) operation, and can exhibit Mpemba-like inversions in their relaxation dynamics. Furthermore, we elucidate how the ETH/MBL transition modulates the prevalence and timescales of these inversions, with the MBL regime enhancing long-lived asymmetry-induced dynamical memory effects.\\

\noindent
The rest of this article is as follows: In Section \ref{sec:EAKC}, we formalize the relationship between entanglement asymmetry and Krylov complexity, highlighting their complementary sensitivities. Section \ref{sec:KCsymmres} outlines in detail the notion of symmetry-resolved Krylov complexity and the required modification of this measure, and presents our main finding: an Mpemba-like crossing in the symmetric complexity. The following section \ref{sec:KCtheta} investigates the initial values of the (a)symmetric complexity and Lanczos coefficients, and analyzes their usefulness as probes of Mpemba inversions. We conclude in Section \ref{sec:conclusion} with a broader discussion of the implications for complexity-theoretic perspectives on nonequilibrium quantum dynamics.

\section{Entanglement Asymmetry and Krylov Complexity}
\label{sec:EAKC}
Entanglement asymmetry has proven to be a valuable probe of the quantum Mpemba effect, particularly in situations where a broken symmetry is approximately restored under time evolution \cite{Ares:2025onj, Ares2023}. Unlike its classical counterpart, however, the quantum version is more nuanced. At least two distinct definitions arise, depending on whether one prioritizes (i) the time required to reach a steady state\footnote{This differs from the classical case, where the final state is strict thermal equilibrium. Quantum systems may not exhibit such equilibration, but for our purposes we assume the existence of a steady state.} or (ii) the extent of symmetry restoration. The strongest form of the QME would demand both: the more asymmetric (“hotter”) state not only crosses the less asymmetric (“colder”) one but also reaches its steady state more rapidly. In this work, however, we adopt a definition that emphasizes the crossing in entanglement asymmetry as the primary diagnostic, since in practice the steady states typically occur on very similar timescales\footnote{This definition is also employed in \cite{ares2025qme}, where the authors state: “In terms of symmetries, the quantum Mpemba effect happens when the more the symmetry is broken, the faster it is restored.”}. A weaker formulation is also conceivable, in which symmetry restoration plays no role and the effect is defined purely in terms of time-to-steady-state. These distinctions arise because, in the classical setting, trajectories converge to a common macroscopic equilibrium, while in the quantum case distinct initial states generally evolve toward different late-time configurations\footnote{Moreover, quantum steady states are subject to fluctuations.}. Under unitary dynamics, the role of the asymmetry operator is to induce an effective temperature that depends on the interplay of the dynamics, the initial state, and the operator itself. Within this framework, entanglement asymmetry serves as a positive semi-definite quantity that measures the distance from a symmetric configuration. Nevertheless, the subtleties of defining and interpreting the QME in the quantum regime require careful consideration.\\

\noindent
Krylov \cite{Parker:2018yvk, Nandy:2024evd, Rabinovici:2025otw, Baiguera:2025dkc}, or more precisely, spread complexity \cite{Balasubramanian:2022tpr} appears, at first sight, to be a natural probe of the QME for several reasons. Unlike entanglement asymmetry, which requires the arbitrary choice of a subsystem, Krylov complexity provides a {\it global} characterization of quantum dynamics: spread complexity measures how the entire wavefunction explores the full accessible Hilbert space. Its evolution is determined solely by the Lanczos coefficients $a_n$, the diagonal entries of the Krylov Hamiltonian which encodes local onsite potentials in the Krylov chain, and $b_n$, the off-diagonals of the Krylov Hamiltonian which encode the microscopic rate of this spreading \cite{Lanczos1950AnIM, viswanath2008recursion}. Moreover, the initial asymmetry quench produces a nontrivial superposition that serves as the seed of the Lanczos recursion. Since each initial state generates a distinct set of Lanczos coefficients and thus a unique trajectory in Krylov space, the corresponding complexity $C_K(t)$ directly encodes the relaxation pathway. As we will show below, this makes Krylov complexity particularly sensitive to Mpemba-like inversions, which are defined by anomalous reorderings of relaxation trajectories between different initial conditions.\\

\noindent
Within the Aubry–André model (\ref{eq:AAHam}), the presence of the ETH/MBL transition offers an ideal setting in which to compare entanglement asymmetry and Krylov complexity as complementary probes of the QME in both thermalizing and localizing regimes. To this end, we examined two extreme initial product states. The fully polarized ferromagnetic state $\ket{\uparrow\uparrow\cdots\uparrow}$, which lies at the edge of the symmetry spectrum ($S_z^{\mathrm{tot}}=N$), explores a sparse region of Hilbert space. Its constrained dynamics make it a natural candidate for exhibiting non-chaotic relaxation and enhanced susceptibility to anomalous effects such as the QME, across all values of the quasiperiodic potential strength $W$. By contrast, the Néel state $\ket{\uparrow\downarrow\cdots\uparrow\downarrow}$ resides in the central sector ($S_z^{\mathrm{tot}}=0$), the largest and densest region of Hilbert space. States in this sector are expected to display strongly chaotic behavior, thereby providing a complementary probe of how the QME manifests in the ergodic regime.
\\ \\
To investigate the QME, it will be necessary to explicitly break the symmetry of the system. Toward this end, we employ the conventional tilt operation \cite{Ares2023,liu2024},
\begin{equation}
\ket{\psi(\theta)} = \exp\left(-i\frac{\theta}{2}\sum_i \sigma_i^y\right)\ket{\psi(0)} \equiv U(\theta)\ket{\psi(0)},
\label{eq:tilt}
\end{equation}
with the initial state $\ket{\psi(0)}$ taken to be either the ferromagnetic or Néel states above. The resulting one-parameter families of states will be referred to as Tilted Ferromagnetic States (TFS) and Tilted Néel States (TNS), respectively. The AA Hamiltonian preserves a global $U(1)$ symmetry associated with the total magnetization, since $[H, \sum_i \sigma_i^z]=0$. However, the tilt operator $U(\theta)$ does not commute with this conserved charge. Consequently, in the presence of tilting, $\theta$ functions as a continuous asymmetry-inducing parameter. Within the interval $\theta \in [0,\pi/2]$, it monotonically increases the initial entanglement asymmetry of the state. For clarity of presentation in all figures, we rescale $\theta$ by a factor of $\pi$, so that $\theta$ denotes a fraction of $\pi$ (e.g., $\theta=0.1$ corresponds to a tilt angle of $0.1\pi$).

\subsection{Entanglement Asymmetry}

The QME in the Aubry–André model has recently been studied in \cite{liu2024}. In our work, the entanglement asymmetry serves as the baseline probe for defining the QME, and reproducing its behavior is essential both for benchmarking against previous results as well as for comparison with Krylov complexity. For completeness, and to verify the accuracy of our numerics, we summarize the key features of the entanglement asymmetry-based analysis here.\\

\begin{figure}[!htbp]
\centering
\subfloat[TFS with $W=2.0$]{\includegraphics[width=7cm]{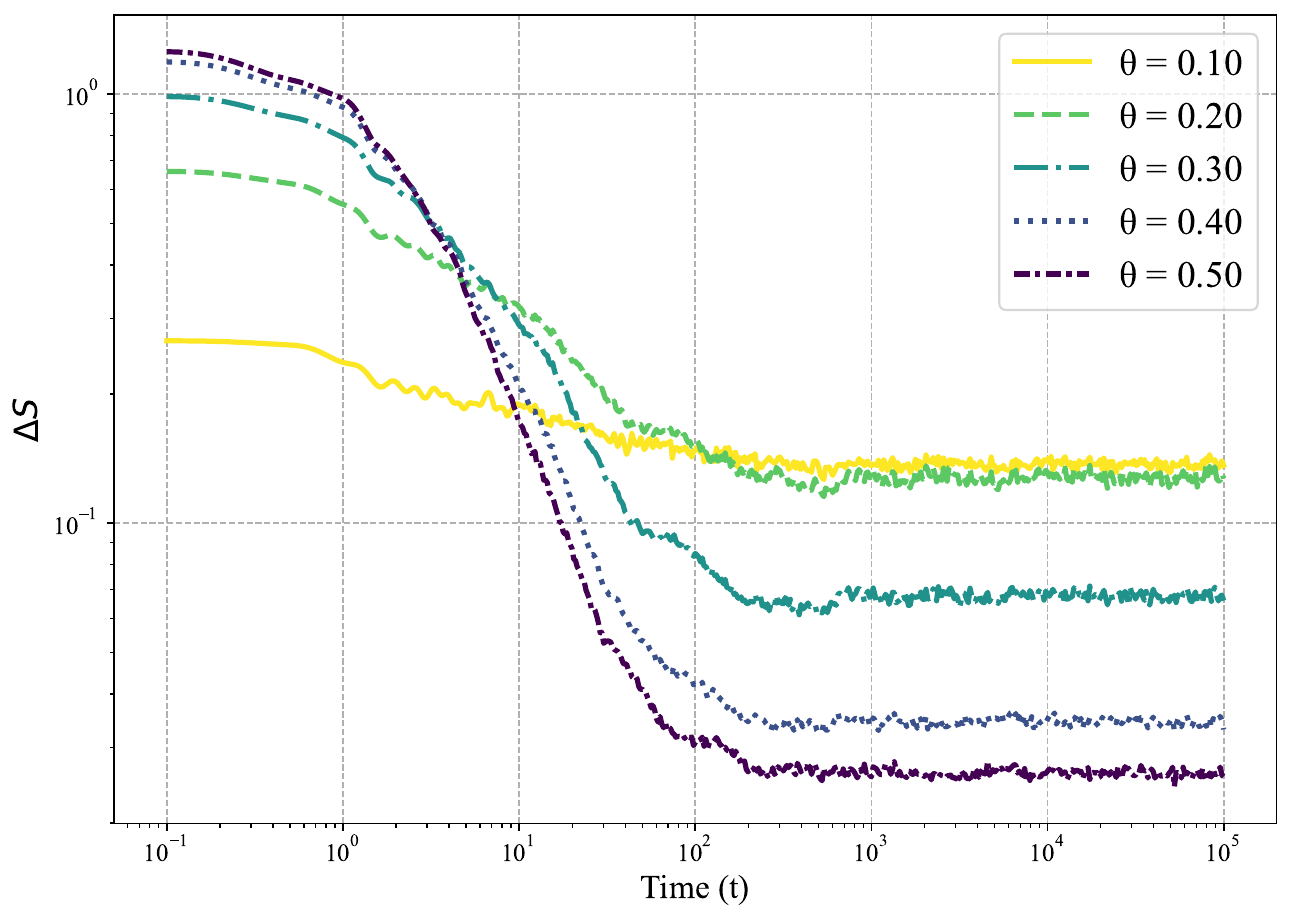}}\hfil
\subfloat[TNS with $W=2.0$]{\includegraphics[width=7cm]{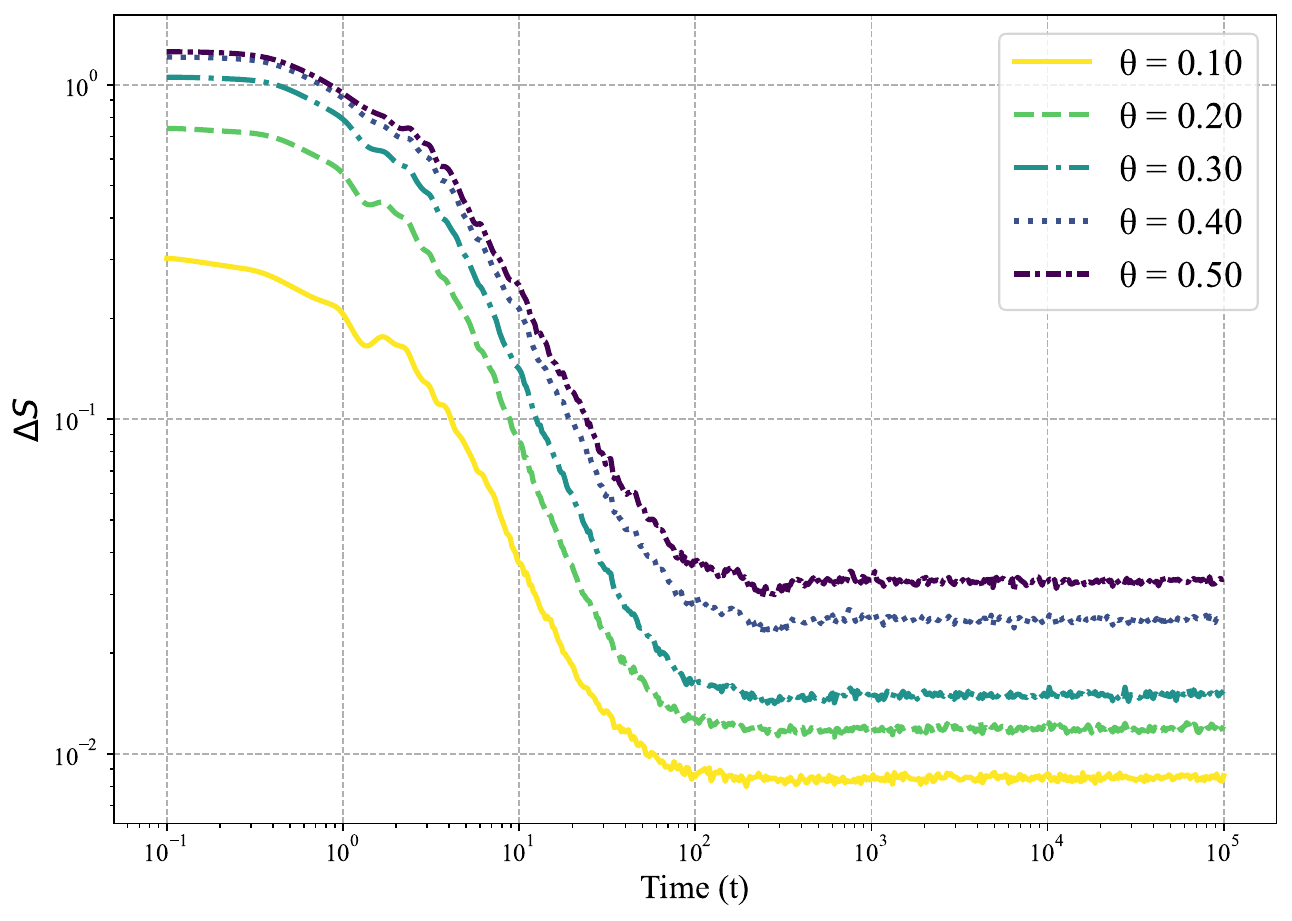}}\hfil 

\subfloat[TFS with $W=2.5$]{\includegraphics[width=7cm]{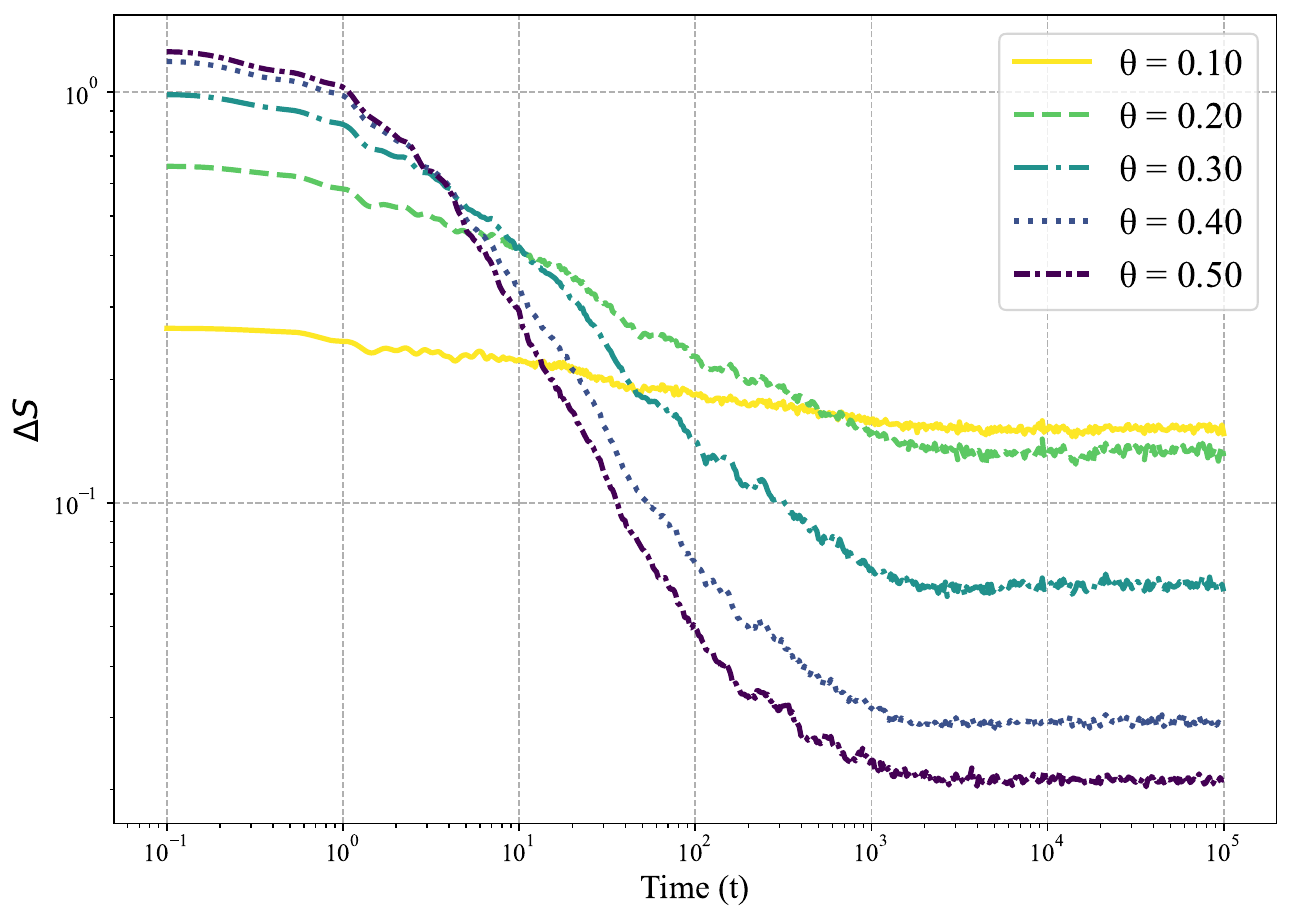}}\hfil
\subfloat[TNS with $W=2.5$ \label{fig:EAsubtleQME}]{\includegraphics[width=7cm]{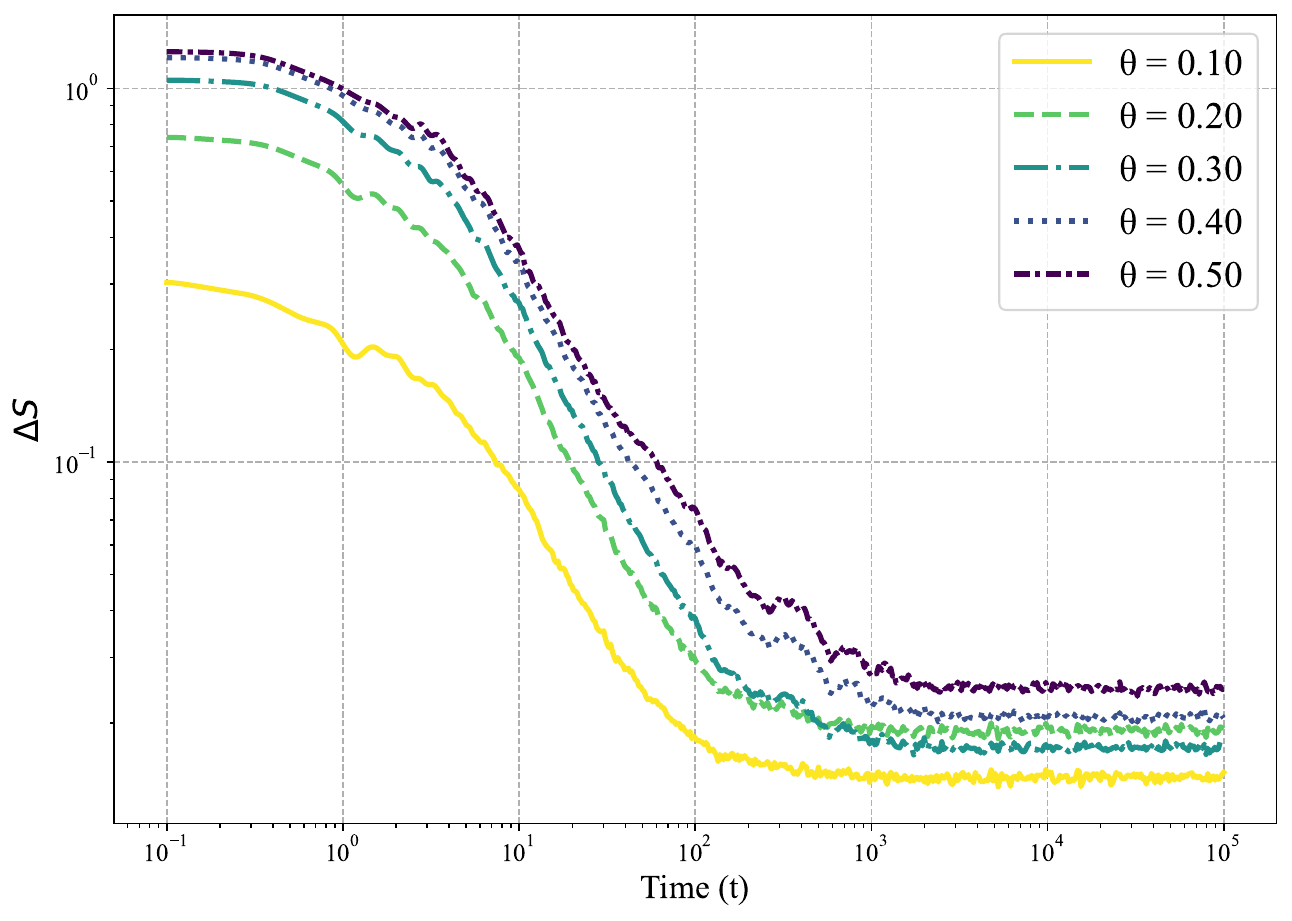}}\hfil 

\subfloat[TFS with $W=5.0$]{\includegraphics[width=7cm]{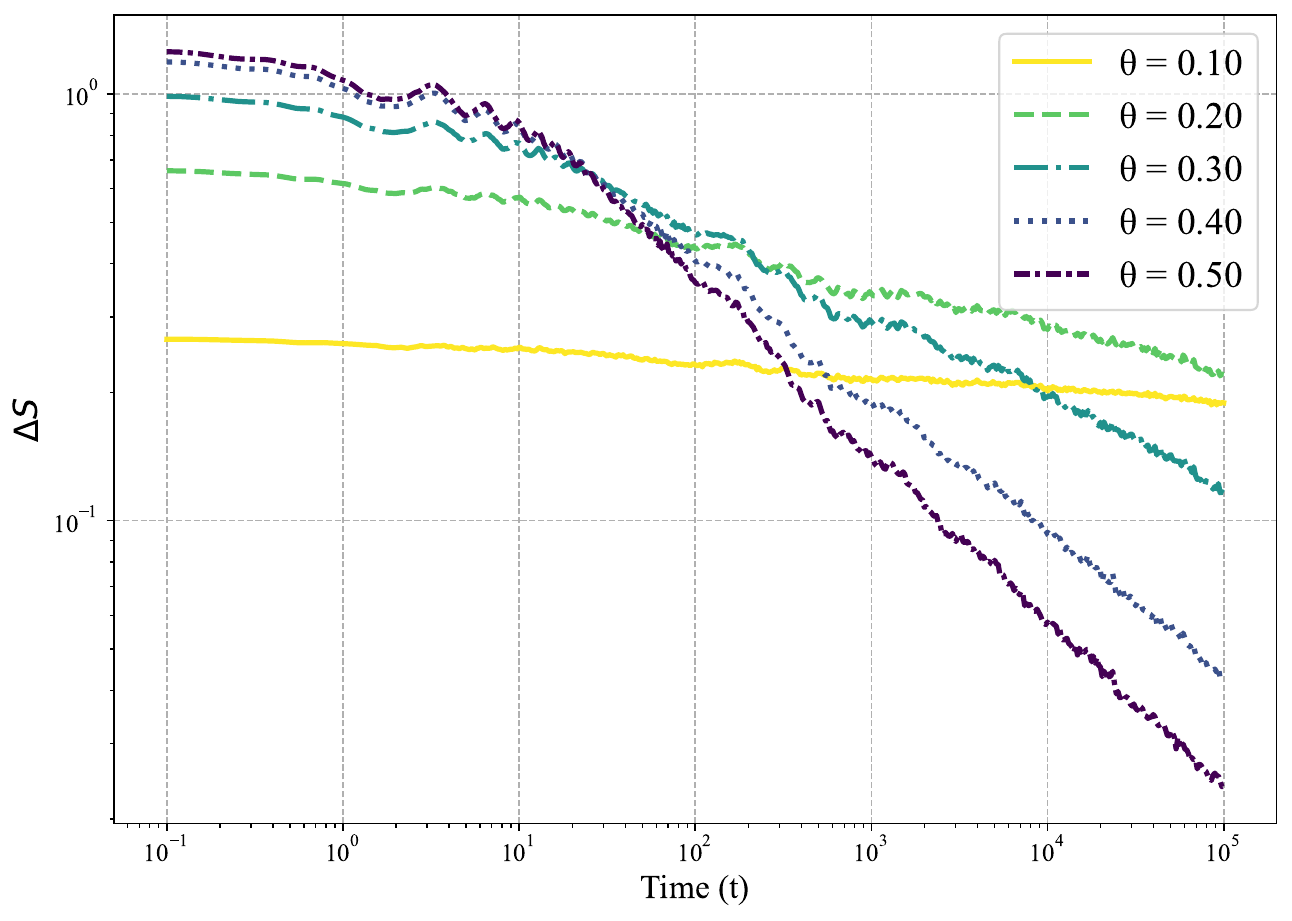}}\hfil
\subfloat[TNS with $W=5.0$]{\includegraphics[width=7cm]{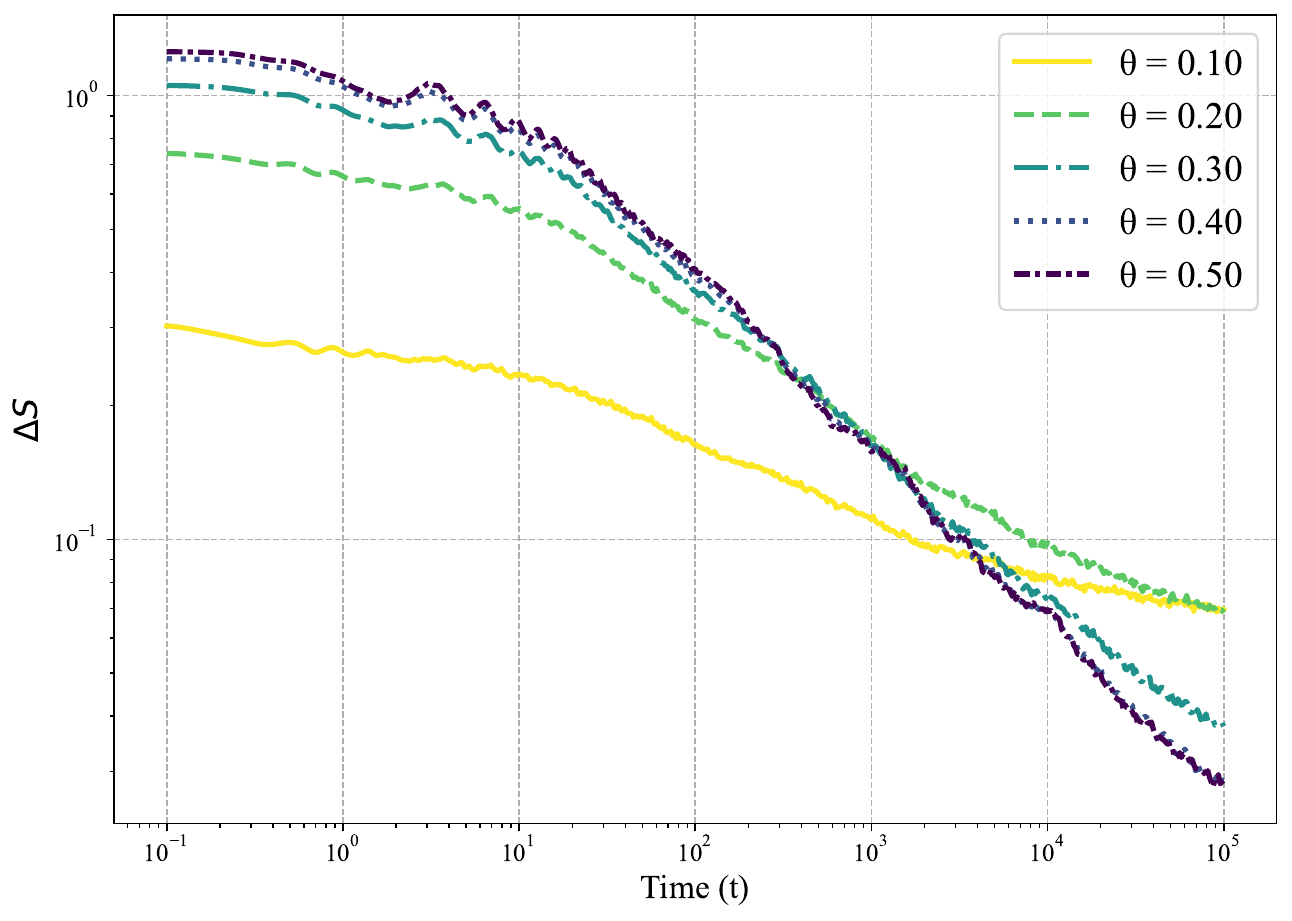}}\hfil 
\caption{Figures demonstrating the difference between the TFS (left) and TNS (right) states, in terms of the entanglement asymmetry for an $N=12$, ($N_A=3$) size (sub)system, averaged over 240 realisations. The QME is always present in the TFS states, but only begins to appear for TNS states at around $W\sim 2.5$ (one can see $\theta=0.3$ EA cross the $\theta=0.2$ EA).}
\label{fig:EA}
\end{figure}

\noindent
We prepare the system in two tilted initial states, $\ket{\psi(\theta_1)}$ and $\ket{\psi(\theta_2)}$, and evolve them in time. A QME is said to occur whenever the entanglement asymmetry curves for the two states exhibit a crossing. Unless otherwise stated, our simulations use system size $N=12$, with subsystem size $N_A=3$ (where required), averaged over 240 realizations of the random phase $\phi$, sampled from the interval $[0,2\pi]$ using stratified sampling to ensure uniform coverage. Figure \ref{fig:EA} demonstrates that, for tilted ferromagnetic states, entanglement asymmetry crossings are observed for all reasonable choices of tilt angles (e.g., $\theta_1=0.2, \theta_2=0.5$), independent of the quasiperiodic potential strength $W$. This robustness extends across the ETH/MBL transition, centered near the critical potential $W_c \approx 3.3$. For $W<W_c$ the system is in the ETH regime, while for $W>W_c$ it resides in the MBL regime.\\

\noindent
In contrast, the behavior of tilted Néel states is more subtle. Deep in the ETH regime (small $W$), no QME is observed. As $W$ increases toward the MBL regime -- but remains below $W_c$ -- crossings appear for some pairs of initial tilts, though not universally (see Fig. \ref{fig:EAsubtleQME}). Once $W$ exceeds the critical value, the QME is observed consistently for all pairs of tilted Néel states. In both ferromagnetic and Néel cases, increasing $W$ delays the approach to steady state; however, given the persistence of the crossings over long time windows (typically $\Delta t \approx 9\times 10^4$), we interpret them as robust indicators of the QME, even without full convergence to late times. To summarise:

\begin{itemize}
    \item For tilted ferromagnetic states, the QME is observed for all $W$.
    \item For tilted Néel states, the QME is observed for all $W>W_c$ (MBL regime).
    \item For $W<W_c$ (ETH regime), the QME may still appear for certain TNS pairs if $W$ is close to criticality, but vanishes for sufficiently small $W \lesssim 2.5$.
\end{itemize}

\noindent
These findings are consistent with the numerical results of \cite{liu2024}, confirming the reliability of our computations. Importantly, the entanglement asymmetry crossings always occur only once (up to small late-time fluctuations) for well-separated tilt values, $0.1\lesssim |\theta_1-\theta_2| $. This implies that the QME can be understood in terms of a reordering of entanglement asymmetry values. If, and only if, the initial ordering at $t=0$ is reversed relative to the late-time ordering, the QME is present. In other words, whenever the more asymmetric state relaxes to become more symmetric than its less asymmetric counterpart, the quantum Mpemba effect is realized.

\subsection{Spread Complexity}

We now turn to the study of Krylov -- or more precisely, spread -- complexity for the tilted ferromagnetic and tilted Néel states. As a brief summary\footnote{For further detail, see \cite{Balasubramanian:2022tpr}.}, spread complexity requires only two inputs: a reference state and the system Hamiltonian. Starting from these, one constructs the Krylov basis
$\{ |K_0\rangle, |K_1\rangle, |K_2\rangle, \dots \}$,
by orthogonalizing and normalizing the sequence
$\{ |\psi\rangle, H|\psi\rangle, H^2|\psi\rangle, H^3|\psi\rangle, \dots \}$.
This procedure yields an orthogonal basis ordered by increasing complexity, spanning the entire subspace explored by the time-evolved reference state.\\

\noindent
The Krylov complexity operator is then defined as
\begin{eqnarray}
\hat{C}_K = \sum_{n} n \, |K_n\rangle\langle K_n|,
\end{eqnarray}
so that the complexity of a target state $|\phi_t\rangle$ is given by the expectation value \\
$C_K(t) = \langle \phi_t | \hat{C}_K | \phi_t \rangle$. In our study, the reference state is taken to be either a tilted ferromagnetic state or a tilted Néel state, while the corresponding target states are their respective time-evolved configurations. In practice, we compute the spread complexity for individual realizations of the Aubry–André Hamiltonian and subsequently average over ensembles of realizations to obtain statistically robust results.\\ 

\begin{figure}[!htbp]

\centering
\subfloat[TFS with $W=2.0$]{\includegraphics[width=7cm]{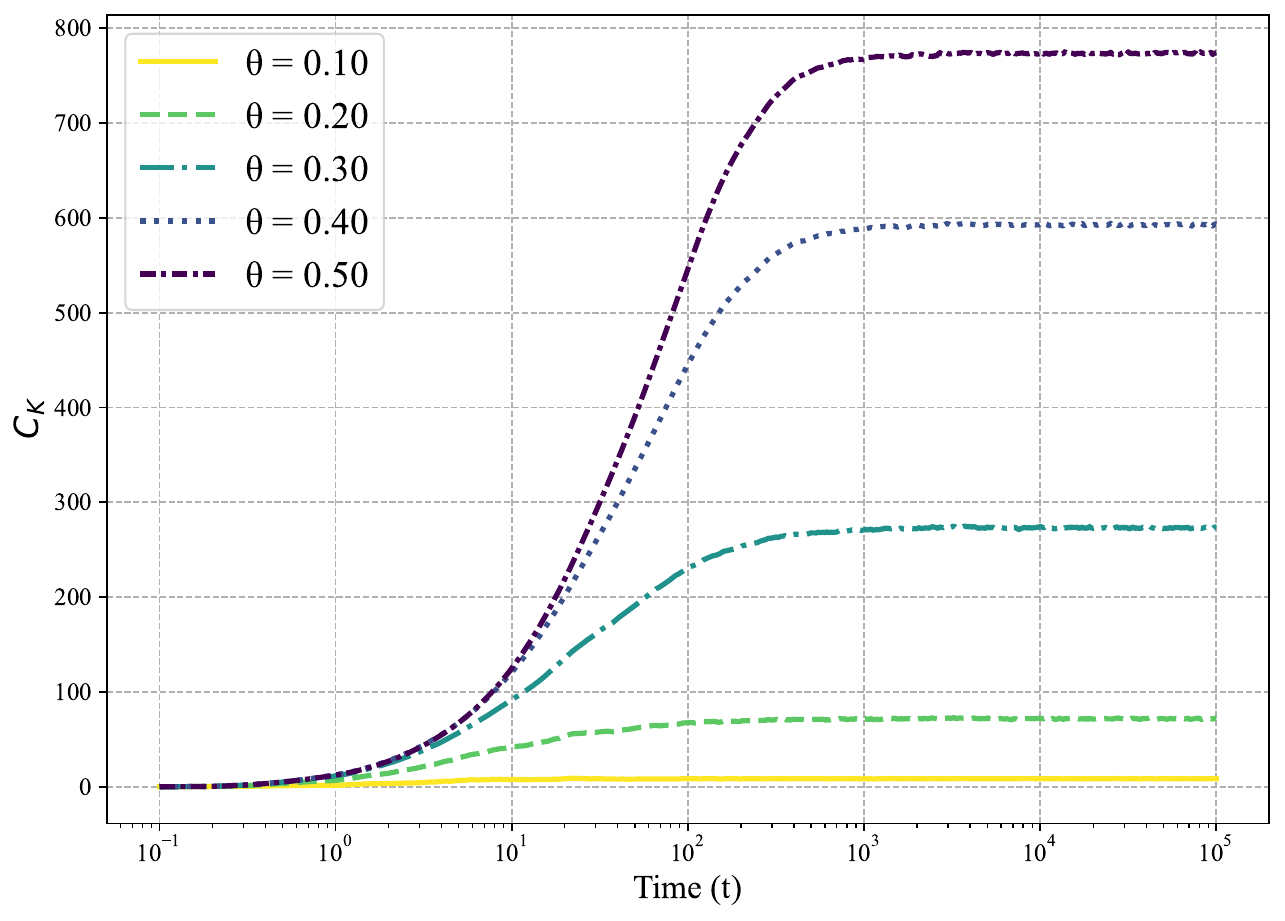}}\hfil
\subfloat[TNS with $W=2.0$]{\includegraphics[width=7cm]{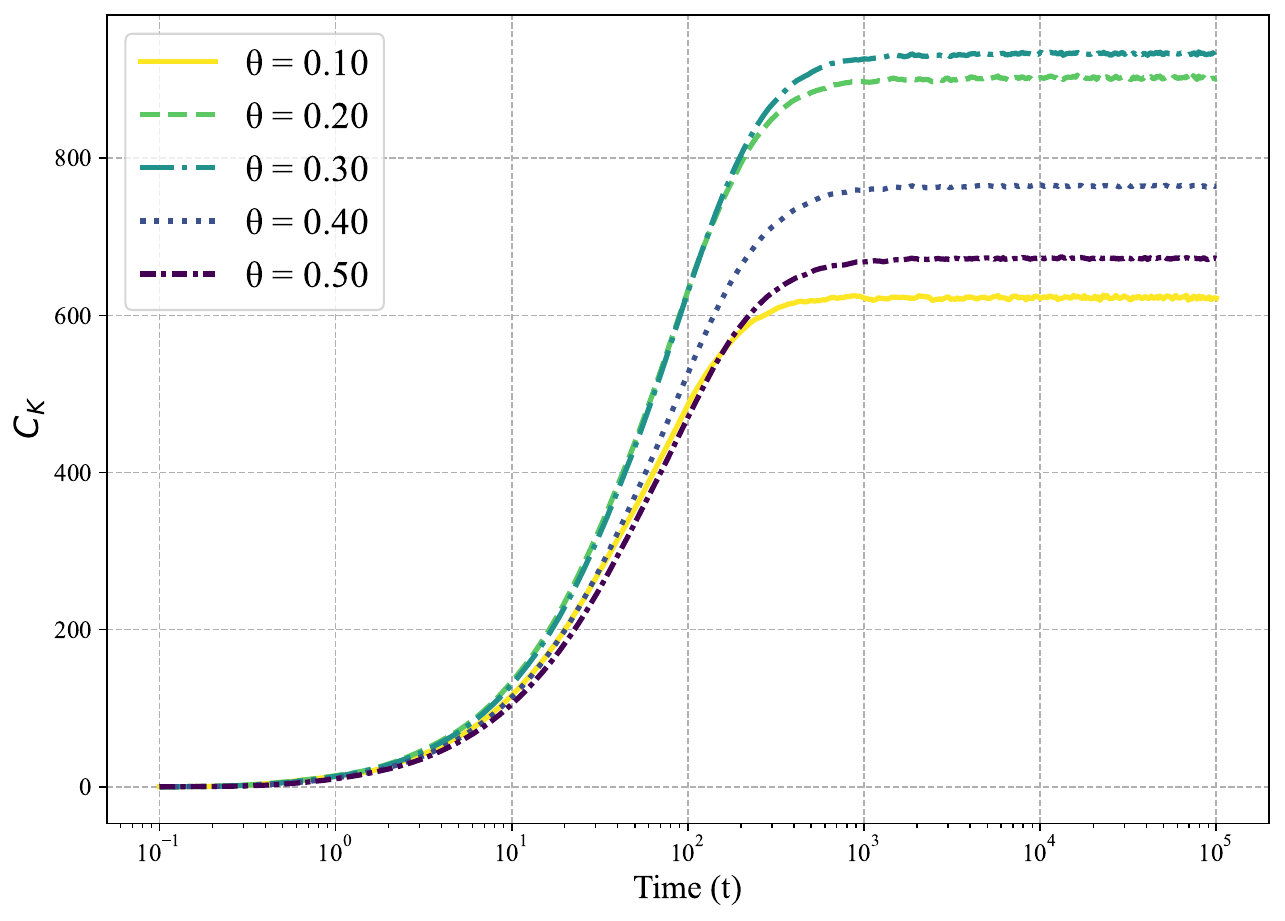}}\hfil 

\subfloat[TFS with $W=2.5$]{\includegraphics[width=7cm]{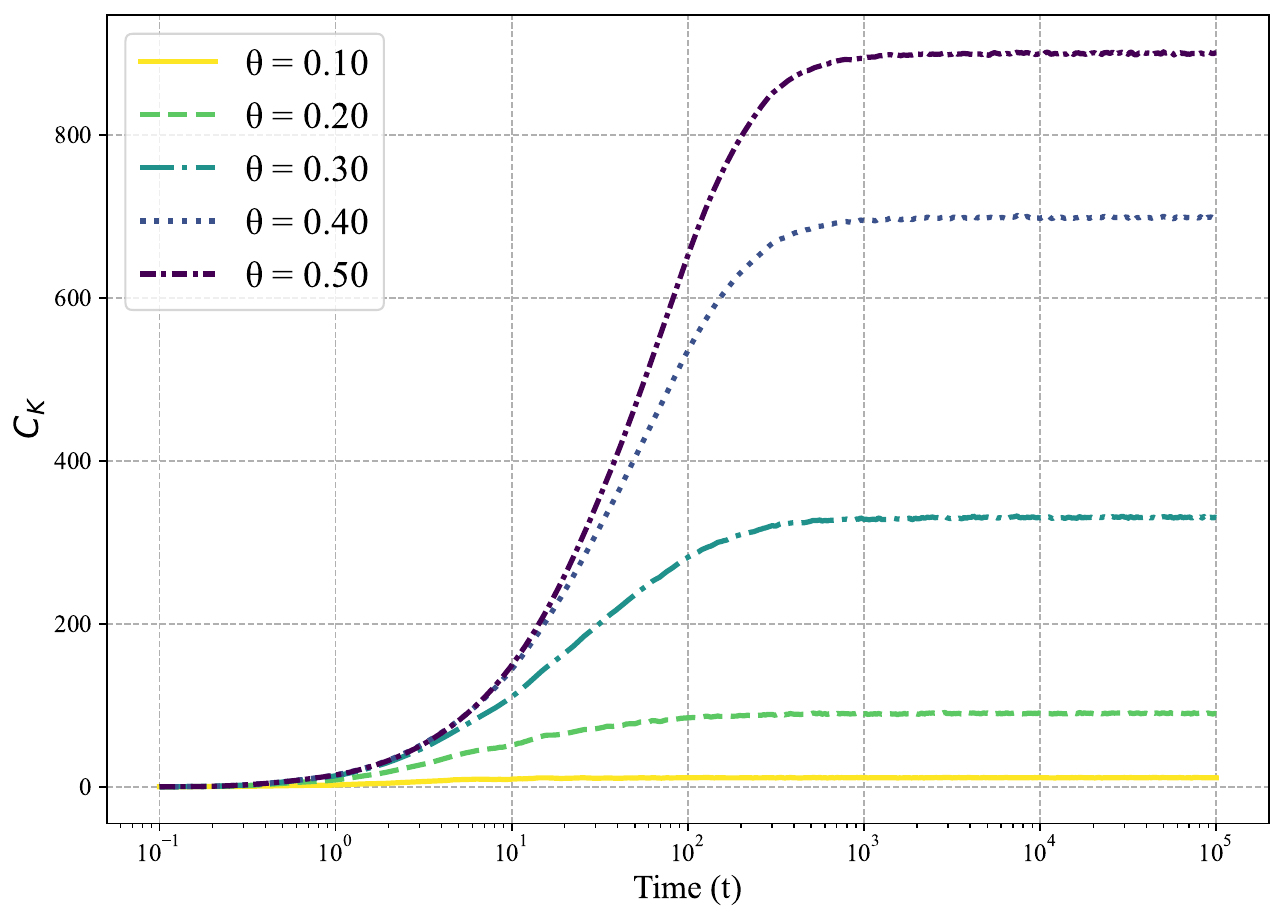}}\hfil
\subfloat[TNS with $W=2.5$ \label{fig:KCsubtleQME}]{\includegraphics[width=7cm]{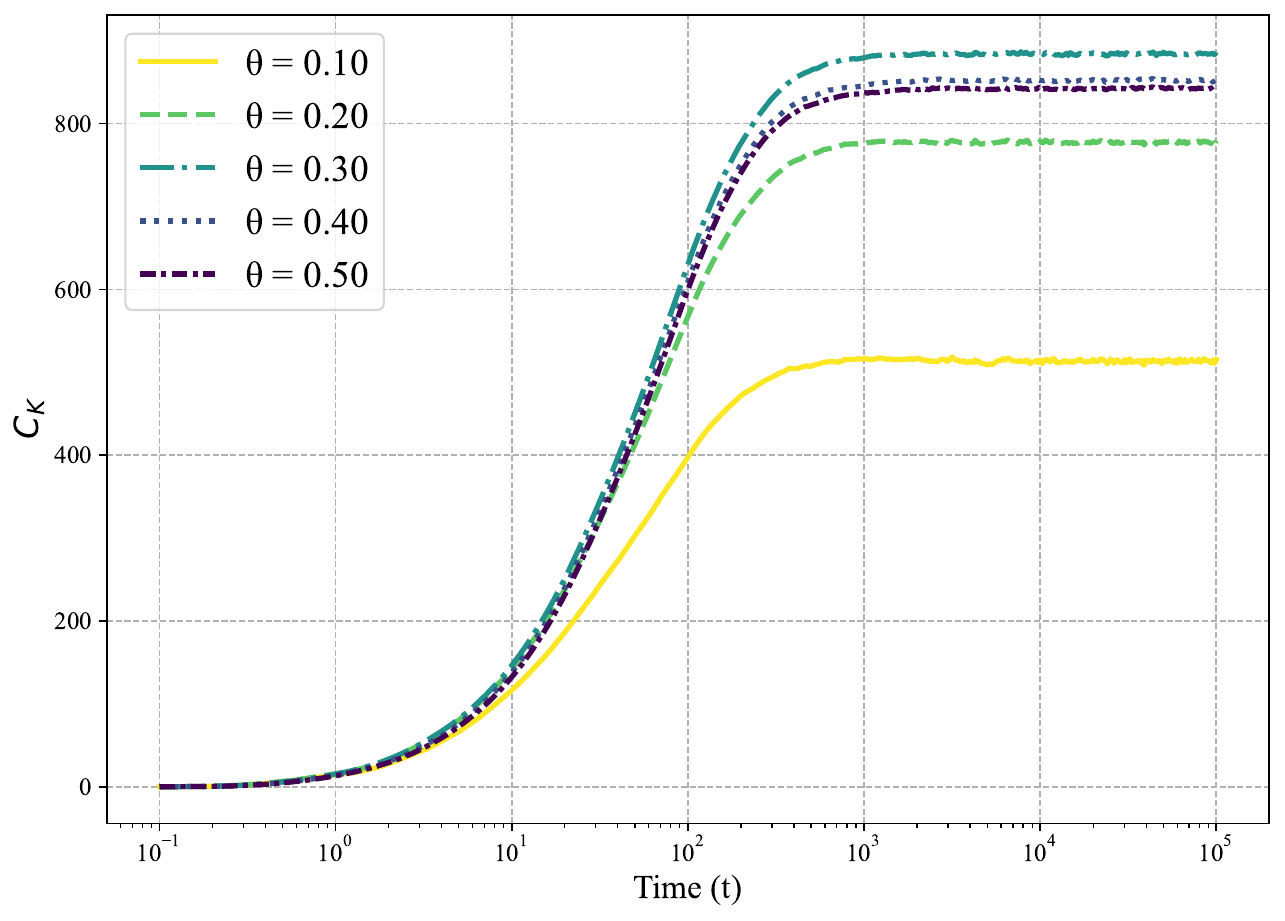}}\hfil 

\subfloat[TFS with $W=5.0$]{\includegraphics[width=7cm]{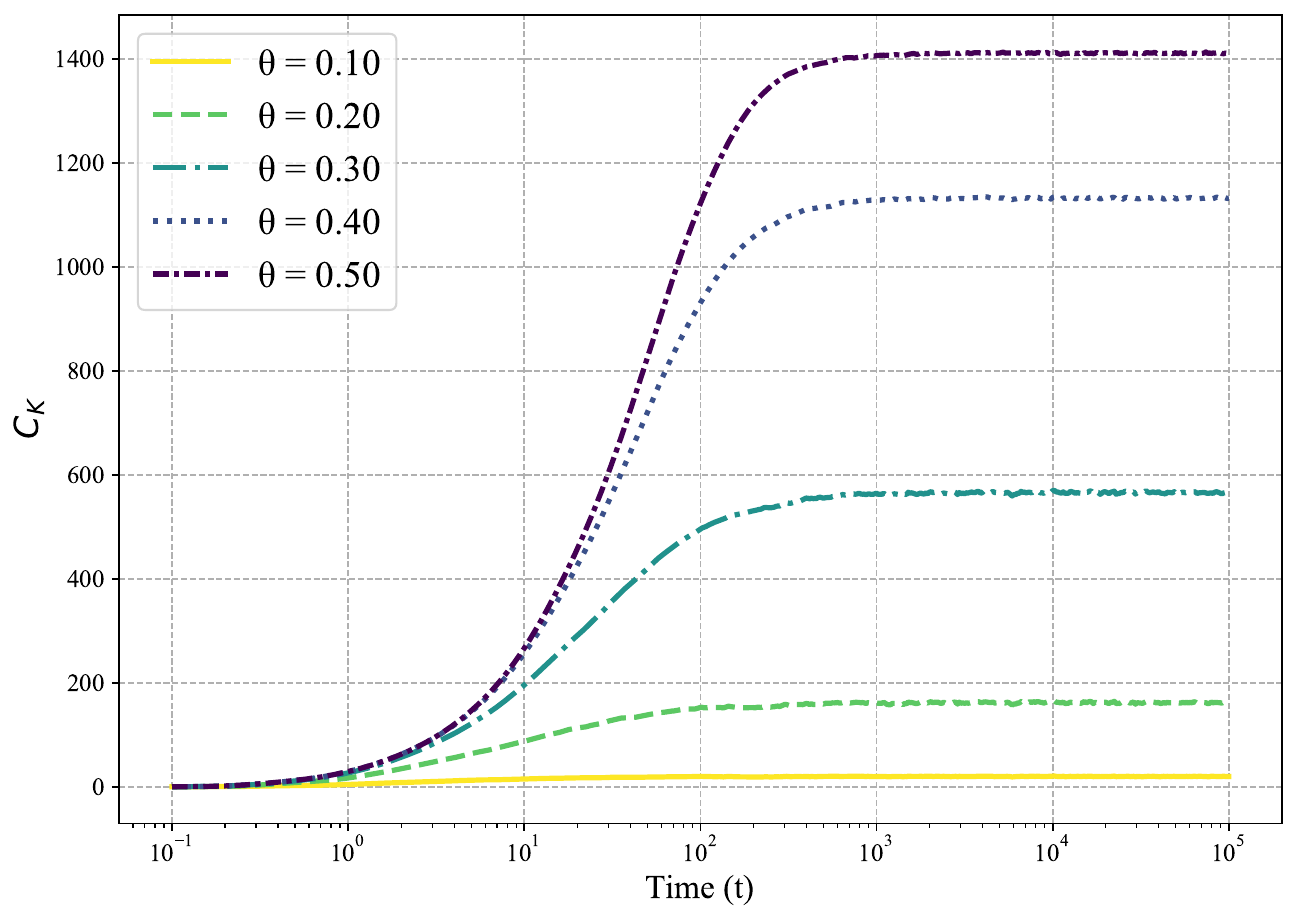}}\hfil
\subfloat[TNS with $W=5.0$]{\includegraphics[width=7cm]{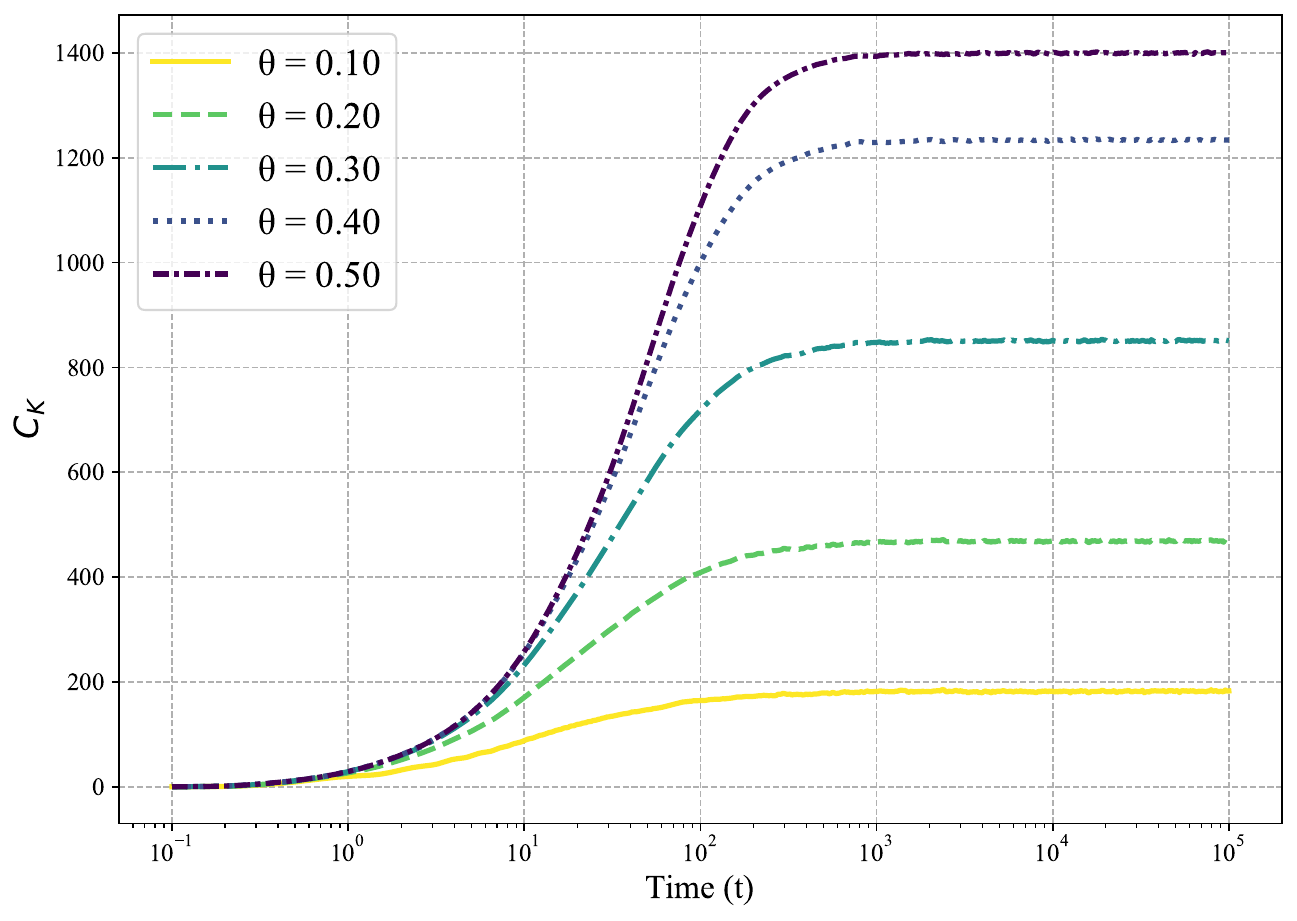}}\hfil 
\caption{Figures demonstrating the difference in Krylov spread complexity between the TFS (left) and TNS (right), for an $N=12$, ($N_A=3$) size (sub)system, averaged over 240 realisations. It is interesting to note that the TFS states always have their KC values ordered by the tilt angle, whereas the TNS states' ordering changes depending on the value of $W$.}
\label{fig:KC}
\end{figure}

\noindent
Our results are presented in Fig. \ref{fig:KC} and should be interpreted alongside the entanglement asymmetry analysis. From the latter, we already know between which pairs of states the QME manifests; the present question is whether, and in what form, this behavior is reflected in the spread complexity of the corresponding time-evolved reference states. A striking feature is the behaviour of the saturation values of the spread complexity. For the TFS states, the saturation values vary significantly across different tilts and are consistently ordered by the tilt parameter. Larger initial tilts yield larger saturation values. In contrast, for the TNS states in the ETH regime ($W < W_c$), the saturation values cluster more closely and are not necessarily ordered by tilt. As $W$ is increased toward and beyond the transition, the ordering by tilt becomes progressively clearer and the spread among saturation values widens. Once the system crosses the critical point, the TNS states display saturation profiles similar to those of their TFS counterparts. In summary, whenever the QME is known to occur (as diagnosed via entanglement asymmetry), we observe that the saturation value of spread complexity is consistently higher for the state with the larger initial tilt ({\it i.e.} the more asymmetric preparation). At this stage, this conclusion is qualitative, but in the following section we will attempt to quantify the relationship between tilt-induced asymmetry and the saturation of spread complexity.   
\\ \\
It is important to recall that $C_K(t)$ measures the average position of the wavefunction along the Krylov chain. Interpreted in the context of the QME, the spread of saturation values indicates the extent to which states are able to explore the accessible Hilbert space. From this perspective, the TFS states behave consistently; they explore comparable relative regions of the Hilbert space across all values of $W$, with the degree of exploration increasing as $W$ grows. This is counterintuitive, since one would normally expect localization at larger $W$ to restrict exploration. The behavior of the TNS states is somewhat more peculiar. As $W$ increases, the variance of the saturation values also increases, again defying the intuition that stronger localization should reduce dynamical spread. However, this enhanced variance does not arise from a uniform increase. Instead, states with smaller tilt angles ($\theta=0.1,0.2,0.3$) exhibit reduced saturation values, indicating restricted exploration, while states with larger tilts ($\theta=0.4,0.5$) follow trends similar to the TFS states. Moreover, for $W<W_c$, the TNS saturation values are not ordered by the tilt parameter, and thus not ordered by the amount of induced asymmetry -- unlike the TFS states, which retain tilt-ordering across all regimes. For the maximum evolution time considered ($t=10^5$), full ordering of the TNS states emerges at $W=4.0$. It is therefore plausible that, with longer time evolution, complete ordering would also appear at lower values of $W$. For all values $W \geq 4.0$, up to at least $W=6.0$, the ordering is robustly established and persists.

% \textbf{\textcolor{red}{NOT SURE WE NEED THIS. Maybe only need section 4}}
% \section{De-diffused Krylov Complexity}
% \label{sec:KCdediff}
% \begin{itemize}
%     \item Same results required as for those above, but with $U_A(\theta) = U(\theta)-U_S(\theta)$, that is, removing the components of the operator that do not change the entanglement asymmetry value.

%     \item In the EA set up, it is conventional to take some state, to act upon it with a symmetry-breaking operator, $U_{\text{full}}$, and then to record the $\Delta S(t)$ as the system is evolved. As previously mentioned, the EA is designed to measure only about coherences between symmetry sectors, not evolution within a sector. Krylov complexity, on the other hand, accounts for both types of changes -- spread within a symmetry sector and hopping between sectors (provided that the state is not an eigenstate localised to a particular symmetry sector, in which case only the spread inside the sector can be captured). We will perform a slightly different procedure: we will act on the initial state, $\ket{\psi}$, with the operator after removing the symmetrically transformed component. More precisely, we write the full operator as a sum of its symmetric and asymmetric parts, 
%     \begin{equation}
%         U_{F} = U_{\text{S}}+U_{A},
%     \end{equation}
% \end{itemize}
% and then compute the Krylov basis generated by $U_A\ket{\psi_0}$. Measuring the Krylov complexity of the fully-transformed state in this basis, we find that the saturation time is significantly reduced by a factor of $\sim100$.

\section{Decomposition of Krylov Complexity}
\label{sec:KCsymmres}
The analysis of the previous section indicates that spread complexity may be (at least qualitatively) sensitive to the presence of the QME, but we do not yet have a quantitative handle on this. To pursue this, we will resolve spread complexity in a more detailed way that highlights the symmetry-restoring nature of the AA Hamiltonian.  To this end, note that, for a generic operator evolving under a symmetry-restoring Hamiltonian, we may expect that $$ O(t) - \sum_q \Pi_q \hat{O}(t) \Pi_q \sim 0\,,$$ at late times.   In the context of entanglement asymmetry the operator is, of course, the density matrix.  In our study another natural candidate is the Krylov complexity operator.  We thus define the \textit{projected symmetric complexity} operator as 
\begin{equation}
\hat{C}_{K,S} = \sum_q \sum_n n \Pi_q |K_n \rangle \langle K_n| \Pi_q    \label{Cproj} 
\end{equation}
and the \textit{projected asymmetric complexity} operator as 
\begin{eqnarray}
\hat{C}_{K,A} &=& \hat{C}_K - \hat{C}_{K,S}     \nonumber \\
& = & \sum_{q \neq r} \sum_n   n \Pi_q |K_n\rangle \langle K_n| \Pi_r  
\end{eqnarray}
We note two important features of the projected asymmetric complexity\footnote{Hereafter, we will refer to the projective (a)symmetric complexity as simply the `(a)symmetric complexity', for short.}.  Firstly, as already alluded to, the matrix elements of $\hat{C}_{K,A}$ are zero if $\hat{C}_K(t)$ is block-diagonal in the charge sectors associated with the symmetry i.e. $\left[ \hat{C}_K(t), \Pi_q  \right] = 0$.  We would expect that in a system where symmetry is approximately restored with time-evolution that the asymmetric complexity should thus approach zero at late times.  Secondly, the expectation value of $\hat{C}_{K,A}$ with respect to the reference state at $t=0$ is negative.  To see this, it is useful to decompose the reference state, $|\psi\rangle$ as 
$$ |\psi\rangle = \sum_{q} \Pi_q |\psi\rangle \equiv \sum_{q} c_q |\psi_q\rangle $$
where $\sum_{q} |c_q|^2 = 1$.  By definition, the reference state is the zero complexity state so that
$$ \langle \psi| \hat{C}_K |\psi\rangle = 0$$
However, the components obtained from projecting the reference state onto the symmetry sectors do not have zero complexity, even at $t=0$ i.e. 
$$
\langle \psi_q | \hat{C}_K |\psi_q\rangle \geq 0   
$$
with equality only when the reference state spans a single symmetry sector.  This can be understood intuitively since the state $|\psi\rangle$ is the \textit{only} state in the Krylov subspace with zero complexity.  More explicitly, there are states built from the same components as the reference state, but that are orthogonal to it 
$$ |\psi'\rangle = \sum_{q} k_q |\psi_q\rangle \ \ \ ; \ \ \ \langle \psi | \psi'\rangle = 0.$$
A state such as $|\psi'\rangle$ will typically be generated at some stage in the Lanczos algorithm.  As such, it has positive complexity.  The individual components $|\psi_q\rangle$ are a linear combination of the zero complexity reference state and states of positive complexity.  In terms of the projectors this leads to 
\begin{equation}
\langle \psi | \Pi_q \hat{C}_K \Pi_q | \psi \rangle = \langle \psi | \hat{C}_{K,S} | \psi \rangle = -\langle \psi | \hat{C}_{K,A} | \psi \rangle  \geq 0 .  \label{posProperty}
\end{equation} 
As argued, the following expectation value will approach $0$ for late times for symmetry-restoring Hamiltonians
\begin{equation}
C_{K,A}(t) = \langle \psi | e^{i t H} \hat{C}_{K,A} e^{-i t H}| \psi \rangle  \rightarrow 0  .  \label{symmRestore}
\end{equation}
These two properties suggest that the asymmetric complexity may be usable as a probe of the quantum Mpemba effect, like entanglement asymmetry.  Three important questions remain:
\begin{enumerate}
\item Is $C_{K,A}(t)$ zero only when symmetry has been restored?
\vspace{0.13cm}
\item Is $C_{K,A}(t)$ non-positive?
\vspace{0.13cm}
\item Is $C_{K,A}(t)$ monotonically increasing?
\end{enumerate}
If these properties are met, then $-C_{K,A}(t)$ satisfies the same important properties that make entanglement asymmetry such a useful measure for the purposes of observing the QME.  We will return to these questions shortly.  \\ \\
Before proceeding we note that in a recent work \cite{Caputa:2025mii}, the authors introduced the notion of \textit{symmetry-resolved Krylov complexity} which considers the growth of operators in contexts where the operator and the Hamiltonian are subject to symmetry constraints.  Since the setting for our questions are closely related to those of \cite{Caputa:2025mii}, we would like to emphasise that the quantity $C_{K,A}(t)$ we will be using is not related to the symmetry-resolved K-complexity, $\bar{C}(t)$, in a simple way.  This goes deeper than the observation that the reference \cite{Caputa:2025mii} is focussed on operator dynamics, since it is straightforward to generalise their construction to quantum states (as we do now).  \\ \\
To briefly summarise symmetry-resolved Krylov complexity, it is sourced by a Hamiltonian that is block diagonal with respect to some symmetry generator
$$ H = \sum_{q} H_q,$$
acting on a reference state that may be decomposed similarly\footnote{When considering operators as opposed to states, one would at this point specialise to operators that are also block-diagonal with respect to the symmetry generator.  Any state can, however, be decomposed in this way.} 
$$ |\psi\rangle = \sum_{q}  c_q |\psi_q\rangle.   $$
Written in terms of projection operators, $\Pi_q$, these are the usual
\begin{eqnarray}
H_q & = & \Pi_q H \Pi_q,   \nonumber \\
|\psi_q\rangle &= & \Pi_q |\psi\rangle. \nonumber
\end{eqnarray}
By acting on $|K_0\rangle = |\psi\rangle$ with increasing integer powers of $H$ and performing a Gram-Schmidt process, one obtains the usual Krylov basis
$$ H |K_n\rangle = b_{n+1}|K_{n+1}\rangle + a_n |K_n\rangle + b_n |K_n\rangle,$$
while acting on $|K_0^{(q)}\rangle = |\psi_q\rangle$ with all integer powers of $H_q$ and orthogonalising one obtains \textit{several different} Krylov bases, each restricted to a symmetry sector
$$ H |K_n^{(q)}\rangle = b_{n+1}^{(q)}|K_{n+1}^{(q)}\rangle + a_n^{(q)} |K_n^{(q)}\rangle + b_{n} |K_n^{(q)}\rangle.$$
The usual Krylov complexity operator is given by 
\begin{equation}
\hat{C}_K = \sum_{n} n |K_n\rangle \langle K_n| 
\end{equation}
while the symmetry-resolved Krylov complexity operator is given by
\begin{equation}
\hat{\bar{C}} = \sum_{q}\sum_{n=1}^{d_q}  n  \rho_q   |K_n^{(q)}\rangle \langle K_n^{(q)}|   \label{CSR}
\end{equation}
where $$\rho_q = |\langle K_0| \Pi_q | K_{0}\rangle|^2.  $$
As demonstrated in \cite{Caputa:2025mii}, the time-evolved state takes a very simple form 
\begin{equation}
|\psi(t)\rangle = \sum \sqrt{\rho_q} |\psi_q(t)\rangle
\end{equation}
so that the evolution of the state is always restricted to the same sectors with the same probability weightings as the reference state.    
Based on this, one may naively expect that it should be possible to identify $|K_n^{(q)}\rangle  \sim \Pi_q |K_n\rangle$ which would imply that (\ref{CSR}) and (\ref{Cproj}) may be equal or (at the very least) related by updating the sector weightings.  However, this expectation is incorrect.  There are two simple ways to see this.  First, we have that
\begin{eqnarray}
\langle K_0| \hat{C}_{K,S} |K_0\rangle & \geq & 0, \quad\text{but}  \nonumber \\
\langle K_0| \bar{C} |K_0\rangle & = & 0. \nonumber
\end{eqnarray}
Secondly, we have that $|K_n^{(q)}\rangle$ can be non-zero for values of $n$ up to the dimension of the symmetry sector $n_q-1$, while $\Pi_q |K_n\rangle$ can, in principle, be non-zero for $n$ up to the full Hilbert space dimension, $n_D-1=(\sum_q n_q)-1$.  The states involved in describing the state dynamics are thus not related in a simple way.  \\ \\ 
From the perspective of studying the spread of the reference state in the different symmetry sectors, the quantity $\bar{C}$ is more natural since it starts at $t=0$ and captures the full dynamics of the time-evolved reference state with fewer vectors.  However, from the perspective of studying the quantum Mpemba effect, the property (\ref{symmRestore}) makes the asymmetric complexity the more natural quantity to consider. We now proceed by computing the decomposed contributions of the Krylov complexity for the AA model.
\\

\subsection{Separating Projected Symmetric and Asymmetric Contributions}
\label{sec:sepKC}
\begin{figure}[!htbp]
\centering
\subfloat[TFS, Proj. Symm. Complexity]{\includegraphics[width=0.47\linewidth]{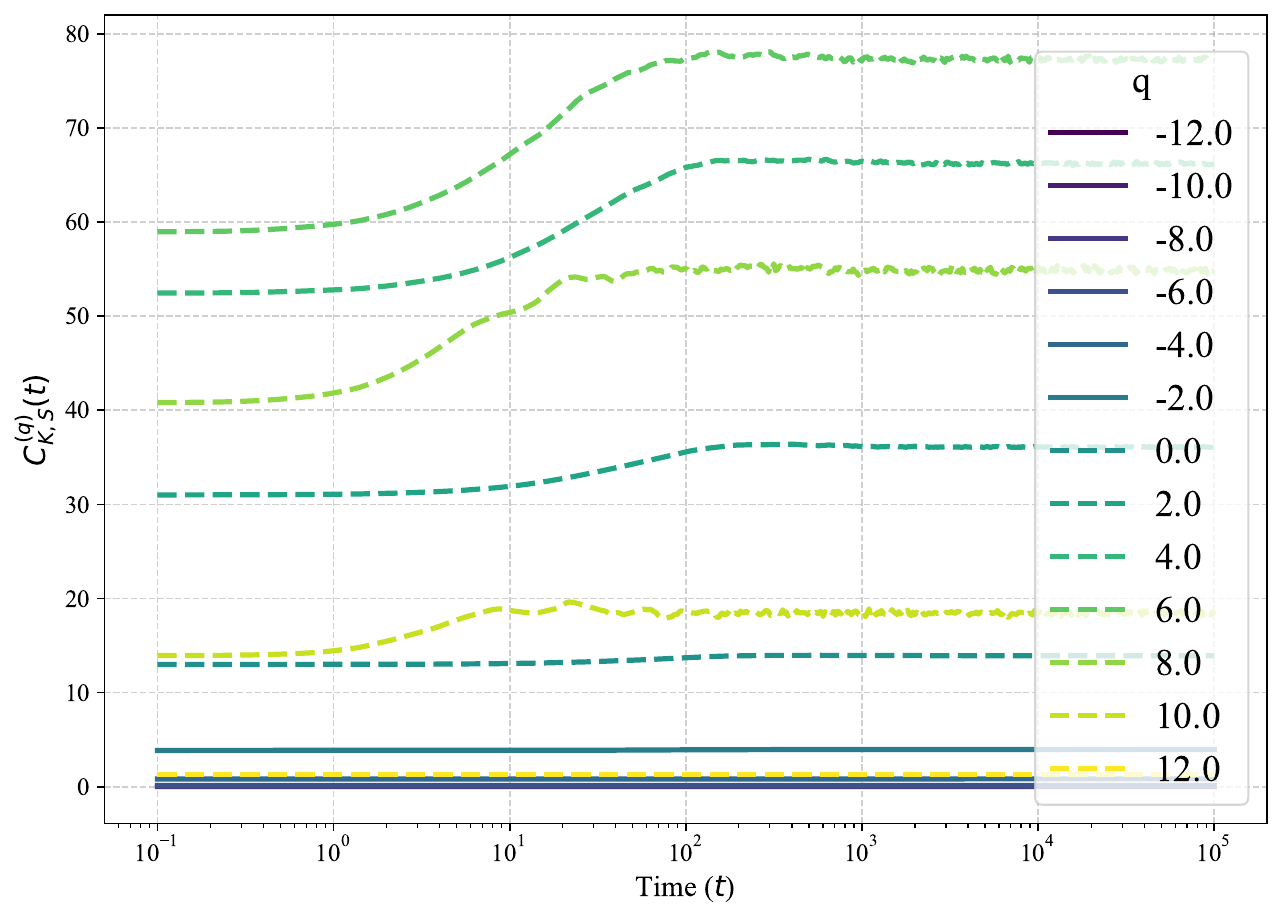}}\hfil
\subfloat[TNS, Proj. Symm. Complexity]{\includegraphics[width=0.47\linewidth]{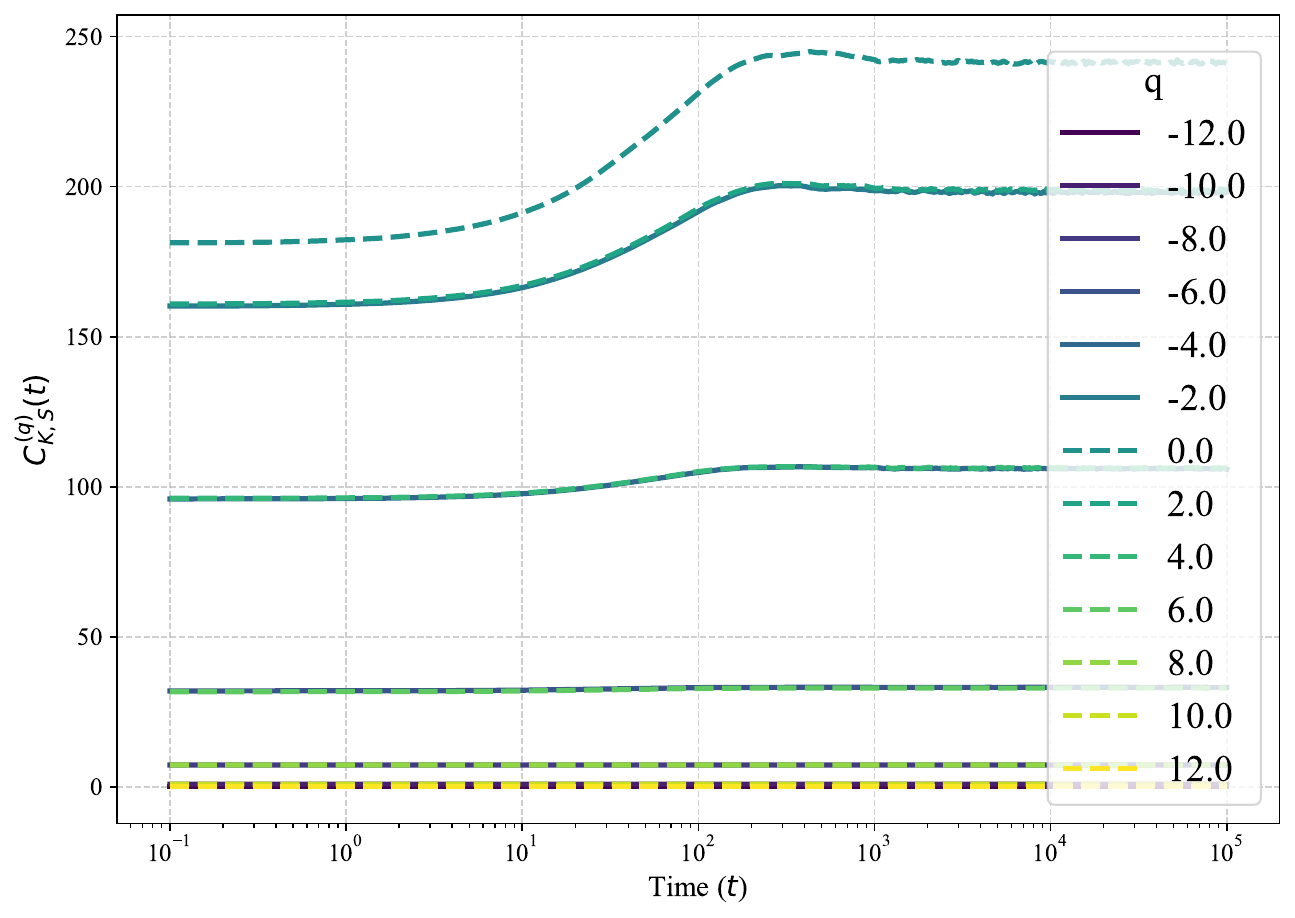}}\hfil 
\subfloat[TFS, Proj. Asymm. Complexity]{\includegraphics[width=0.47\linewidth]{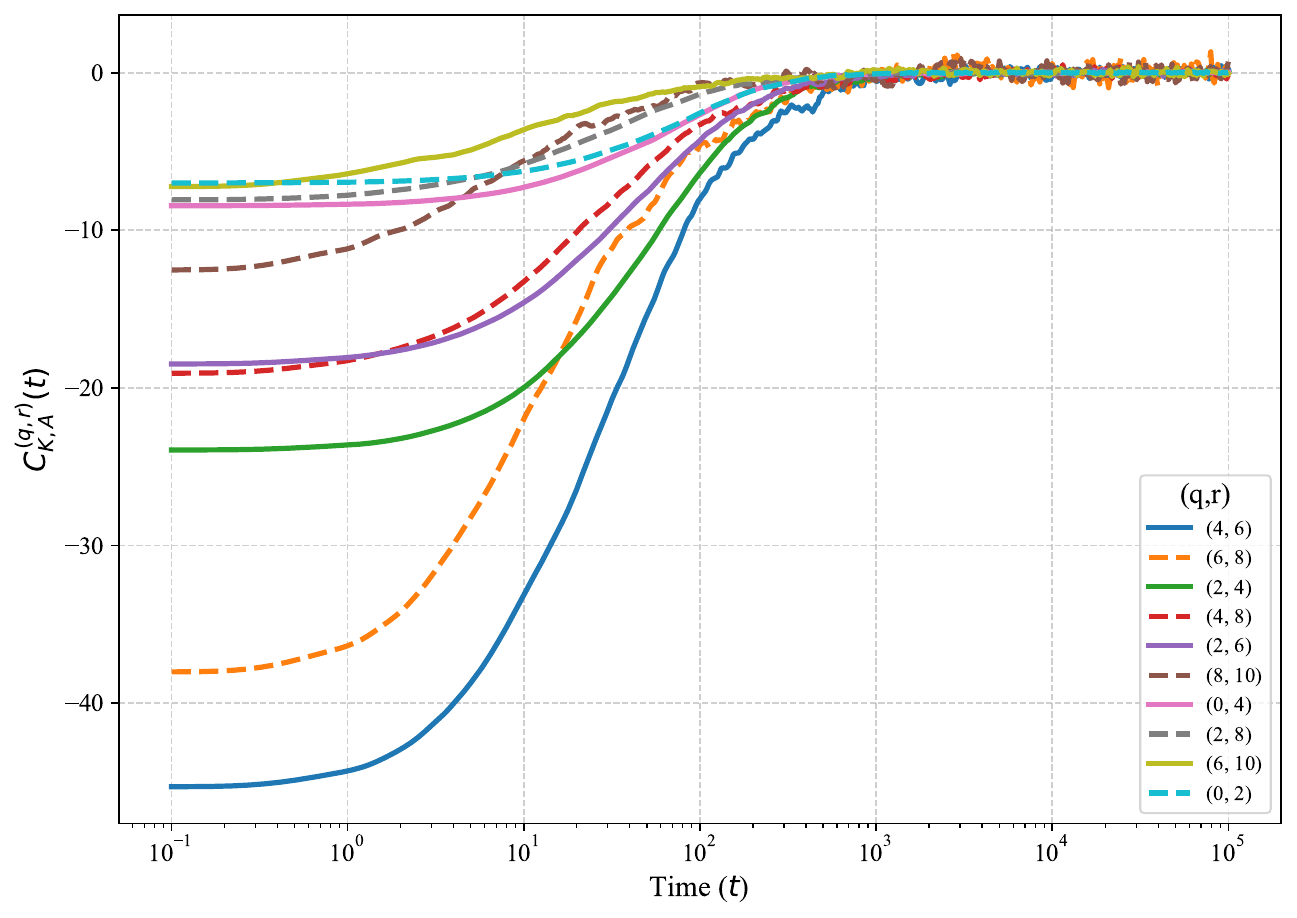}}\hfil
\subfloat[TNS, Proj. Asymm. Complexity]{\includegraphics[width=0.47\linewidth]{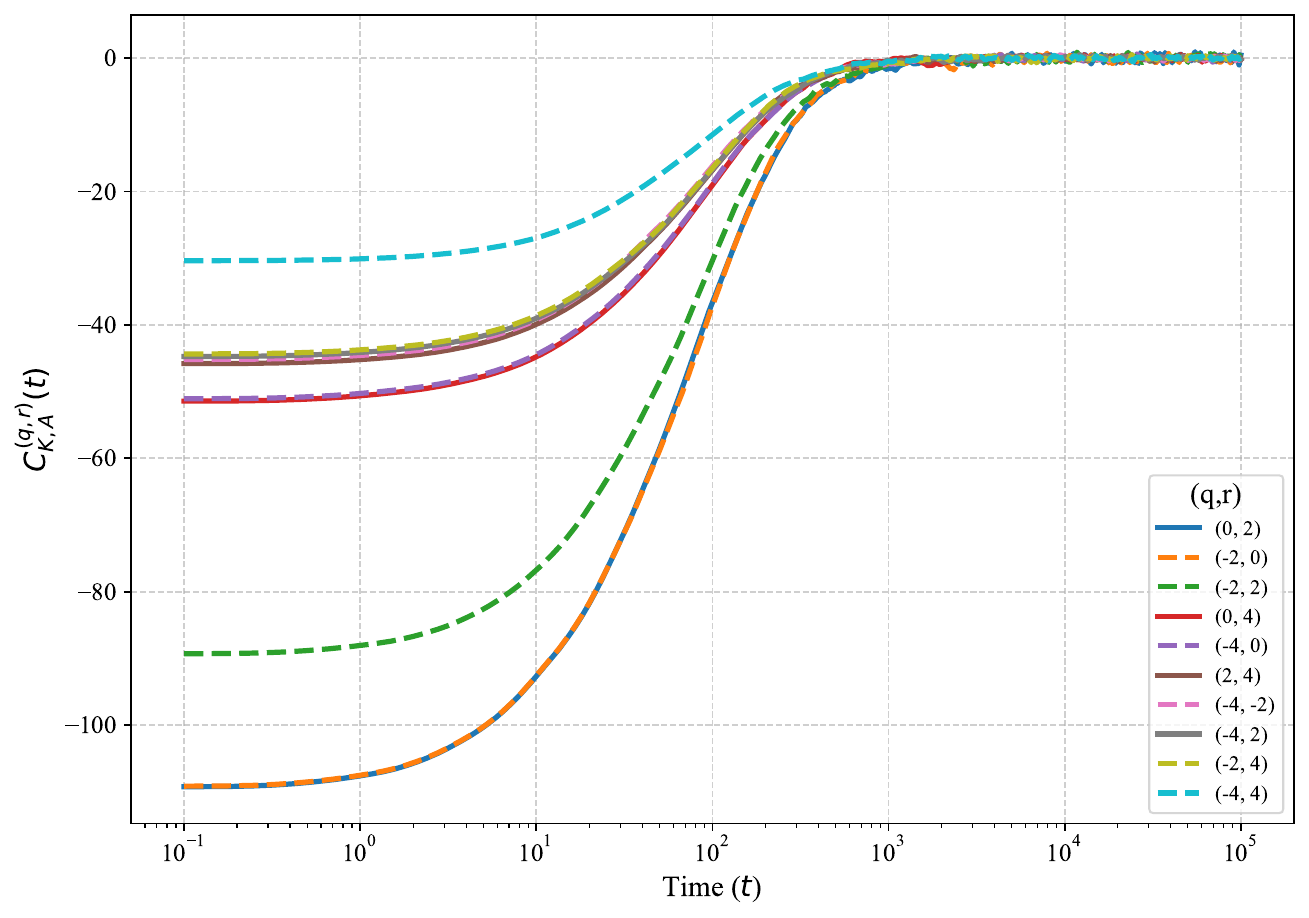}}\hfil 
\caption{Figures demonstrating the positivity of the projected symmetric complexity (above) and the initial negativity of the largest (by magnitude at $t=0$) asymmetric complexity values (below), for an $N=12$ size system, averaged over 240 realisations, and with $\theta=0.3,W=2.0$. The contributions to the symmetric complexity values (top) are labelled by their symmetry sector $q$, and the asymmetric complexity contributions are labelled by the symmetry sectors between which there is coherence $(q,r)$. We have left out a large number of smaller asymmetric contributions to improve the presentation of the data.}\label{fig:DiffNMix}
\end{figure}

\begin{figure}[!htbp]
\centering
\subfloat[$W=1.0,\theta=0.1$]
{\includegraphics[width=0.47\linewidth]{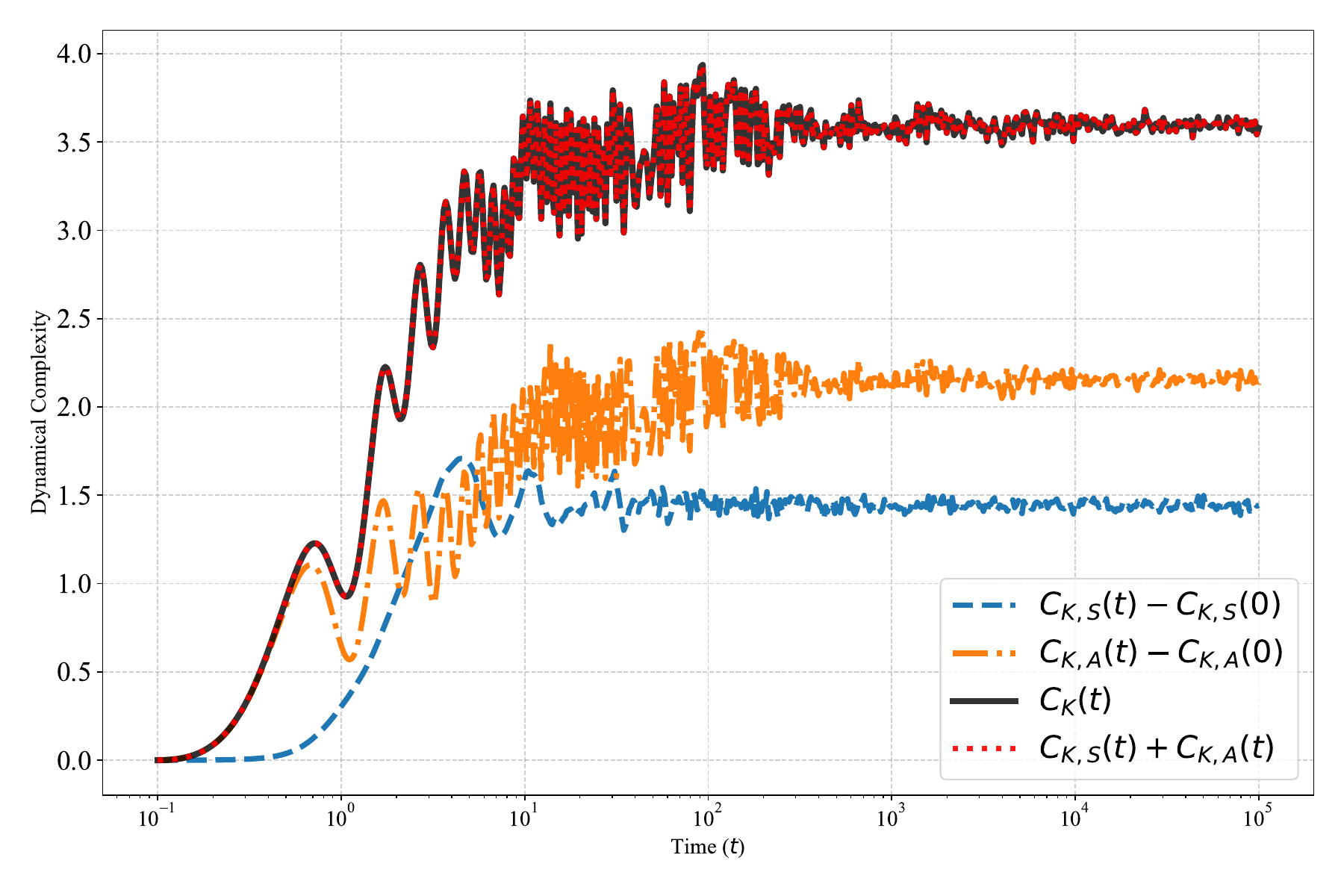}}\hfil
\subfloat[$W=1.0,\theta=0.5$]{\includegraphics[width=0.47\linewidth]{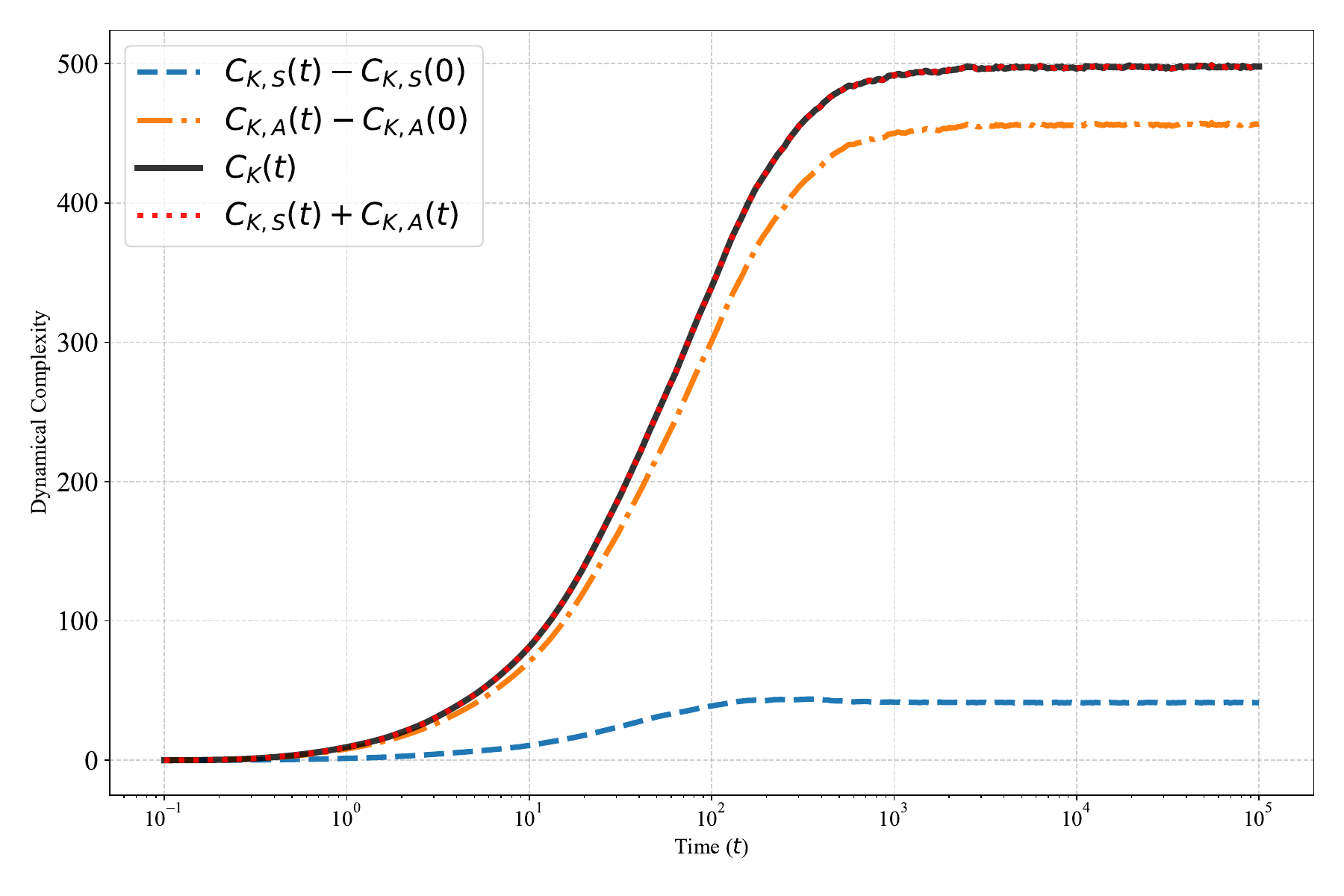}}\hfil 
\subfloat[$W=5.0,\theta=0.1$]{\includegraphics[width=0.47\linewidth]{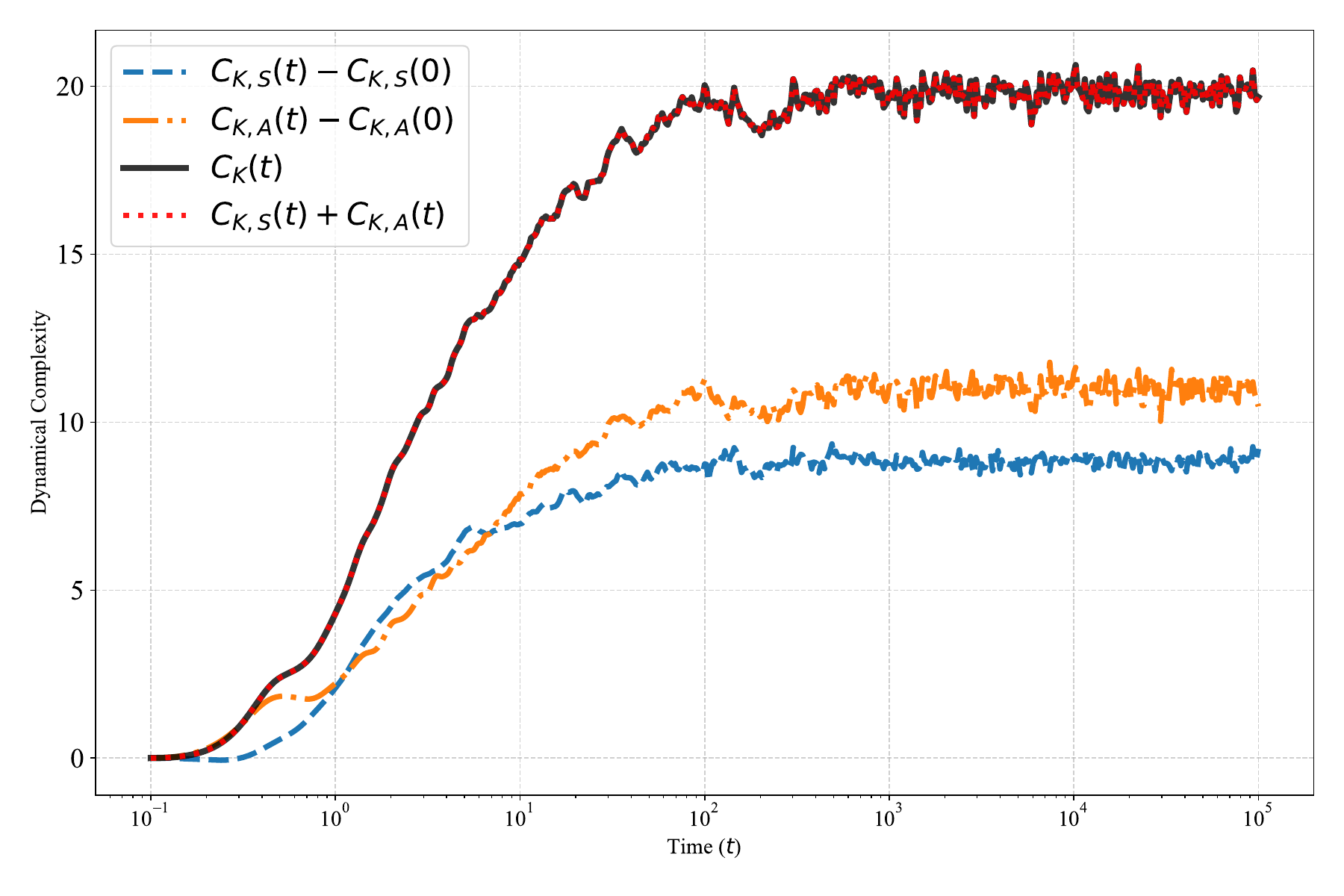}}\hfil
\subfloat[$W=5.0,\theta=0.5$]{\includegraphics[width=0.47\linewidth]{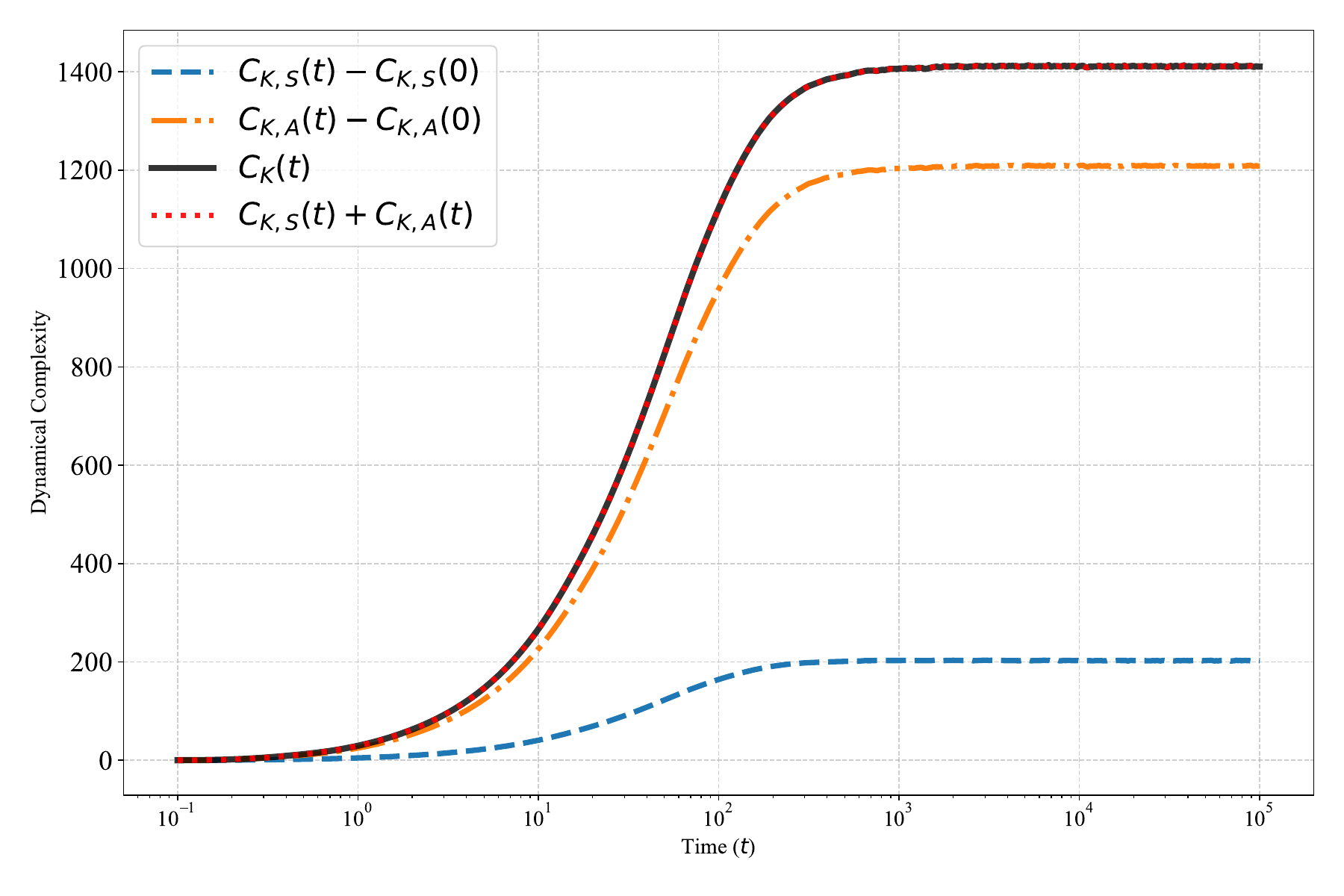}}\hfil 
\caption{Figures verifying the decomposition of the projected complexity for the TFS states at different potential strengths and different tilt angles. Similar computations for the TNS states may be found in appendix \ref{app:TNSdecomposition}.}
\label{fig:SumOfParts}
\end{figure}

In the previous section, we analysed the properties of the symmetric and asymmetric complexities by focusing on the projection operators applied to the Krylov complexity operator.  For the purposes of computation it is simpler to leave the Krylov operator untouched, and project the time-evolved states themselves onto the symmetry sectors, since
\begin{equation}
\begin{split}
\bra{\psi}\hat{C}_\text{K,S}\ket{\psi}&=\sum_q\bra{\psi}\hat\Pi_q \hat{C}_{K} \hat\Pi_q\ket{\psi}=\sum_q \bra{\psi_q}\hat{C}_K\ket{\psi_q},\\
\bra{\psi}\hat{C}_\text{K,A}\ket{\psi}&=\sum_{q\neq r}\bra{\psi}\hat\Pi_q \hat{C}_K \hat\Pi_r\ket{\psi}=\sum_{q\neq r} \bra{\psi_q}\hat{C}_K\ket{\psi_r},
\end{split}
\label{eq:srkcproj}
\end{equation}
where we have abbreviated the projection onto symmetry sector $q(r)$ as $\hat\Pi_{q(r)}$, and the projected state as $\ket{\psi_{q(r)}}$. 
\\ \\
%Looking closely at \eqref{eq:srkcproj}, there are two interesting observations that can be made. First, by considering the form of the diffusive KC, it is clear that what it measures is the Krylov complexity of the projected state in the basis generated by the initial state and, in particular, this means that it will not start at zero. 
%The second observation follows naturally from the first: given that the diffusive KC is not generally 0 at the initial time, then it must be true that the mixing KC is also \textit{not} generally 0 at the initial time, but instead must be \textit{negative} -- this is enforced by \eqref{eq:KCbreakdown}. These observations are highlighted by figure \ref{fig:DiffNMix}.
In Fig. \ref{fig:DiffNMix} we have plotted the contributions to the projected symmetric and asymmetric complexities from the various symmetry sectors.  We note that the contributions to the asymmetric complexity for all sector pairs starts at a negative value and tends towards zero as time increases.  This is an important test of the properties (\ref{posProperty}) and (\ref{symmRestore}).  However, we note that the contributions from the various sectors can fluctuate about zero and attain positive values at late time (see appendix \ref{app:latetimefluctuations}).  This would suggest that $C_{K,A}(t)$ is not strictly negative and monotonically increasing.  As such, $-C_{K,A}(t)$ does not satisfy all the convenient properties that make entanglement asymmetry a useful measure in the context of the QME.  We note, however, that the fluctuations are small compared to the initial values $C_{K,A}(0)$.  This would suggest that a small modification of this measure may yield a quantity that does satisfy these useful properties.  Indeed, entanglement asymmetry is derived from the Shannon entropy which is strictly positive, but in analysis of the QME it is often replaced by Renyi entropy which is not.  In this analogy, we expect $C_{K,A}(t)$ to be comparable to a Renyi entropy.  
\FloatBarrier
We could also study only the dynamical aspects of asymmetric and symmetric complexities shown in Fig. \ref{fig:DiffNMix}, by simply subtracting off the zero-time value of each component from all of the individual components -- this does not affect the sum of the contributions, as they are necessarily balanced at $t=0$. As a consistency check these components are calculated separately (i.e. \textit{not} making use of the relation $C_K(t)=C_{K,S}(t)+C_{K,A}(t)$), and their contributions are verified to sum to the full KC, shown in Fig. \ref{fig:SumOfParts}.

% \subsection{Comparison with Symmetry Resolved Krylov Complexity}

\begin{figure}[!htbp]
\centering
\subfloat[$W=1.0,\theta=0.1$]{\includegraphics[width=0.47\linewidth]{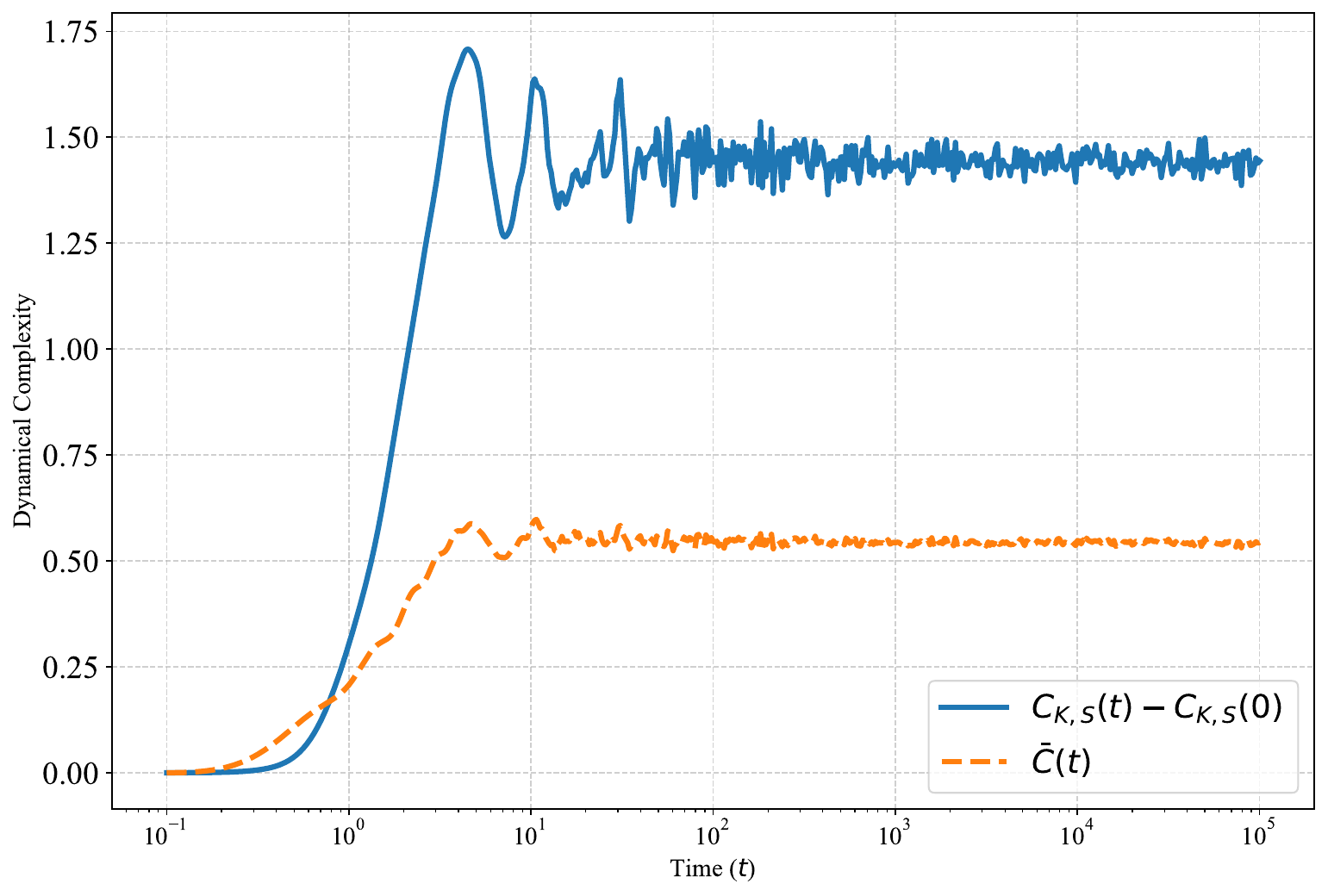}}\hfil
\subfloat[$W=1.0,\theta=0.5$]{\includegraphics[width=0.47\linewidth]{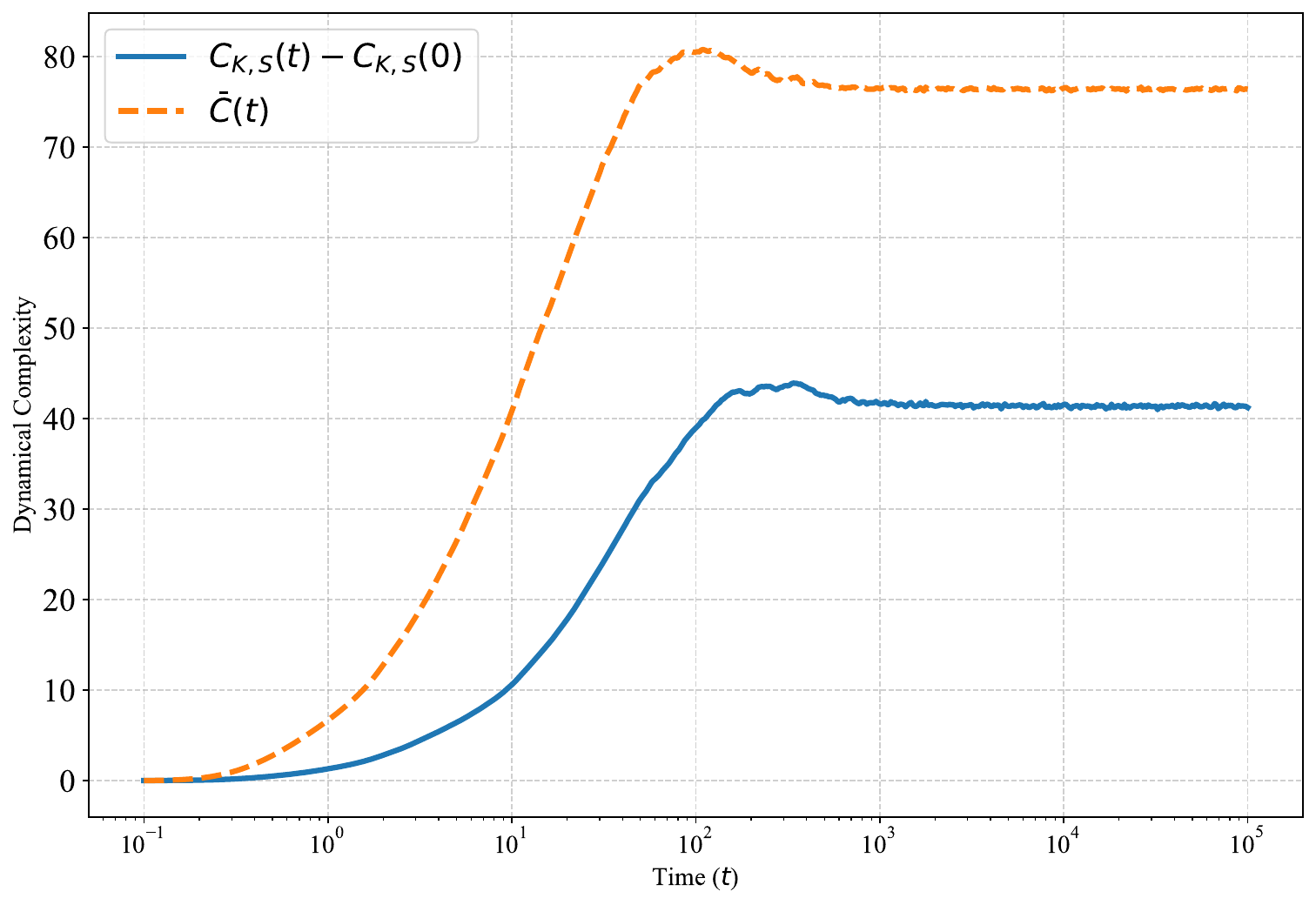}}\hfil 
\subfloat[$W=5.0,\theta=0.1$]{\includegraphics[width=0.47\linewidth]{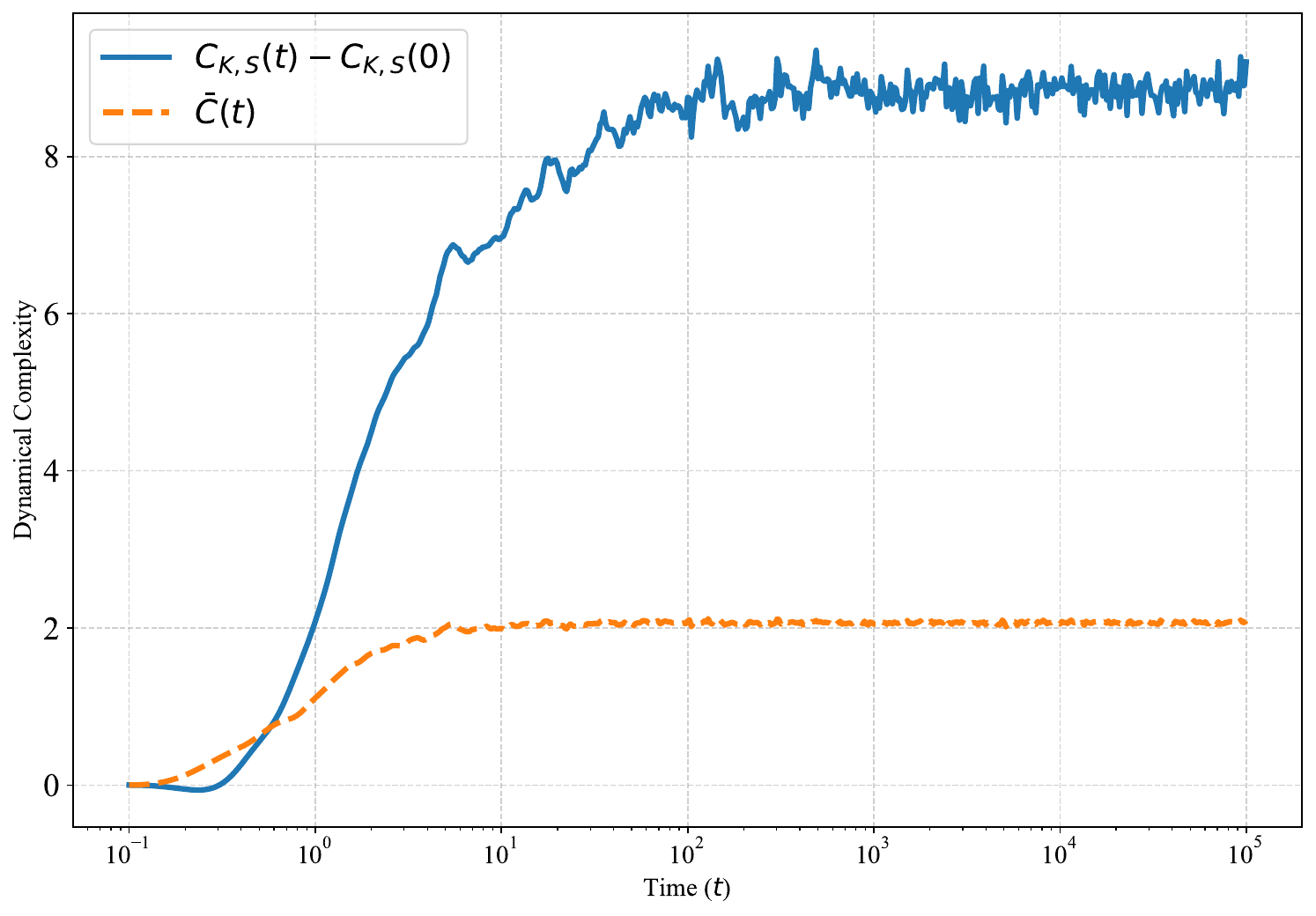}}\hfil
\subfloat[$W=5.0,\theta=0.5$]{\includegraphics[width=0.47\linewidth]{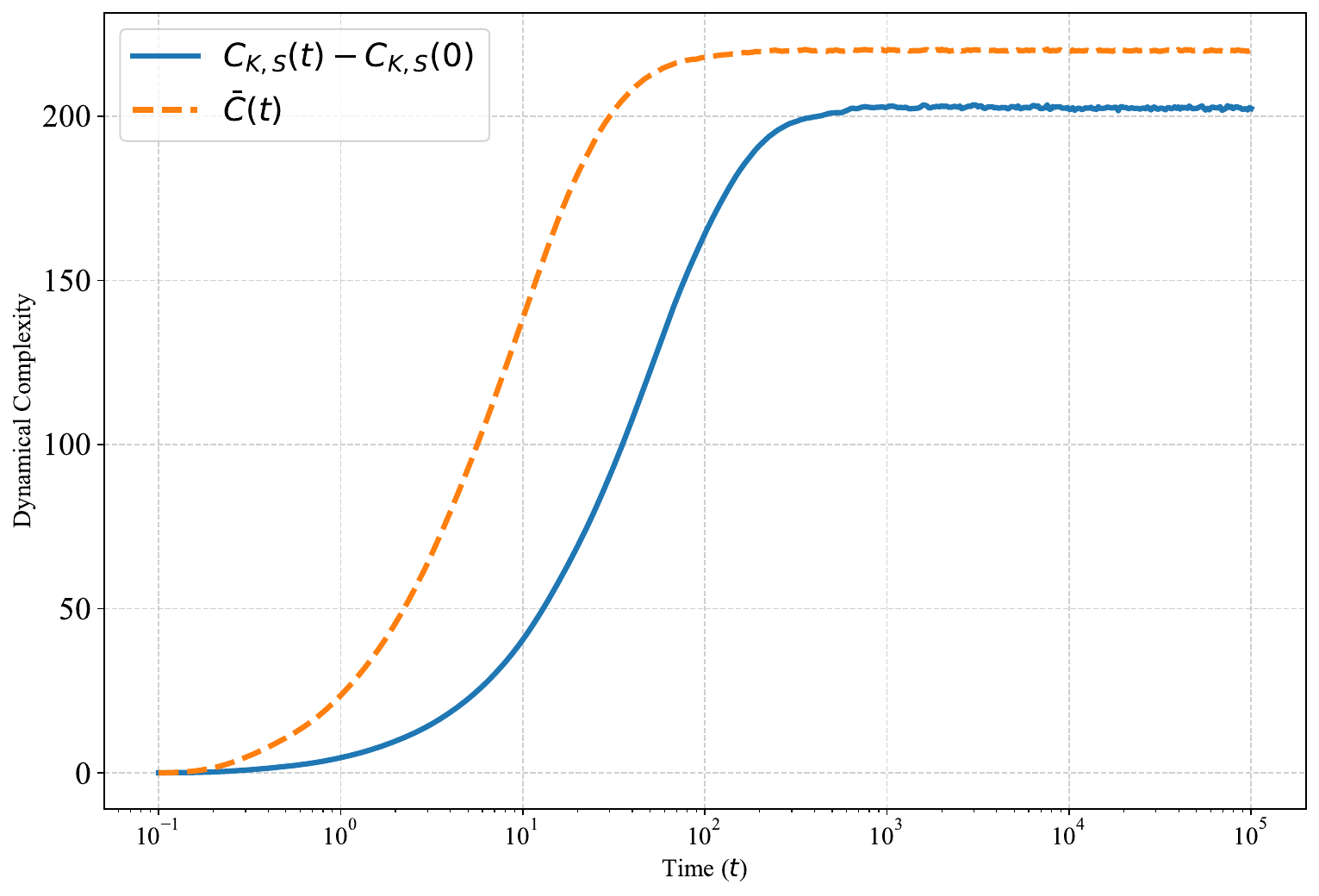}}\hfil 
\caption{Figures showing the discrepancy between the different intrasector diffusive measures for the TFS states at different tilt angles and different potential strengths. We compare the total discrepancy using the $\bar{C}$ measure described in \cite{caputa2025block} (dashed lines), and the projective symmetric complexity introduced in this work. See the appendix \ref{app:diffcomp} for the sector-by-sector comparisons and the TNS results.}
\label{fig:discrepancy}
\end{figure}

In Fig. \ref{fig:discrepancy} we have compared the symmetric complexity (shifted to start at zero) with the symmetry-resolved Krylov complexity for the system.  This demonstrates that these quantities are different.  Perhaps not surprisingly, many of the same dynamical features are captured by both measures, though their values and the relative scales of various features are clearly different.  \\ \\

\subsection{Asymmetric complexity as a probe of the QME}

\begin{figure}[!htbp]
\centering
\subfloat[TFS, $W=1.0$]{\includegraphics[width=0.47\linewidth]{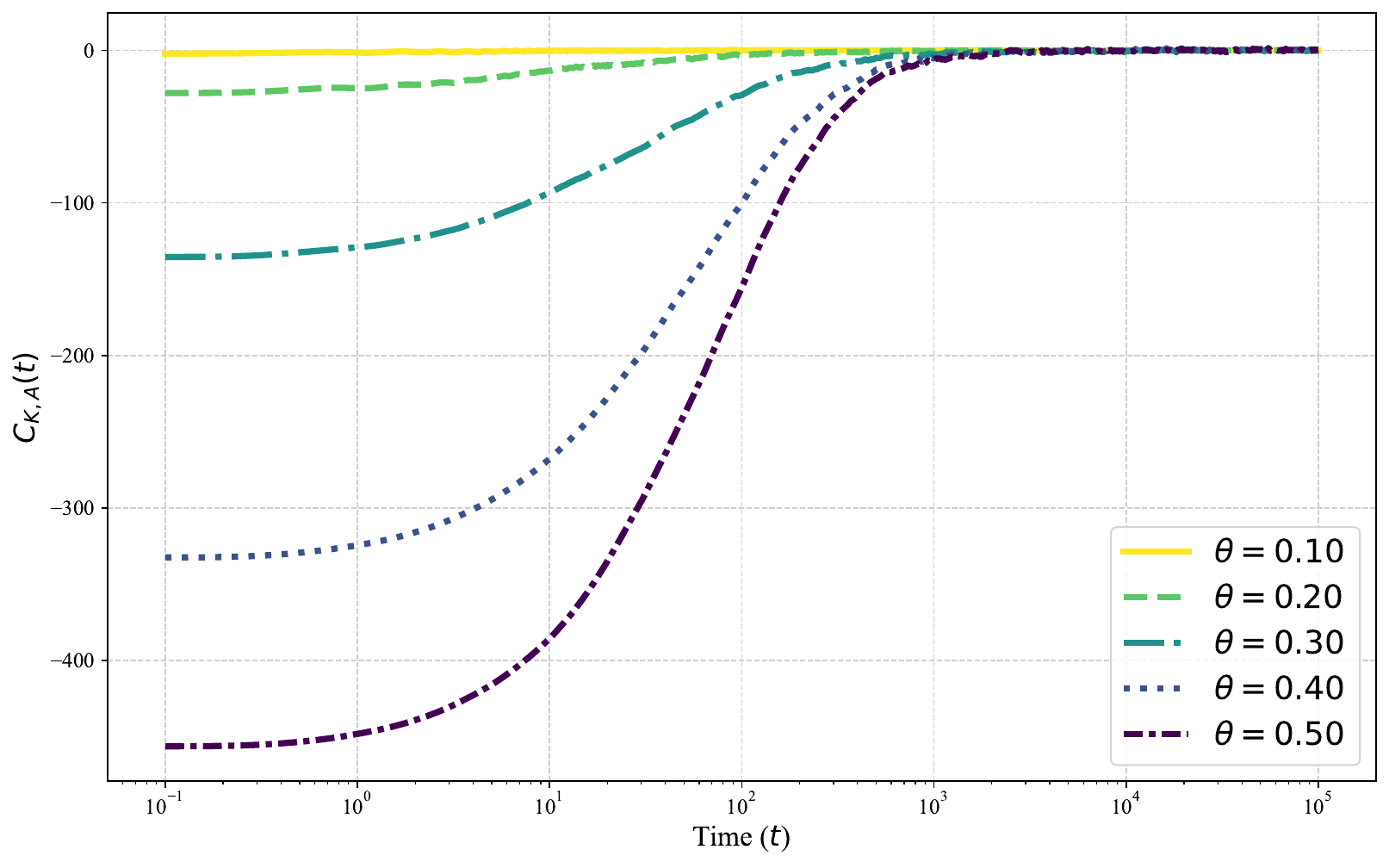}}\hfil
\subfloat[TNS, $W=1.0$]{\includegraphics[width=0.47\linewidth]{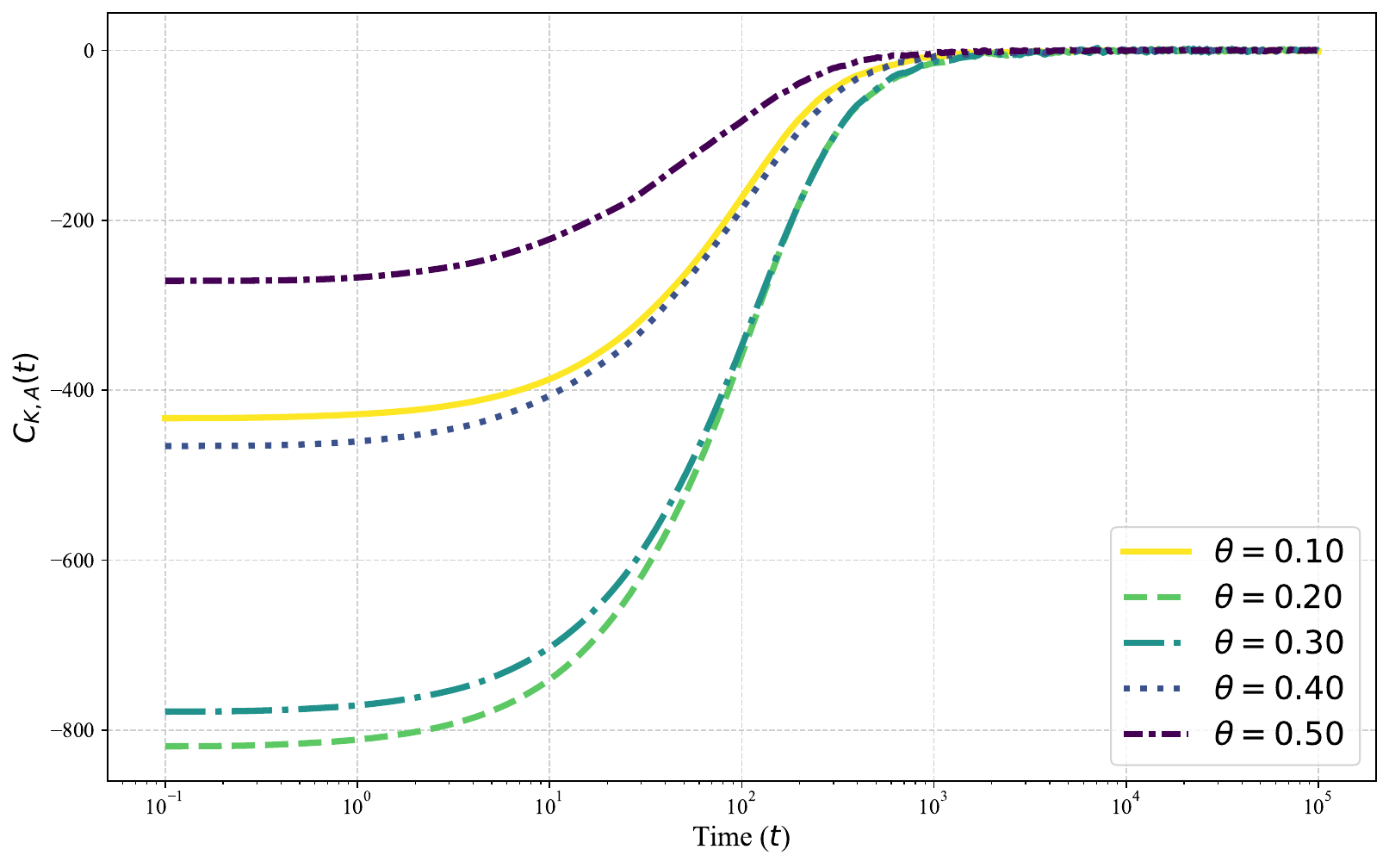}}\hfil 
\subfloat[TFS, $W=2.5$]{\includegraphics[width=0.47\linewidth]{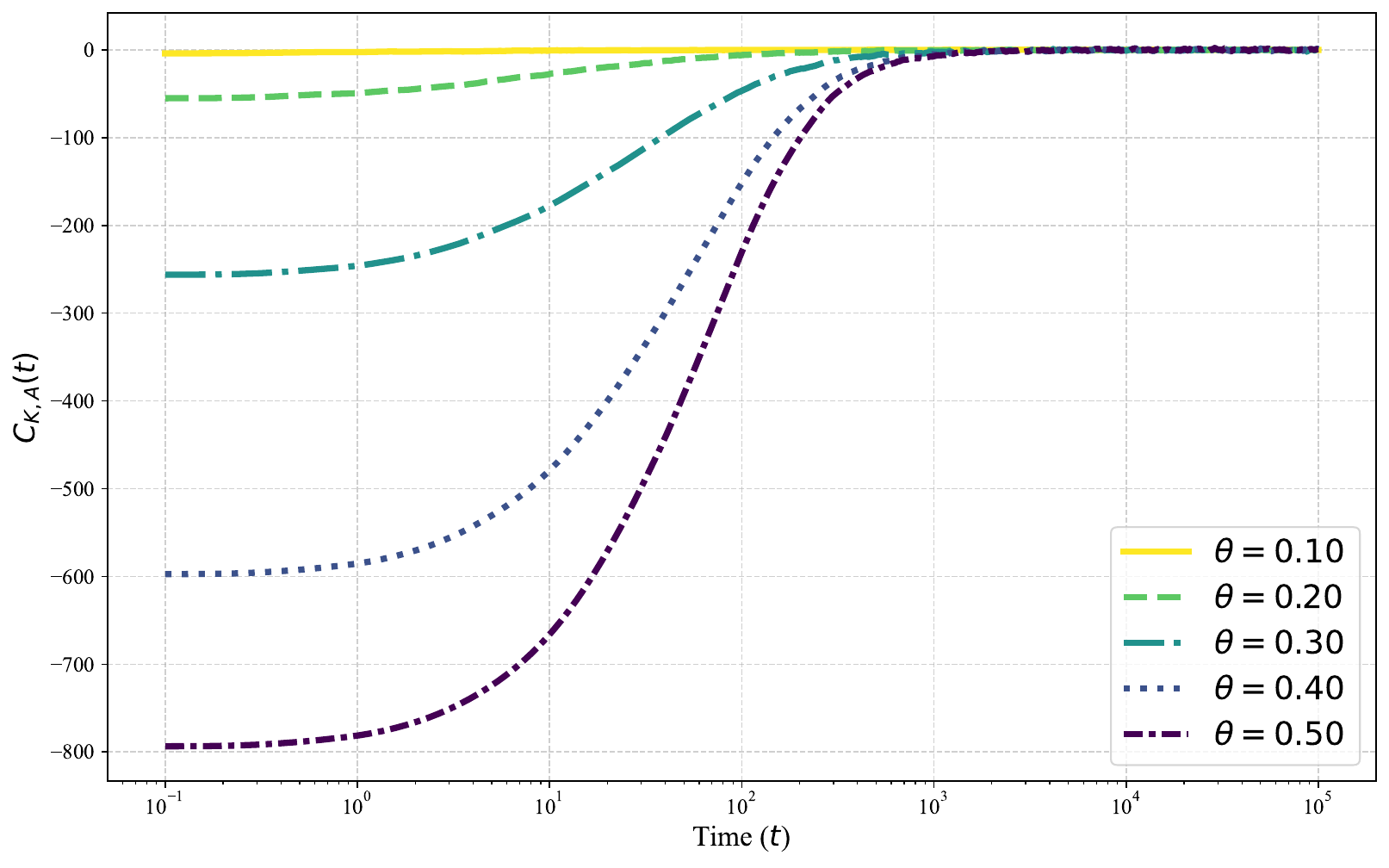}}\hfil
\subfloat[TNS, $W=2.5$ \label{fig:monomixing}]{\includegraphics[width=0.47\linewidth]{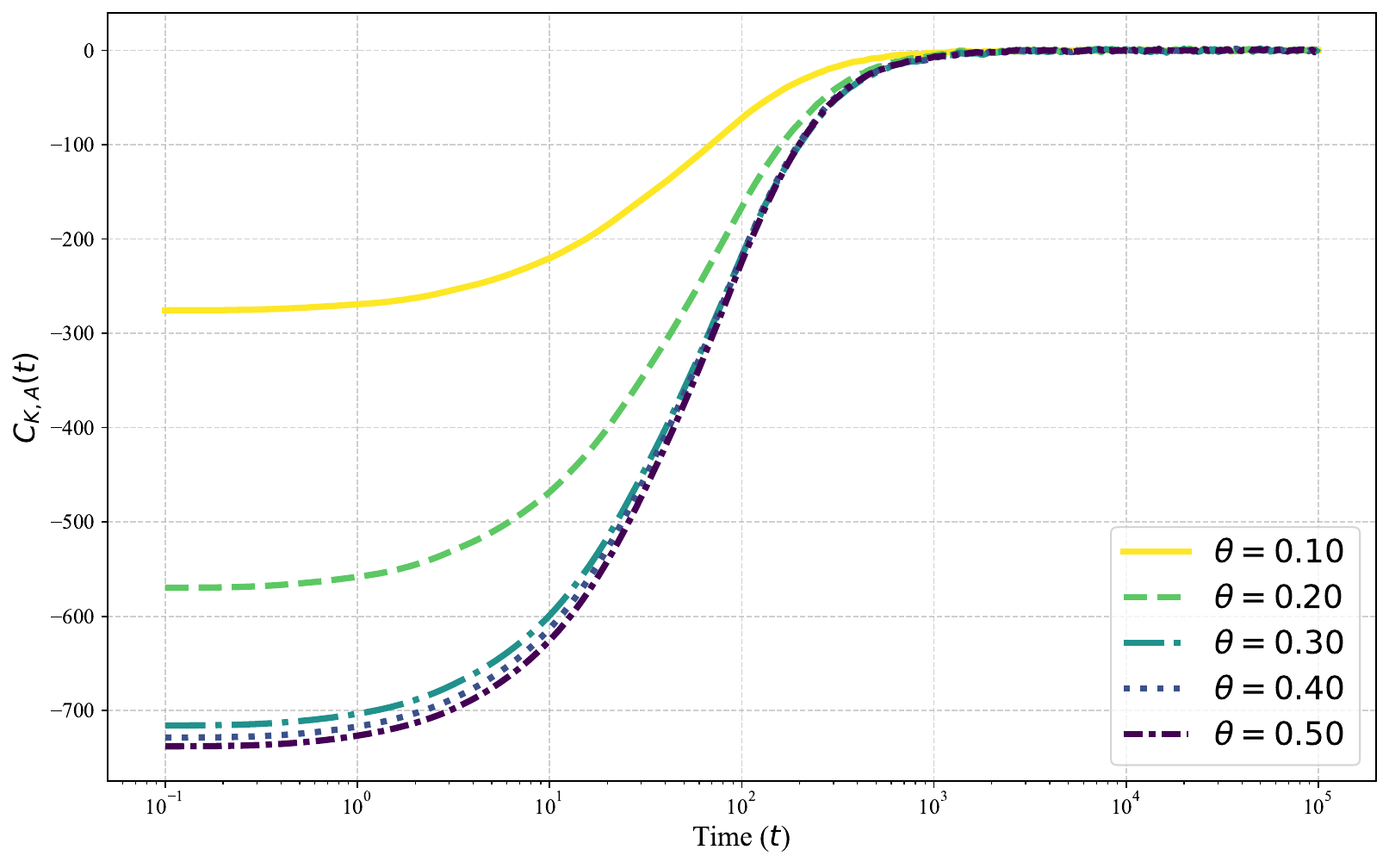}}\hfil 
\subfloat[TFS, $W=5.0$]{\includegraphics[width=0.47\linewidth]{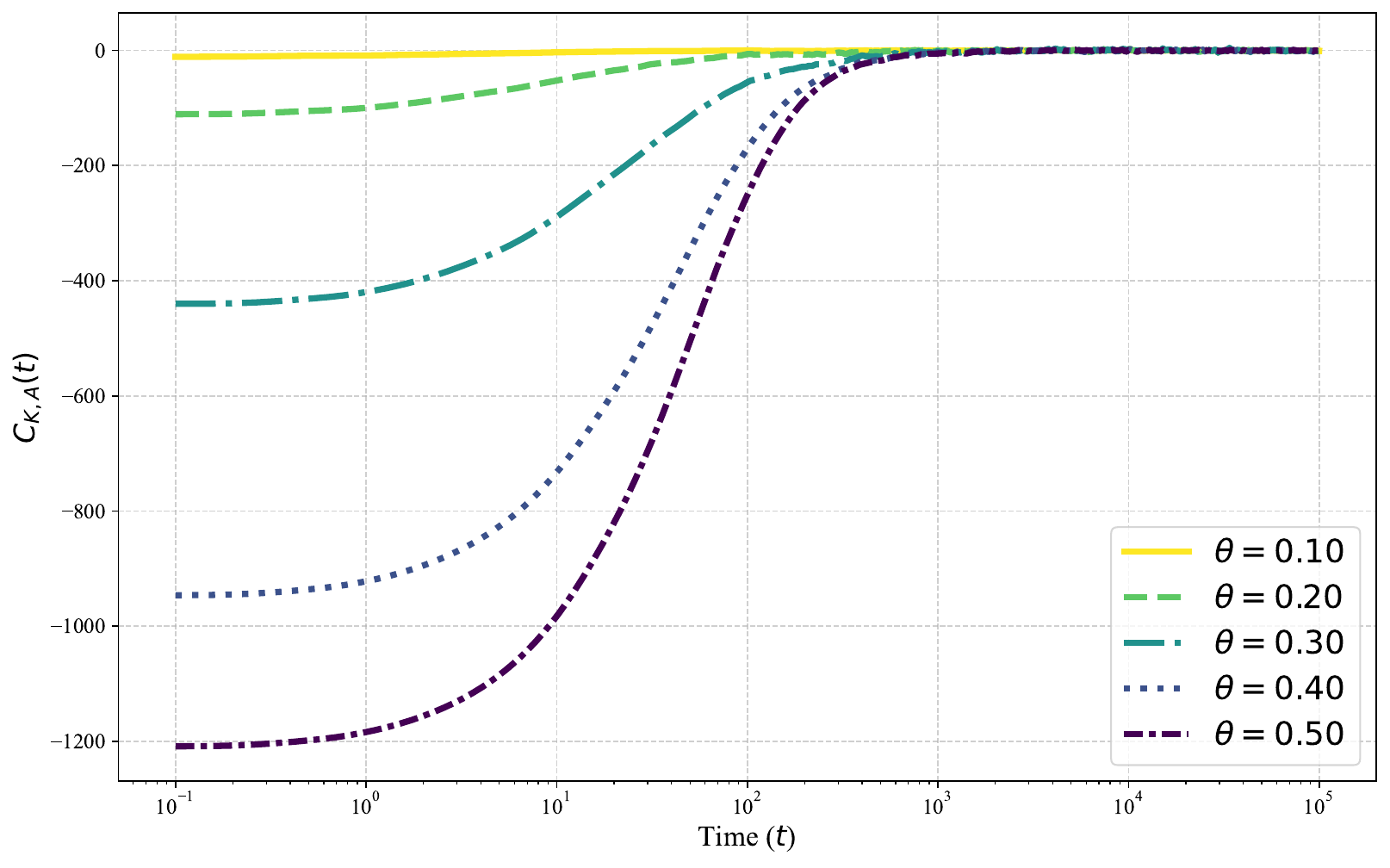}}\hfil
\subfloat[TNS, $W=5.0$]{\includegraphics[width=0.47\linewidth]{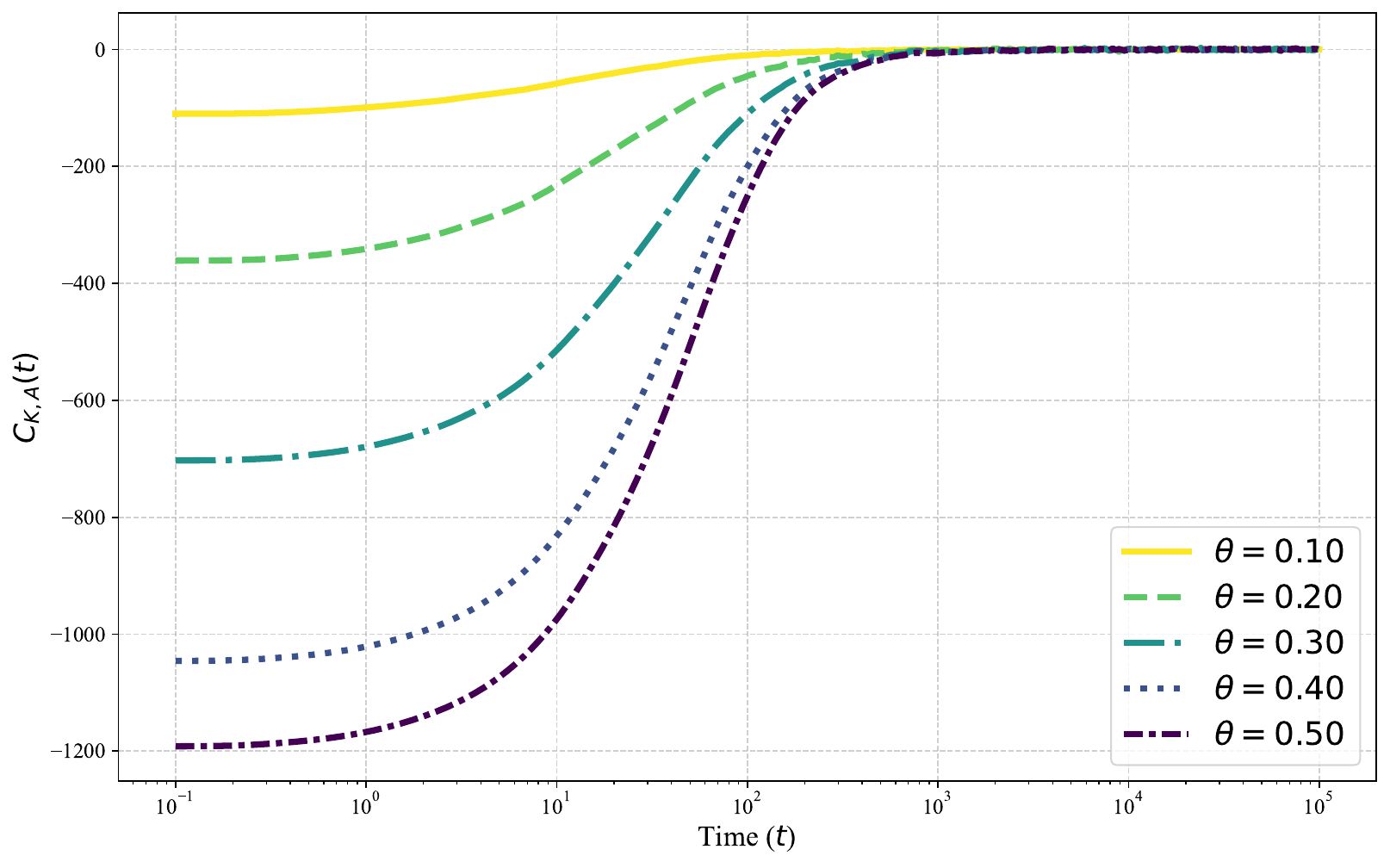}}\hfil 
\caption{Figures showing the total projective asymmetric complexity for the TFS and TNS states for different tilt angles, and increasing potential strength (top to bottom).}
\label{fig:mixingKC}
\end{figure}

This begs the question: {\it Can the asymmetric complexity be used to identify a quantum Mpemba effect?}  In Fig. \ref{fig:mixingKC} we have plotted the asymmetric complexity for various values of $\theta$, $W$ and for both the tilted ferromagnetic states and tilted Néel states.  We identify a clear distinction between the cases where the QME occurs and where it does not.  When the QME occurs, we observe a clear separation between the initial values, $C_{K,A}(0)$, as we change $\theta$.  Specifically, the initial value always decreases (i.e. increases in magnitude) as we increase $\theta$ so that the initial values are ordered in line with the tilt value.  In contrast, when the the QME does not occur we find either only a small spacing or, at low values of $W$, an increase in the initial value as we increase $\theta$.  \\ \\
Though this is clearly a distinguishing feature that separates pairs of states that are known to exhibit the Mpemba effect and those that do not, the direct interpretation of this is more difficult.  In particular, the asymmetric complexities of states do not cross, even when a pair of states is known to exhibit the Mpemba effect.  This is, on the face of it, rather odd.  Like entanglement asymmetry, the projected asymmetric complexity can be viewed as a measure of how far the Krylov complexity operator is from a symmetric operator.  As such, one may have expected that two states that exhibit a Mpemba effect (i.e. whose EAs cross) should also show a crossing in their asymmetric complexity.  \\ \\ 

As a counterpoint to this reasoning, the asymmetric complexities of all states reach zero at about the same time and, since some states start off with a greater amount of asymmetry, it may still be possible to argue a sort of ``crossing" if this is normalised correctly.  For this reason it is useful to consider the timescale associated to the asymmetric complexity, and to this end we use the half-life $t_{1/2}$ -- the time it takes for half of all the total initial asymmetric complexity to be lost (see Table \ref{tab:halflife}). First, note that while the asymmetric complexities have different initial magnitudes, even after normalizing these data based on their $t=0$ values, their half-lives will remain unchanged. In the context provided by the table, the decay rates are slower for larger initial asymmetric complexities, thus the hypothesized crossing in the decay rate is \textit{not} exhibited. We observe the following general trends: half-life increases with increasing potential strength, $W$; and, in the regimes where the QME is observed ($2.5\lesssim W$ for the Néel case, always for ferromagnetic case), these half-lives are ordered monotonically by their $\theta$ value\footnote{The $W=2.5$ Néel state data is an edge case, and it fails only `slightly' to be monotonic.}.
\\ \\
\begin{table}[!htbp]
    \centering 

    \begin{tabular}{|c||c|c|c|c|c||c|c|c|c|c|}
        \hline
        \textbf{$W \ \setminus \ \theta$} & \textbf{0.1} & \textbf{0.2} & \textbf{0.3} & \textbf{0.4} & \textbf{0.5} & \textbf{0.1*} & \textbf{0.2} & \textbf{0.3} & \textbf{0.4} & \textbf{0.5} \\
        \hline \hline

        \textbf{1.0} & 73.45 & 82.25 & 84.35 & 70.39 & 46.86 & 0.62 & 8.77 & 26.16 & 44.22 & 55.05 \\
        \hline
        \textbf{2.0} & 53.40 & 60.05 & 62.55 & 57.14 & 53.63 & 1.28 & 10.84 & 25.22 & 43.07 & 52.74 \\
        \hline
        \textbf{2.5} & 41.11 & 47.14 & 50.35 & 50.06 & 51.57 & 1.28 & 9.96 & 24.07 & 40.17 & 49.01 \\
        \hline
        \textbf{3.3} & 26.02 & 33.08 & 37.91 & 43.35 & 47.84 & 1.94 & 9.69 & 21.22 & 36.68 & 45.16 \\
        \hline
        \textbf{5.0} & 11.21 & 18.04 & 26.08 & 35.12 & 39.49 & 4.08 & 8.89 & 19.05 & 32.01 & 38.73 \\
        \hline
    \end{tabular}
    \caption{Table showing the half-lives $t_{1/2}$ of the asymmetric complexities for given $W$ values (rows) and $\theta$ values (columns) for the Néel state (left) and ferromagnetic state (right) averaged over 240 realisations of $\phi$ for an $N=12$ size system. \\ \textbf{*}: The magnitude of the asymmetric complexity at $\theta=0.1$ for the ferromagnetic state is small and prone to large oscillatory behavior, whose influence is reduced as $\theta$ and/or $W$ increases. Hence, this column (and the $W=1.0$, $\theta=0.2$ ferromagnetic entry) should be treated with caution.}
        \label{tab:halflife}
\end{table}

Perhaps what is most interesting about the half-lives of the projected asymmetric complexity is that they largely demystify the QME: we observe the QME when the projected asymmetric complexity is ordered in magnitude by the magnitude of the tilt. States that are more asymmetric by this measure require more time to approach a symmetric one. Quite counter-intuitively (based on the EA), the more interesting states are those in the thermalizing regime with larger tilts (higher initial EA) but shorter half-lives. Again, this is not a mystery from the perspective of the projective asymmetric complexity, as the states have different orderings of their initial asymmetry magnitudes, and these initial orderings appear to always be preserved. 
\\ \\
Before proceeding with the analysis of the symmetric complexity, we comment on the vanishing of the asymmetric complexity. In the MBL phase, coherences persist within a single system, while they vanish in the ETH phase due to decoherence. This distinction is crucial when interpreting metrics like the asymmetric complexity, which may appear to predict vanishing coherence in both phases when averaged over many realizations. The vanishing of coherences in the ETH regime is an intrinsic process: the chaotic spectrum results in rapidly oscillating phase factors $e^{-i(E_n-E_m)t}$ and, when computing the expectation value of an observable, the sum over these fluctuating phases averages to 0. This is understood as the system acting as its own bath. Conversely, a single realisation of the system in the MBL phase can preserve coherences indefinitely due to localization. While each realisation of the system may preserve some coherence, the phase of that coherence is random. When performing the average over an ensemble of system realisations, these random phases lead to destructive interference, resulting in the coherences present in an individual realisation (and any phase-sensitive metrics contained therein) to vanish. Therefore, the vanishing of the realisation-averaged asymmetric complexity is not a signal of thermalization, but rather a direct result of taking an ensemble average of localizing systems.

\subsection{Shifted Projected Symmetric Complexity}
\label{sec:symmetricQME}
\begin{figure}[!htbp]
\centering
\subfloat[TFS, $W=3.0$]{\includegraphics[width=0.47\linewidth]{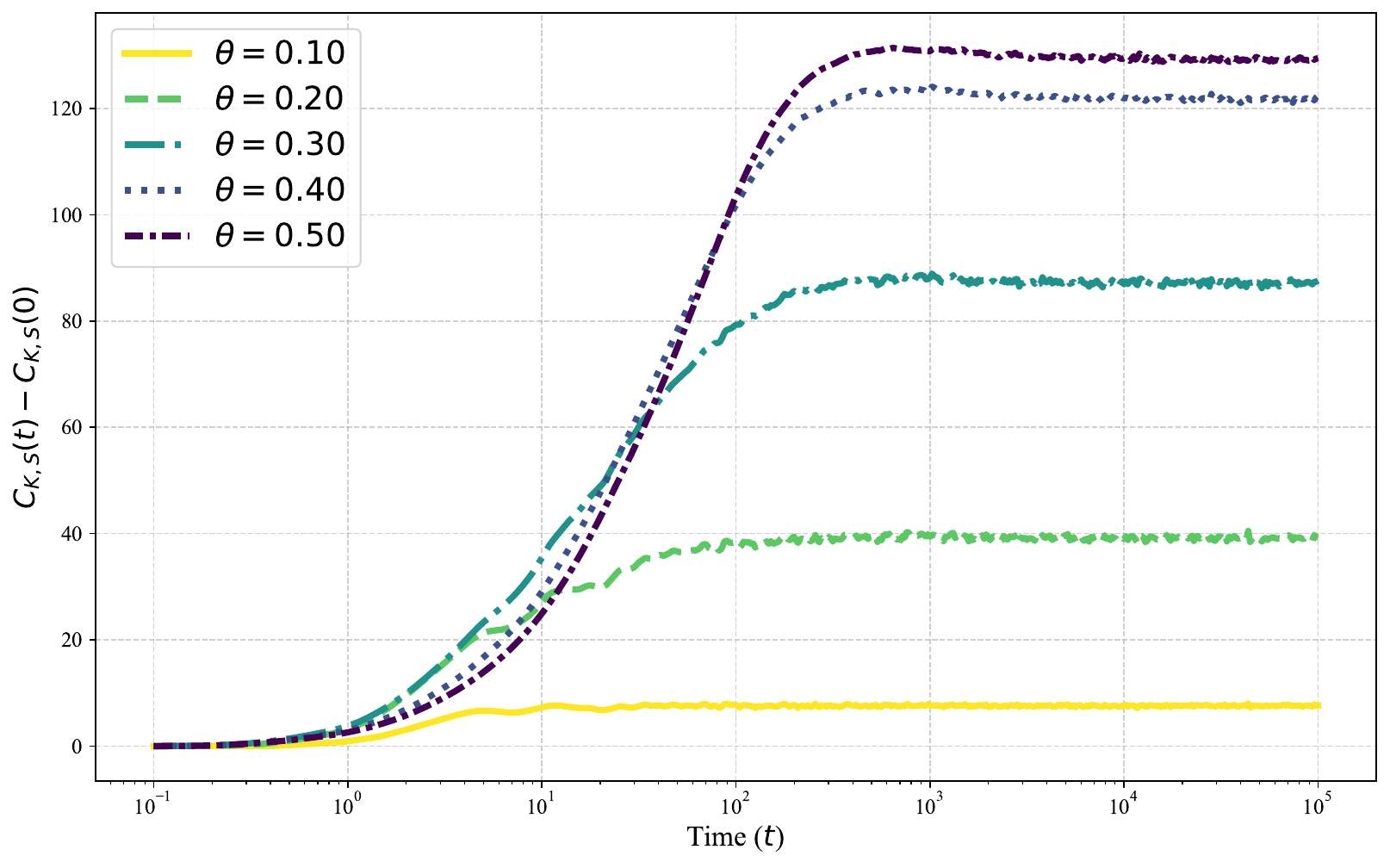}}\hfil
\subfloat[TNS, $W=3.0$ \label{fig:diffKCbelow}]{\includegraphics[width=0.47\linewidth]{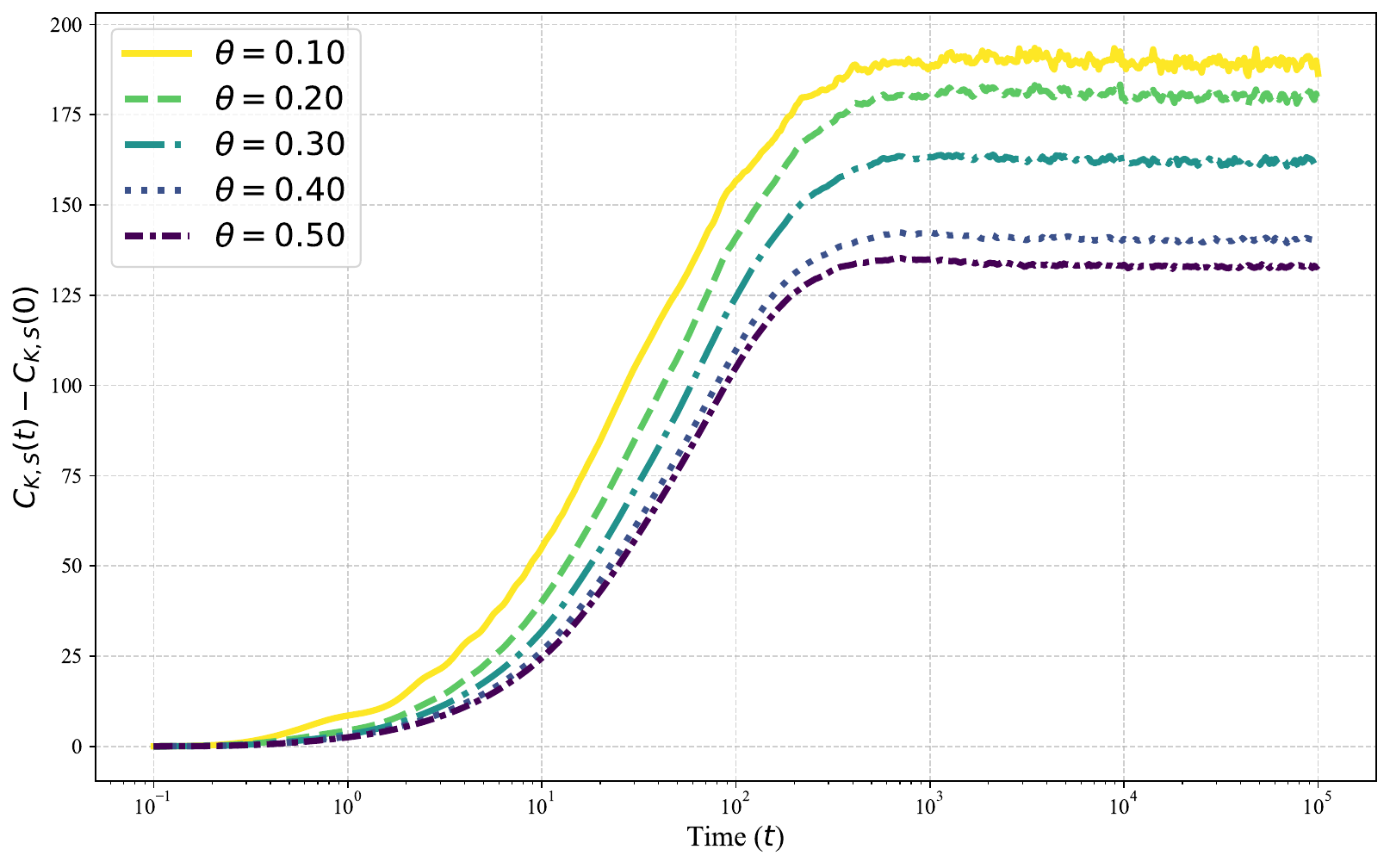}}\hfil 
\subfloat[TFS, $W=3.3$]{\includegraphics[width=0.47\linewidth]{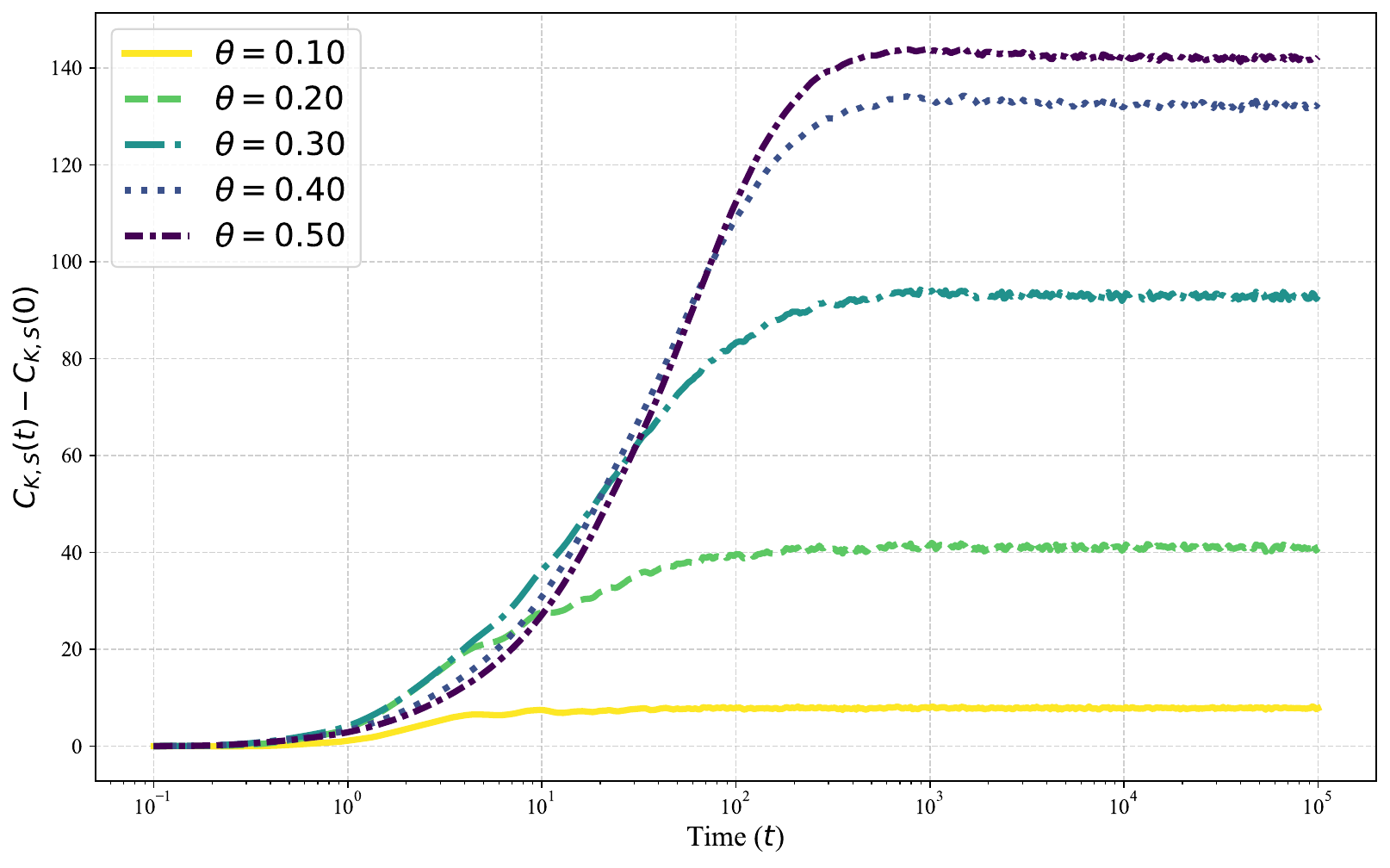}}\hfil
\subfloat[TNS, $W=3.3$\label{fig:diffKCcrit}]{\includegraphics[width=0.47\linewidth]{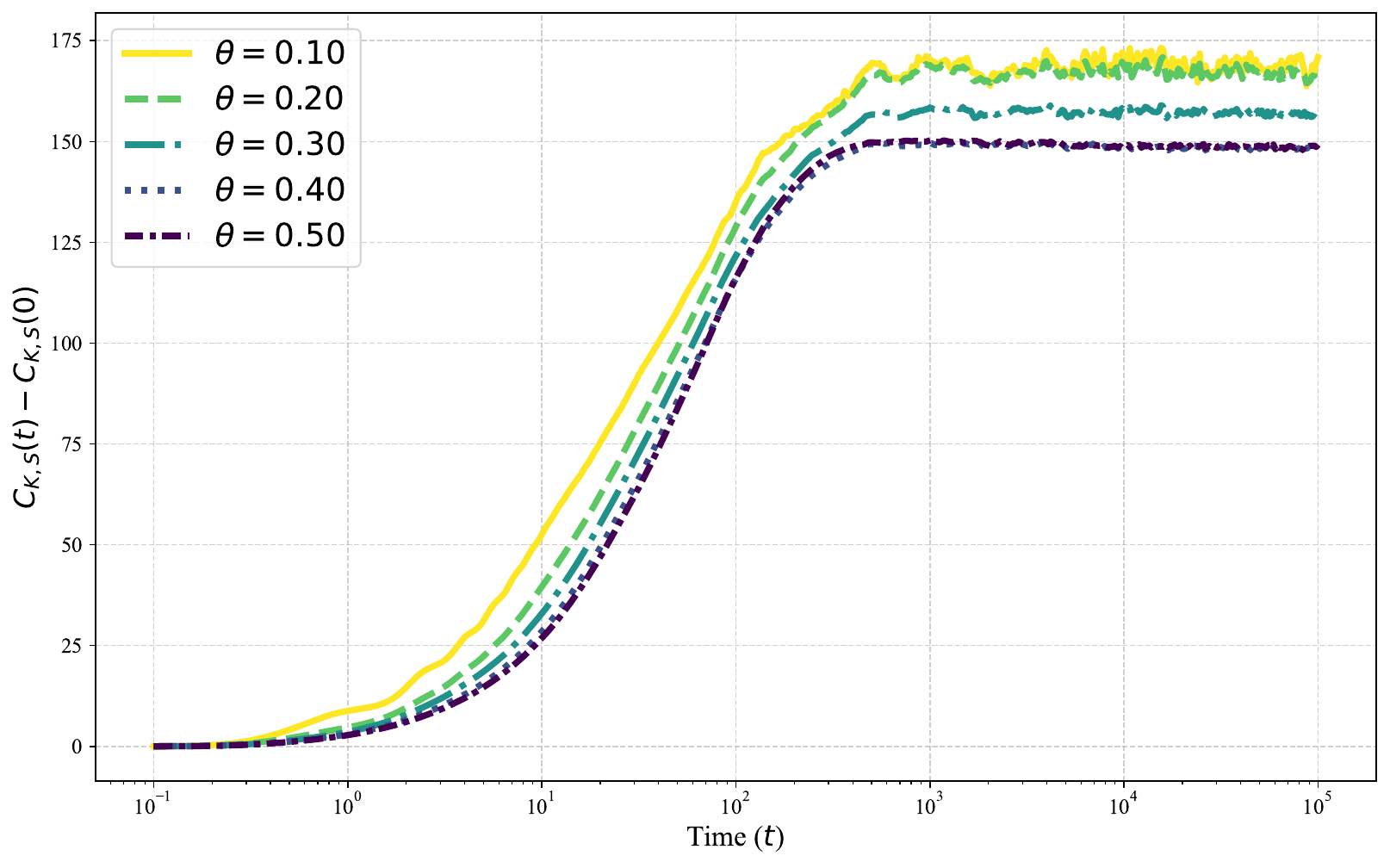}}\hfil 
\subfloat[TFS, $W=3.6$]{\includegraphics[width=0.47\linewidth]{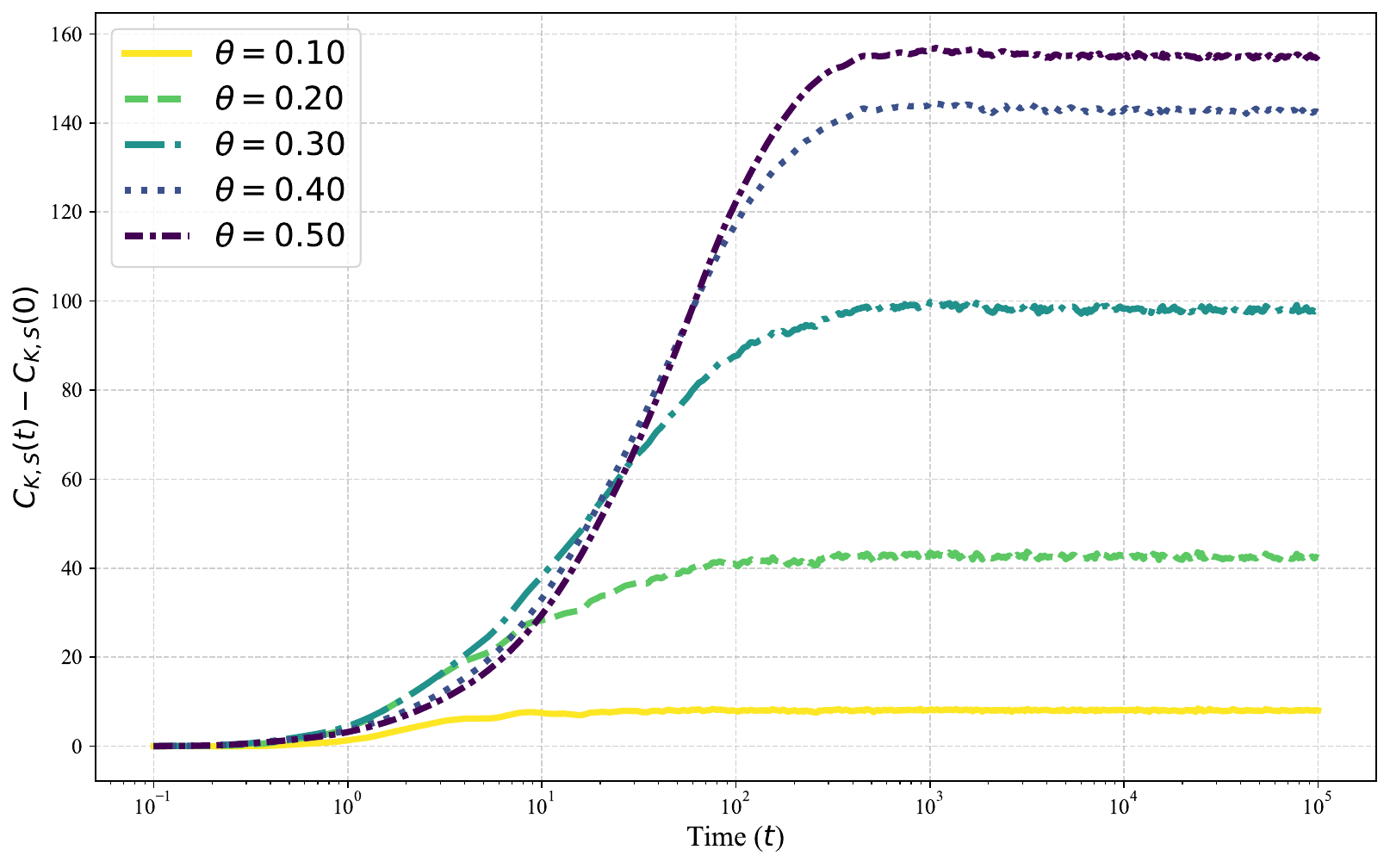}}\hfil
\subfloat[TNS, $W=3.6$ \label{fig:diffKCabove}]{\includegraphics[width=0.47\linewidth]{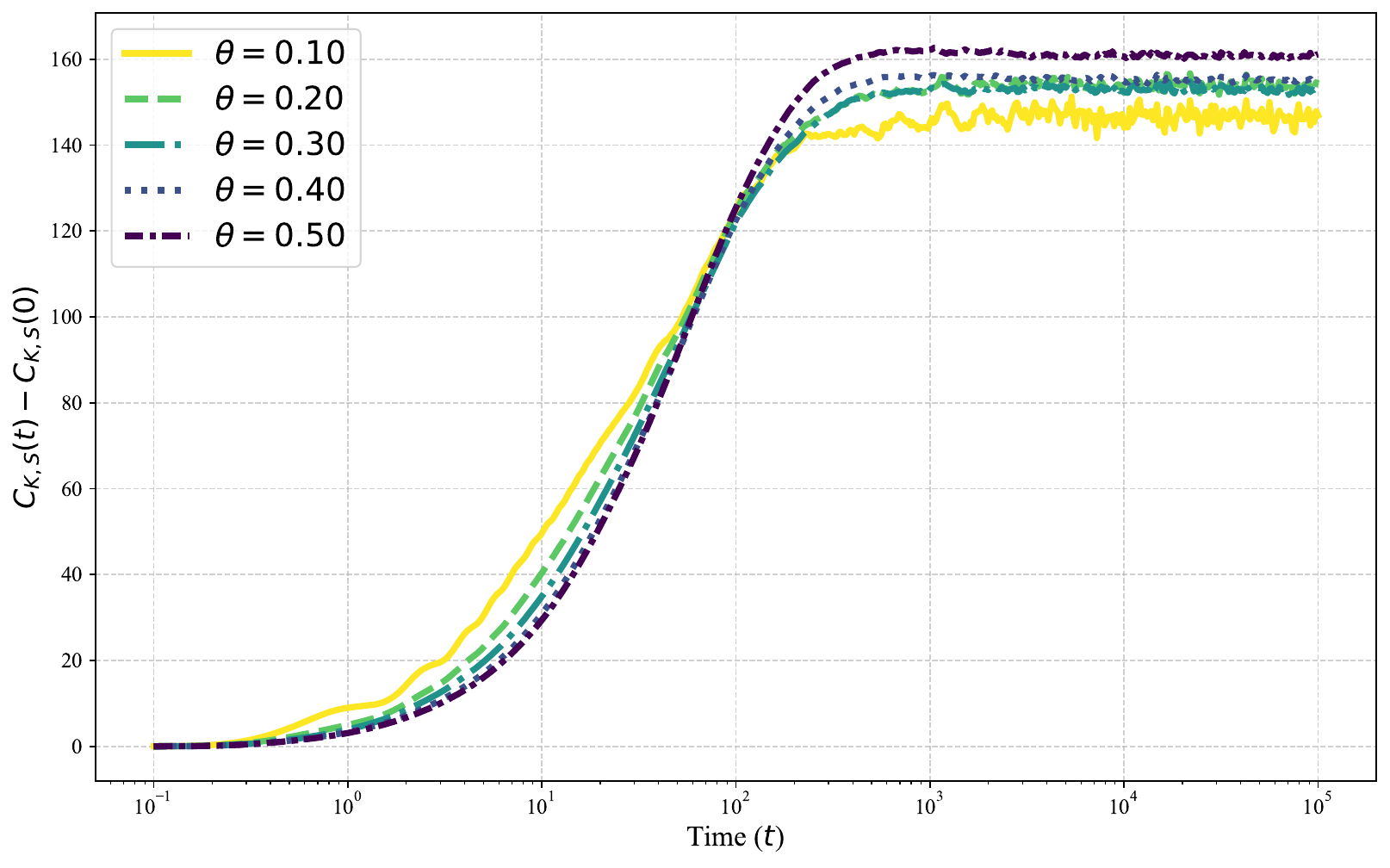}}\hfil 
\caption{Figures showing the dynamical projective symmetric complexity for the TFS (left) and TNS (right) states for different tilt angles, and increasing potential strength (top to bottom). We choose values close to the ETH/MBL critical value ($W_c\approx3.3$). In appendix \ref{app:SRKCQME} we plot the equivalent figures for the symmetry-resolved Krylov complexity.}
\label{fig:diffusiveKC}
\end{figure}

The projective symmetric components of the complexity will counterbalance the large negative values of the asymmetric complexity. While it is interesting to observe the values of symmetric complexities at $t=0$, it seems more appropriate to attribute the non-zero values at $t=0$ to the asymmetric behavior of the system -- if the complexity were initially block diagonal, then the asymmetric complexity would be zero, and, therefore, so too would be the symmetric contribution. For this reason, in Fig. \ref{fig:diffusiveKC}, we plot the \textit{dynamical} part of the symmetric complexity, $C_{K,S}(t)-C_{K,S}(0)$, to isolate the growth of complexity in these components.
Contrary to the asymmetric components which do \textit{not} show a clear dependence on the ETH/MBL phase change -- the ordering in magnitude of $|C_{K,A}|$ being monotonically ordered by tilt angle first occurs at $W\approx2.5$, significantly before the phase transition at $W\approx3.3$ -- the dynamical symmetric complexity appears to be acutely sensitive to the phase transition for the Néel states. For $W=3.0<W_c$ (Fig. \ref{fig:diffKCbelow}), the symmetric complexity exhibits slower initial growth and lower saturation values for larger values of $\theta$. As the system is considered at increasing values of $W\geq W_c$, while the initial slow growth is always present for the states of larger tilts, the saturation values grow such that at $W=3.3$ (Fig. \ref{fig:diffKCcrit}) the state tilted by $0.5\pi$ plateaus just higher than the $0.4\pi$ tilted state, and, when one increases the potential strength a little further to $W=3.6$ (Fig. \ref{fig:diffKCabove}) the saturation values have completely reversed the ordering they had before the phase transition.\\

\noindent
These results are both striking and challenging to interpret, reflecting the subtleties of studying the Aubry–André model. Broadly speaking, the behavior of the symmetric Krylov complexity aligns with that observed for entanglement asymmetry: crossings are consistently present for tilted ferromagnetic states at all values of $W$ studied, while for tilted Néel states they emerge only once the system approaches or crosses the ETH/MBL critical point. The complication lies in interpretation. The ETH/MBL transition, while of intrinsic interest, acts as a confounding factor: it remains unclear whether the observed crossings are uniquely attributable to the quantum Mpemba effect, or whether they reflect a more generic sensitivity of operator dynamics to phase structure. If one nevertheless regards the observed behavior as Mpemba-like, an immediate puzzle arises. Based on entanglement asymmetry, one would naturally expect crossings to manifest in the asymmetric component of the Krylov decomposition, since this quantity encodes inter-sector coherences and explicitly tracks symmetry restoration. Instead, the crossings appear in the dynamical symmetric complexity. This suggests a more nuanced mechanism at play. One possible interpretation is that coherence functions as a resource in the quantum setting \cite{Chitambar_2019}. In the decomposition procedure, the projection onto symmetry sectors effectively acts as a resource-destroying map, separating the free (symmetric) complexity from the resource-rich (asymmetric) contribution. The negative initial value of the asymmetric complexity at $t=0$ can then be understood as the “cost” of preparing a coherent superposition, which is absorbed into the initial offset of the symmetric complexity. Under this view, it is only the dynamical growth of the symmetric component that directly reflects the system’s relaxation, and hence it is this part that exhibits Mpemba-like crossings.\\

\noindent
Crucially, the fact that the data in each panel of Fig. \ref{fig:diffusiveKC} correspond to a fixed potential strength $W$ implies that the observed crossings cannot be dismissed as trivial consequences of comparing across phases. Rather, they represent re-orderings of relaxation trajectories under the same dynamical conditions, closely paralleling the entanglement-asymmetry definition of the QME. Taken together, these results strongly support our interpretation of the observed crossings in symmetric complexity as a genuine manifestation of the quantum Mpemba effect, albeit one whose appearance is deeply entangled with the ETH/MBL transition.

\section{Static ($t=0$) Predictors of the Mpemba Effect}
\label{sec:KCtheta}
The decomposition of Krylov complexity into its projected symmetric and projected asymmetric parts provides a method for identifying a set of states where a Mpemba effect will be observed. The remarkable feature is that the spread of the asymmetric complexity at time $t=0$ and the value of its tilt appear to contain enough information to make the prediction of whether a pair of states will exhibit the Mpemba effect or not.  In this section we would like to further investigate this.

\subsection{Structural Complexity}
\begin{figure}[!htbp]
\centering
\subfloat[TFS]{\includegraphics[width=0.47\linewidth]{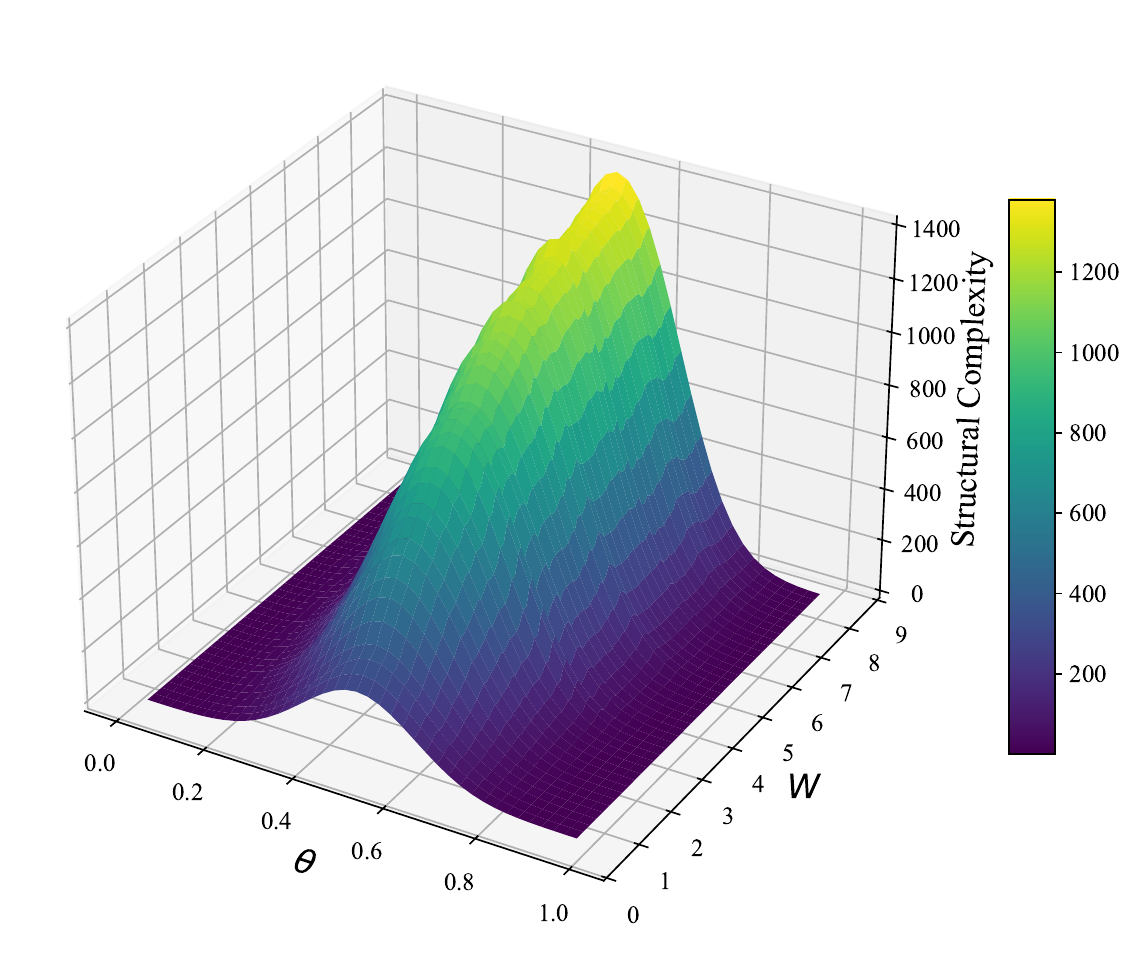}}\hfil
\subfloat[TNS \label{fig:structuralTNS}]{\includegraphics[width=0.47\linewidth]{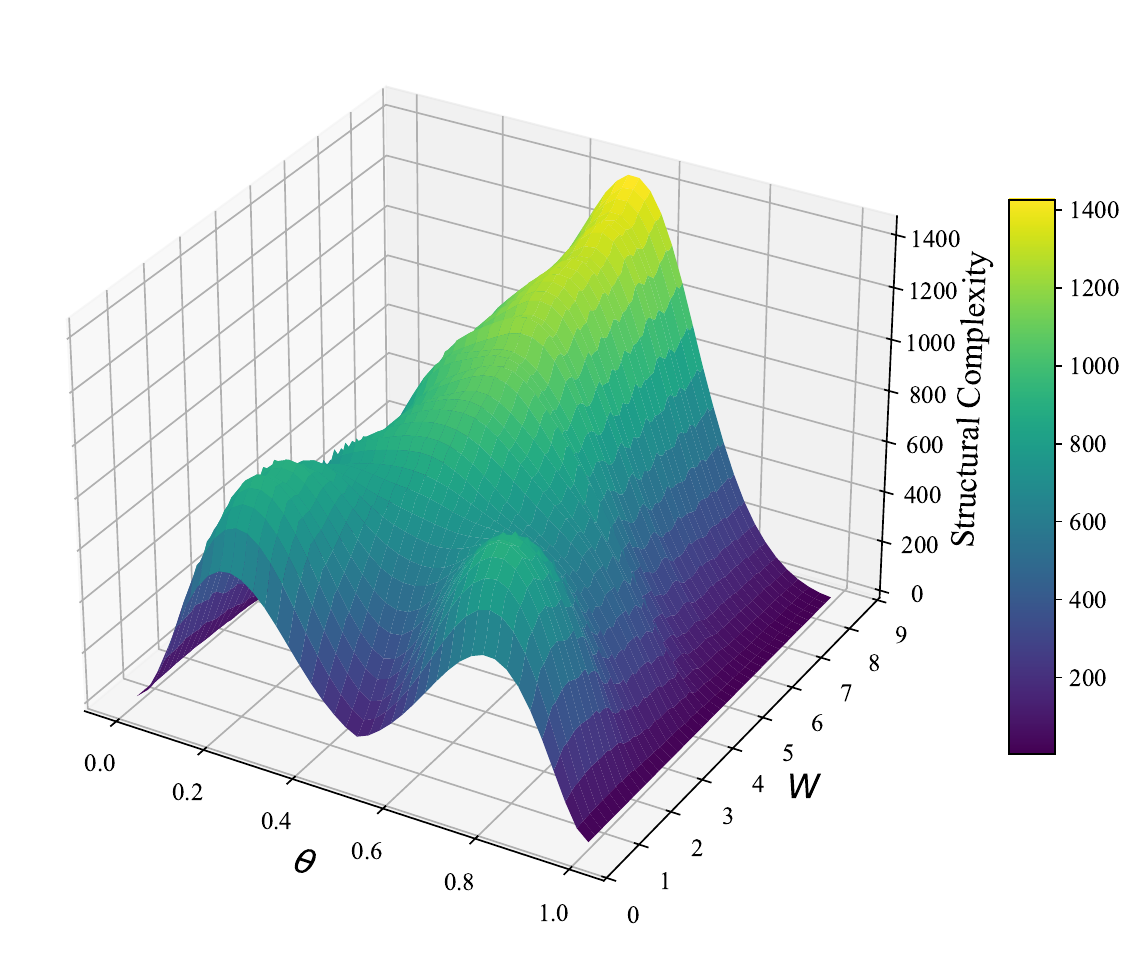}}\hfil 
\caption{Figures showing the structural complexity surface described by the initial tilt angle, $\theta$, and the quasiperiodic potential strength, $W$ for a single realisation of the system ($\phi=0$). Note the single dominant peak in the ferromagnetic state plot and the transition from two peaks to one peak as $W$ increases for the Néel state plot.}
\label{fig:structuralKC}
\end{figure}

The decomposition, presented in sec. \ref{sec:KCsymmres}, of Krylov complexity into symmetric and asymmetric components leads to an interesting observation: there is an initial \textit{structural} complexity of the asymmetric states. The $t=0$ symmetric complexity is a positive semi-definite quantity, and is necessarily balanced by the semi-definite negativity of the asymmetric complexity\footnote{This is specifically a $t=0$ statement. There is no obvious obstruction to the asymmetric complexity becoming positive, however, the dynamics we observe show the asymmetric complexity typically decays to fluctuations around zero.}. To be precise, we will define the structural complexity as
\begin{equation}
    C_{structural} = |C_{K,A}(0)|=C_{K,S}(0),
\end{equation}
that is, the $t=0$ magnitude of the projective asymmetric complexity, which is, of course, equal to the $t=0$ projective symmetric complexity. We choose this nomenclature as it is a quantity which is non-dynamical\footnote{But it is \textit{informed} of the dynamics via the Lanczos algorithm.}, and it quantifies the amount of non-symmetric `structure' that the system must remove to achieve symmetry. We emphasize that, if the structural complexity begins at zero, this means that there are no coherences -- the projective symmetric complexity is positive semi-definite for all times, and vanishing only if the initial state begins in the block diagonal symmetry sectors, whereafter, given that the Hamiltonian commutes with the generator of the symmetry, the asymmetric complexity will always be zero. Recalling now the candidacy questions of sec. \ref{sec:KCsymmres}, we see that this observable does satisfy property (1) -- it is vanishing only when there is no asymmetry -- and rephrasing property (2) as `non-negative', it satisfies the property of being non-negative as well.  Given that it is a static quantity, property (3) -- requiring monotonicity of the time evolution -- is irrelevant. It is important to note that the asymmetric complexity need only be non-positive at $t=0$, and there is no obstruction preventing these values from evolving to being positive (see late times of Fig. \ref{fig:DiffNMix} and App. \ref{app:latetimefluctuations}). Intuitively, the structural complexity is a measure of the amount of asymmetry (coherences) between symmetry sectors computed using the Krylov basis generated by repeated applications of the Hamiltonian to the tilted state. It is tempting to wonder if this initial structural complexity encodes information about the states that is predictive of the Mpemba effect in the form of some necessary but insufficient condition. 
\\ \\
Figure \ref{fig:structuralKC} highlights both quantitative and qualitative differences in the structural complexity of TFS and TNS states. For the TFS states, the structural complexity exhibits a single peak centered around $\theta = 0.5$, independent of the quasiperiodic potential $W$. In contrast, the TNS states display a transition: at small $W$, the structural complexity is bi-peaked, but as $W$ increases the profile evolves into a single peak. This crossover occurs near $W \approx 2.5$ (see App. \ref{app:crossinganalysis}). Importantly, it is precisely in the regime where the structural complexity becomes singly-peaked that the Mpemba effect—when probed using entanglement asymmetry—can occur.\\

\noindent
This implies that the structural complexity may be used to predict whether a pair of states will exhibit a Mpemba effect.  This, despite the fact that figure  \ref{fig:structuralKC} is not averaged {\it i.e.} it is computed for only a single realisation of the Aubry-André model. Of course, using a single realisation means that the data is susceptible to realisation-dependent noise that would be removed by averaging over realisations. However, the general behaviour demonstrated does appear to be a robust feature of almost all realisations. While structural complexity provides valuable predictive insight, it is important to recognize the computational challenges associated with its evaluation. To generate Fig. \ref{fig:structuralKC}, the full Krylov basis was constructed at every point in phase space, since each point corresponds to a distinct combination of initial tilt and quasiperiodic potential strength. Numerically, this requires full re-orthogonalization at each iteration of the Lanczos algorithm to preserve the fidelity of the basis, a process that is computationally expensive. Moreover, obtaining the complete Krylov basis together with the associated Lanczos coefficients is effectively equivalent to computing the time evolution of the state to arbitrarily late times. This places significant computational cost on the method and may ultimately, and sadly, constrain the practicality of structural complexity as a predictive diagnostic of the quantum Mpemba effect.

\subsection{Lanczos Coefficients}
\label{sec:lanczos}
\begin{figure}[!htbp]
\centering
\subfloat[TFS]{\includegraphics[width=0.47\linewidth]{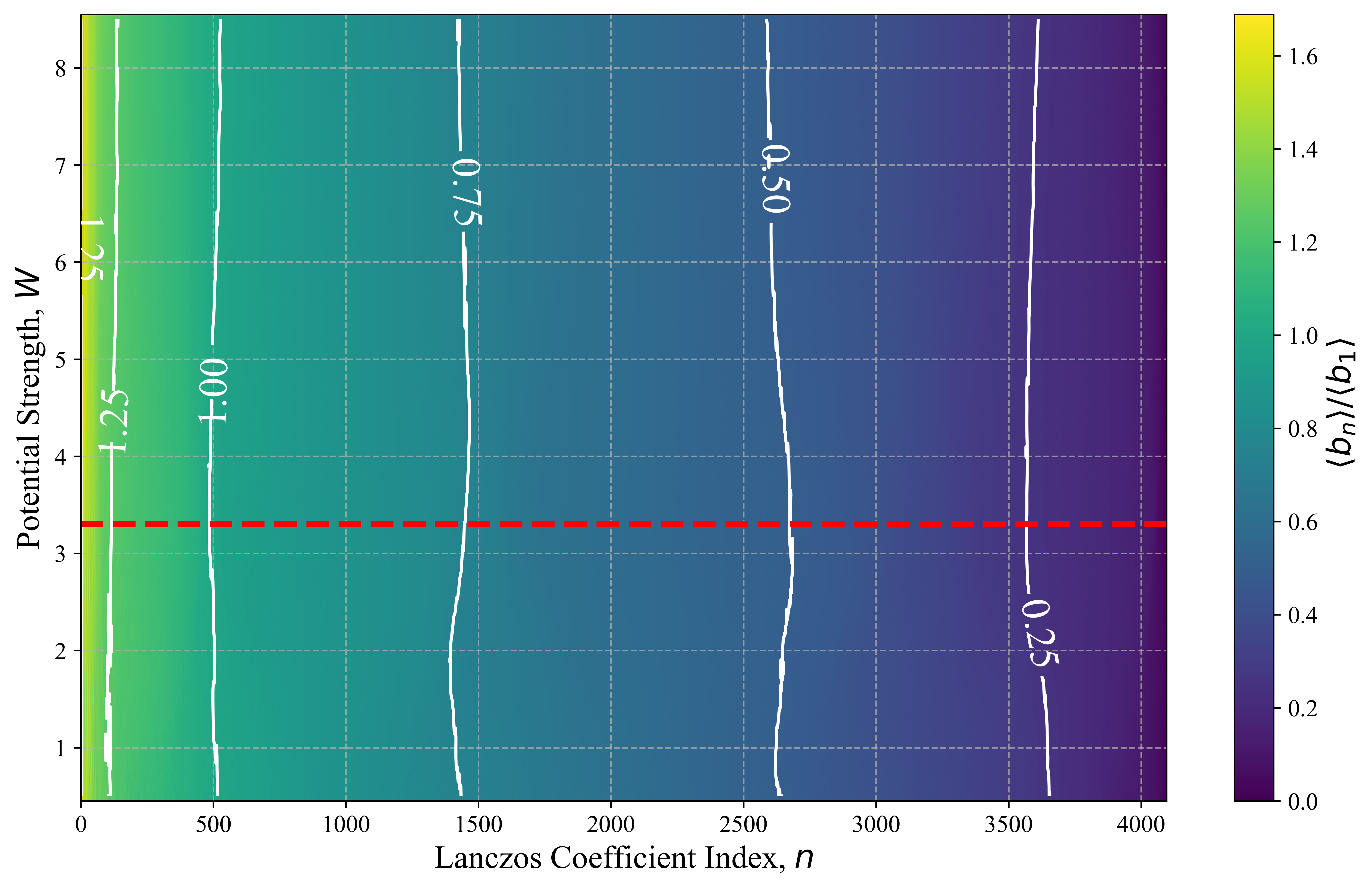}}\hfil
\subfloat[TNS]{\includegraphics[width=0.47\linewidth]{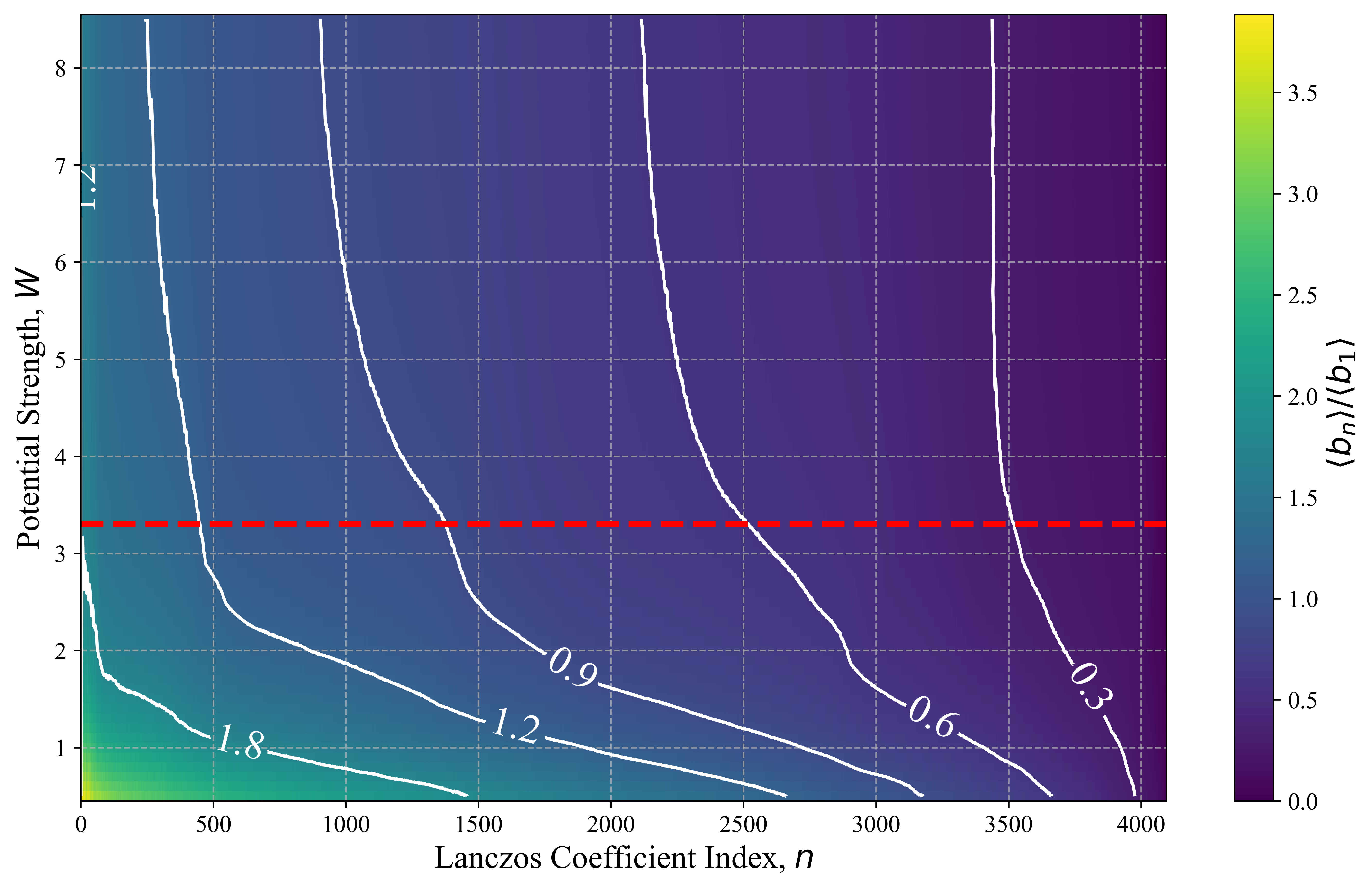}}\hfil 
\caption{Figures showing the microscopic encoding of the dynamics via the normalized Lanczos coefficients for the $\theta=0.5$ tilted state. The red-dashed line indicates the critical potential value, $W_c\approx3.3$. The white lines represent the points at which the Lanczos coefficients take on the same normalized value, with these values labelled on the lines themselves.}
\label{fig:lanczosheatmap}
\end{figure}

As was in section \ref{sec:EAKC}, the Lanczos coefficients provide a microscopic insight into the dynamics of the system. As the quasiperiodic potential strength increases, so too do the initial values of the Lanczos coefficients. Recall, that $b_1^2=\mu_2=\langle H^2\rangle-\langle H\rangle^2$, and, by increasing $W$ we increase the magnitude of the variance of the spectrum of the Hamiltonian. This necessitates a normalization scheme to compare the scaling of the Lanczos coefficients that counteracts this growth. Hence, to make the Lanczos coefficients comparable irrespective of their $W$ value, we use the simple normalization scheme\\ $b_n\rightarrow b_n/b_1$. The results of computing these normalized coefficients is shown in Fig. \ref{fig:lanczosheatmap}.

\begin{figure}[!htbp]
\centering
\subfloat[TFS]{\includegraphics[width=0.47\linewidth]{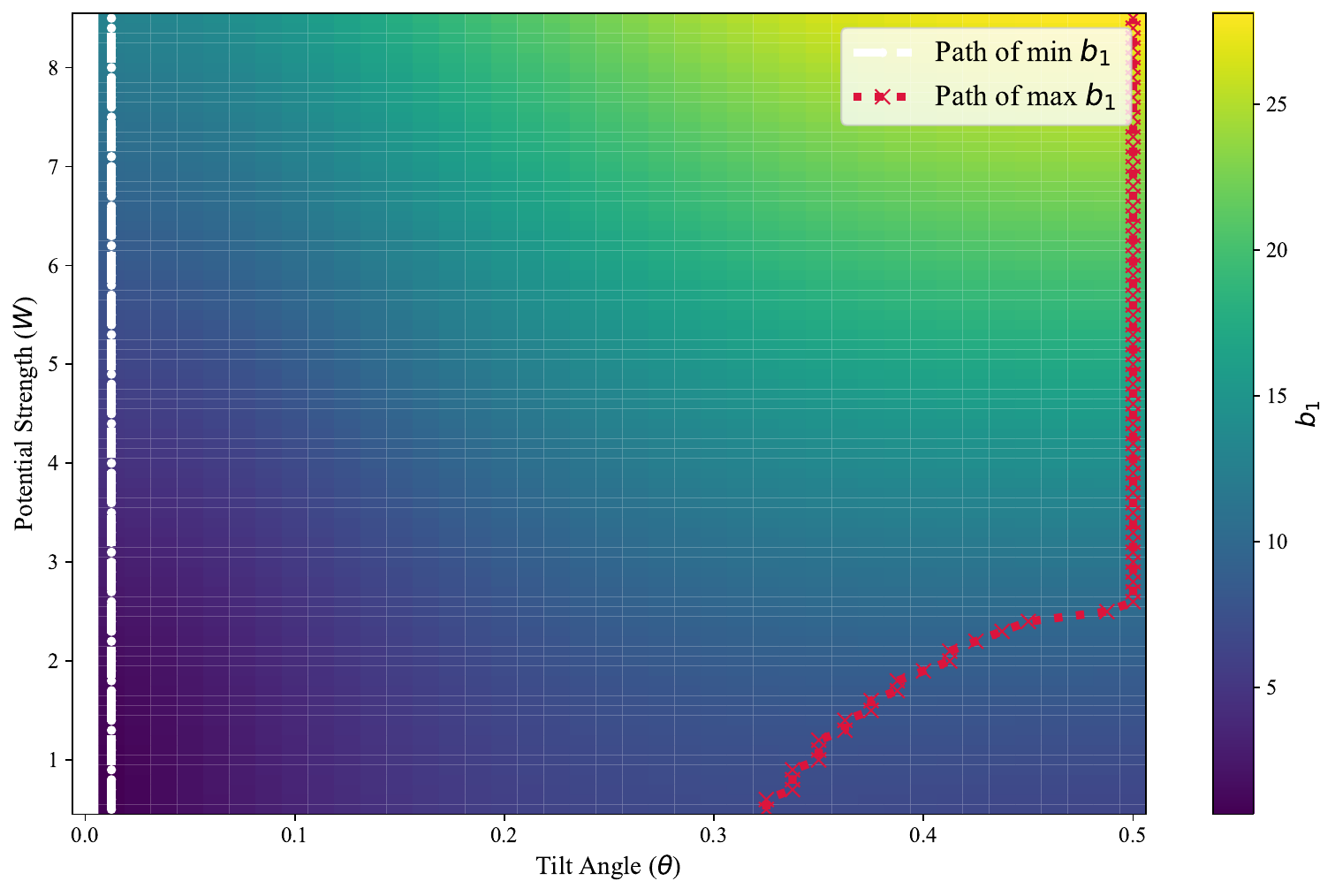}}\hfil
\subfloat[TNS]{\includegraphics[width=0.47\linewidth]{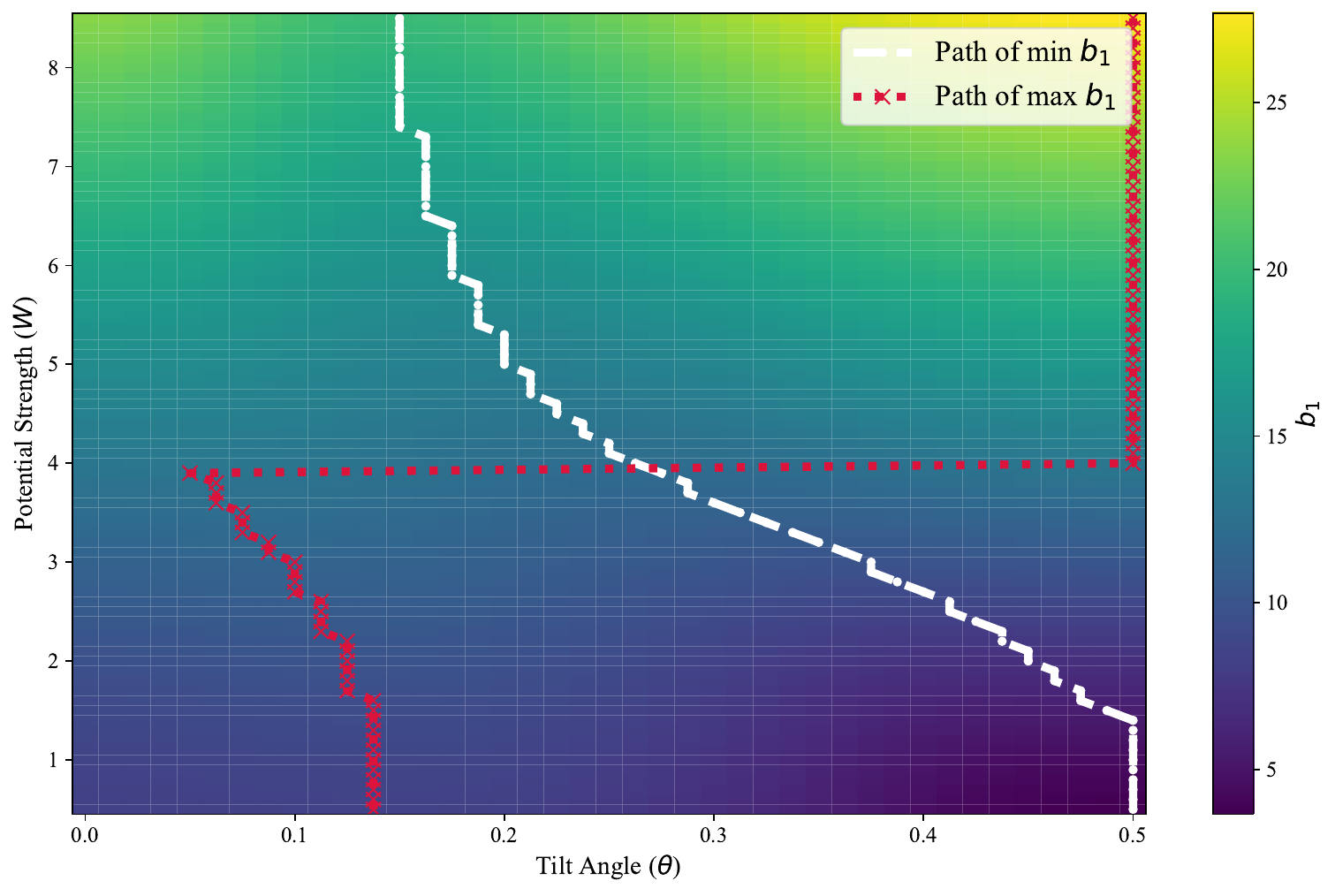}}\hfil 
\caption{Figures showing the first Lanczos coefficient value as a heatmap over the phase space of the potential strength and the tilt angle. The maximum (minimum) $b_1$ value for a given value of $W$ is included as the dotted (dashed) red (white) line. The sudden shift in maximum $b_1$ for the Néel state occurs when transitioning from $W=3.9$ to $W=4.0$.}
\label{fig:b1s}
\end{figure}

Figure \ref{fig:lanczosheatmap} provides insight into the \textit{rate} of spread of the asymmetry-quenched state under the dynamics of the Hamiltonian. For the ETH regime (below the red-dashed lines), the Néel state clearly has a rate of spread that persists well into larger $n$ values. The transition of the contours from skewing-right at $W<W_c$, to being almost vertical at $W>W_c$ shows how MBL effects cause a reduction in the rate of spread of the state. Conversely, the ferromagnetic state never experiences the `encouraged' spreading at $W<W_c$; the contour lines run almost vertically for all choices of $W$, providing an explanation for its dynamics and, in particular, the presence of the QME, being largely independent of the choice of $W$.
\\ \\
While it is true that increasing $W$ generally results in an increase in $b_1$, it is instructive to also consider how it is affected by the tilt angle. To that end, we include Fig. \ref{fig:b1s}.
It is clear that the first Lanczos coefficients always tend to increase with increasing potential strength. That said, these values are not increasing at the same rate for different tilt values. The tilted ferromagnetic states, where the QME is always observed, presents a generally increasing ordering in the $b_1$ values with increasing tilts. The tilted Néel states do not display this behavior; the majority ($0.14<\theta<0.5$) appear reverse-ordered by the tilt value at low values of $W$, but exhibit a sudden reordering when $W$ increases from 3.9 to 4.0. Note that the minimum value for $b_1$ for the tilted ferromagnetic state is always at the lowest tilt, but the tilted Néel state minimum transitions from the maximal tilt to almost align with the value of the lowest potential maximum location (the maximal $b_1$ starts at $\theta\approx0.14$ and the minimum $b_1$ ends at $\theta\approx0.15$ for the values of $W$ tested). The sudden shift in the location of the tilt value of the maximum $b_1$ value can be understood by recognising that, when the phase change takes place, the action of the tilt takes one closer to a (hypothetical) Gibbs distribution of the system (see App. \ref{app:thermomajorization}). The late-time steady state of the system in the MBL regime, however, is not well-approximated by the Gibbs state since the system does not thermalise in this regime. Hence, the $b_1$ values are largest for larger tilts in the MBL regime since the state will evolve to a localized steady state, for which the thermal Gibbs state is a poor approximation.
\\ \\
Given that these values are a first order approximation of the rate of spread of the initial state, they contain surprising predictive power. Including additional coefficients would naturally enhance the predictive capability, since the average rate of spread over longer times can be better approximated. Interestingly, the sudden shift in the maximum is present in the region $W\in[3.6,4.1]$, even when including an average over any number of the first 50 Lanczos coefficients (see App. \ref{app:avgLanczos}). The sudden reorganization of the maximum can be understood as an early indicator of the QME, as states with larger tilts (and therefore larger initial entanglement asymmetries) suddenly acquire comparatively greater impetus to spread than less tilted (lower EA) states. These results indicate that, at least for the Aubry-André model studied in this work, the early Lanczos coefficients have significant predictive power for the quantum Mpemba effect.

\subsection{What is the Quantum Mpemba Effect?\label{sec:KCQME}}
\begin{figure}[!htbp]
\centering
\subfloat[$W=1.0$\label{sprob1}]{\includegraphics[width=0.47\linewidth]{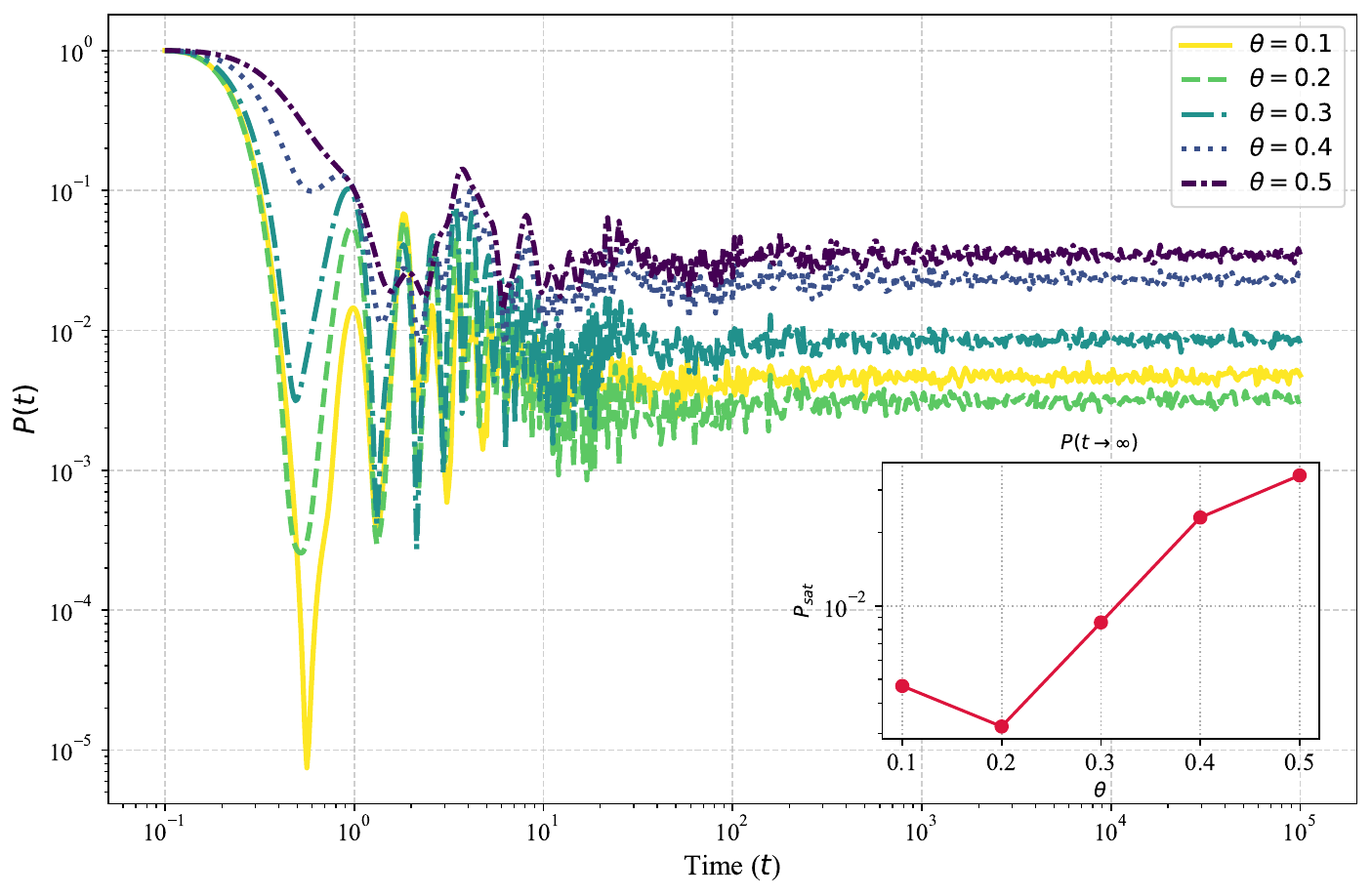}}\hfil
\subfloat[$W=2.0$ \label{fig:sprob2}]{\includegraphics[width=0.47\linewidth]{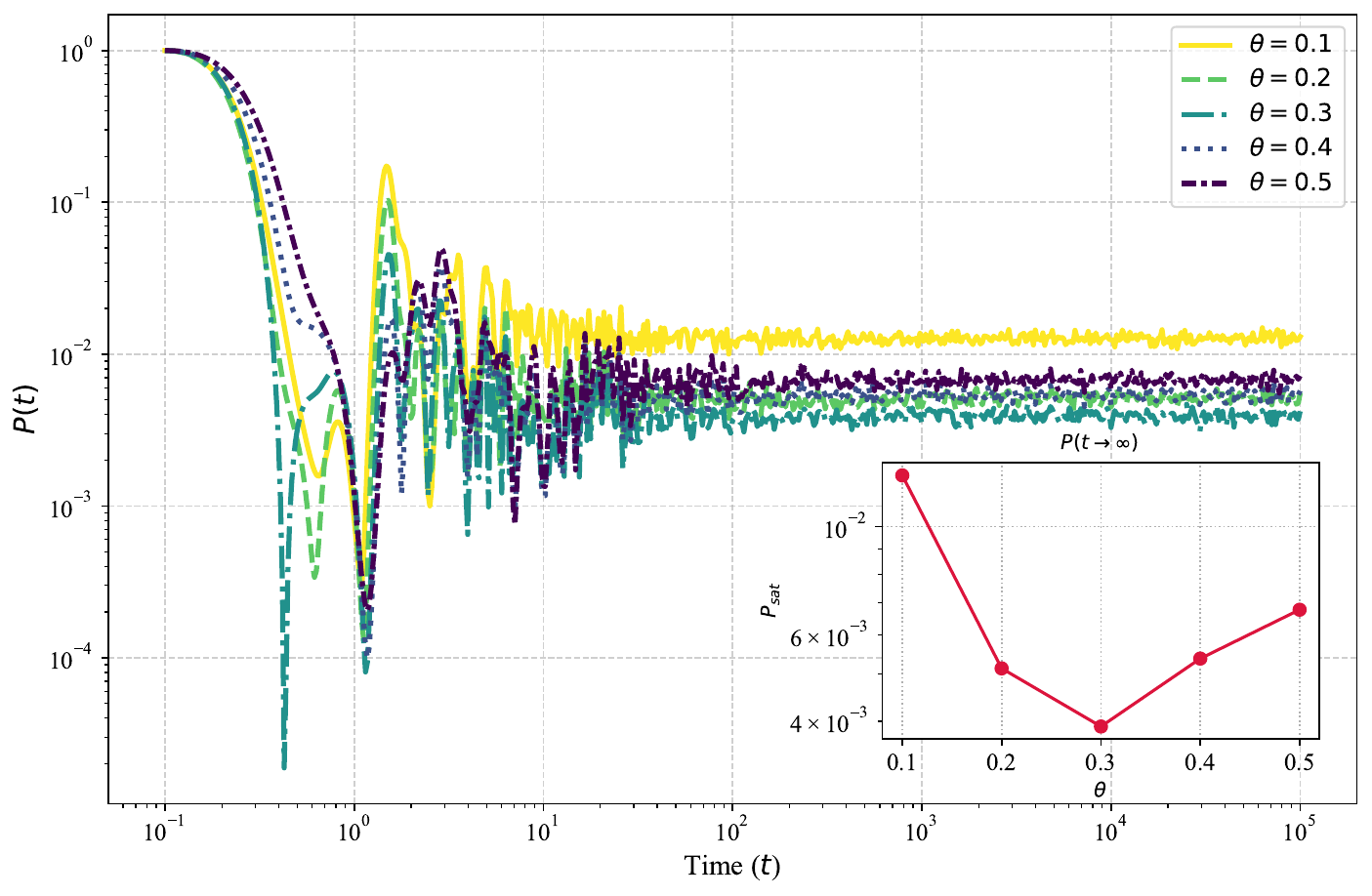}}\hfil 
\subfloat[$W=2.5$]{\includegraphics[width=0.47\linewidth]{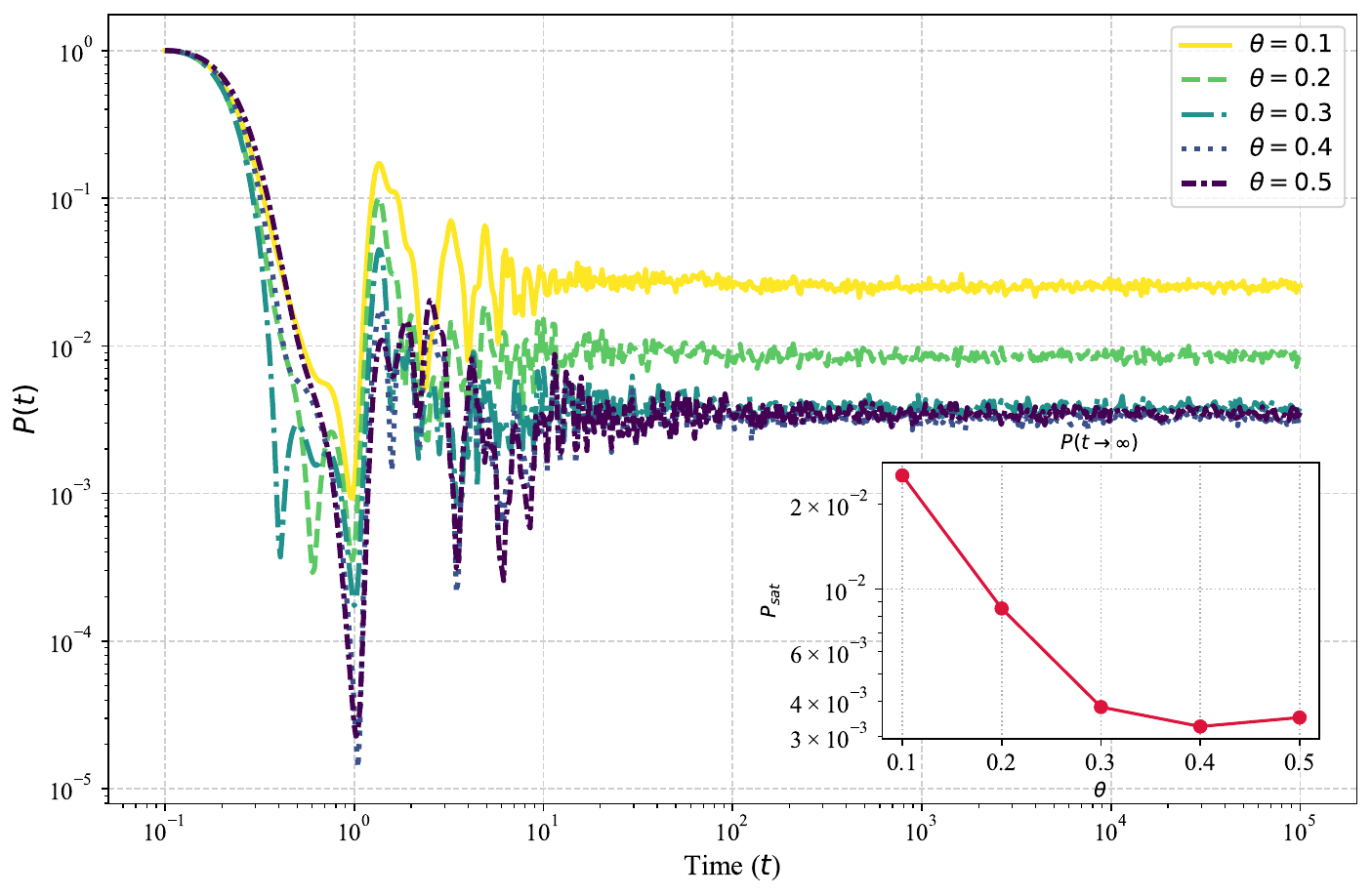}}\hfil
\subfloat[$W=3.0$]{\includegraphics[width=0.47\linewidth]{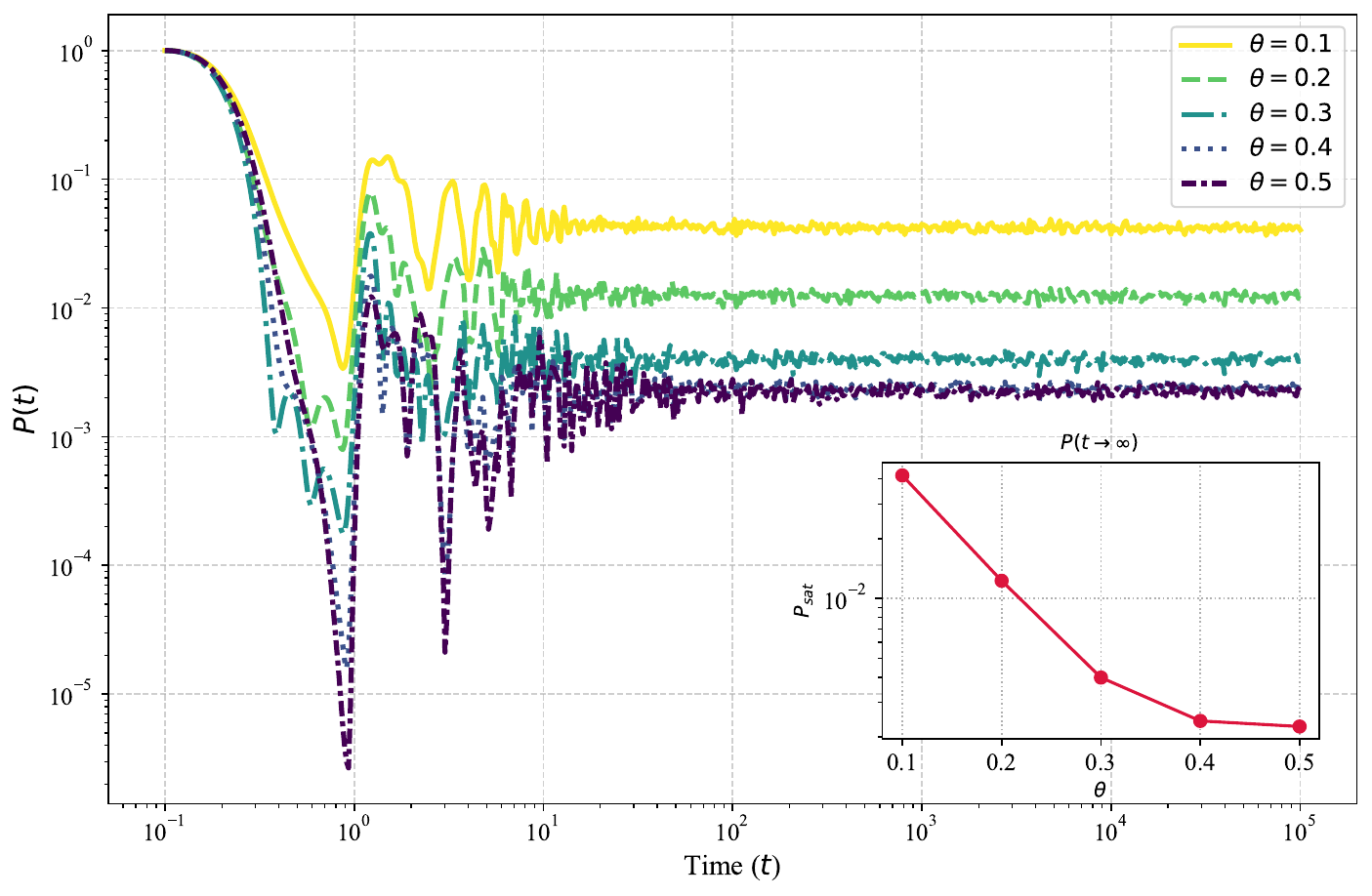}}\hfil 
\subfloat[$W=3.3$]{\includegraphics[width=0.47\linewidth]{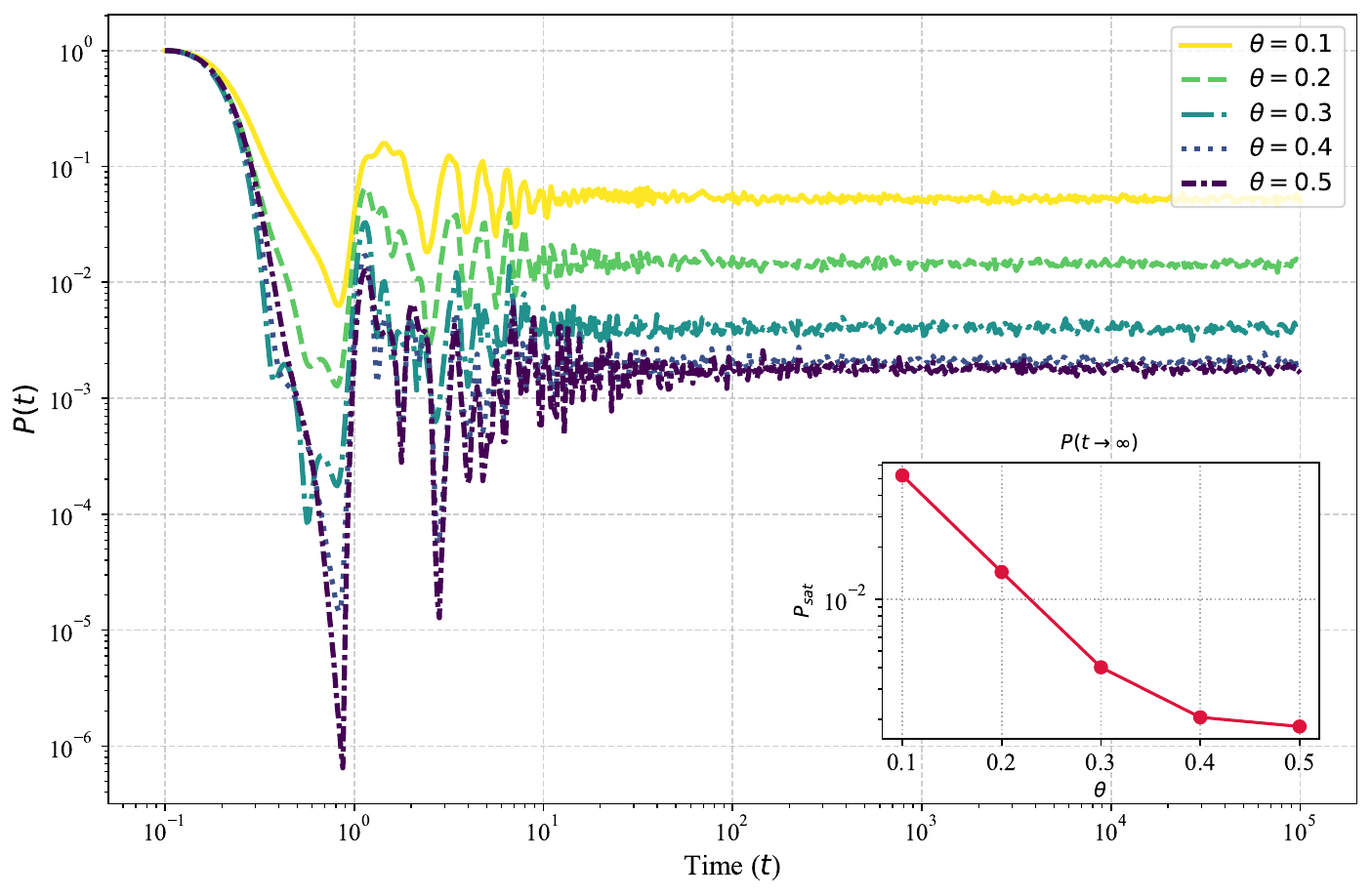}}\hfil
\subfloat[$W=5.0$]{\includegraphics[width=0.47\linewidth]{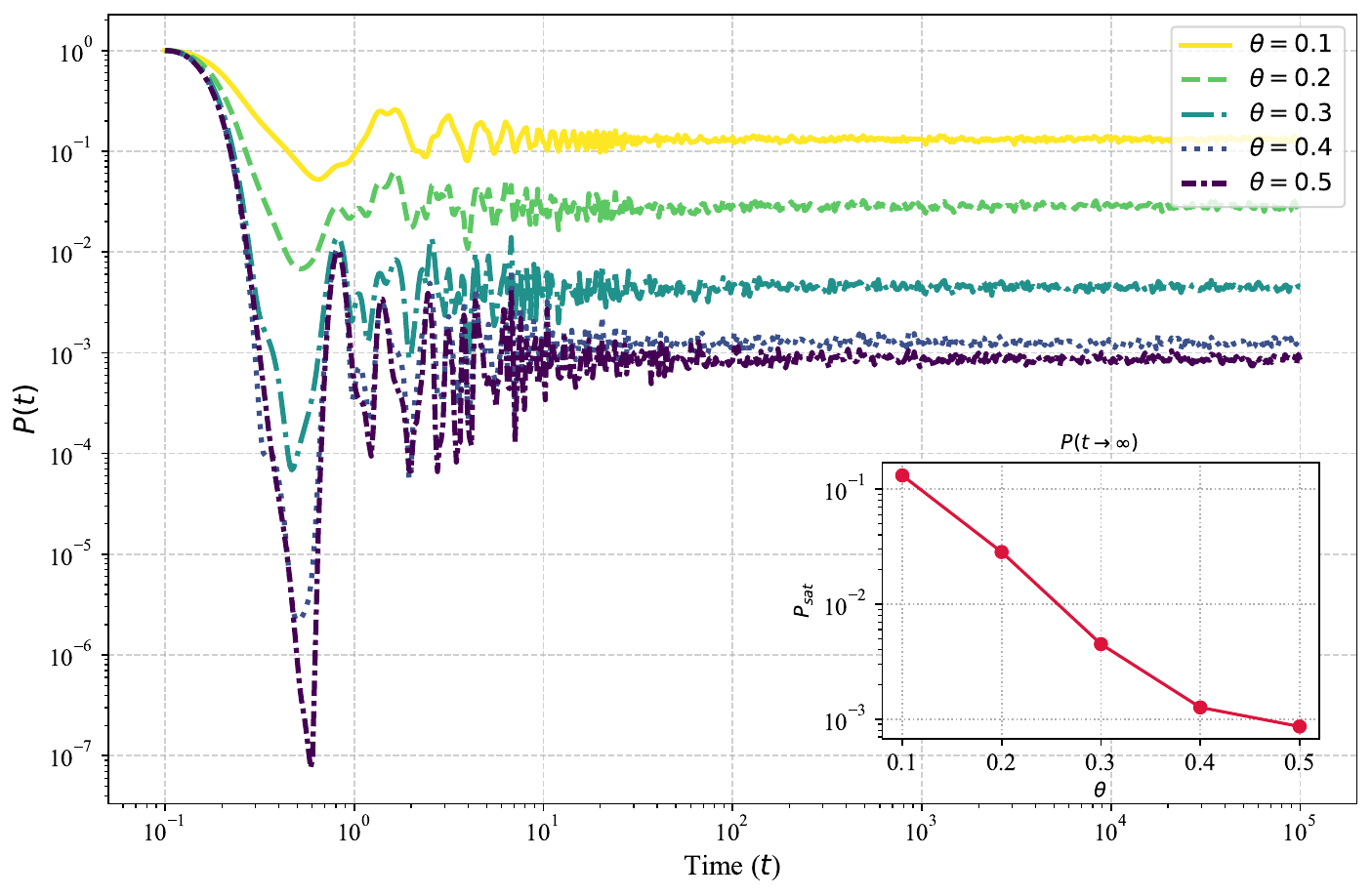}}\hfil 
\caption{Figures showing the survival probability for multiple tilts at different quasiperiodic potential strengths for the TNS states. The inset plot shows the saturation values of the survival probabilities as a function of the initial tilt.}
\label{fig:ISCmpemba}
\end{figure}

Throughout this work, we have taken entanglement asymmetry to be the measure that one uses to define the quantum Mpemba effect. As is well established, this is not the \textit{only} measure of asymmetry that one may use for the purposes of observing the QME\cite{ares2025qme}. Any good probe for the Mpemba effect must faithfully quantify how far a quantum system is from its steady state. In systems where the Hamiltonian possesses some symmetry, a measure for how far that system is from a symmetric configuration can serve as a proxy for this. Quantifying this is, however, ambiguous.  In general, one may consider any positive definite operator and use projection operators to isolate the part that is not commuting with the system Hamiltonian. On the space of operators, there are then a vast number of valid inner products.  In particular, the asymmetric complexity we have defined serves as a measure of how far the system is from a symmetric state. In this light, figure \ref{fig:structuralTNS} informs a challenging question: How can a system with a larger entanglement asymmetry ($W=1.0$, $\theta=0.5$) have a smaller structural complexity if both of these measures may be interpreted as quantifiers of how much symmetry is to be restored? The answer, we think, is found by considering the survival probabilities of the initial tilted states, shown in Fig. \ref{fig:ISCmpemba}.\\

\noindent
As shown in Fig. \ref{fig:EA}, the asymmetry operator monotonically increases the asymmetry of the system when quantified via entanglement asymmetry. However, in the strong ETH phase, its effect is more subtle. Beyond a certain value of the tilt angle, $\theta_{\text{peak}}$, the asymmetry operator also increases the overlap of the initial state with the thermal state to which it approximately converges (see insets of Figs. \ref{sprob1}, \ref{fig:sprob2}, and App. \ref{app:thermomajorization}). Thus, although the asymmetry operator is strictly monotonic with respect to entanglement asymmetry, it is not monotonic in terms of reducing thermal overlap. In this regime, increasing $\theta$ simultaneously induces larger asymmetry and enhances the state’s proximity to the eventual thermal configuration.\\

\noindent
There is an additional subtlety that needs to be kept in mind: despite being related by a unitary transformation, two states may not be relaxing to a comparable equilibrium.  In particular, the temperature of the equilibrium configuration may be different.  This can have a detrimental impact when trying to realise the calculated QME experimentally.  It may be necessary to restrict the discussion only to states that will reach (approximately) the same equilibrium state\footnote{One natural way this can be achieved is to consider an open quantum system where the bath is held at a fixed temperature. See also App. \ref{app:thermomajorization}.}.   These concerns can be summarised as follows:
\begin{enumerate}
\item Entanglement asymmetry may not always provide the most appropriate measure of distance from a symmetric configuration for the states under consideration; and
\item The choice of asymmetry operator is critical: it should not only increase asymmetry monotonically, but must also drive the initial state further away from the state to which it will eventually converge.
\end{enumerate}
\noindent
Although point (2) directly informs point (1), the broader caution here is that for any choice of asymmetry measure, asymmetry itself may not constitute the most faithful notion of distance from the steady state. Care must therefore be taken in selecting both the diagnostic and the operator when interpreting the presence or absence of Mpemba-like behavior.

% \section{Quantum Ruelle-Pollicott Resonances?}
% Hunch that these are actually what describe the QME, and that these can be used to predict movement towards equilibrium... see ``Theory of Irreversibility in Quantum Many-Body Systems" at https://arxiv.org/abs/2501.06183.

\section{Conclusion\label{sec:conclusion}}
\noindent
In this work, we set out to explore the utility of Krylov complexity as a probe of the Quantum Mpemba Effect. We focused on the Aubry–André model, which is particularly well-suited to this investigation; not only does it display the QME when quantified using entanglement asymmetry, but it also undergoes an ETH/MBL transition, providing a rich backdrop against which to test the sensitivity and universality of complexity-based diagnostics. Our first line of inquiry asked whether Krylov (spread) complexity is responsive to the presence of the QME already established in the Aubry–André model through entanglement asymmetry. We found that it is indeed sensitive, though it does not yield the kind of “clean” Mpemba crossings characteristic of entanglement asymmetry. Instead, the appearance of the QME in Krylov complexity is tied to the monotonicity of the saturation values under tilts. Indeed, when increasing the asymmetry parameter leads to consistently larger saturation values of Krylov complexity, the QME manifests robustly. However, when monotonicity breaks down, more subtle forms of the QME still persist, as our results demonstrated. This indicates that Krylov complexity encodes genuine information about Mpemba-like behavior, but requires a more refined analysis than entanglement asymmetry.\\

\noindent
Motivated by this observation, we next introduced a formal decomposition of Krylov complexity into symmetric and asymmetric components, achieved through projective operations onto symmetry sectors. This decomposition isolates the contributions from block-diagonal versus off-block-diagonal coherences and provides a sharper diagnostic of symmetry restoration. The resulting symmetric and asymmetric complexities display unconventional features -- non-zero, and in some cases negative, initial values at $t=0$. Still, they form a valid decomposition of Krylov complexity and yield clear interpretive power: the asymmetric complexity $|C_{K,A}|$ quantifies the distance from symmetry, and its monotonic ordering under tilt resolves cases where the total complexity alone appears ambiguous. Consequently, this decomposition reveals a unique advantage of Krylov complexity over entanglement asymmetry in that, unlike the latter, complexity does not require arbitrary subsystem choices (size and location), and avoids the information loss inherent in reduced density matrices.\\

\noindent
Finally, we asked whether the initial values of the symmetric and asymmetric complexities -- which we referred to as the {\it structural complexity} -- carry predictive information for the quantum Mpemba effect. Here the answer is unambiguously yes. Structural complexity provides early signals of Mpemba-like behavior and even highlights the limits of entanglement asymmetry. In the ETH regime of the model, we observed that certain choices of asymmetry operator induce both larger asymmetries and larger overlaps with the thermal state, leading to non-monotonic responses that entanglement asymmetry alone would miss. Krylov complexity, in contrast, is sensitive to this subtlety at t=0, underscoring its potential as a predictive tool.\\

\noindent
This study opens several concrete and feasible directions for further work. First, it would be valuable to test the universality of the structural complexity diagnostic by applying it to other models that exhibit anomalous relaxation, such as the XXZ spin chain or random circuit models, to determine whether the predictive signatures we observed persist. Second, extending the Krylov complexity decomposition framework to non-Abelian or higher-form symmetries could clarify how different symmetry structures condition the quantum Mpemba effect. Third, investigating the role of complexity in open or non-Hermitian quantum systems, particularly near exceptional points, offers a natural extension, since such systems often display enhanced sensitivity to initial conditions. Fourth, having established Krylov (a)symmetric complexity as a faithful probe of the quantum Mpemba effect, it would be useful to investigate the effect when the quenched initial states all converge to the same final state. This is perhaps most easily realised by studying a system coupled to a bath at a fixed temperature so that all initial configurations evolve towards the same final state. This would strengthen the analogy with the classical Mpemba effect. Finally, experimental realizations in quasiperiodic cold-atom setups, trapped ions, or superconducting circuits could provide a direct test of whether structural complexity can serve as a measurable predictor of Mpemba-like dynamics. Together, these directions highlight the promise of Krylov complexity not only as a theoretical construct but also as a practical diagnostic of anomalous relaxation in quantum matter.

\section*{Acknowledgements}
JM and HJRVZ
are supported in part by the “Quantum Technologies for Sustainable Development” grant
from the National Institute for Theoretical and Computational Sciences of South Africa
(NITHECS). CB is supported by the Oppenheimer Memorial Trust Research Fellowship
and the Harry Crossley Research Fellowship. JM and CB would like to acknowledge support from the ICTP, Trieste through the Associates Programme and from the Simons Foundation through grant number 284558FY19. Computations were performed using facilities provided by the University of Cape Town’s ICTS High Performance Computing team: hpc.uct.ac.za  -- \url{https://doi.org/10.5281/zenodo.10021612}.

\bibliographystyle{JHEP}
\bibliography{biblio}

\clearpage
\appendix
\section{Identifying the Critical Potential}
\label{app:criticalW}
As a verification of the basic methodology used to produce the results herein, we reproduce the result highlighted in \cite{liu2024} for the critical value of the quasiperiodic potential strength. To do so, we make use of a finite-size scaling analysis.
\subsection{Finite-Size Scaling Analysis}
\begin{figure}[!htb]
    \centering
    \includegraphics[width=0.9\linewidth]{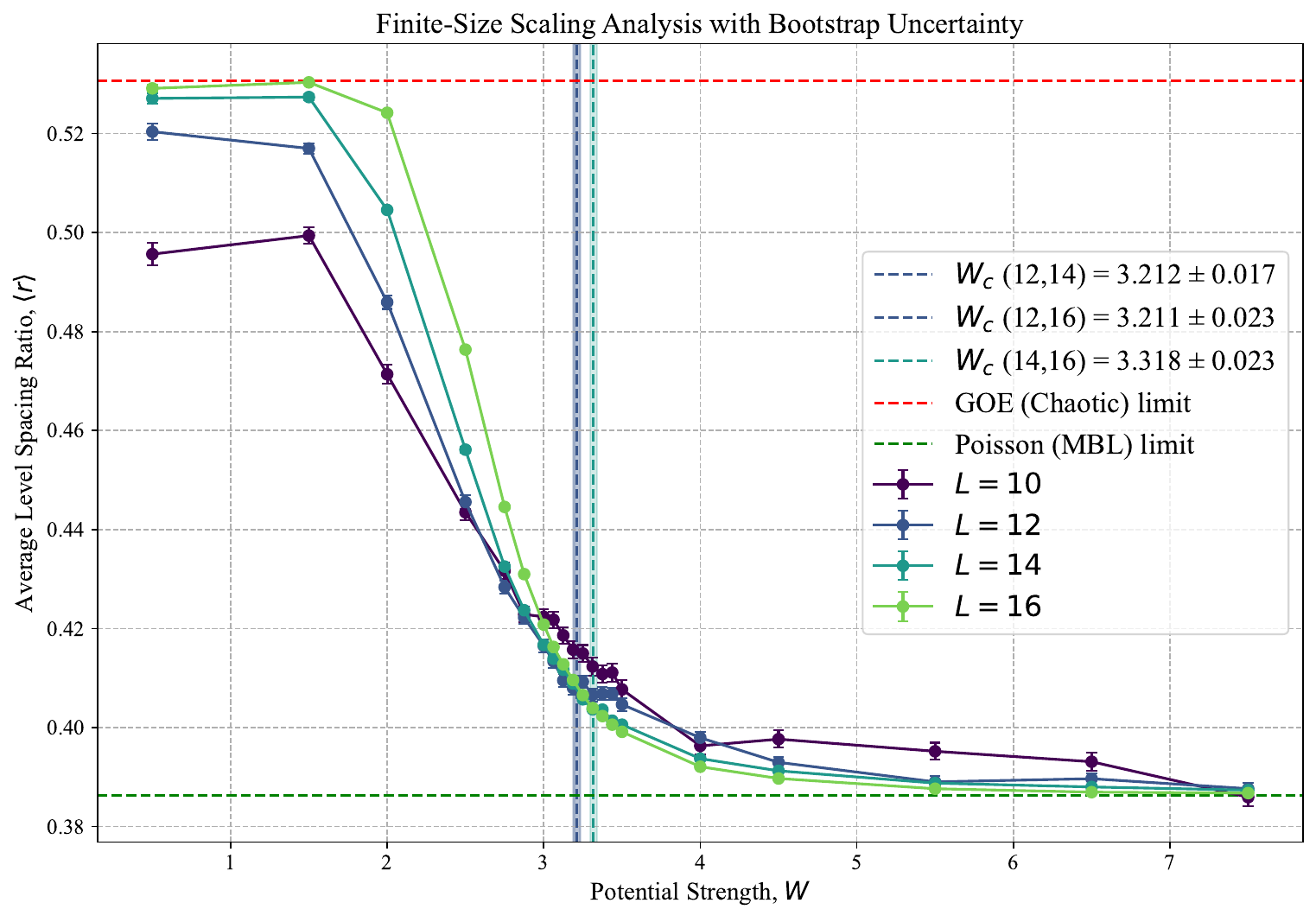}
    \caption{Finite-size scaling analysis to detect the ETH/MBL transition via the critical potential $W_c$. 240 realisations were used, and a higher density of points were collected near the transition region. The estimate for the transition is obtained by approximating the crossing point between the $N=14$ and $N=16$ data. As the system size grows, there is some drift in the crossing point --  this is an expected finite-sized effect, and provides a reason fo the lack of agreement between the two estimates. The error on the crossing is calculated using a bootstrap analysis.}
    \label{fig:fsa}
\end{figure}

\begin{figure}[!htb]
    \centering
    \includegraphics[width=0.95\linewidth]{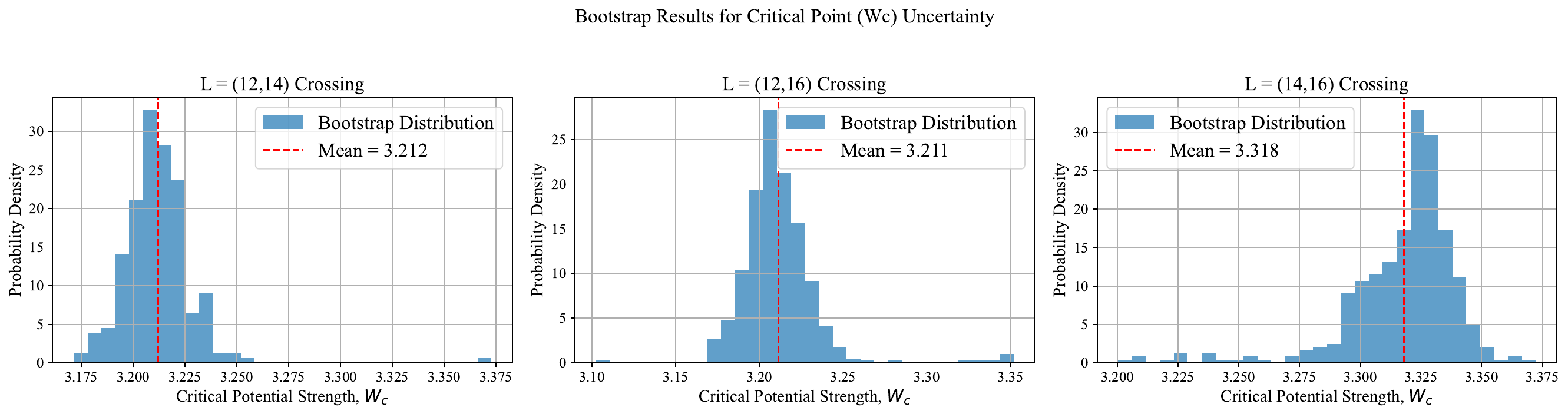}
    \caption{Bootstrap analysis of the crossing points for different crossing pairs of $N$. A minimum value filter was used of (left-to-right) 3.1, 3.1, and 3.2. There were (left-to-right) 232, 498, and 424 valid samples (not disqualified by the minimum filter) for these analyses, respectively.}
    \label{fig:bootstrap}
\end{figure}

At the point of the ETH/MBL transition, there is a scale-invariant point $W_c$, at which we would expect all lines to cross. In these simulations, we restrict ourselves to the 240 realisations that are averaged over, and we analyse only the largest symmetry sector (the $\langle S_Z\rangle=0$ sector), restricting the analysis to only the central 80\% of terms in this sector. For a system of size $N=2k$, there are $(2k)!/(k!k!)$ states that belong to this sector ($N=10\leftrightarrow252, N=12\leftrightarrow924, N=14\leftrightarrow3432, N=16\leftrightarrow12870$). The results of the finite-size scaling analysis are shown in Fig. \ref{fig:fsa}, and their corresponding bootstrap analysis is shown in Fig. \ref{fig:bootstrap}. It should be noted that this value is in agreement with the value calculated in \cite{liu2024}.
The bootstrap method is a simple resampling procedure that we use to inform the uncertainties of the finite-size scaling analysis. The bootstrap method assumes that the 240 randomly sampled $\phi$ values are representative of the true distribution of the $\phi$ values. As was mentioned in the text, random stratified sampling was used to preserve randomness and enhance uniformity; the latter property often requiring unmanageably large sample sizes to converge to -- especially considering that each realisation will be simulated as an $N=12$ size system.
Bootstrapping samples the 240 realisations \textit{with} replacement. From this data, an average level spacing ratio is computed for the $N=12,14,16$ data. The estimated crossing of a pair of curves is then computed and stored. This process is repeated $N_{bootstrap}$ times and, when complete, the population of crossing values is representative of the true crossing value, provided that the original data is representative. In this implementation, the bootstrap sampling procedure was attempted 500 times. These samples were used to generate a population of crossing values between pairs $(N_1,N_2)$ of the system sizes $N_1$ and $N_2$.  Additionally, we make use of a minimum filter of $3.1<W_c$ for the (12,14) and (12,16) pairs, and of $3.2<W_c$ for the (14,16) pair.

\clearpage
\section{Late Time Fluctuations of Asymmetric Complexity}
\label{app:latetimefluctuations}
\begin{figure}[!htbp]
\centering
\subfloat[TFS $W=2.0$]{\includegraphics[width=7cm]{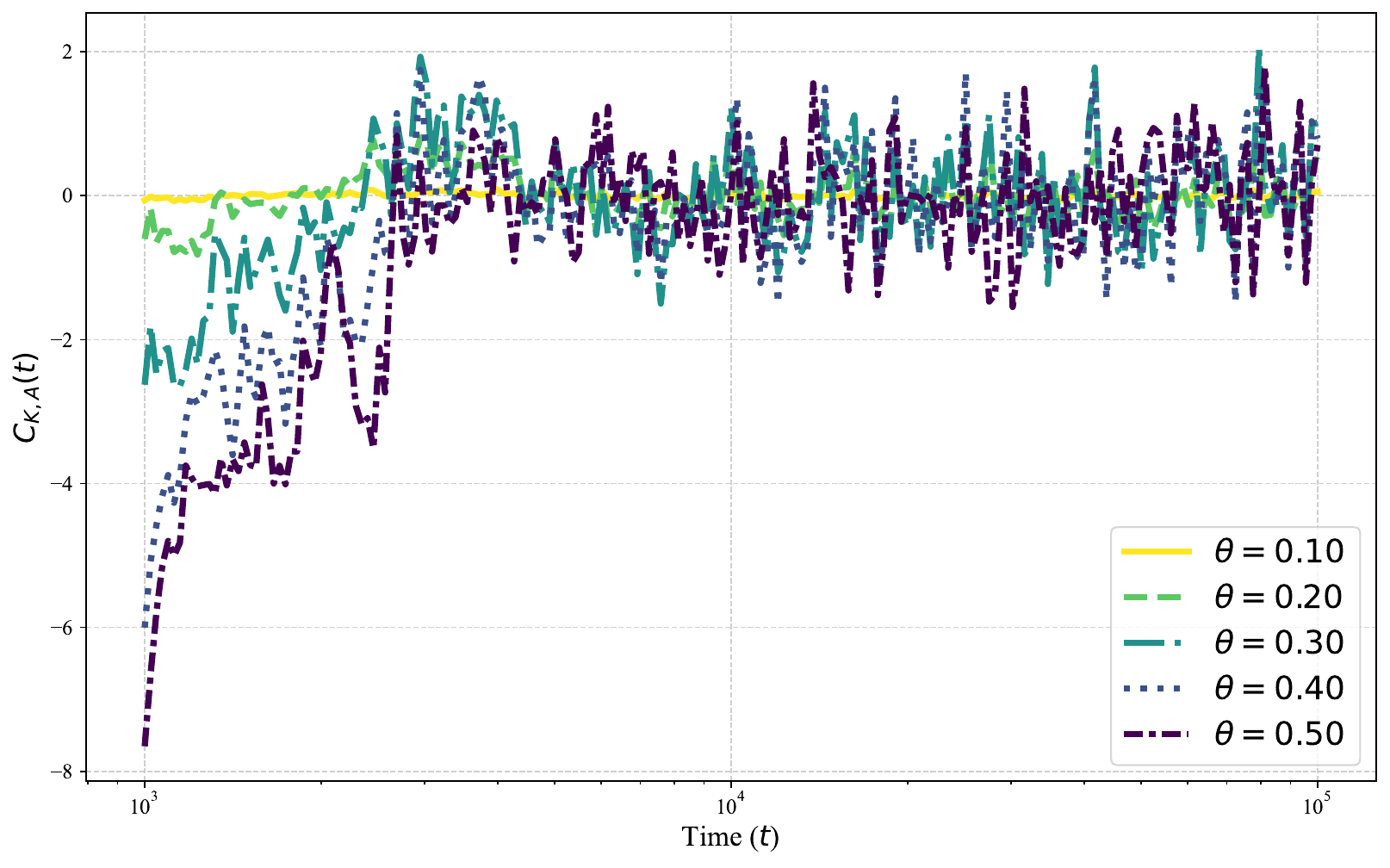}}\hfil
\subfloat[TNS $W=2.0$]{\includegraphics[width=7cm]{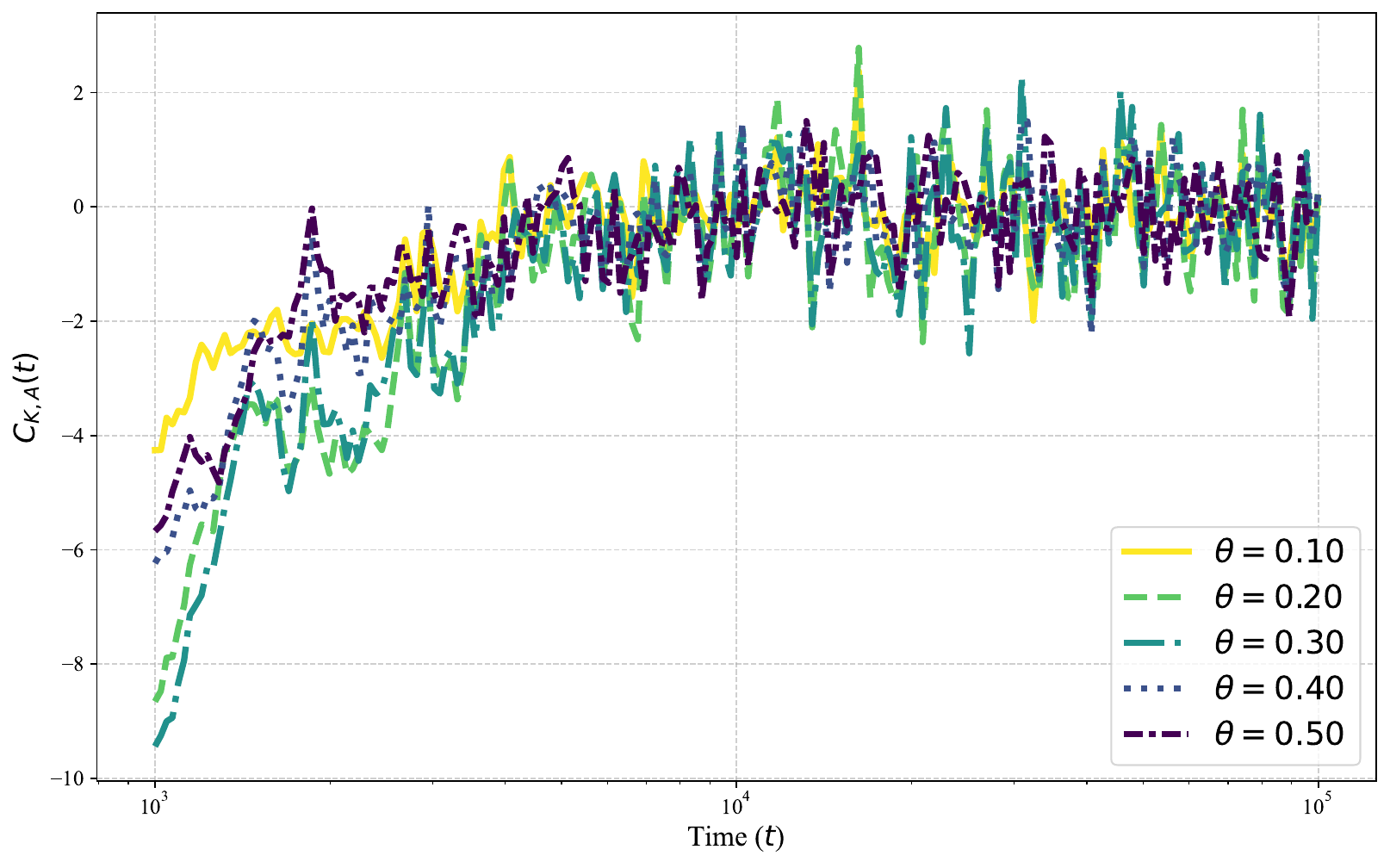}}\hfil 

\subfloat[TFS $W=2.5$]{\includegraphics[width=7cm]{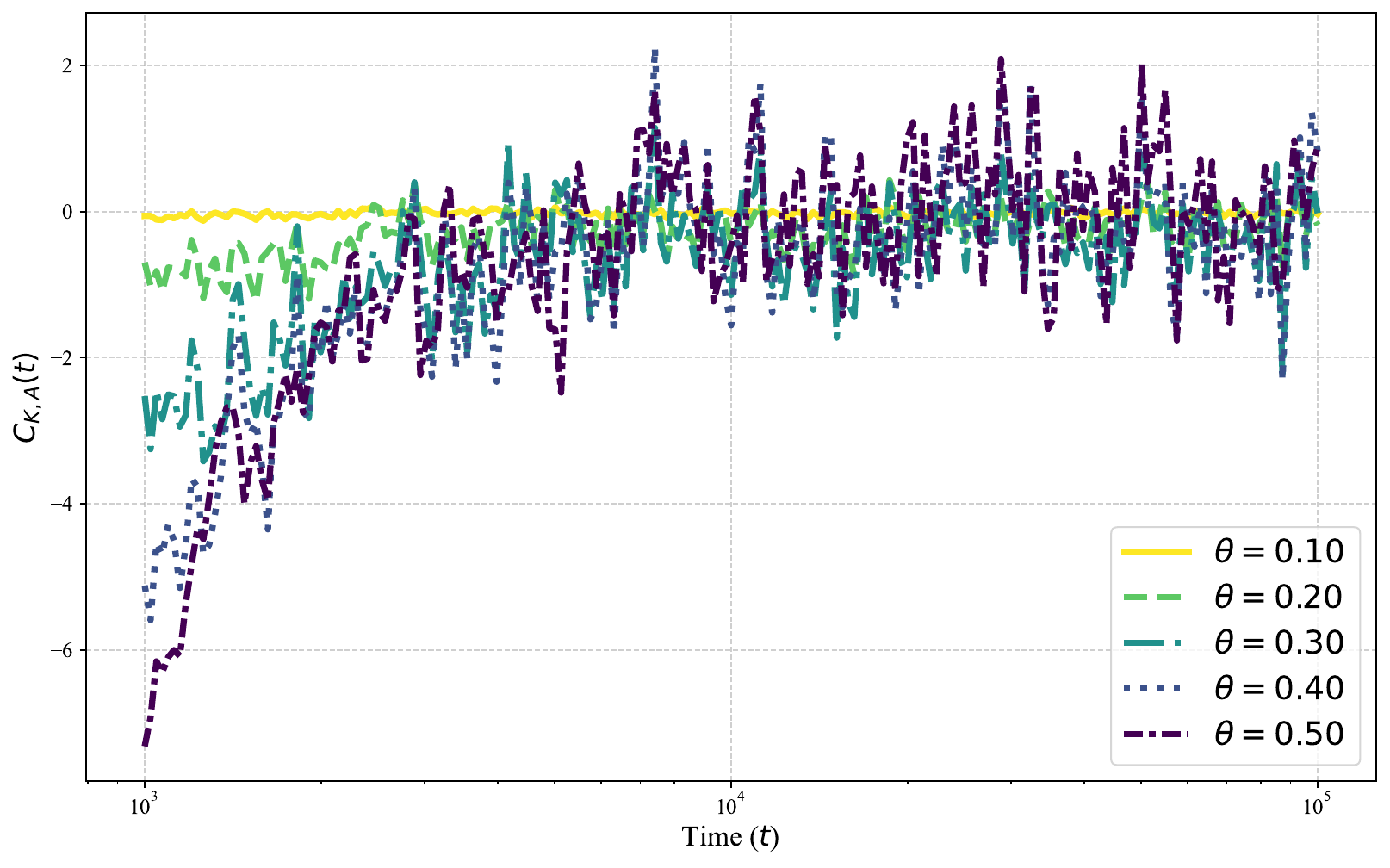}}\hfil
\subfloat[TNS $W=2.5$]{\includegraphics[width=7cm]{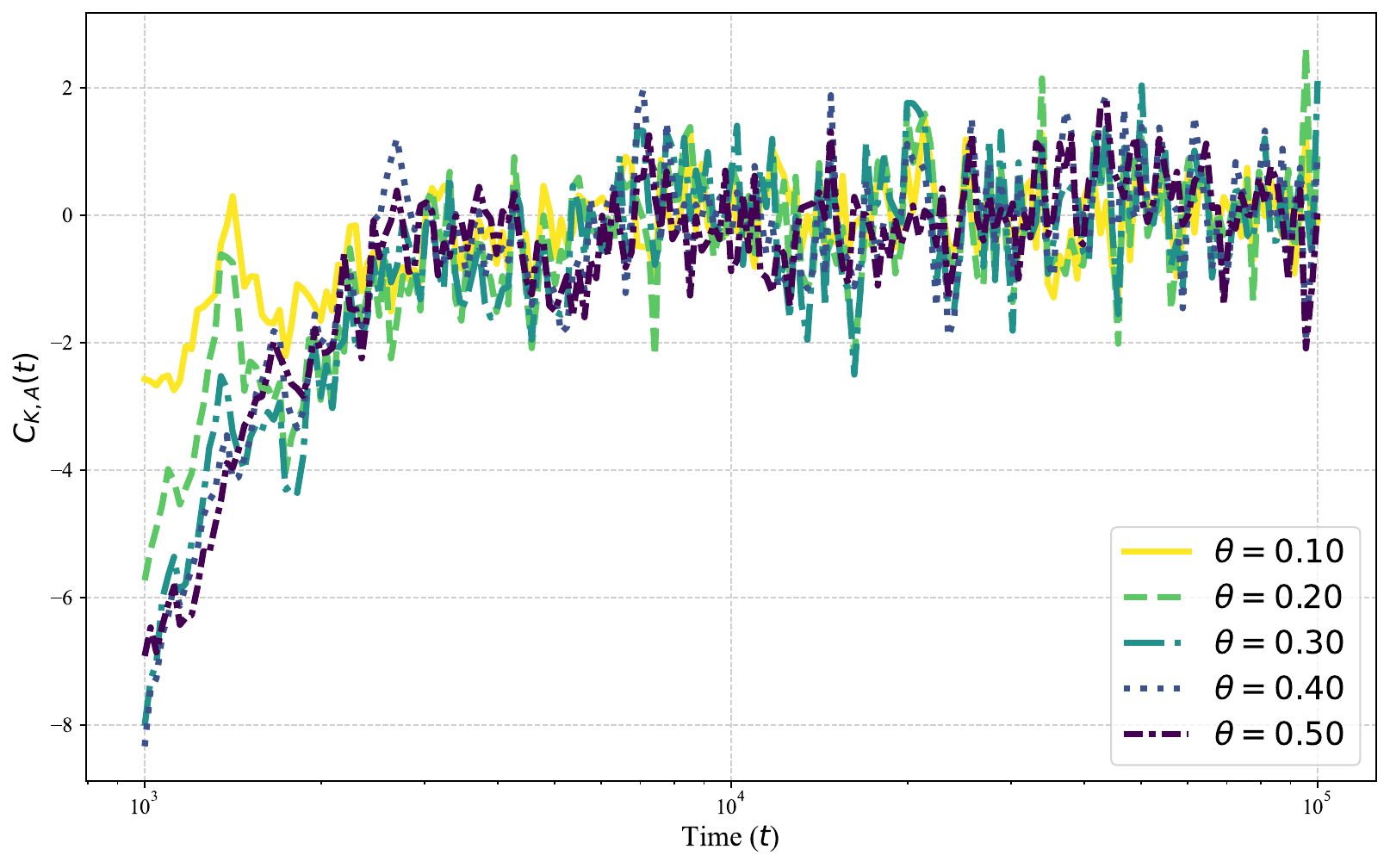}}\hfil 

\subfloat[TFS $W=5.0$]{\includegraphics[width=7cm]{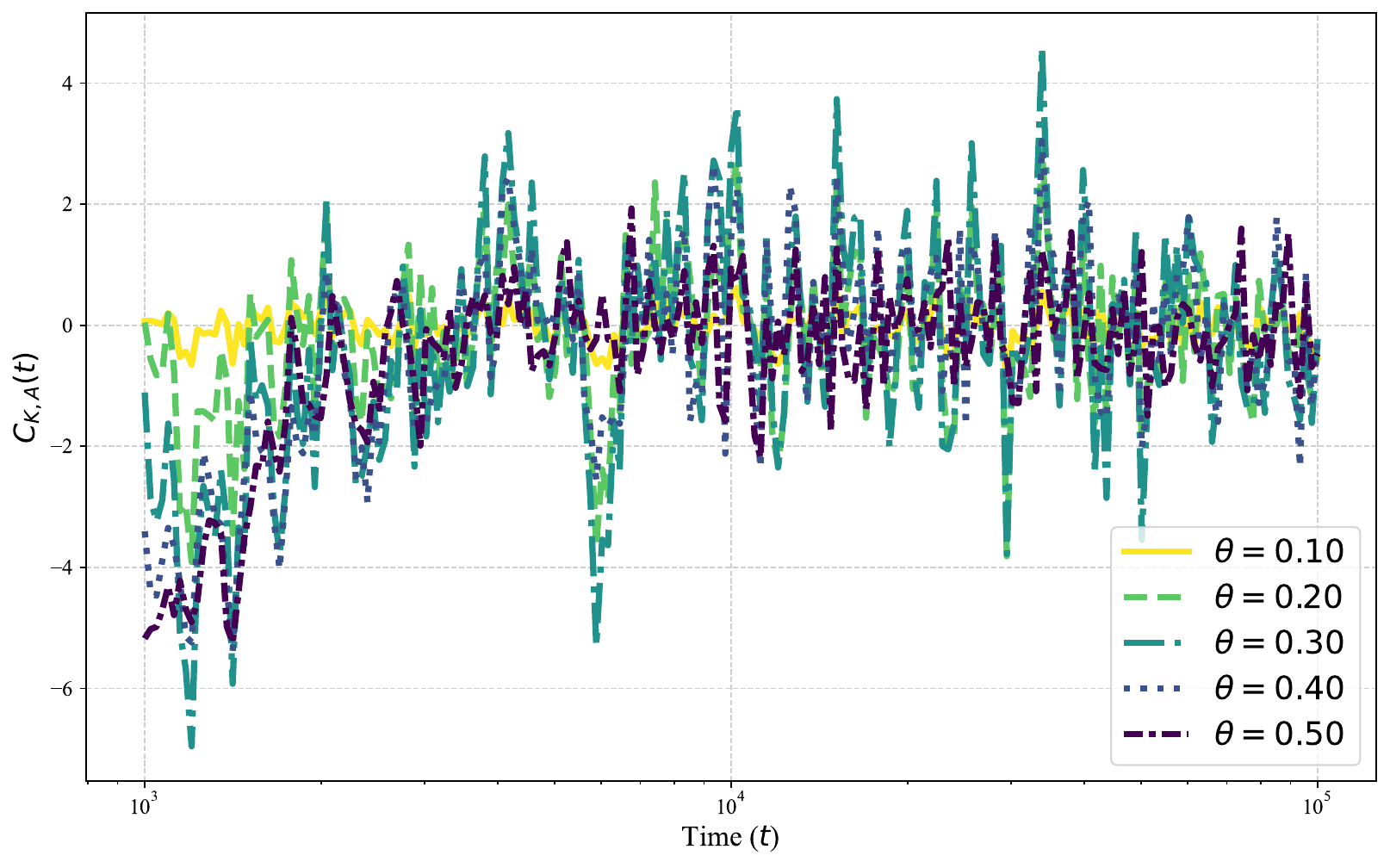}}\hfil
\subfloat[TNS $W=5.0$]{\includegraphics[width=7cm]{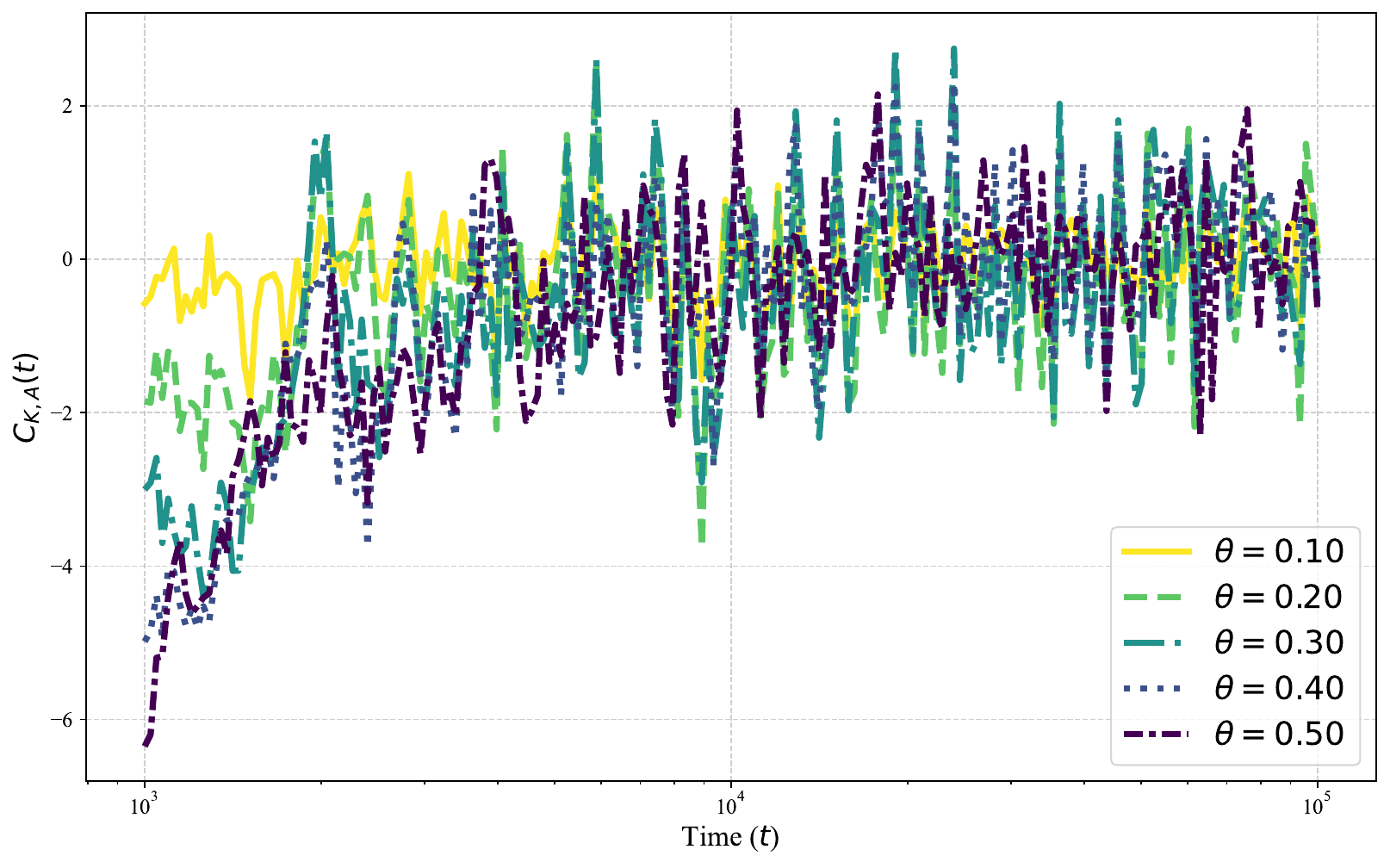}}\hfil 
\caption{Figures showing the late time fluctuations of the asymmetric complexity for the TNS and TFS states, for an $N=12$ size system, averaged over 240 realisations of $\phi$.}
\label{fig:latetimefluctuations}
\end{figure}

As discussed in the main text, the projected asymmetric complexity starts at negative values and evolves to being centred around 0, with both positive and negative fluctuations at late times. We isolate the late time behavior of Fig. \ref{fig:mixingKC} in Fig. \ref{fig:latetimefluctuations} to visually confirm the statement in a  more obvious way, since the fluctuation magnitudes are typically small compared to the initial values for the projected asymmetric complexities.

\clearpage
\section{Other ETH/MBL Observables}
In addition to the data presented in the main text, we also collected data for the imbalance, entanglement entropy, and survival probabilities. These observables are often used in studying the ETH/MBL transition, and we include this data here for completeness.

\subsection{Imbalance}
\begin{figure}[!htbp]

\centering
\subfloat[$\theta=0.1$]{\includegraphics[width=7cm]{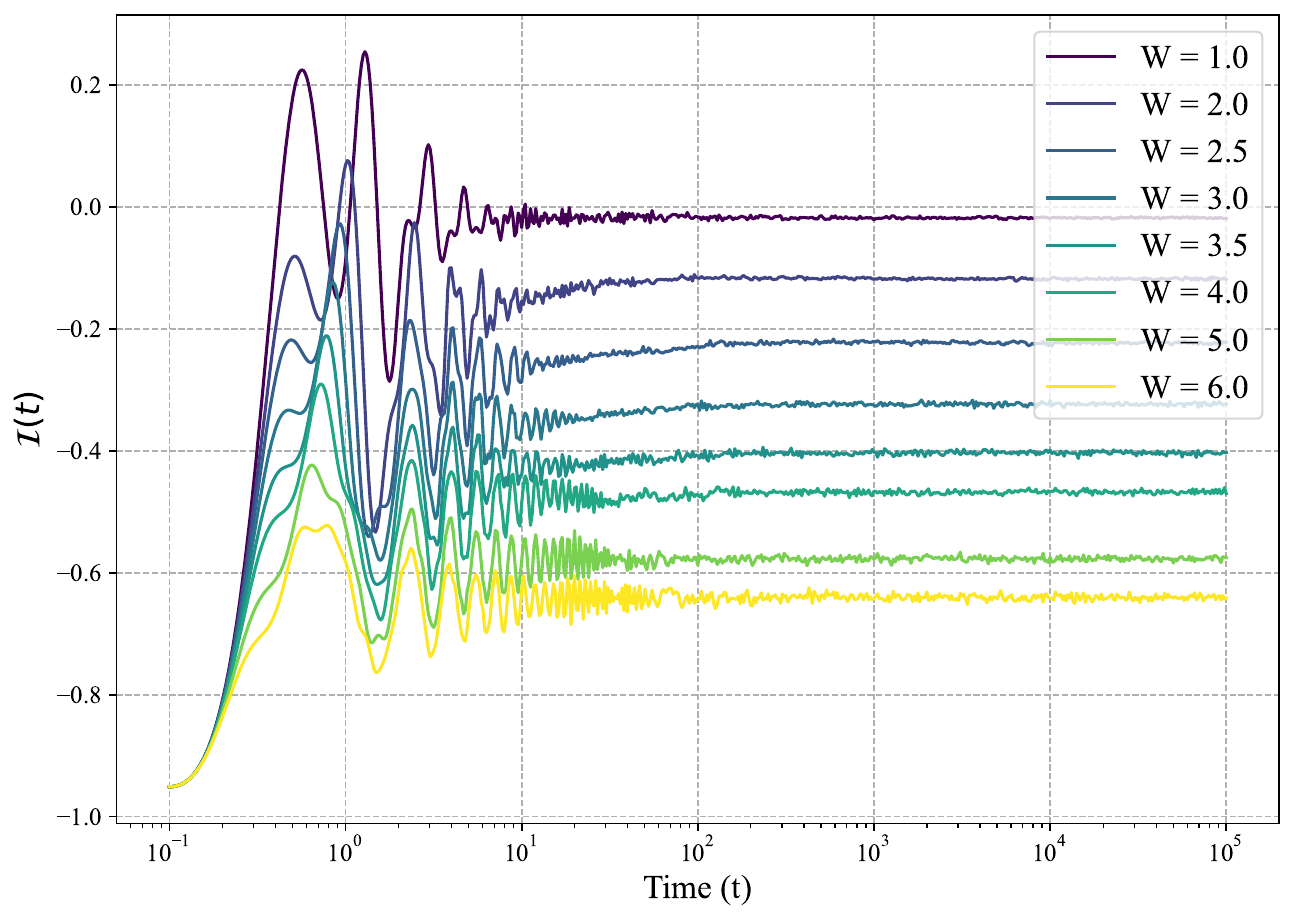}}\hfil
\subfloat[$\theta=0.2$]{\includegraphics[width=7cm]{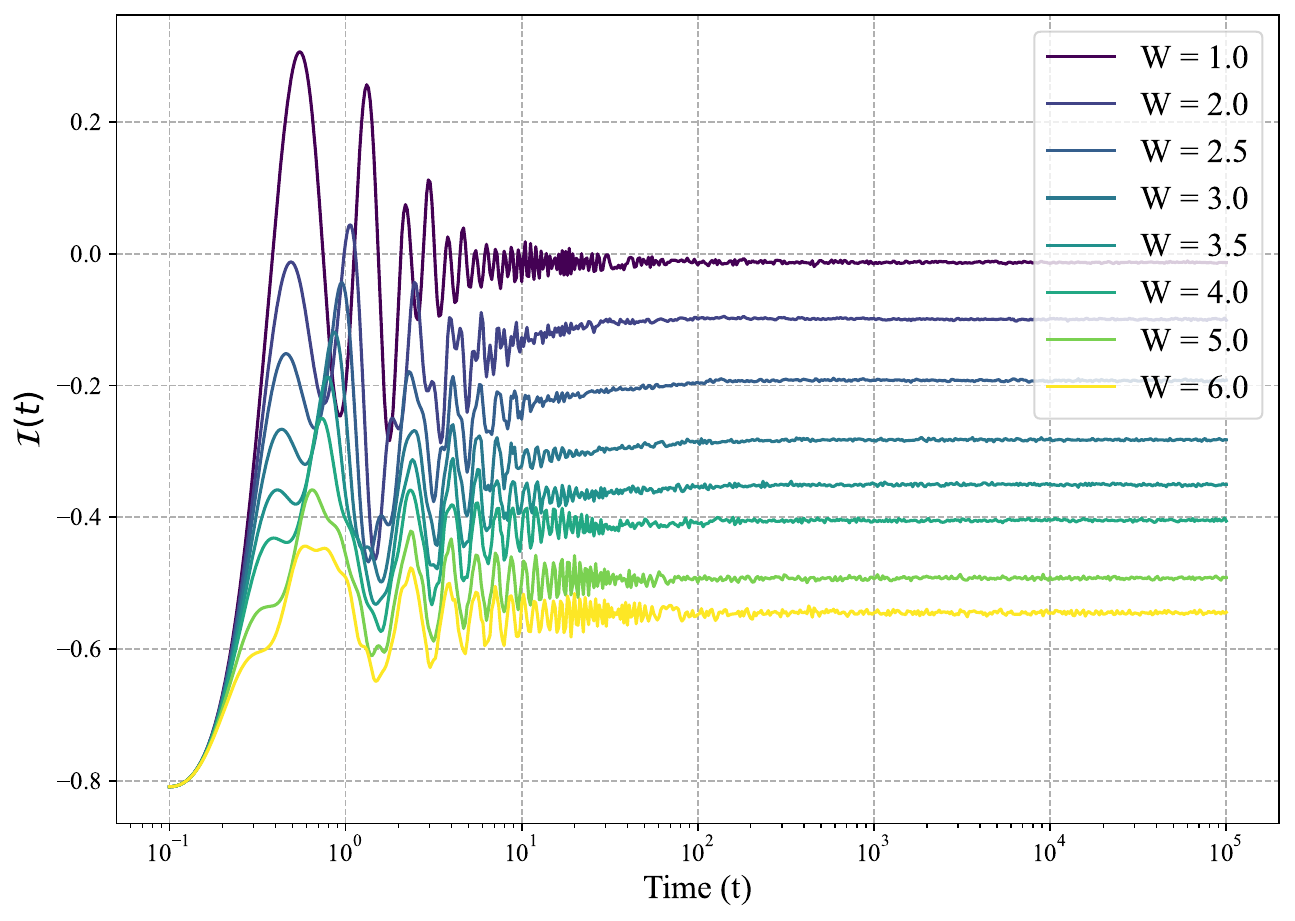}}\hfil 

\subfloat[$\theta=0.3$]{\includegraphics[width=7cm]{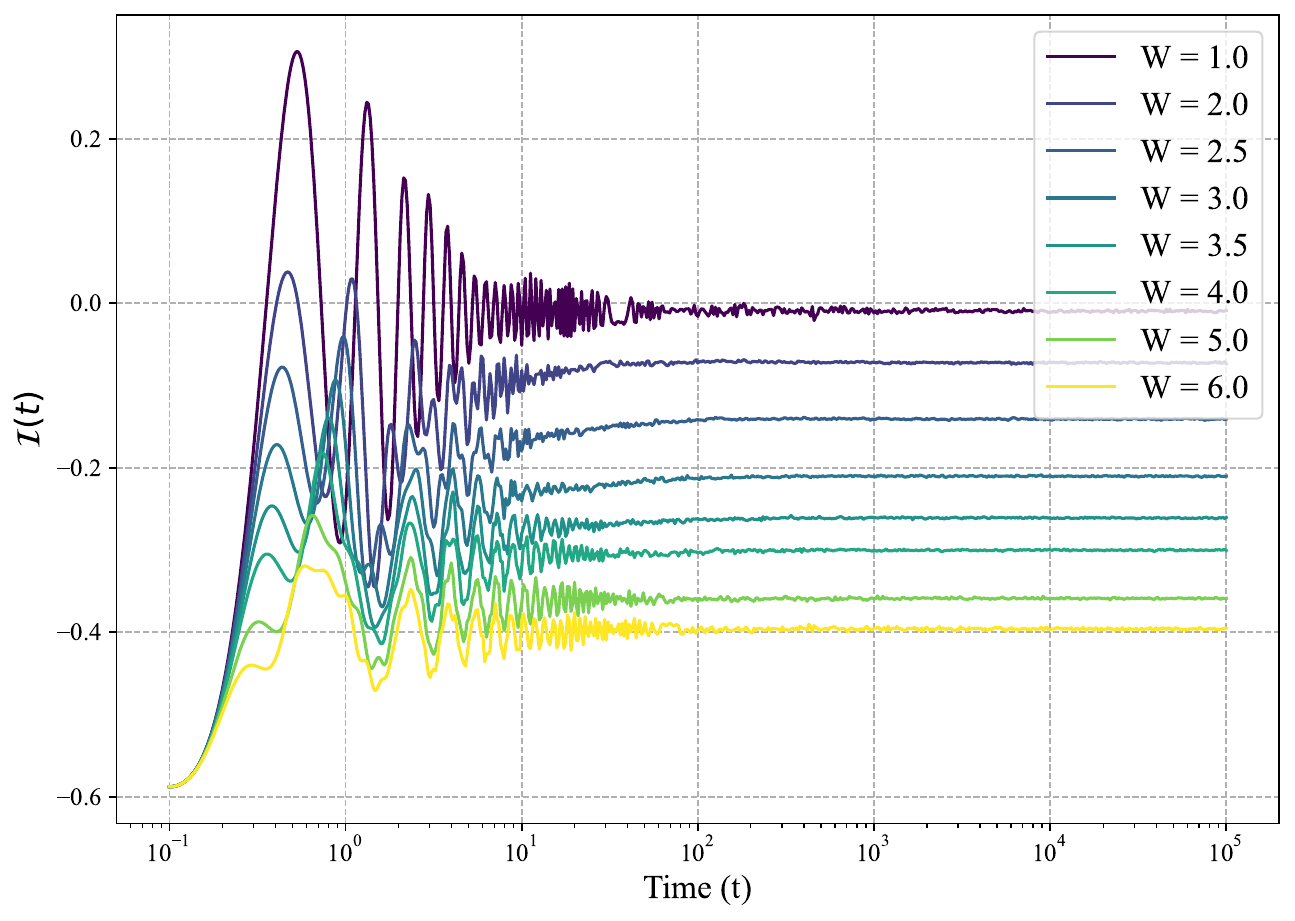}}\hfil
\subfloat[$\theta=0.4$]{\includegraphics[width=7cm]{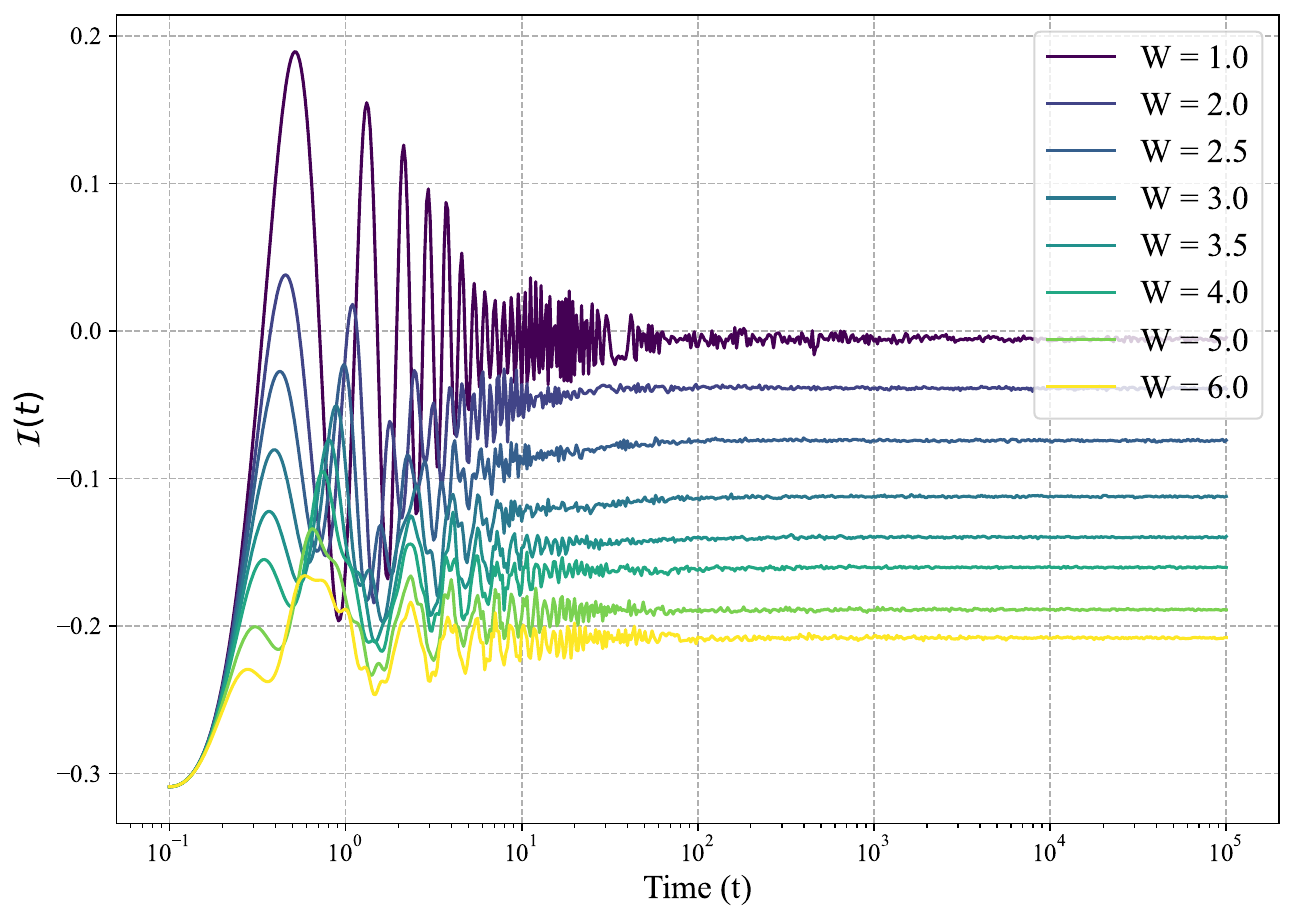}}\hfil 

\subfloat[$\theta=0.5$]{\includegraphics[width=7cm]{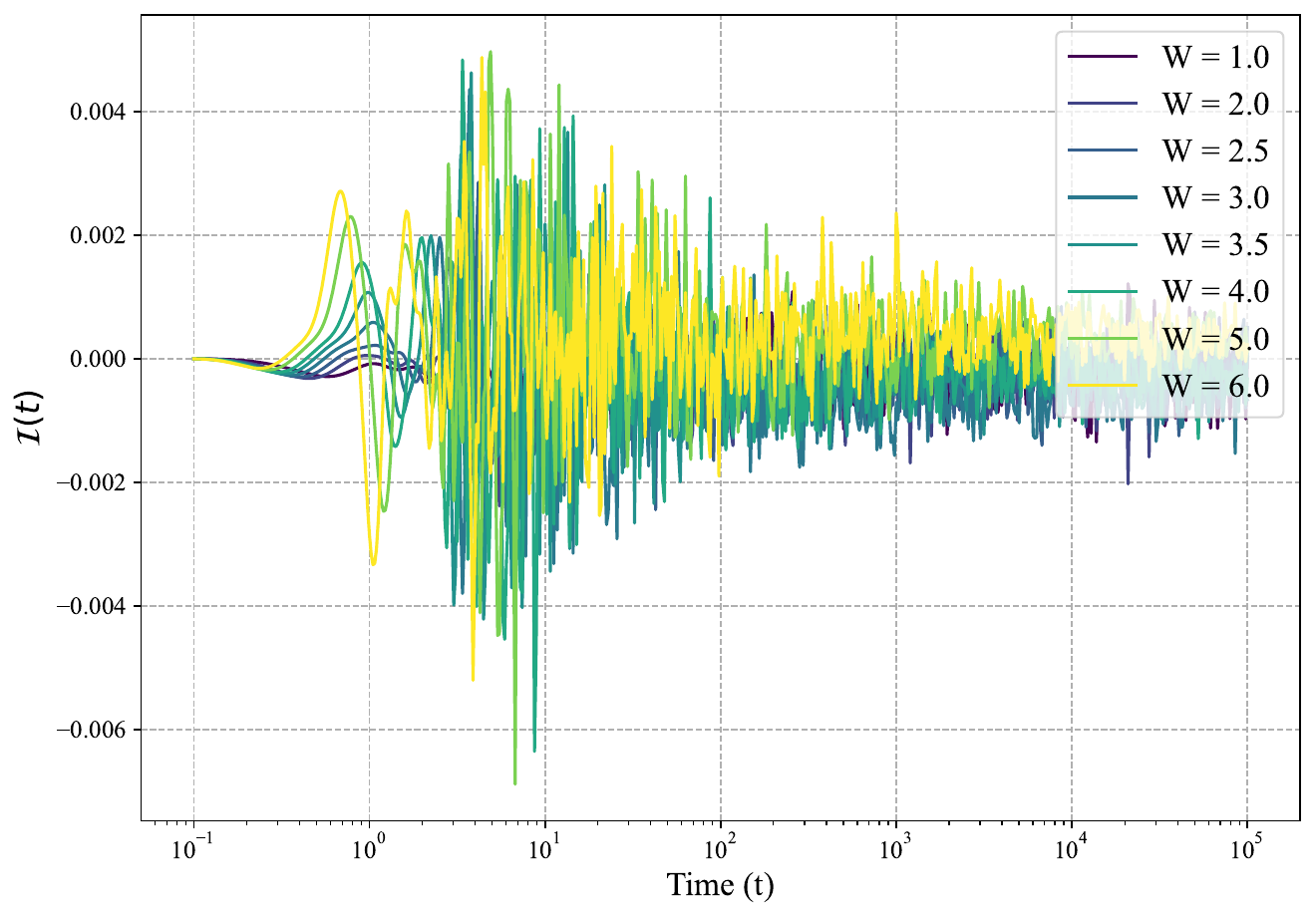}}\hfil
\caption{Figures showing the imbalance for the TNS states with different initial tilts for an $N=12$ system, averaged over 240 realisations over a range of quasiperiodic potentials, $W$. Note that all plots would start at $|\mathcal{I}|=1$ if the states were \textit{not} tilted, and larger tilts act to lower the magnitude of the initial imbalance. When the tilt is $0.5\pi$, all the spins are perpendicular to the imbalance axis, so the initial imbalance becomes 0. Typically, $\mathcal{I}(t)\rightarrow0$ is used as an indicator of thermalization.}
\label{fig:imbalance}
\end{figure}

The imbalance operator,
\begin{equation}
    \mathcal{I}(t)=\frac{1}{N}\sum_{i=1}^L(-1)^i\bra{\psi(t)}\sigma^z_i\ket{\psi(t)},
    \label{eq:imbalance}
\end{equation}
is an observable that can be used to quantify the ability of the system to retain som `memory' of an initial staggered (Néel-like) spin pattern. The Néel state (and its globally-flipped counterpart) are the unique states with $|\mathcal{I}(0)|=1$. The ferromagnetic state, having all spins aligned, has $\mathcal{I}(0)=0$, and thus the imbalance is not conventionally used to study the memory/localisation effects for the ferromagnetic states. It is expected that in the ETH regime this observable should decay to 0, while in the MBL regime some of the initial information about the staggered spins should be preserved \cite{RevModPhys.91.021001,Nandkishore_2015}. In Fig. \ref{fig:imbalance}, we present the results of computing the imbalance for the Néel state for different initial tilt values, and at different potential strengths.
The plots begin at different initial values ($\mathcal{I}(0)\neq1$) due to the initial tilts reducing the magnitude of the imbalance, with the rotation of $\theta=0.5$ causing all spins to point in directions orthogonal to the $z$-direction. For all the figures, there is a pleasing correlation between $W$ and the value at which $\mathcal{I}(t)$ stabilizes. It is interesting to note that being in the ETH regime ($W<3.3$) is \textit{not} a sufficient condition for the imbalance to decay to zero (all plots exhibit at least a slight deviation from 0), but the effect of the potential strength as an `enhancer' of localisation is clearly perceivable. 

\subsection{Entanglement Entropy}
\begin{figure}[!htbp]

\centering
\subfloat[TFS with $\theta=0.1$]{\includegraphics[width=7cm]{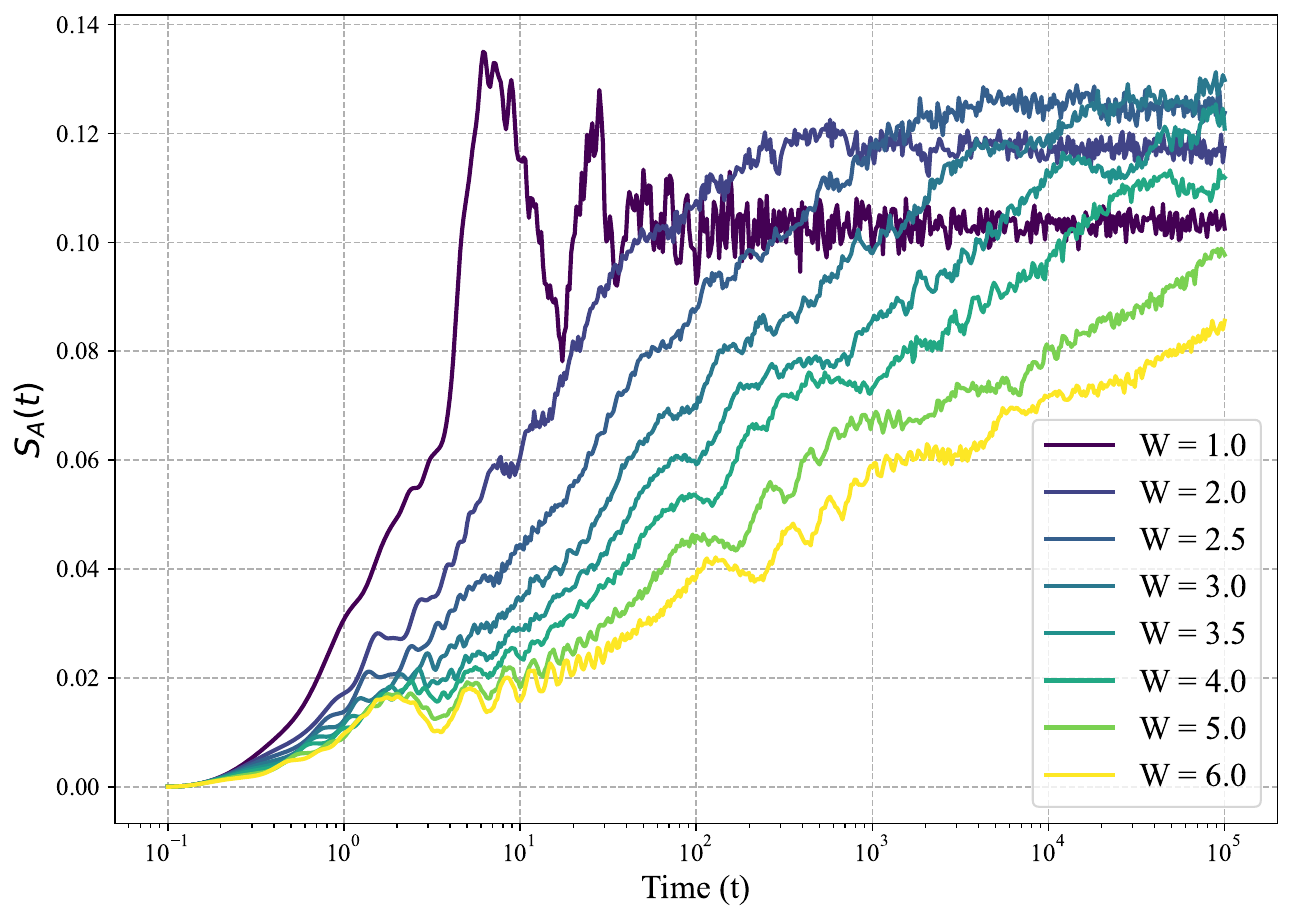}}\hfil
\subfloat[TNS with $\theta=0.1$ \label{fig:normalEE}]{\includegraphics[width=7cm]{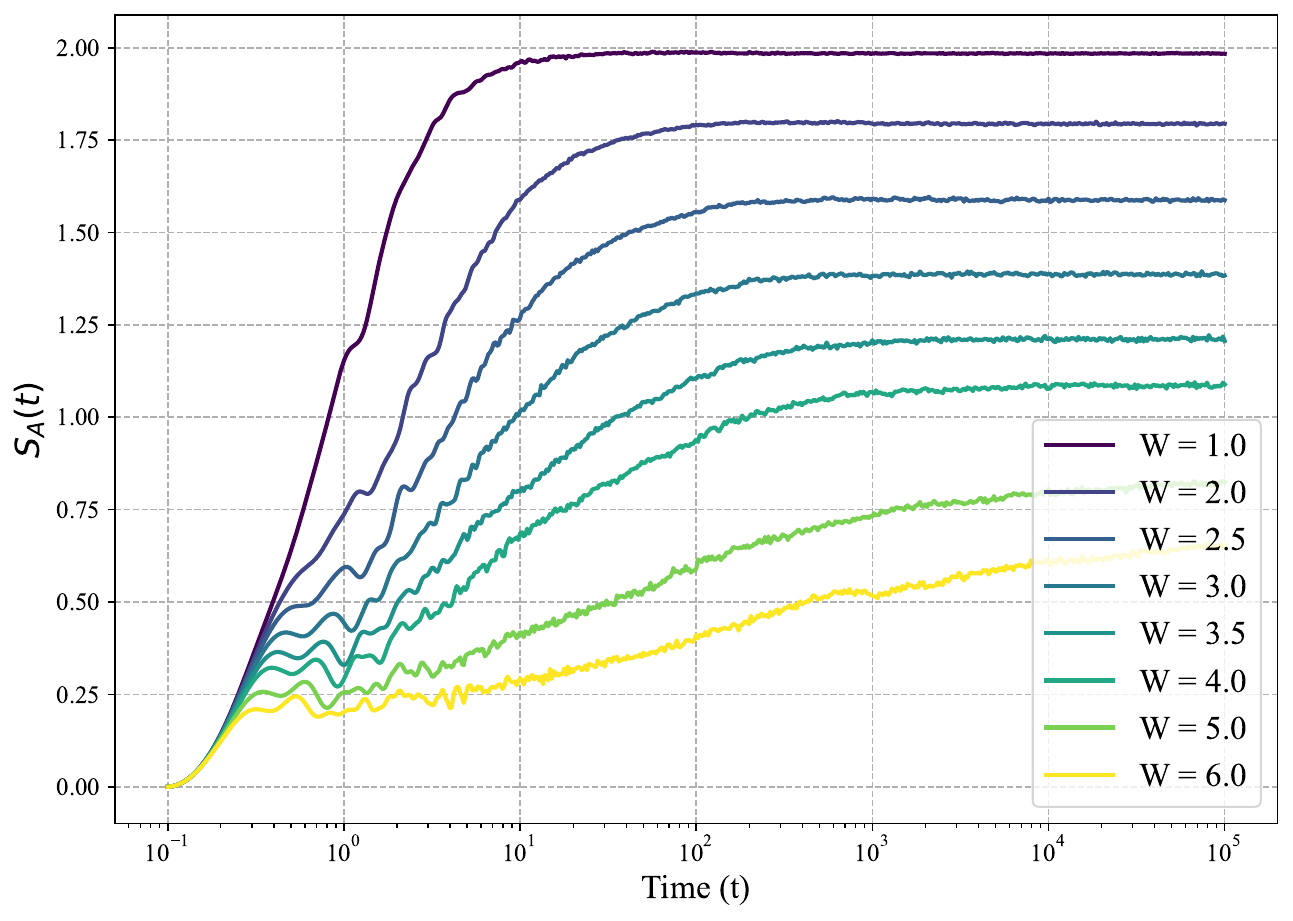}}\hfil 

\subfloat[TFS with $\theta=0.3$]{\includegraphics[width=7cm]{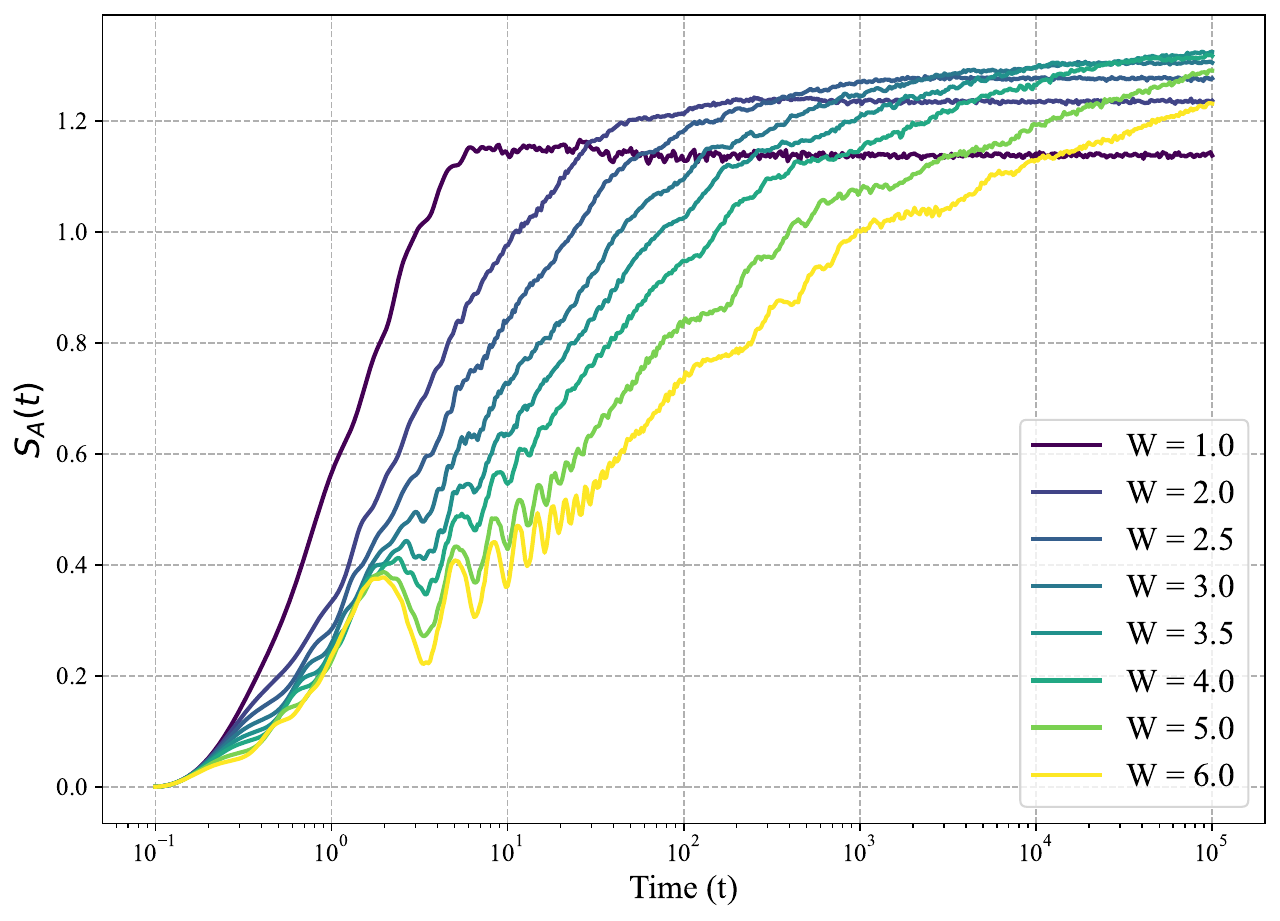}}\hfil
\subfloat[TNS with $\theta=0.3$]{\includegraphics[width=7cm]{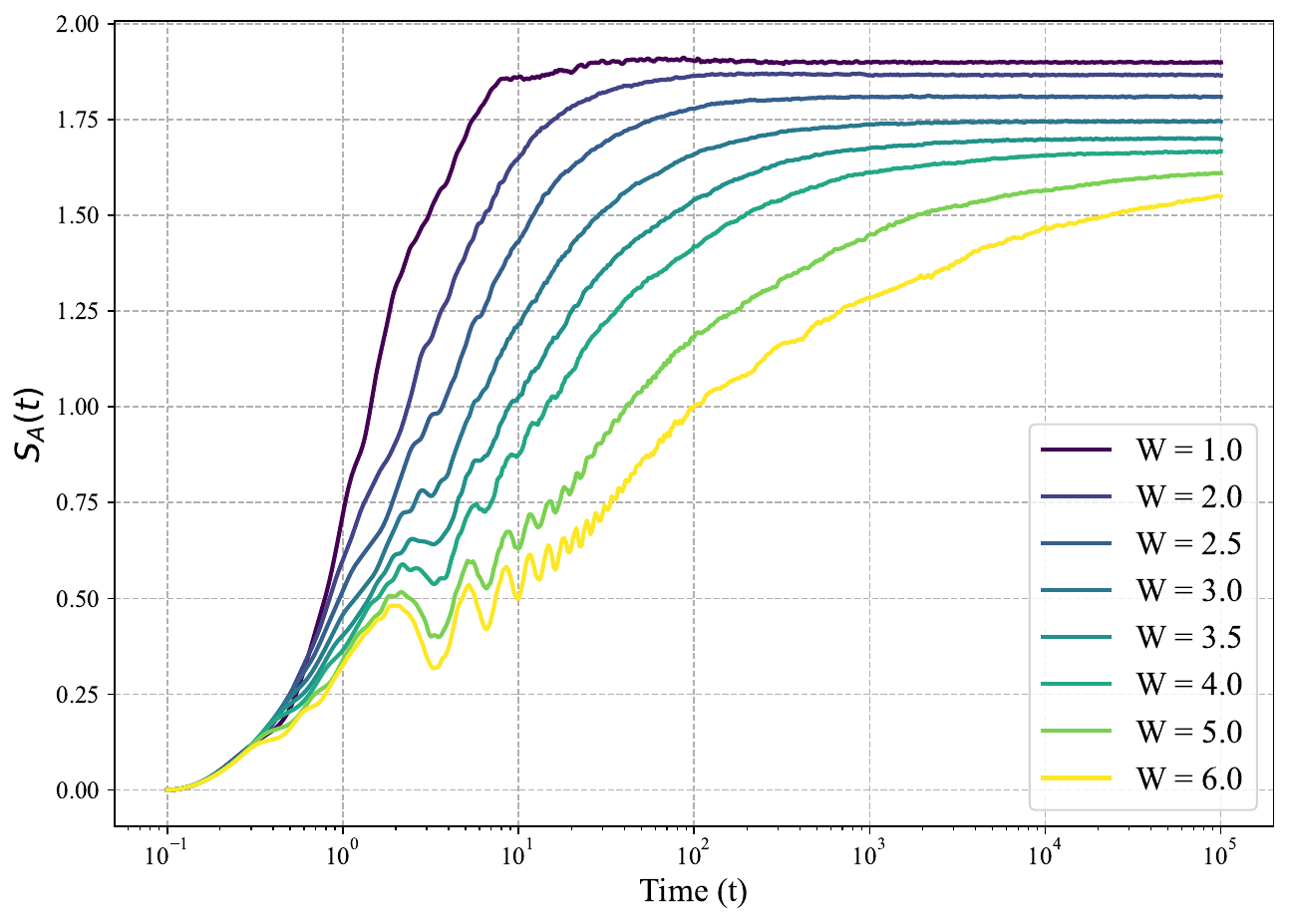}}\hfil 

\subfloat[TFS with $\theta=0.5$]{\includegraphics[width=7cm]{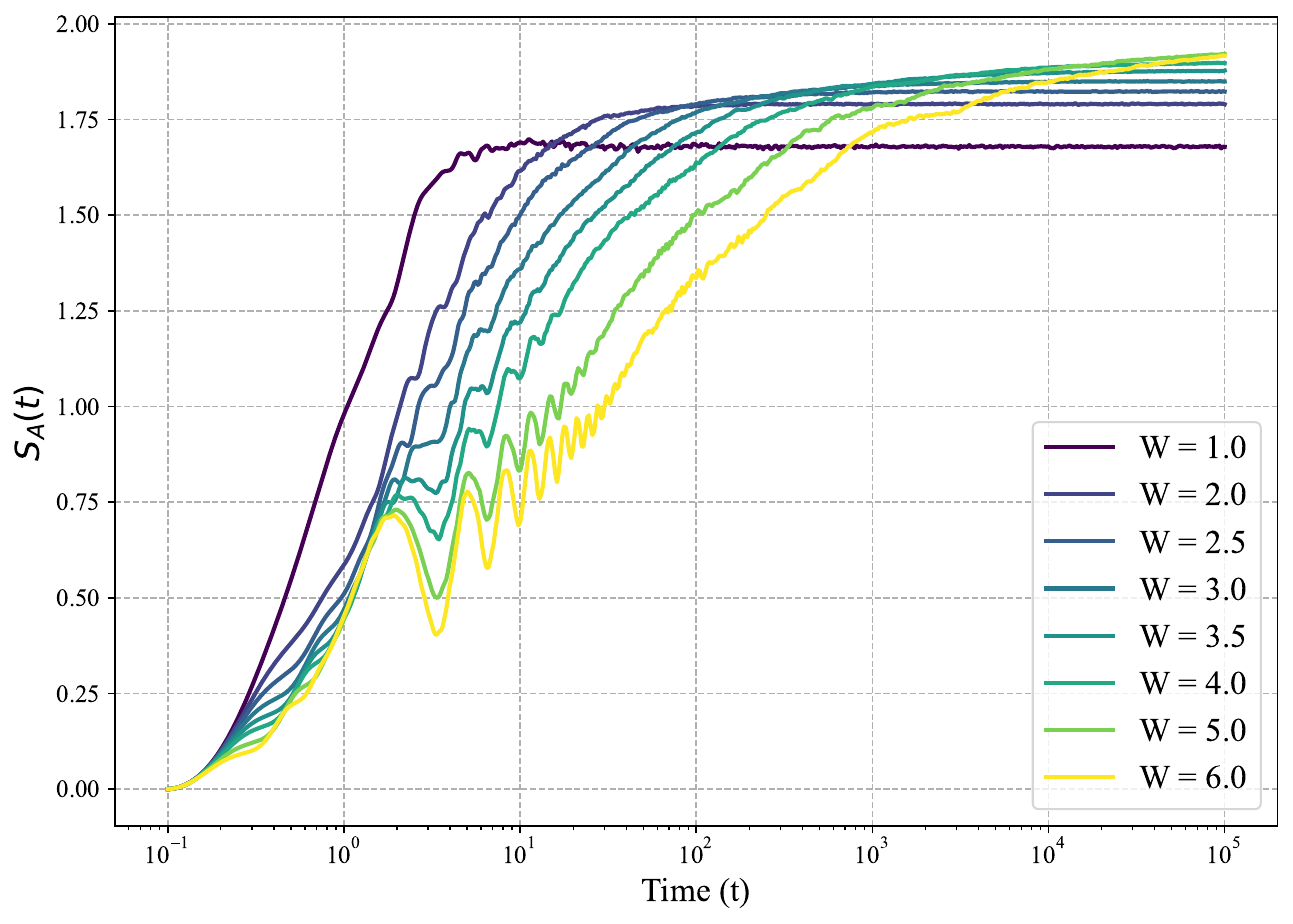}}\hfil
\subfloat[TNS with $\theta=0.5$]{\includegraphics[width=7cm]{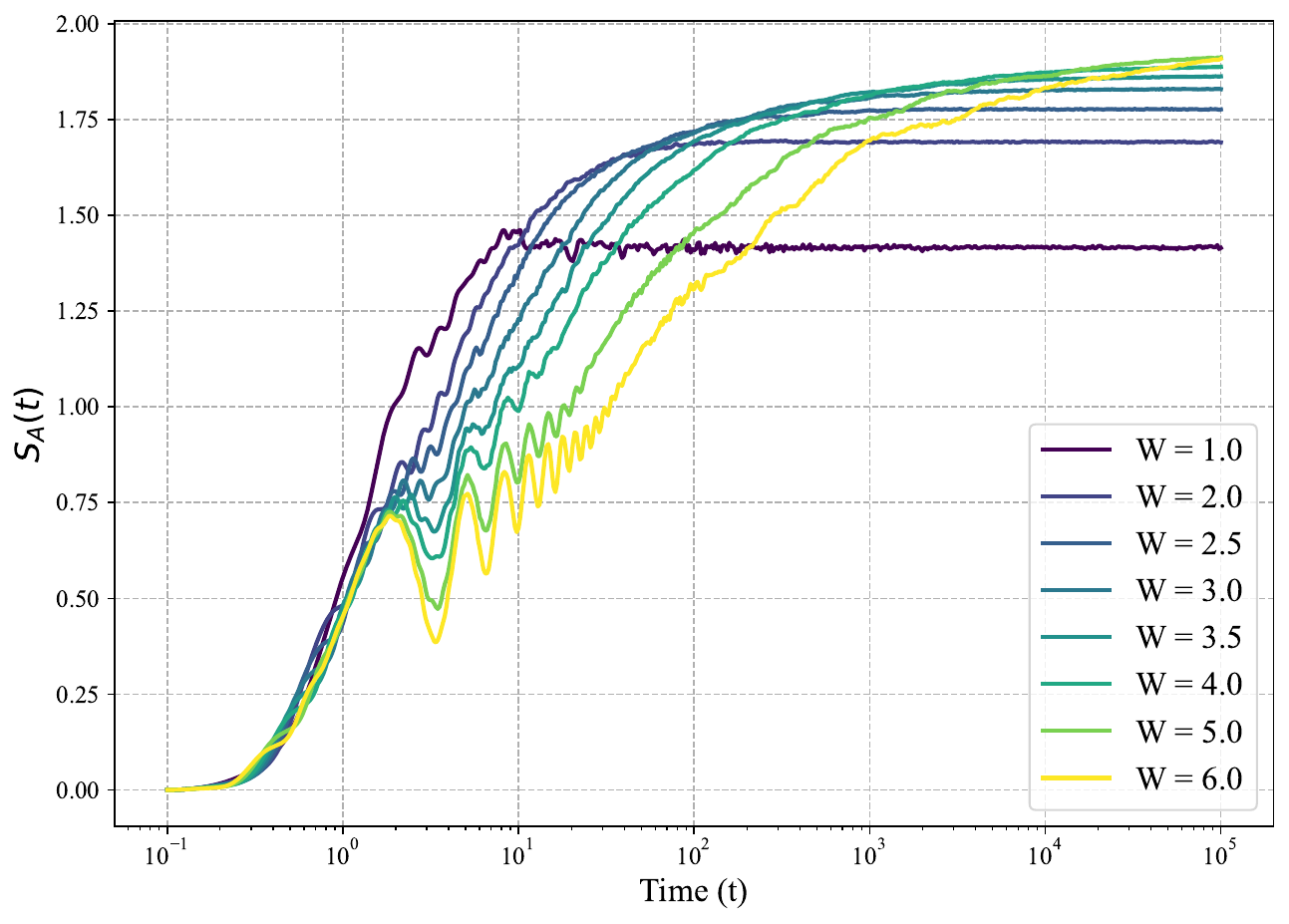}}\hfil 
\caption{Figures showing the relationship between the entanglement entropy and the quasiperiodic potential strength $W$ as we increase the asymmetry parameter $\theta$ (top to bottom), of the the TFS (left) and TNS (right) states, for an $N=12$, ($N_A=3$) size (sub)system, averaged over 240 realisations.}
\label{fig:EE}
\end{figure}

The entanglement entropy quantifies the entanglement between the subsytem of interest (the $N_{sub}=3$ subsystem) and the remainder of the system,
\begin{equation}
    S_A = -\text{Tr}_B(\rho\ln\rho),
    \label{eq:ee}
\end{equation}
and is the \textit{von Neumann} entropy of the subsystem being considered. In the ETH regime, $S_A$ it is found to exhibit fast linear growth, reaching its saturation value $S_A\propto\text{vol(subsystem)}$ quickly when compared to the MBL regime, where growth is typically logarithmic in time and $S_A\propto \text{area(subsystem)}$ \cite{RevModPhys.91.021001}. The maximal entropy for an $n=3$ size subsystem is $\ln(2^3)\approx2.08$. The entanglement entropy results for the AA model are shown in Fig. \ref{fig:EE}.
One can see that the saturation always occurs faster for the ETH regime states, but it does not always saturate to a higher value. The (non-Mpemba) crossing allows for additional insights into the role of the asymmetry transformation. Focusing on the TNS states, Fig. \ref{fig:normalEE} shows the conventional ETH/MBL style transition; at low potential strength (the ETH regime), the system saturates rapidly and very nearly achieves the maximal value (note, there is a non-zero tilt present), but the saturation value decreases for the ETH regime as we increase $\theta$, whereas the saturation values tend to increase with increasing $\theta$. Note that the rate of growth (at early times) also tends to increase with increasing $\theta$ in both the ETH and MBL regimes. The asymmetry (tilt) operation thus gives access to larger late time entanglement entropies -- while the states in the MBL regime might not grow as rapidly as those in the ETH regime, they are able to exceed the conventional area-law predictions of the MBL phase. %\textcolor{red}{A surprisingly interesting result. Not sure if you guys want to move it to the main text or not? Not sure if this is a known result or not.} 

\subsection{Survival Probability}
\begin{figure}[!htbp]

\centering
\subfloat[TFS with $\theta=0.1$]{\includegraphics[width=7cm]{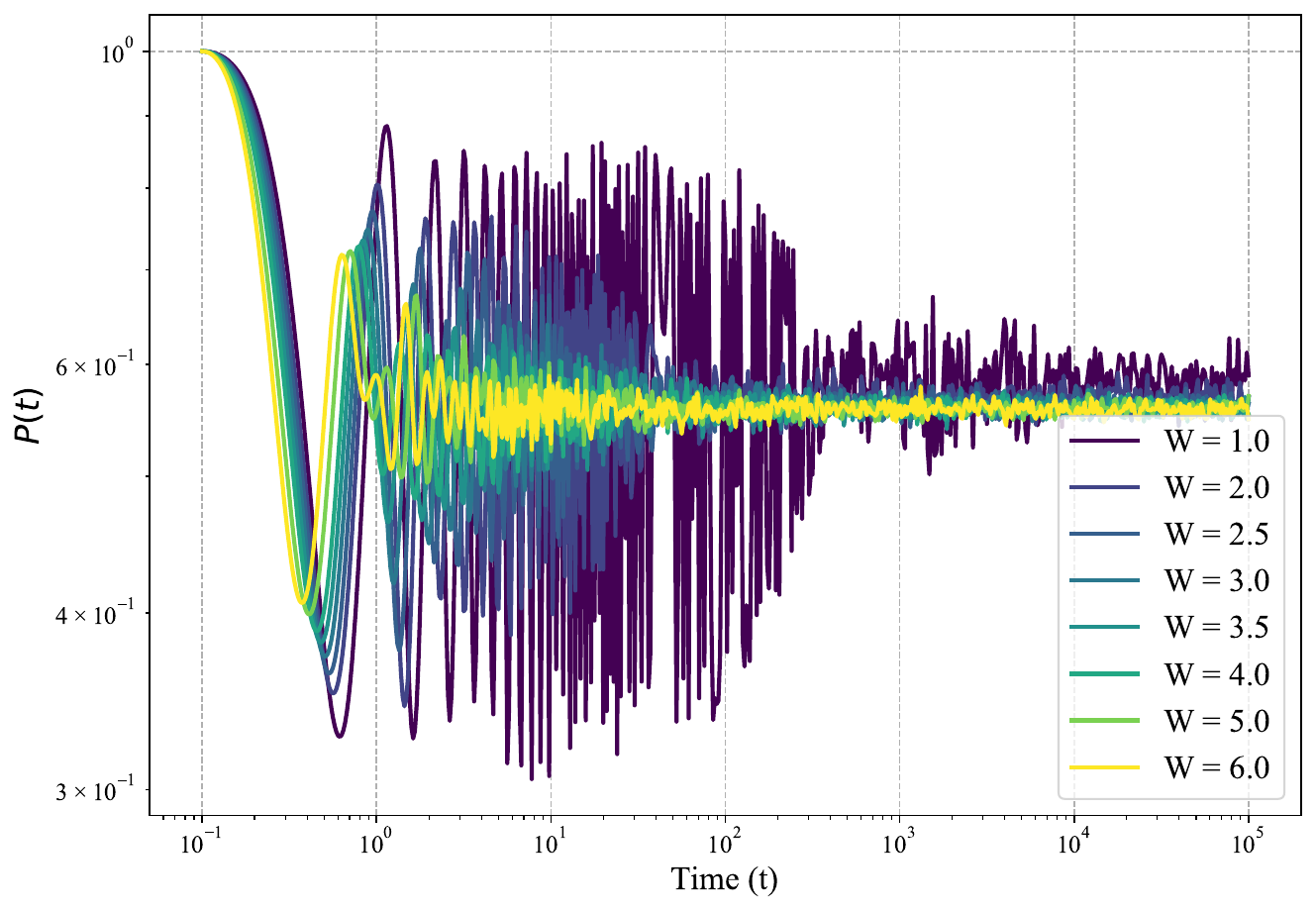}}\hfil
\subfloat[TNS with $\theta=0.1$ \label{fig:normalsurvival}]{\includegraphics[width=7cm]{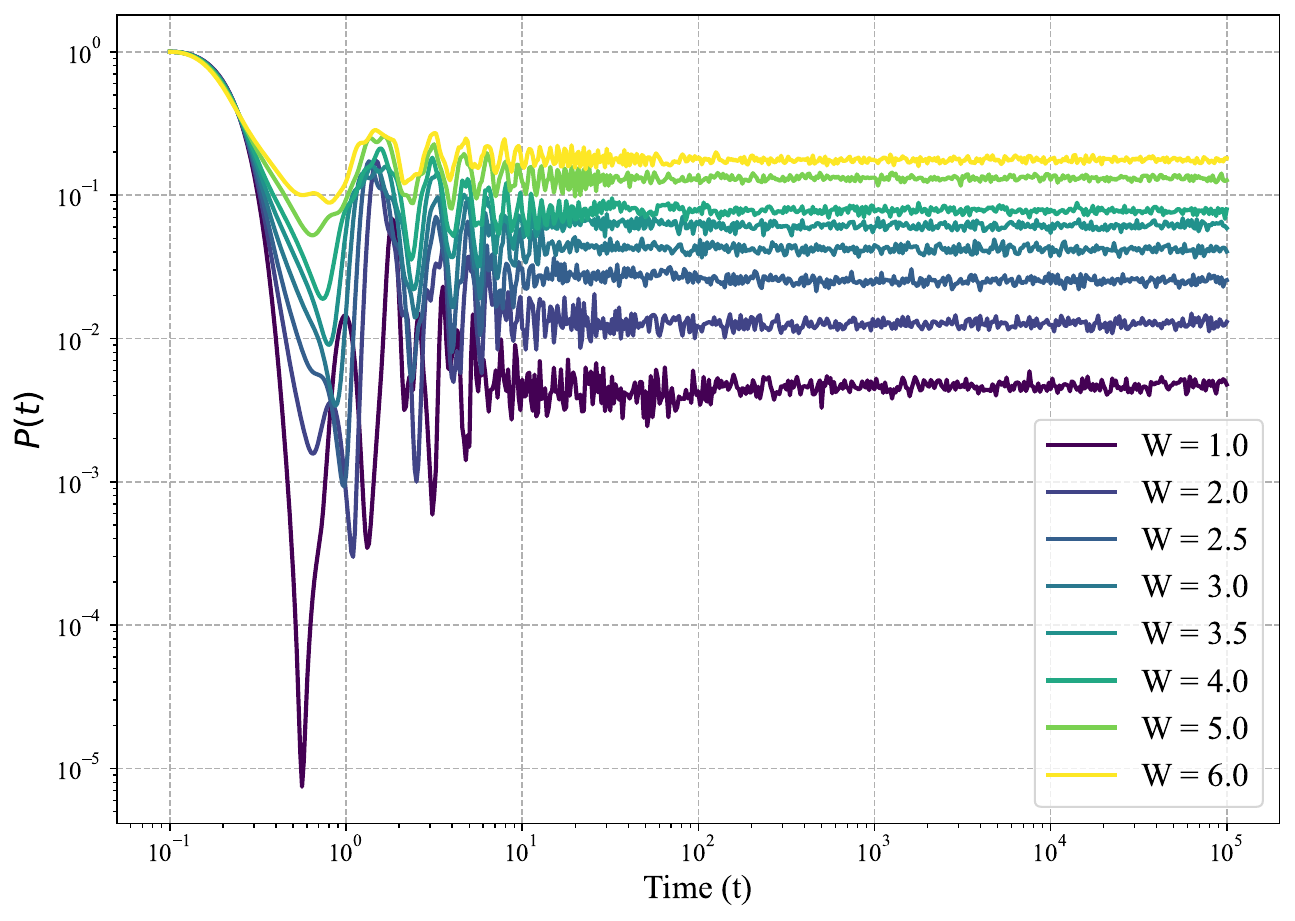}}\hfil 

\subfloat[TFS with $\theta=0.3$]{\includegraphics[width=7cm]{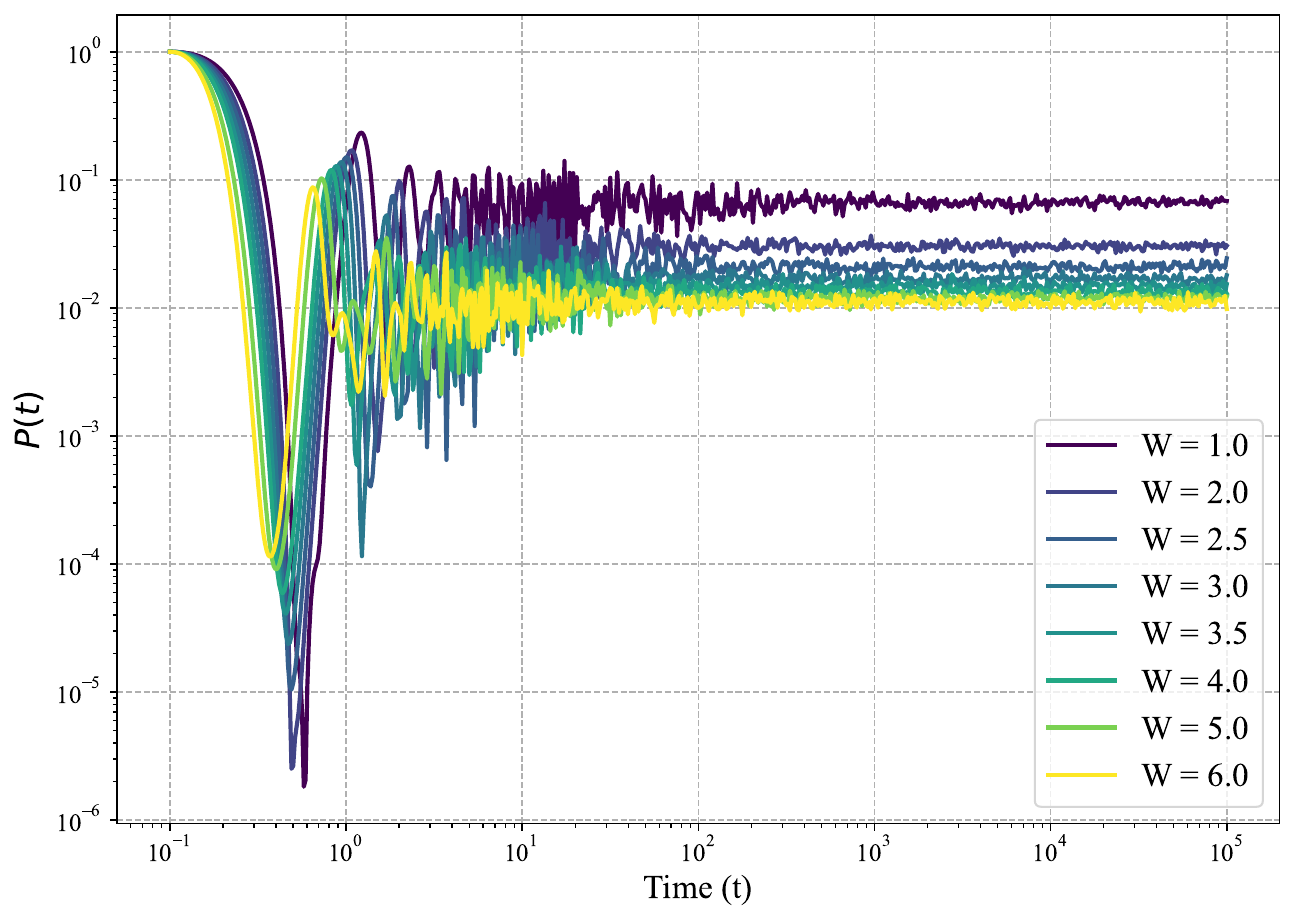}}\hfil
\subfloat[TNS with $\theta=0.3$ \label{fig:midtiltsurvival}]{\includegraphics[width=7cm]{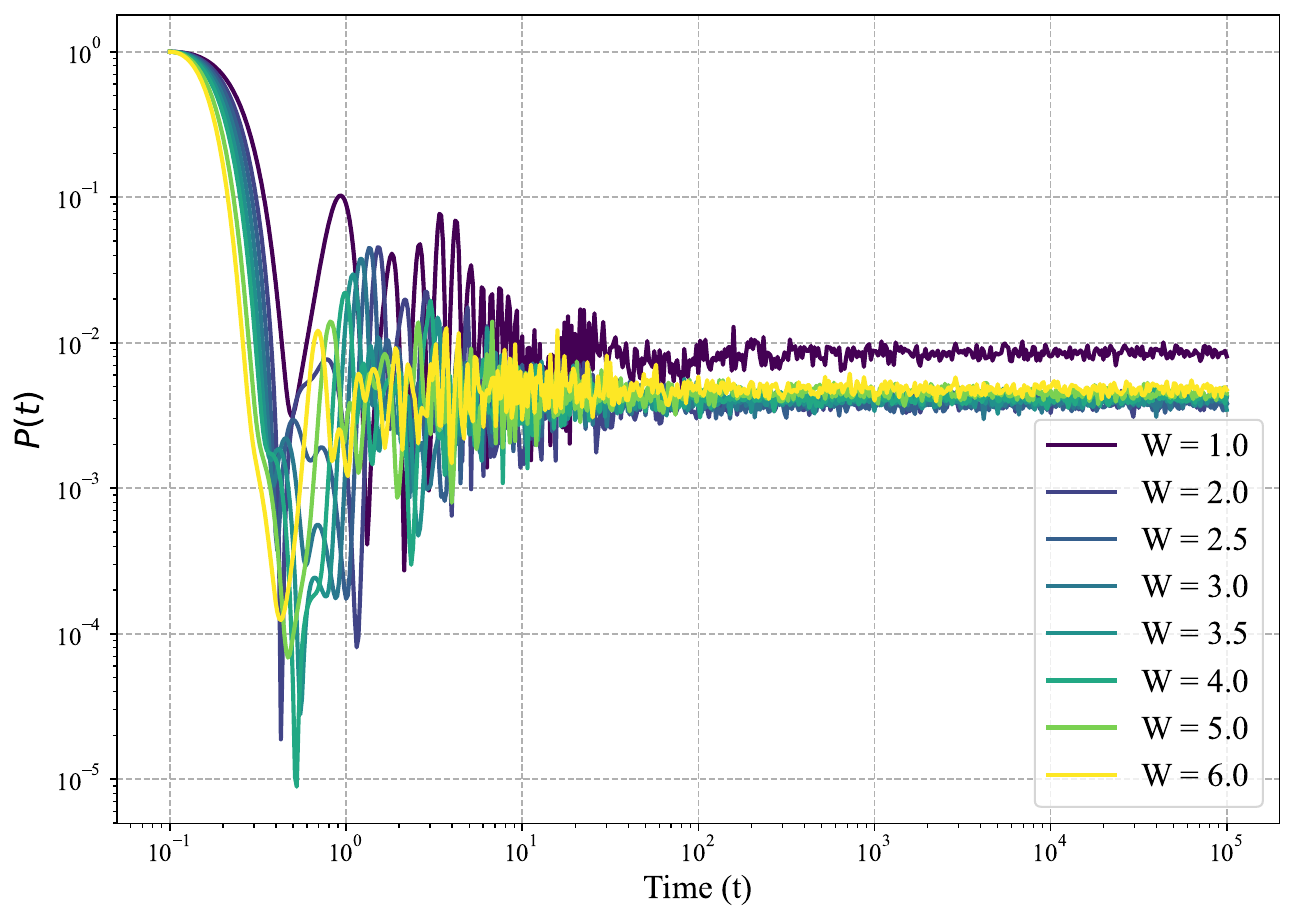}}\hfil 

\subfloat[TFS with $\theta=0.5$]{\includegraphics[width=7cm]{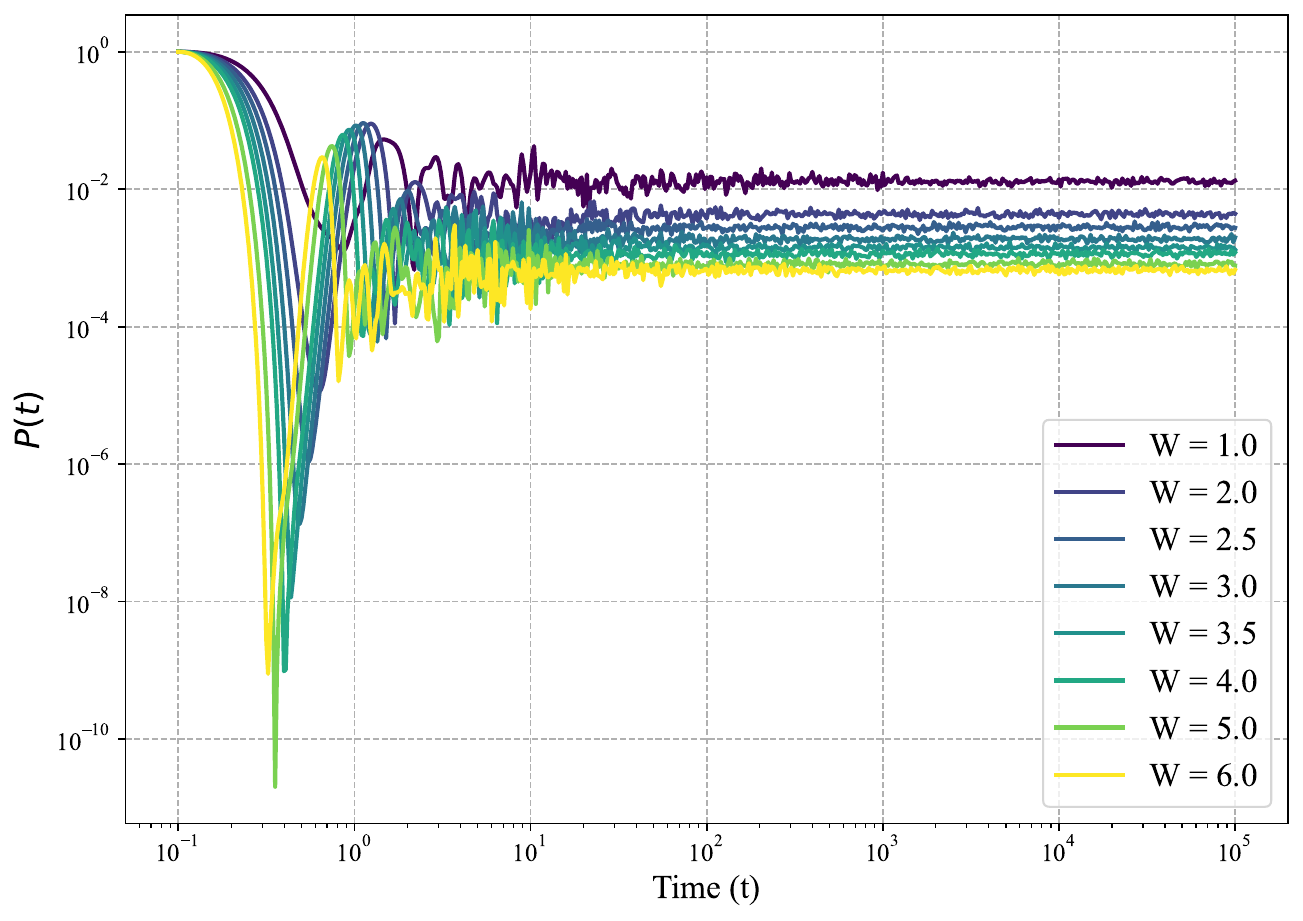}}\hfil
\subfloat[TNS with $\theta=0.5$ \label{fig:maxtiltsurvival}]{\includegraphics[width=7cm]{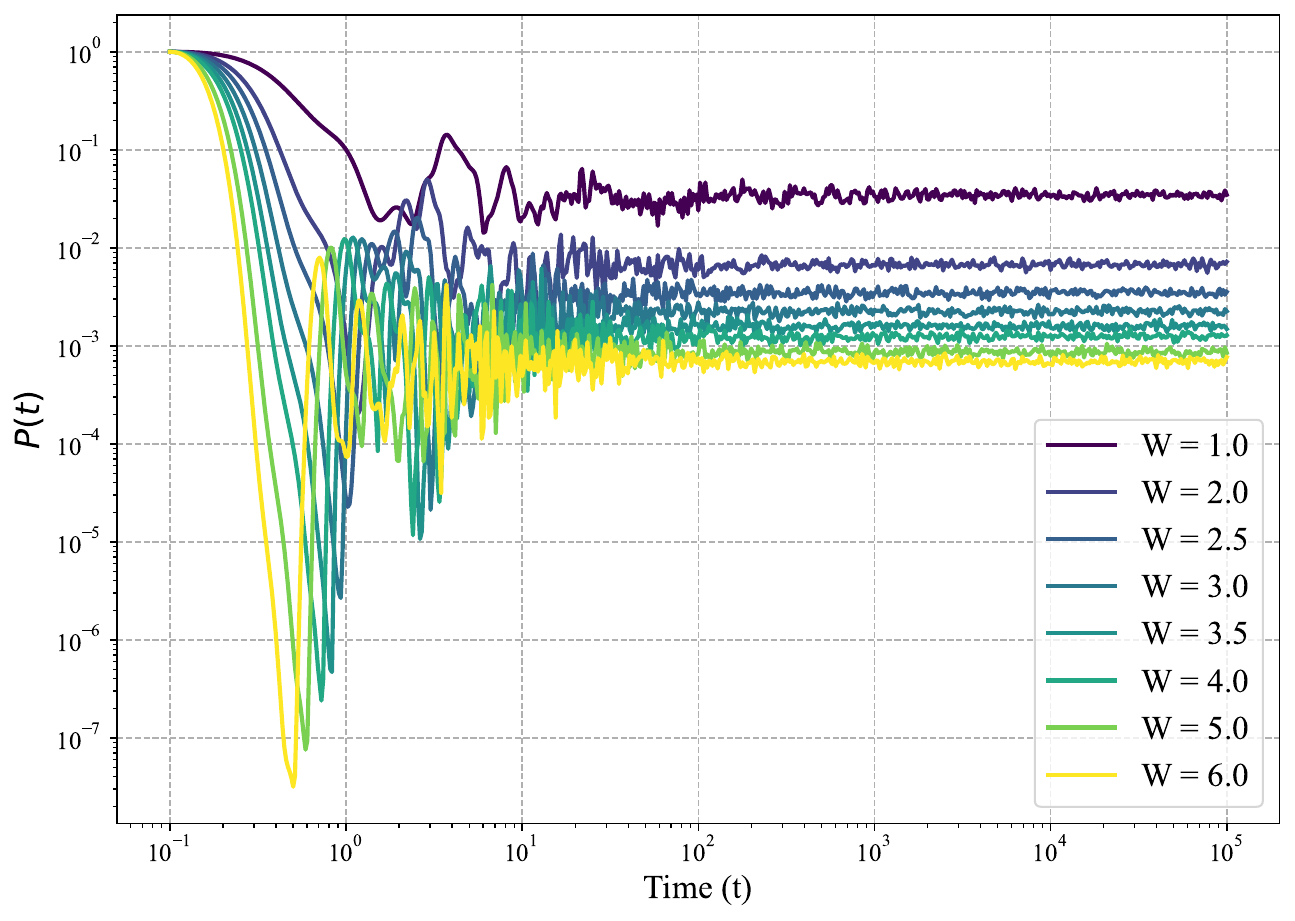}}\hfil 
\caption{Figures showing the relationship between the survival probability and the quasiperiodic potential strength $W$ as we increase the asymmetry parameter $\theta$ (top to bottom), of the the TFS (left) and TNS (right) states, for an $N=12$ size system, averaged over 240 realisations. Take careful note of the different values for the logarithmic scales on the probability axes.}
\label{fig:survivalprob}
\end{figure}

Finally, we comment on the survival probability of the initial tilted states. This quantity is simply the probability of finding that the time-evolved state is the same as the initial state,
\begin{equation}
    P(t) = |\braket{\psi(0)|\psi(t)}|^2.
    \label{eq:survivalprob}
\end{equation}
It is a simple quantifier of `memory' of initial conditions. In the chaotic (ETH) regime, it is expected that this probability should decay rapidly to zero (up to fluctuations), while in the MBL regime one expects a reduction in the decay rate, and an increase in the saturation value (i.e. the state should have a non-trivial overlap with the initial state). The plots of these computations for the TFS and TNS states are presented in Fig. \ref{fig:survivalprob}.
The interplay between the tilt angles and the potential strength is very interesting. As the tilt angle is increased (reading top to bottom in the figures), the survival probability saturates at lower values; that is, less information about the initial state is preserved, a `thermal-like' effect. However, the ordering of the late-time saturation values -- that one would expect to be ordered by the potential strength (as in Fig. \ref{fig:normalsurvival}) -- begins to show violations in this ordering as we increase the tilt (see Fig.c\ref{fig:midtiltsurvival}), until the ordering is completely reversed at the maximal tilt (Fig. \ref{fig:maxtiltsurvival})\footnote{Similar behavior is observable in the TFS states (left-hand side of Fig. \ref{fig:survivalprob}), but the TNS states demonstrate it most clearly.}. %\textcolor{red}{Another really interesting result??? Again, not sure if we want this in the main text?}

\clearpage
\section{Additional Plots}
Given the large numbers of parameters at play -- 2 states $\times$ 5 tilt angles $\times$ 13 different potential strengths -- it is necessary to place some additional figures outside of the main text. We have tried to avoid being overzealous in adding figures, but include figures where we believe it will enhance the understanding of the main text.

\subsection{Decomposition of Complexity: Néel}
\label{app:TNSdecomposition}
\begin{figure}[!htbp]
\centering
\subfloat[$W=1.0,\theta=0.1$]
{\includegraphics[width=0.47\linewidth]{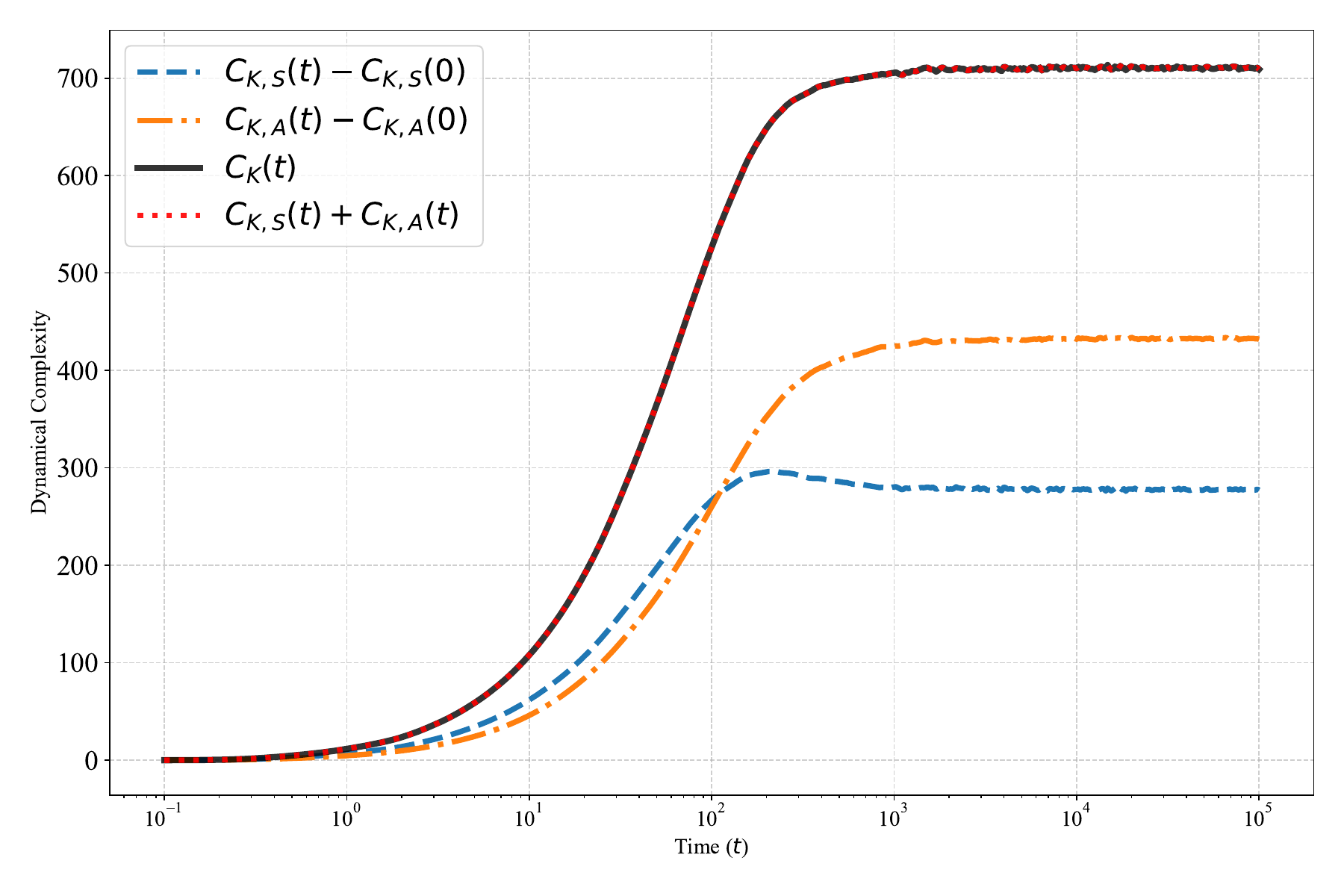}}\hfil
\subfloat[$W=1.0,\theta=0.5$]{\includegraphics[width=0.47\linewidth]{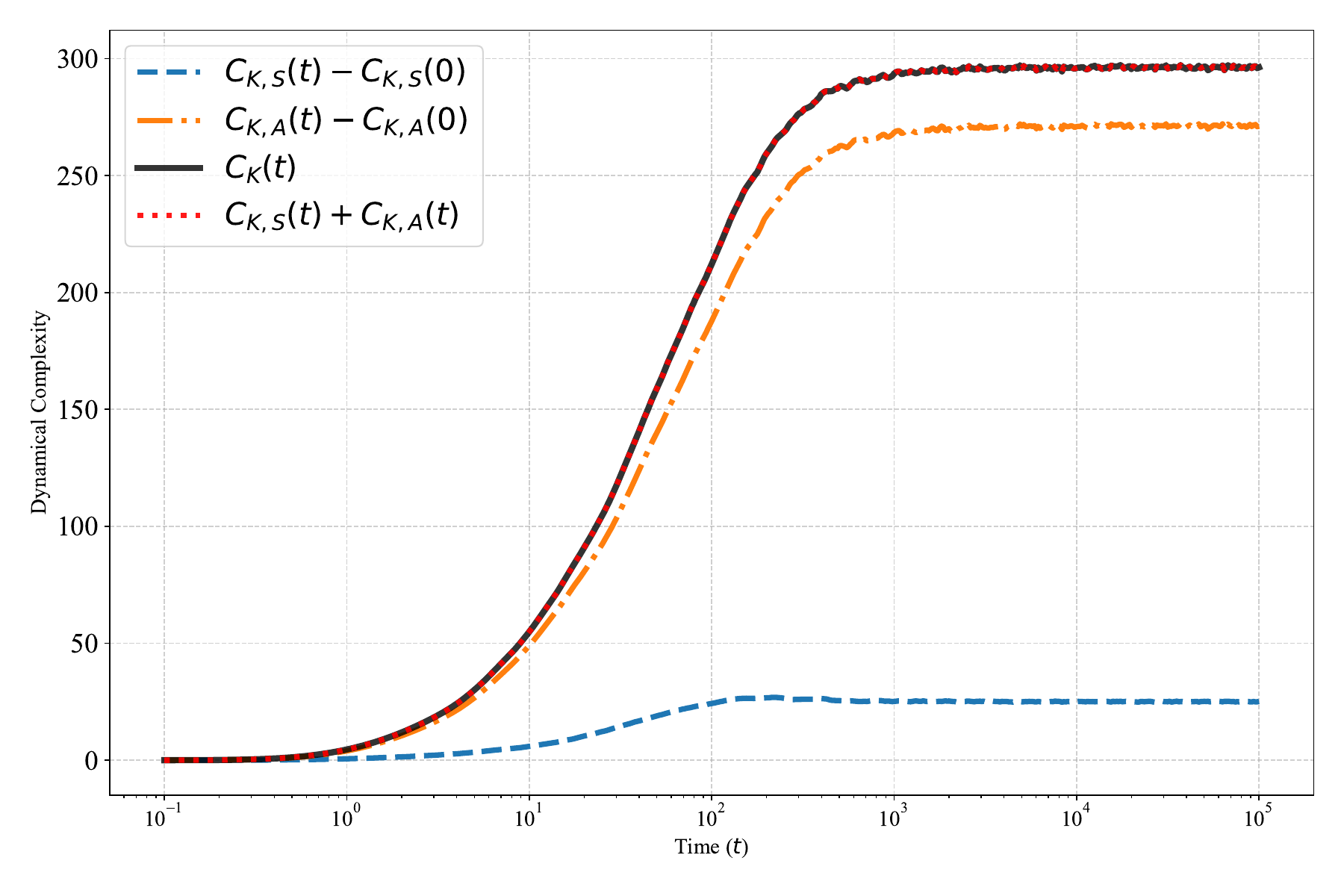}}\hfil 
\subfloat[$W=5.0,\theta=0.1$]{\includegraphics[width=0.47\linewidth]{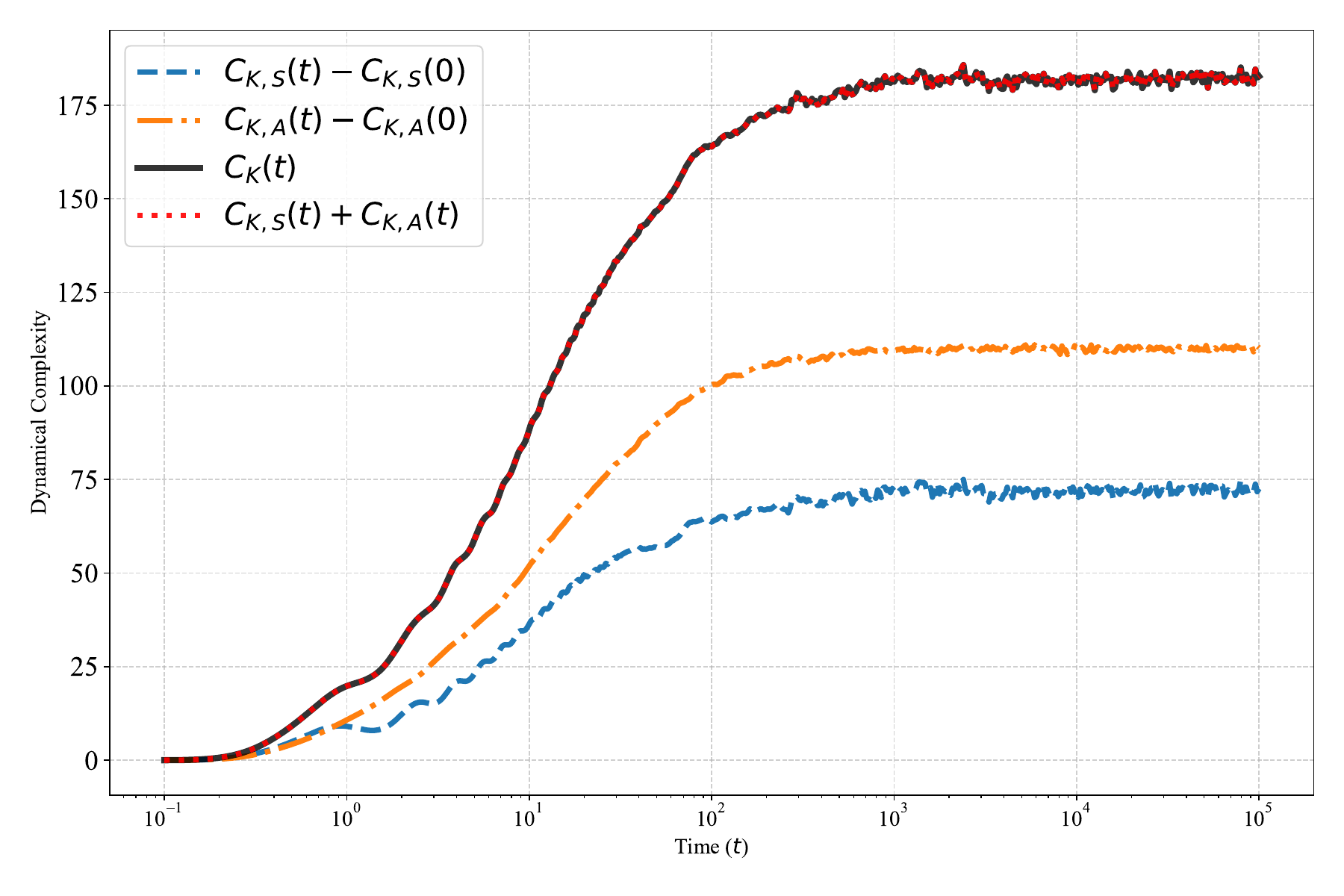}}\hfil
\subfloat[$W=5.0,\theta=0.5$]{\includegraphics[width=0.47\linewidth]{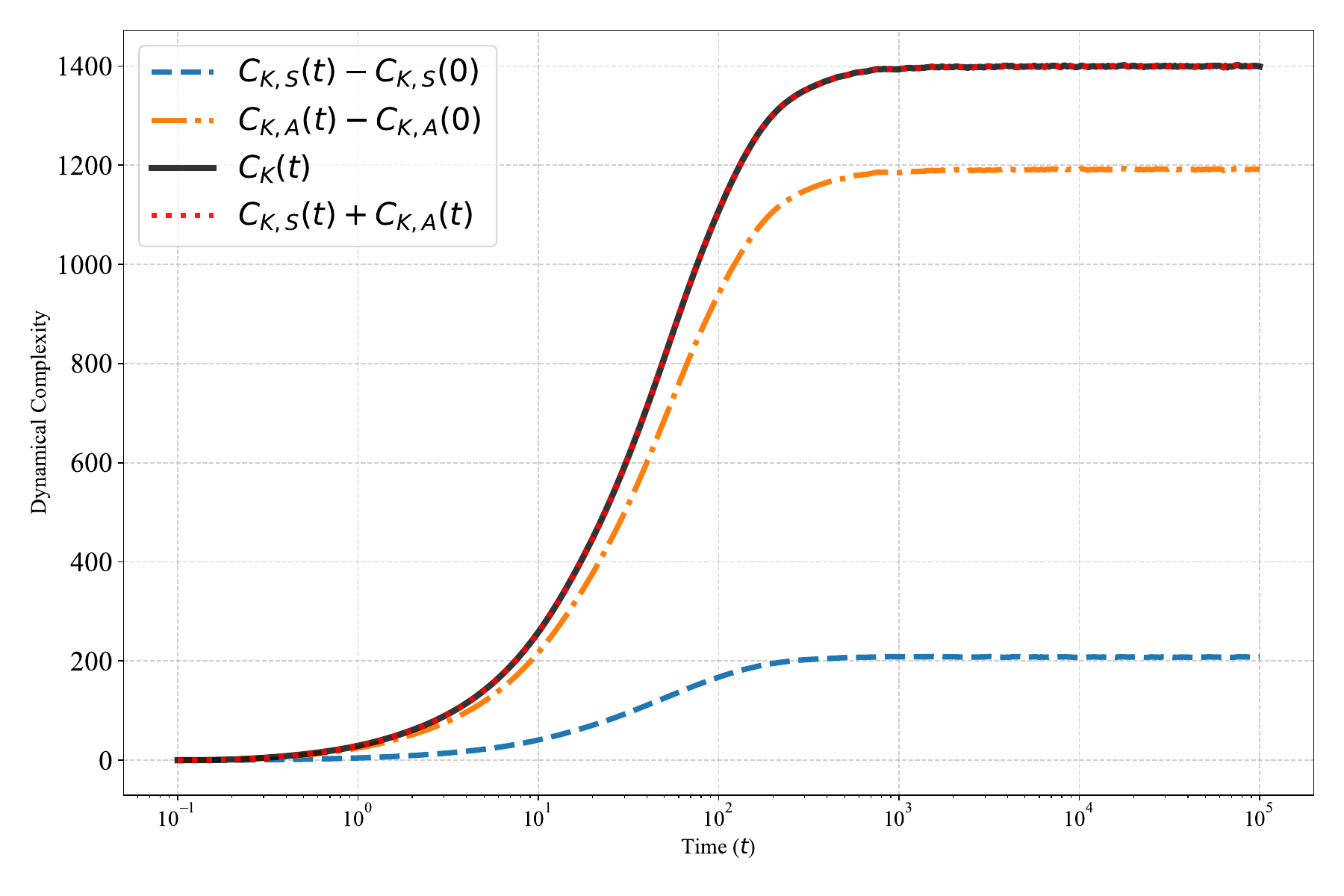}}\hfil 
\caption{Figures verifying the decomposition of the projected complexity for the TNS states at different potential strengths and different tilt angles.}
\label{fig:SumOfPartsApp}
\end{figure}

In Fig. \ref{fig:SumOfParts}, we show that the decomposition of the Krylov complexity into symmetric and asymmetric components is consistent, and faithfully reproduces the total consistency. This served as both a consistency check of both the formalism and the numerics. For completeness, we include Fig. \ref{fig:SumOfPartsApp}, which is the same decomposition plot as was shown for the TFS states in the main text, but for the TNS states. These data faithfully reproduce the total Krylov complexity curves.

\clearpage
\subsection{Comparison of Diffusive Measures}
\label{app:diffcomp}
\begin{figure}[!htbp]
\centering
\subfloat[$W=1.0,\theta=0.1$]{\includegraphics[width=0.47\linewidth]{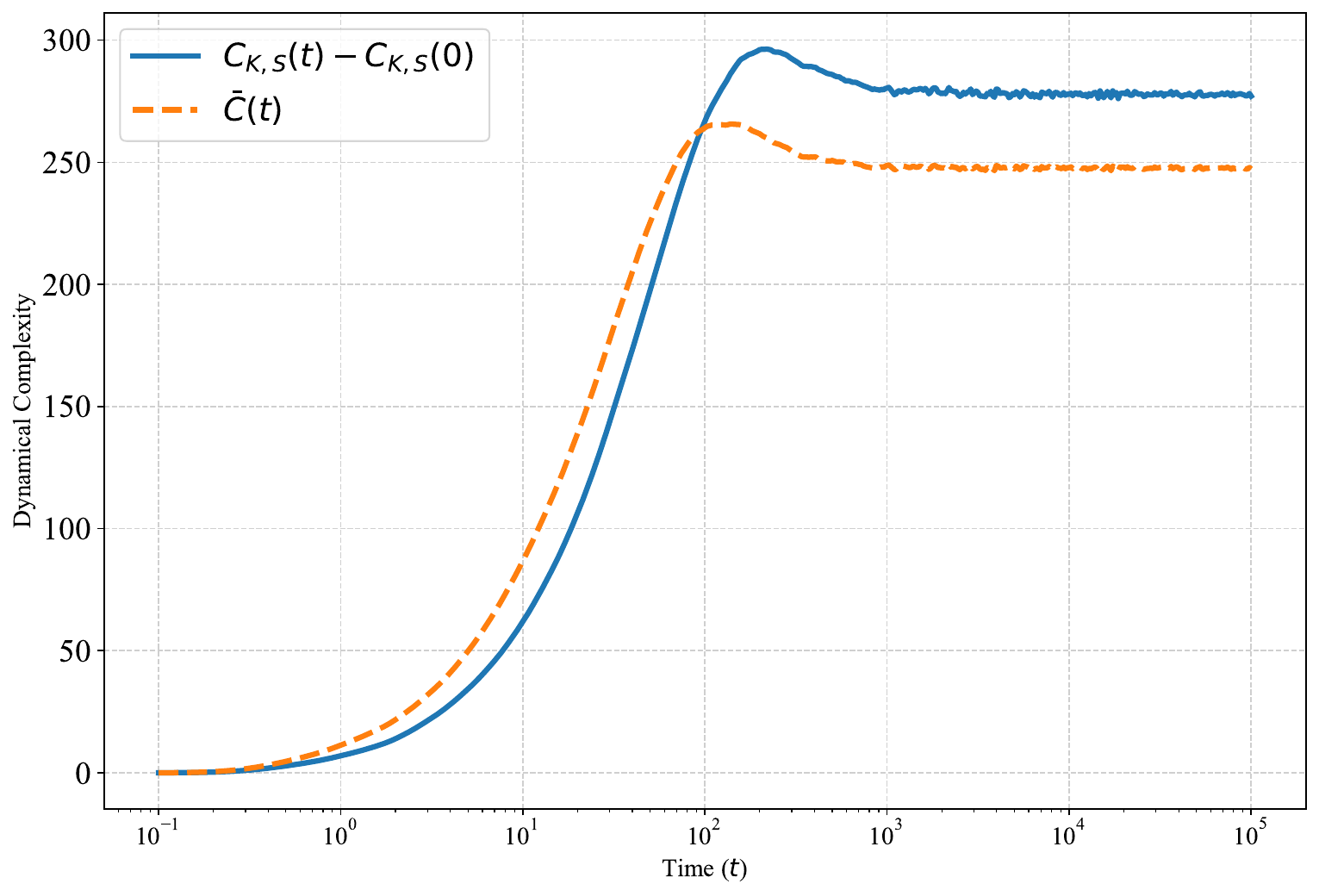}}\hfil
\subfloat[$W=1.0,\theta=0.5$]{\includegraphics[width=0.47\linewidth]{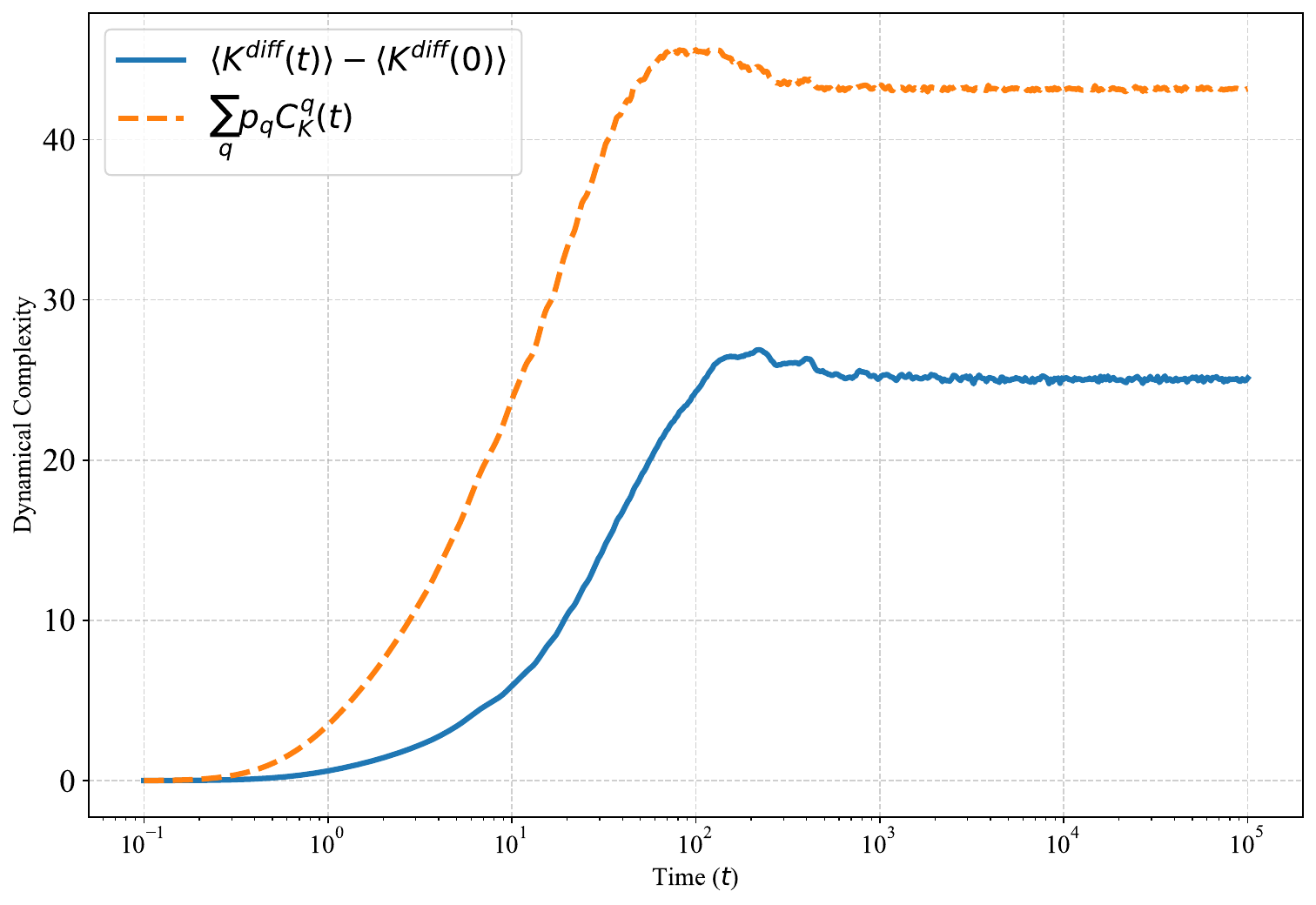}}\hfil 
\subfloat[$W=5.0,\theta=0.1$]{\includegraphics[width=0.47\linewidth]{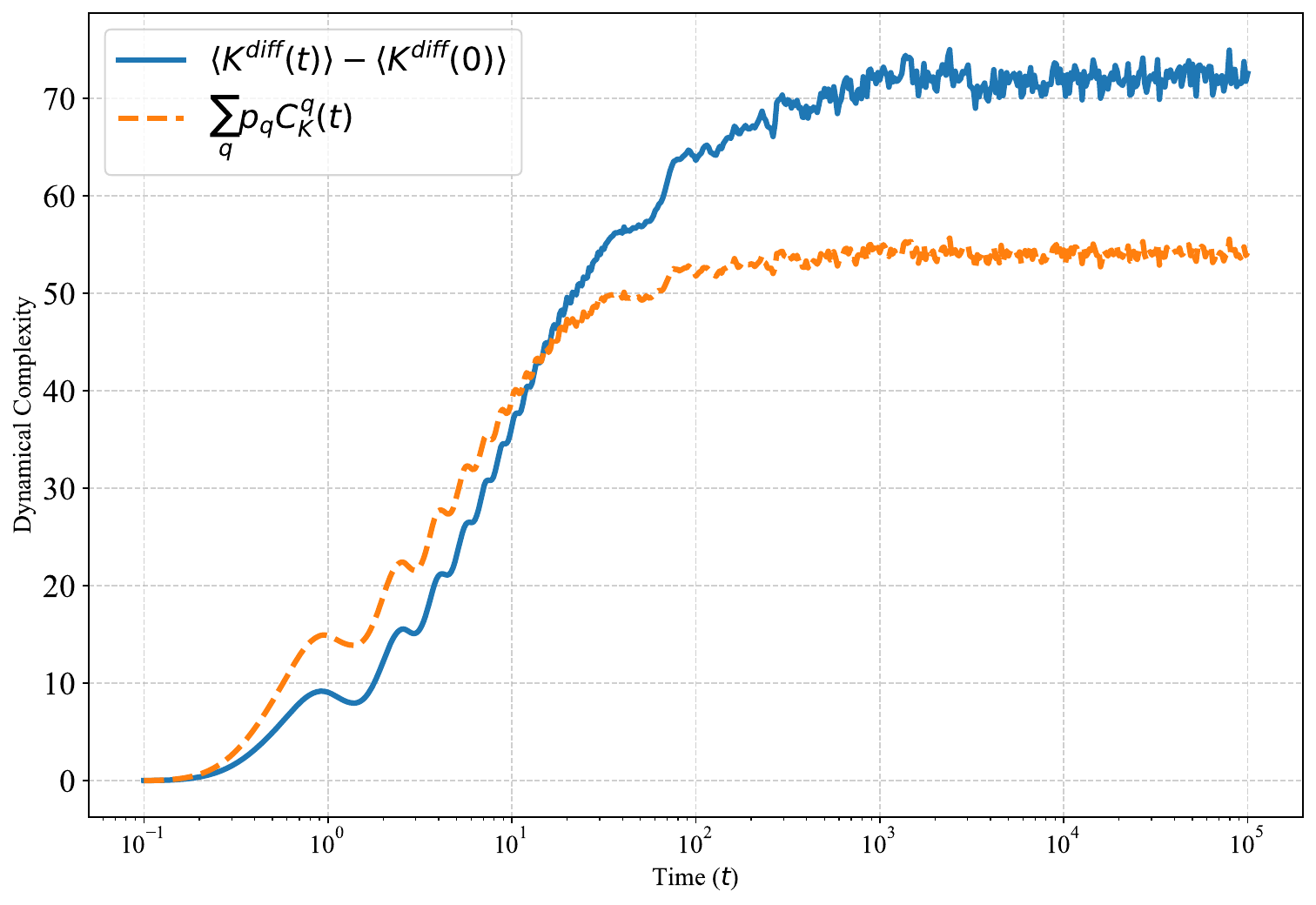}}\hfil
\subfloat[$W=5.0,\theta=0.5$]{\includegraphics[width=0.47\linewidth]{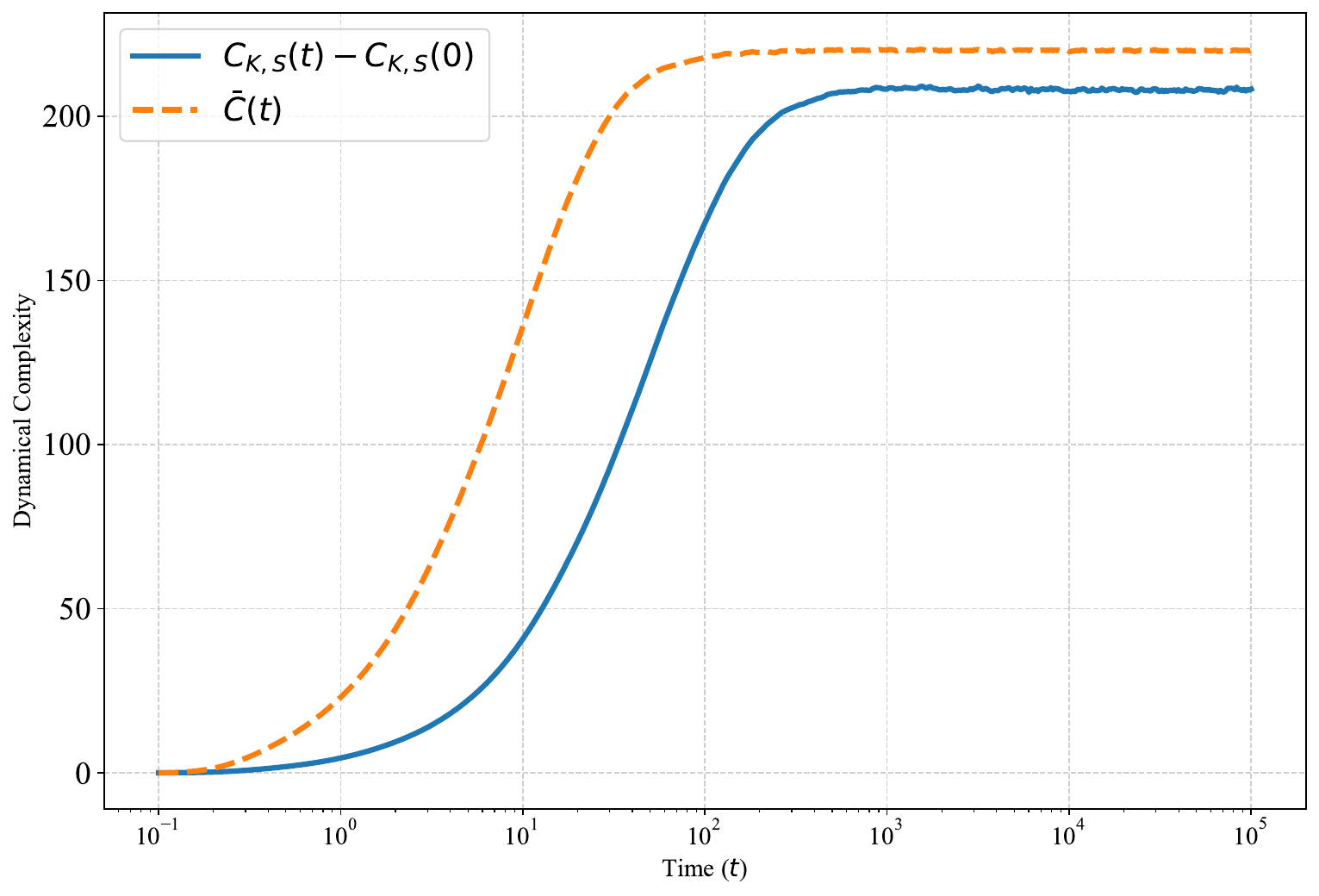}}\hfil 
\caption{Figures showing the discrepancy between the different intrasector diffusive measures for the TNS states at different tilt angles and different potential strengths. We compare the total discrepancy using the $\bar{C}$ measure described in \cite{caputa2025block} (dashed lines), and the projective symmetric complexity introduced in this work.}
\label{fig:discrepancyapp}
\end{figure}

\begin{figure}[!htbp]
\centering
\subfloat[$W=1.0,\theta=0.1$]{\includegraphics[width=0.47\linewidth]{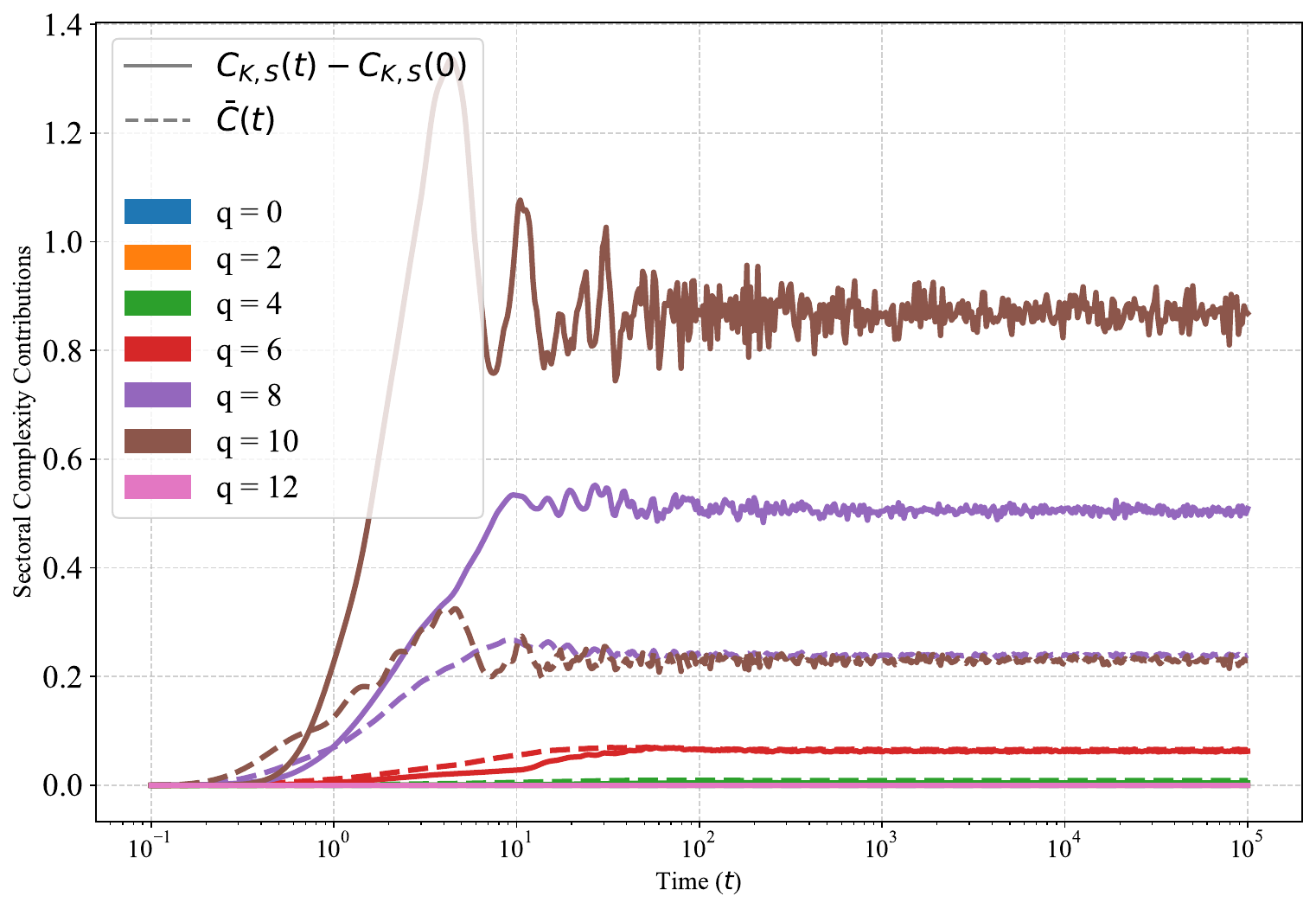}}\hfil
\subfloat[$W=1.0,\theta=0.5$]{\includegraphics[width=0.47\linewidth]{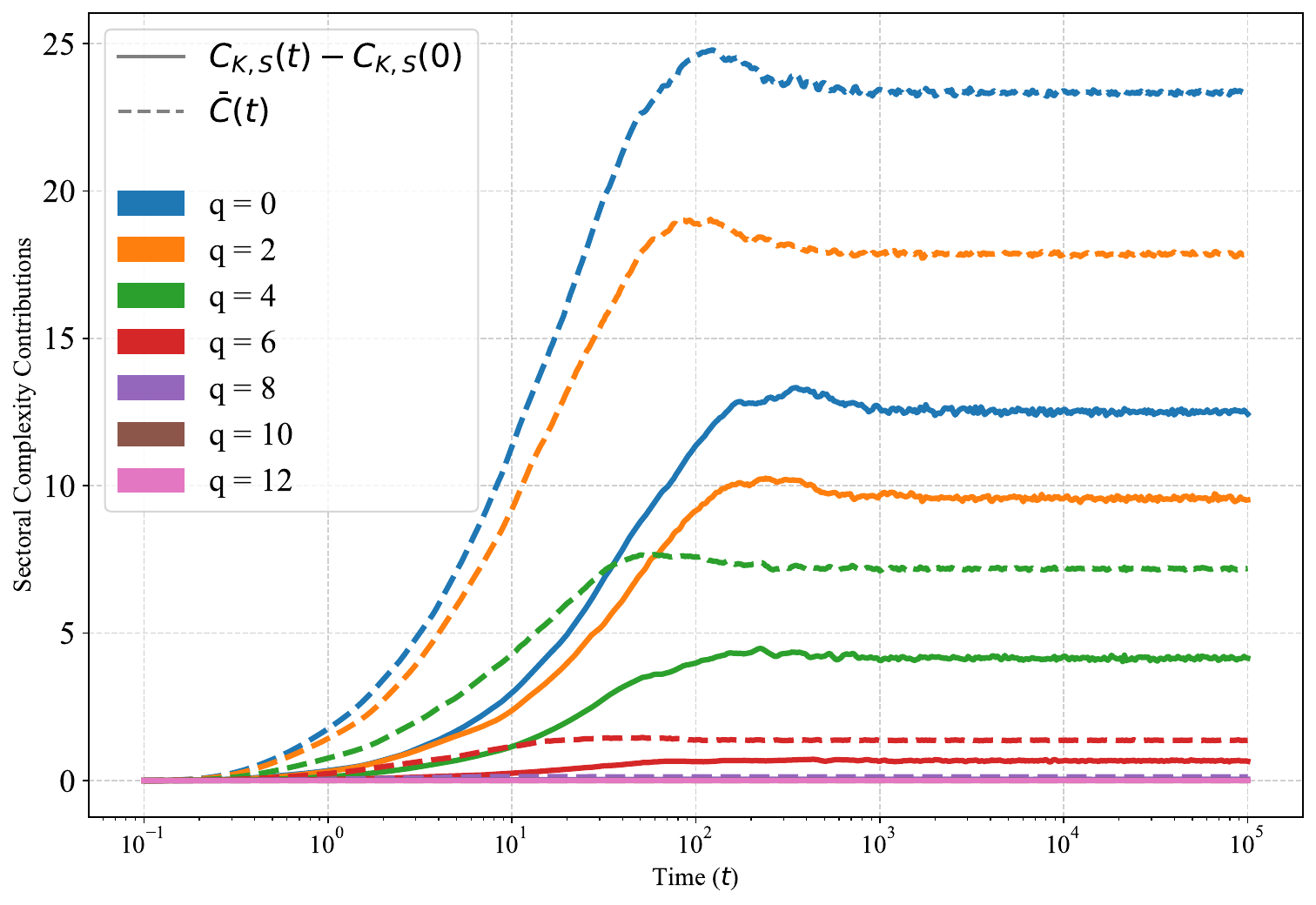}}\hfil 
\subfloat[$W=5.0,\theta=0.1$]{\includegraphics[width=0.47\linewidth]{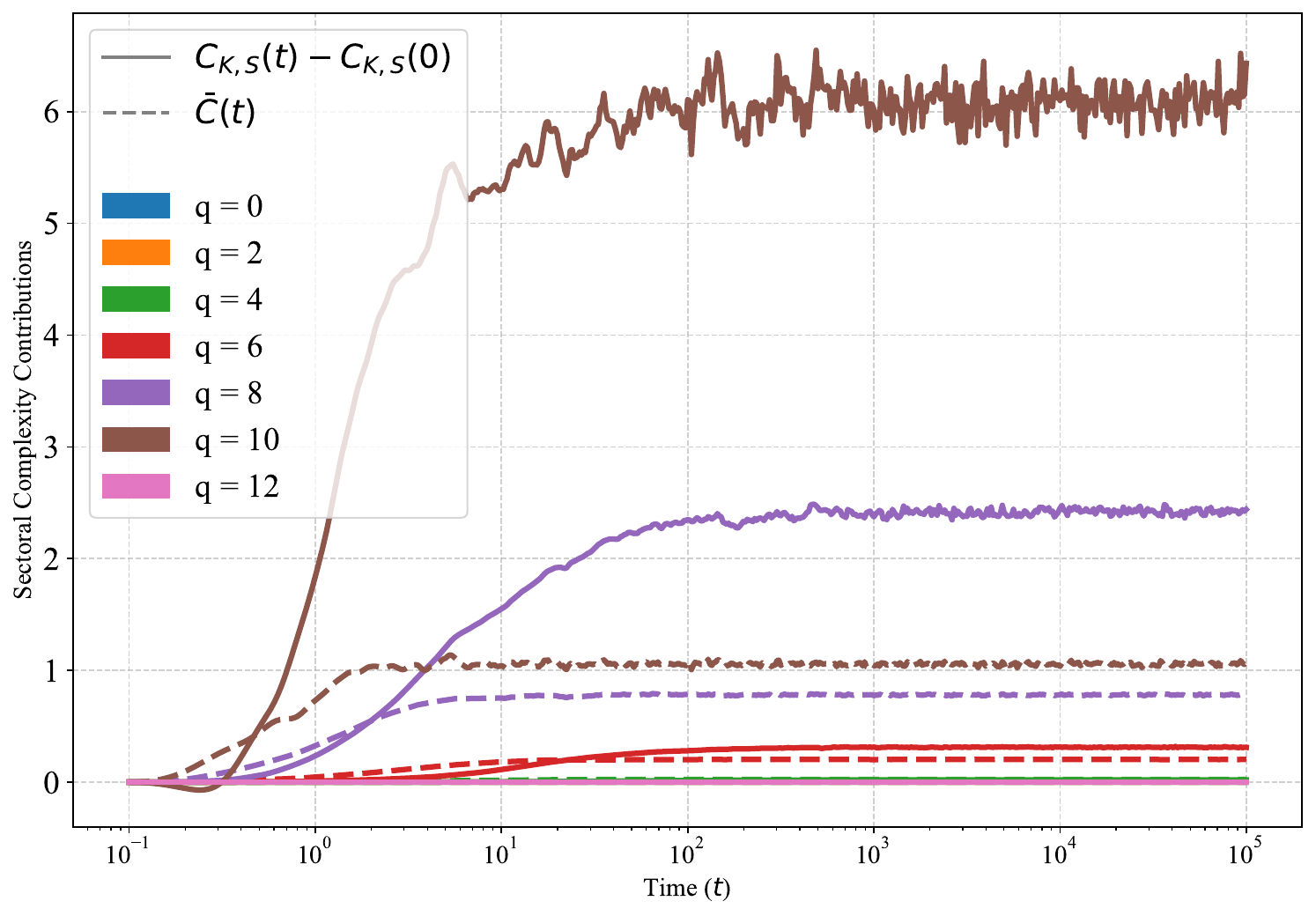}}\hfil
\subfloat[$W=5.0,\theta=0.5$]{\includegraphics[width=0.47\linewidth]{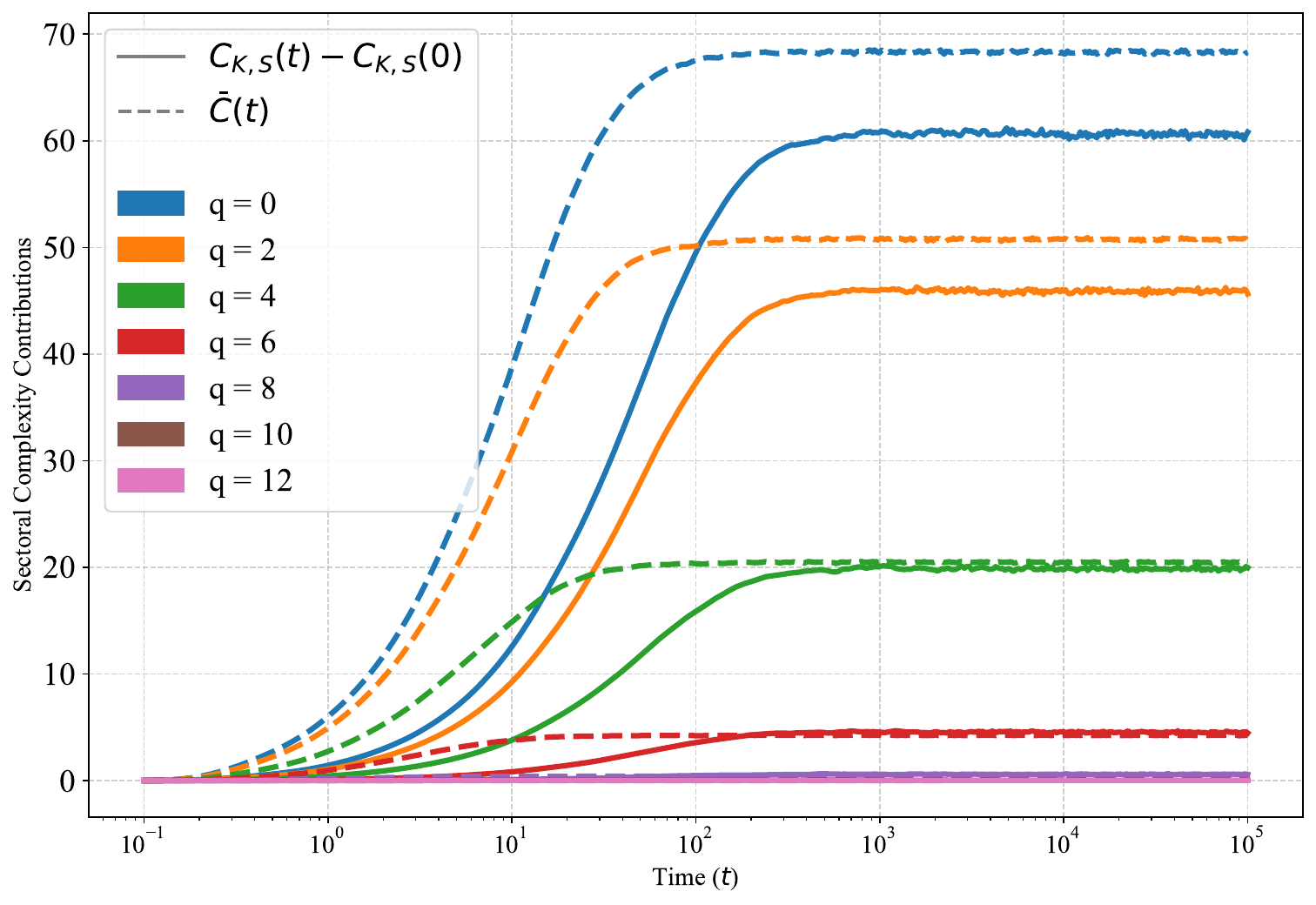}}\hfil 
\caption{Figures showing the discrepancy between the different intrasector diffusive measures for the TFS states at different tilt angles and different potential strengths. We compare the individual sector contributions to the diffusive complexity using an unsummed form of the $\bar{C}$ measure described in \cite{caputa2025block} (dashed lines), and the unsummed projective symmetric complexity introduced in this work.}
\label{fig:sectoraldiscrepancytfs}
\end{figure}

\begin{figure}[!htbp]
\centering
\subfloat[$W=1.0,\theta=0.1$]{\includegraphics[width=0.47\linewidth]{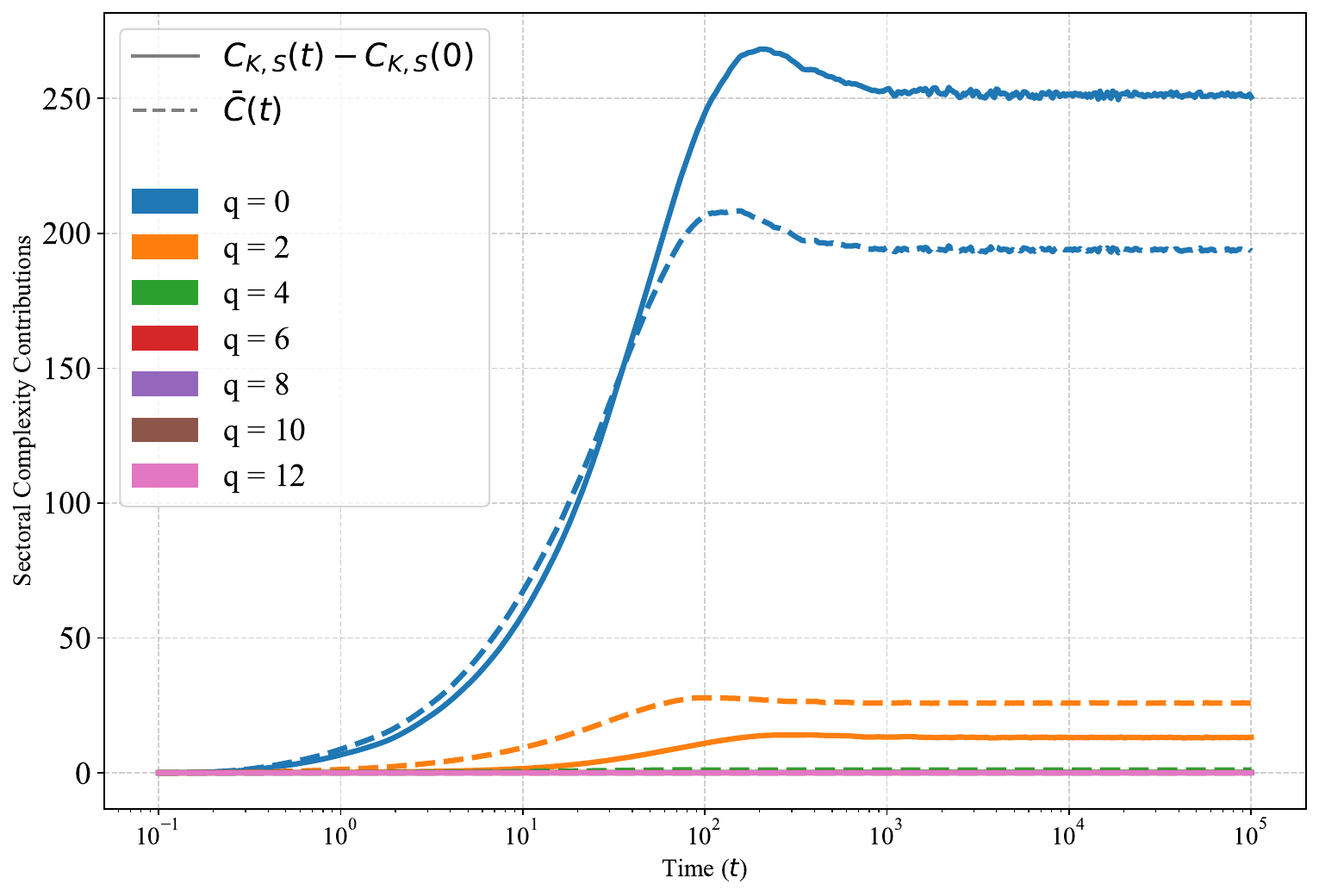}}\hfil
\subfloat[$W=1.0,\theta=0.5$]{\includegraphics[width=0.47\linewidth]{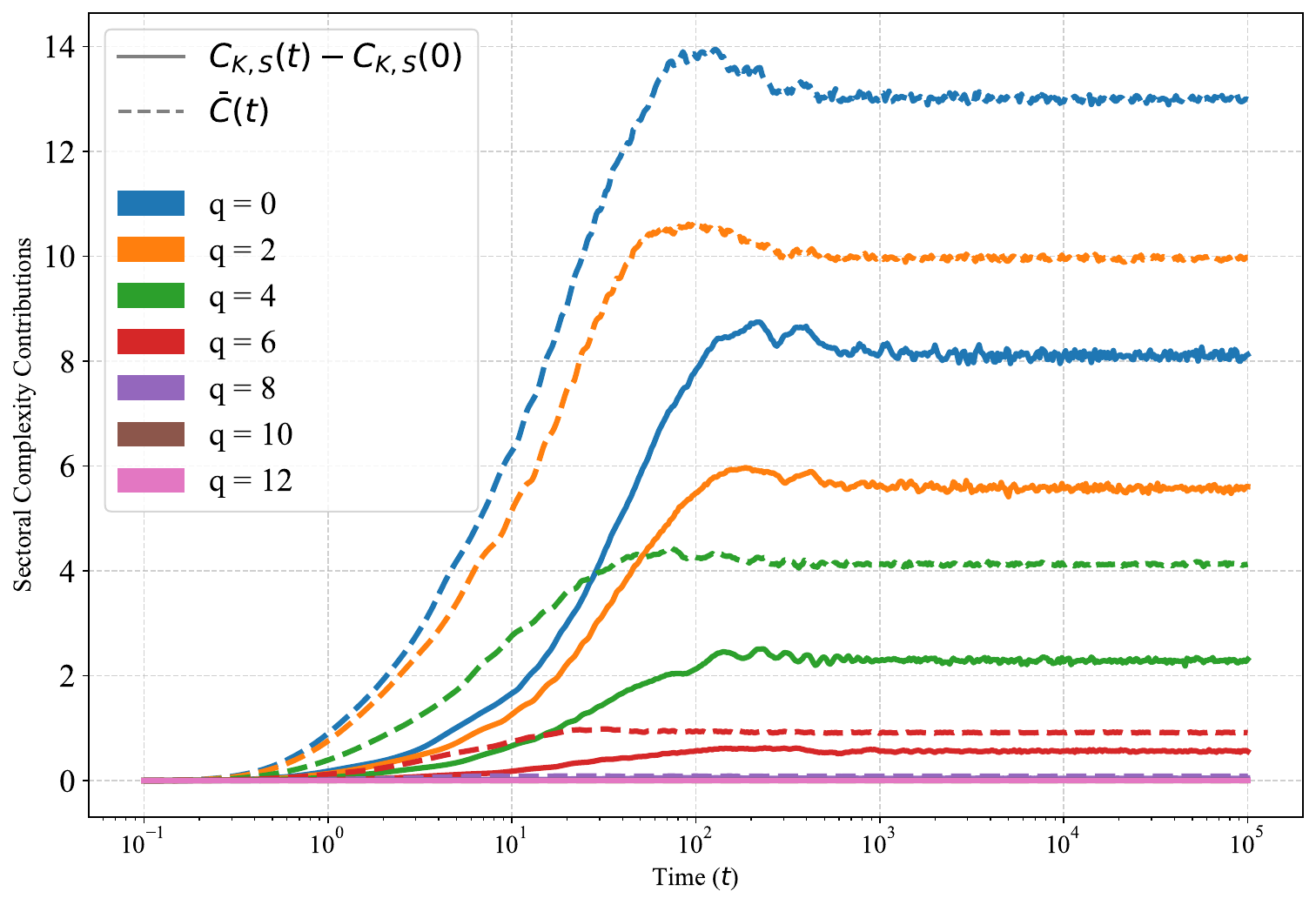}}\hfil 
\subfloat[$W=5.0,\theta=0.1$]{\includegraphics[width=0.47\linewidth]{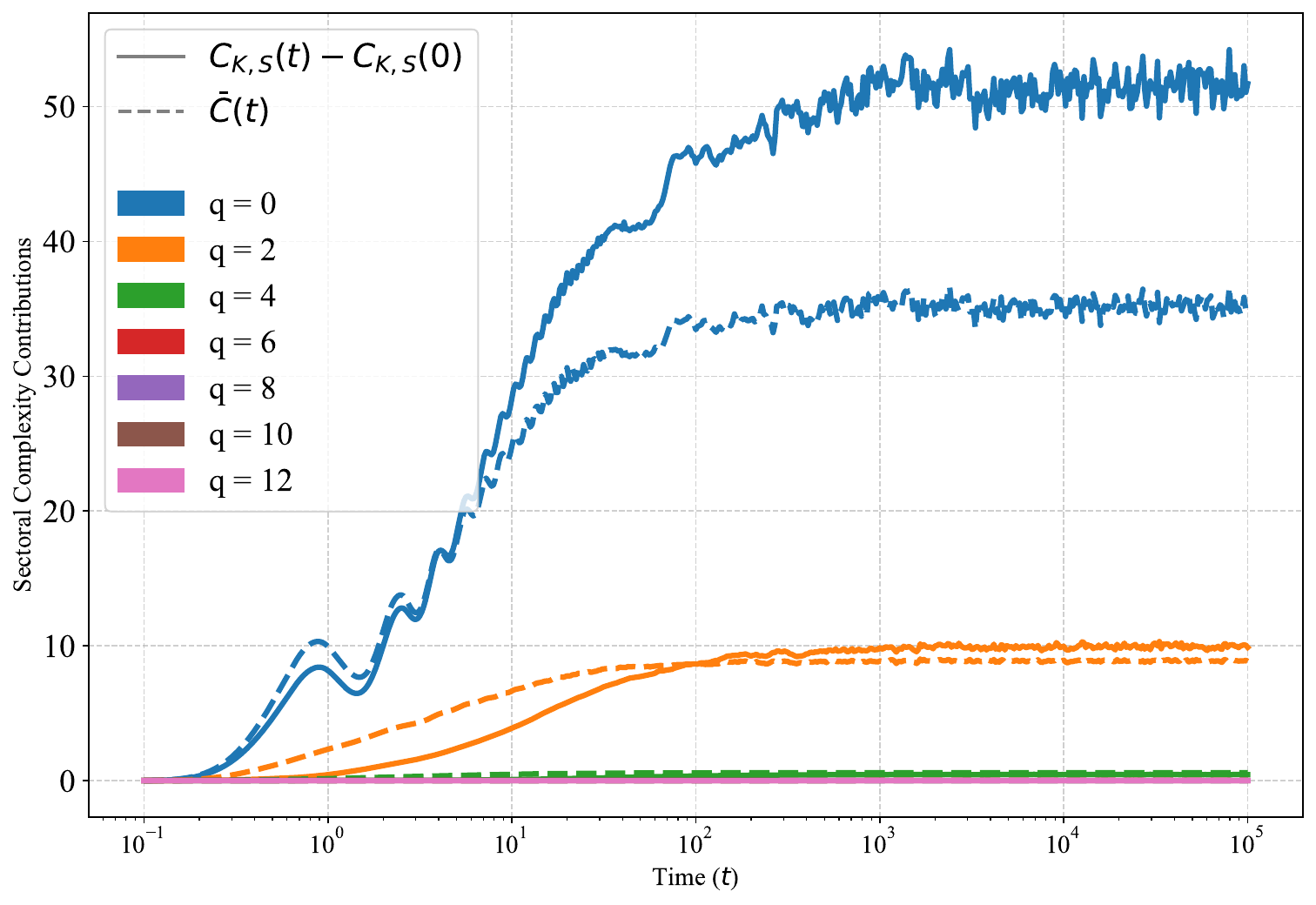}}\hfil
\subfloat[$W=5.0,\theta=0.5$]{\includegraphics[width=0.47\linewidth]{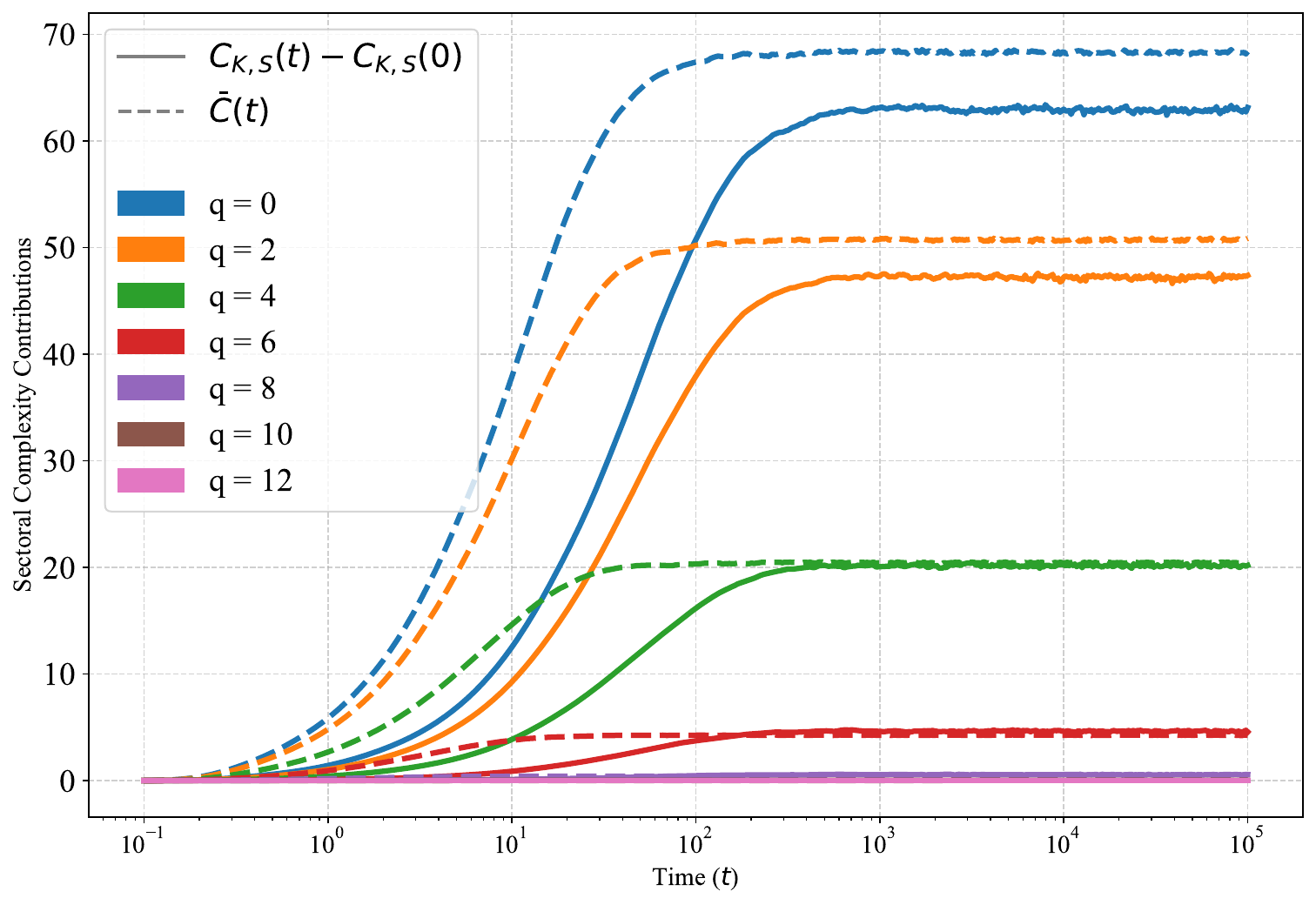}}\hfil 
\caption{Figures showing the discrepancy between the different intrasector diffusive measures for the TNS states at different tilt angles and different potential strengths. We compare the individual sector contributions to the diffusive complexity using an unsummed form of the $\bar{C}$ measure described in \cite{caputa2025block} (dashed lines), and the unsummed projective symmetric complexity introduced in this work.}
\label{fig:sectoraldiscrepancytns}
\end{figure}

In the main text, we compare the projective symmetric complexity $C_{K,S}(t)$ to the symmetry-resolved Krylov complexity measure $\bar{C}(t)$ presented in \cite{caputa2025block}. We call these `diffusive' measures because both are intended to capture the diffusive growth of spread complexity within a particular symmetry sector. This appendix serves to elaborate on the comparisons by providing further plots for comparison.
\\ \\
First, in Fig. \ref{fig:discrepancyapp} we present the partner plots of Fig. \ref{fig:discrepancy}; these are identical parameter choices but for the TNS states, which are not presented in the text.\\ \\
\textbf{TNS Diffusive Discrepancy}
\FloatBarrier
In addition to measuring the summed contributions for all symmetry sectors, we include a comparison of the individual contributions for the TFS states in Fig. \ref{fig:sectoraldiscrepancytfs} and perform an identical comparison for the TNS states in Fig. \ref{fig:sectoraldiscrepancytns}.
\clearpage
\textbf{TFS Diffusive Measures by Sector}
\FloatBarrier

\clearpage
\textbf{TNS Diffusive Measures by Sector}

\subsection{Symmetric Complexity's Chaos `Bump'}
\label{app:symmBump}
\begin{figure}[!htbp]
\centering
\subfloat[TFS $W=1.0$]{\includegraphics[width=7cm]{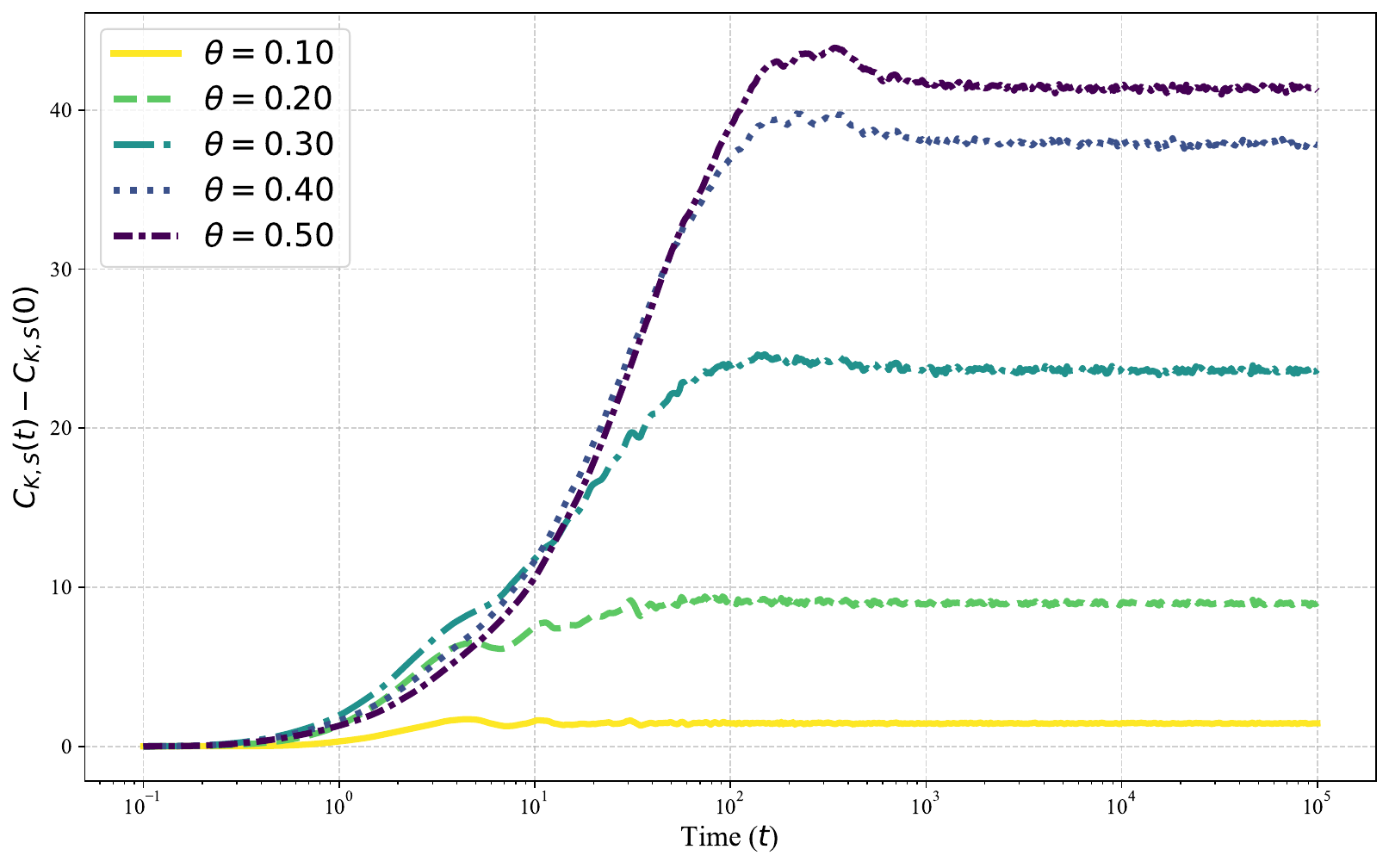}}\hfil
\subfloat[TNS $W=1.0$]{\includegraphics[width=7cm]{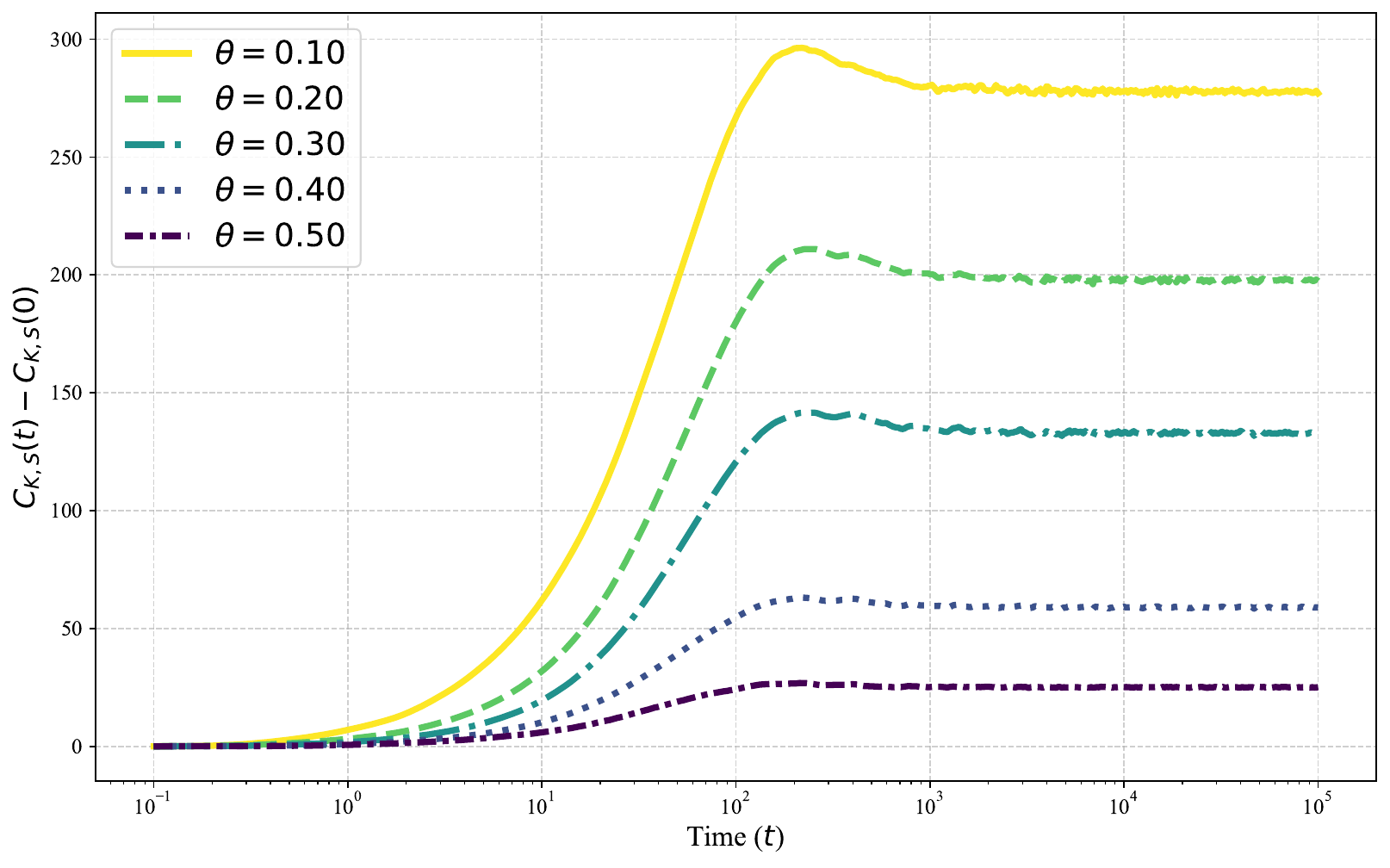}}\hfil 

\subfloat[TFS $W=2.0$]{\includegraphics[width=7cm]{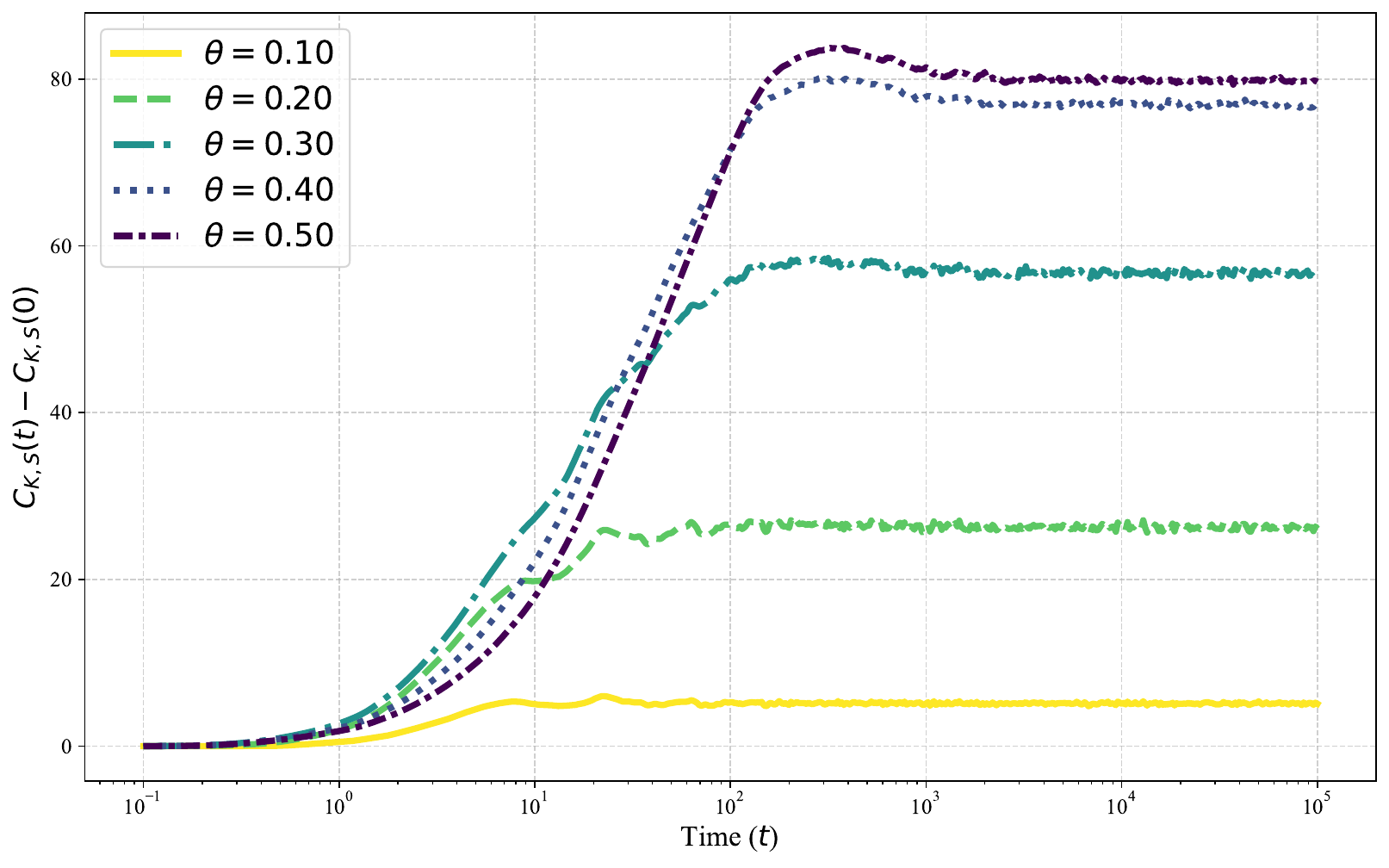}}\hfil
\subfloat[TNS $W=2.0$]{\includegraphics[width=7cm]{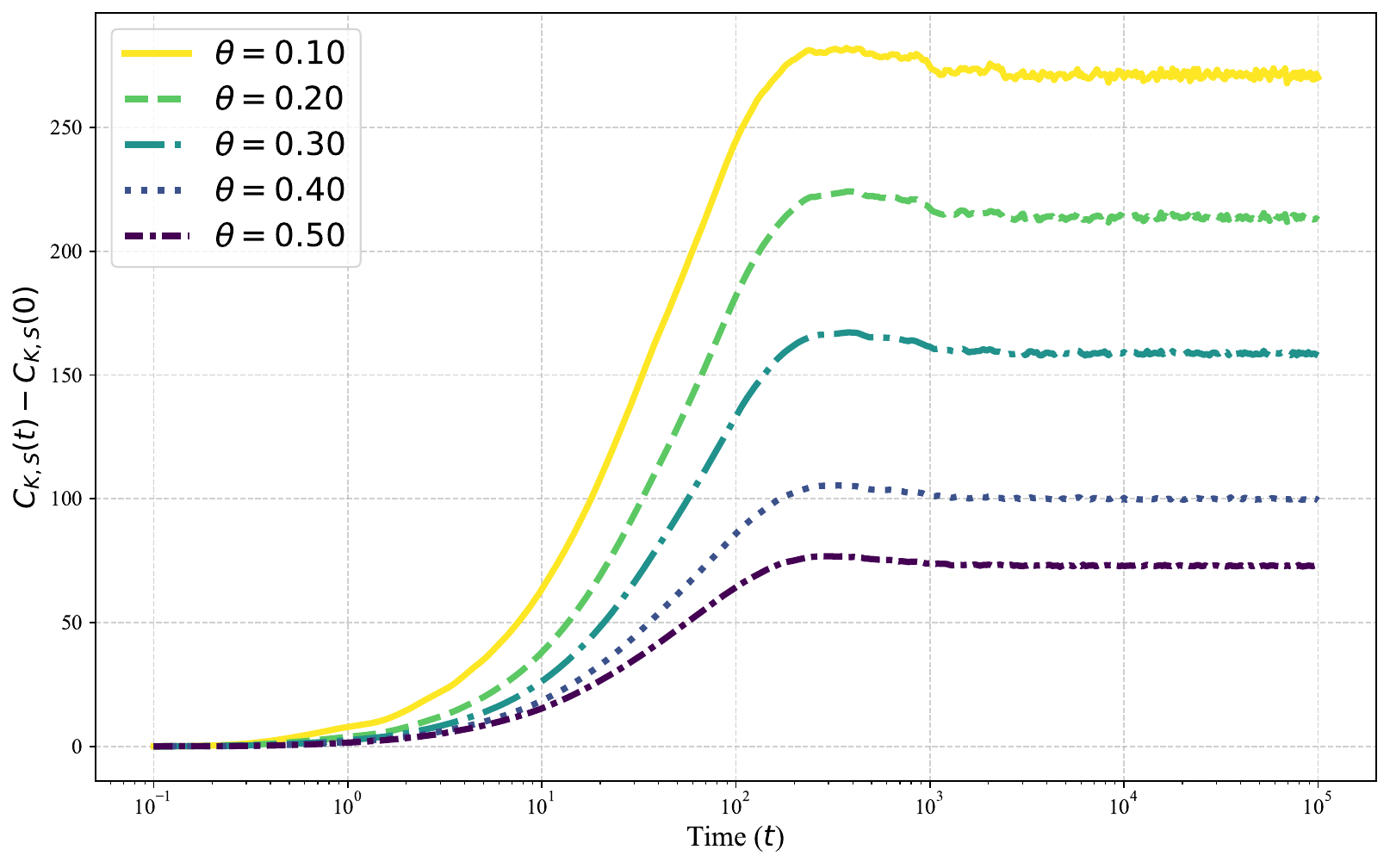}}\hfil 

\subfloat[TFS $W=2.5$]{\includegraphics[width=7cm]{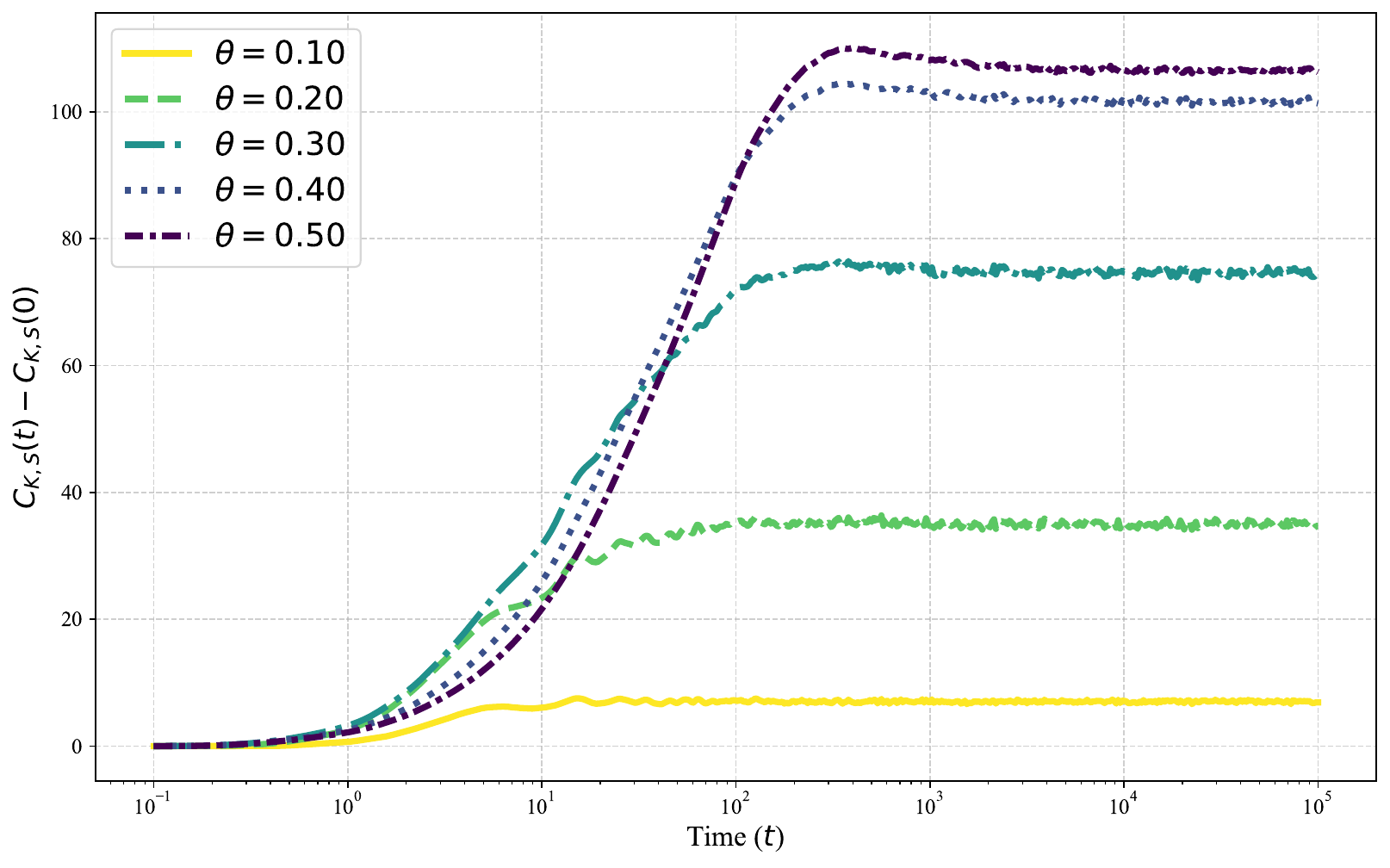}}\hfil
\subfloat[TNS $W=2.5$]{\includegraphics[width=7cm]{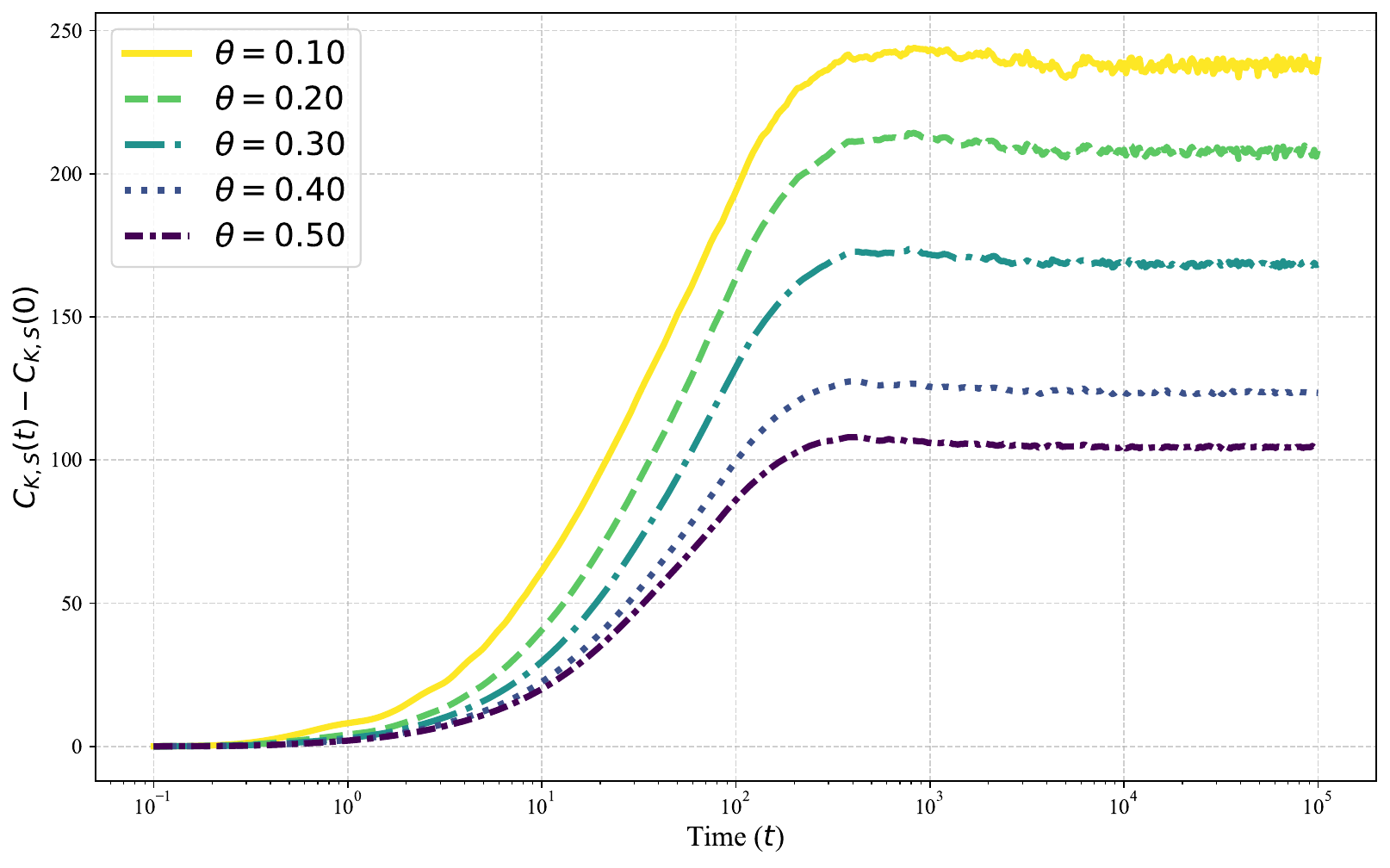}}\hfil 
\caption{Figures showing the enhanced sensitivity of the symmetric complexity to the ETH phase of the system.}
\label{fig:appsymmkc}
\end{figure}

In sec. \ref{sec:KCsymmres}, we justify choosing to identify the crossing of the symmetric complexities as a QME. In this justification, we make reference to the fact that Krylov complexity demonstrates sensitivity to the phase of the system being studied via different mechanisms. As was mentioned, chaotic systems can exhibit a `bump'; a pre-plateau peak that is higher than the saturation value of the complexity \cite{Balasubramanian:2023kwd, Baggioli:2024wbz}. The symmetric complexity produces such a peak far away from the transition region in the ETH regime. Interestingly, the symmetric complexity appears to exhibit this pre-plateau peak even when it is absent for the total Krylov complexity. This result is shown in Fig. \ref{fig:appsymmkc}. The peak becomes less-pronounced as $W$ is increased, and is present  (but not overtly so) at $W=2.5$. These should be contrasted with Fig. \ref{fig:KC} in the main text, which exhibits no clear pre-plateau peak whatsoever.

\subsection{Symmetry-Resolved Krylov Complexity and the Quantum Mpemba Effect}
\label{app:SRKCQME}
\begin{figure}[!htbp]
\centering
\subfloat[TFS $W=3.0$]{\includegraphics[width=7cm]{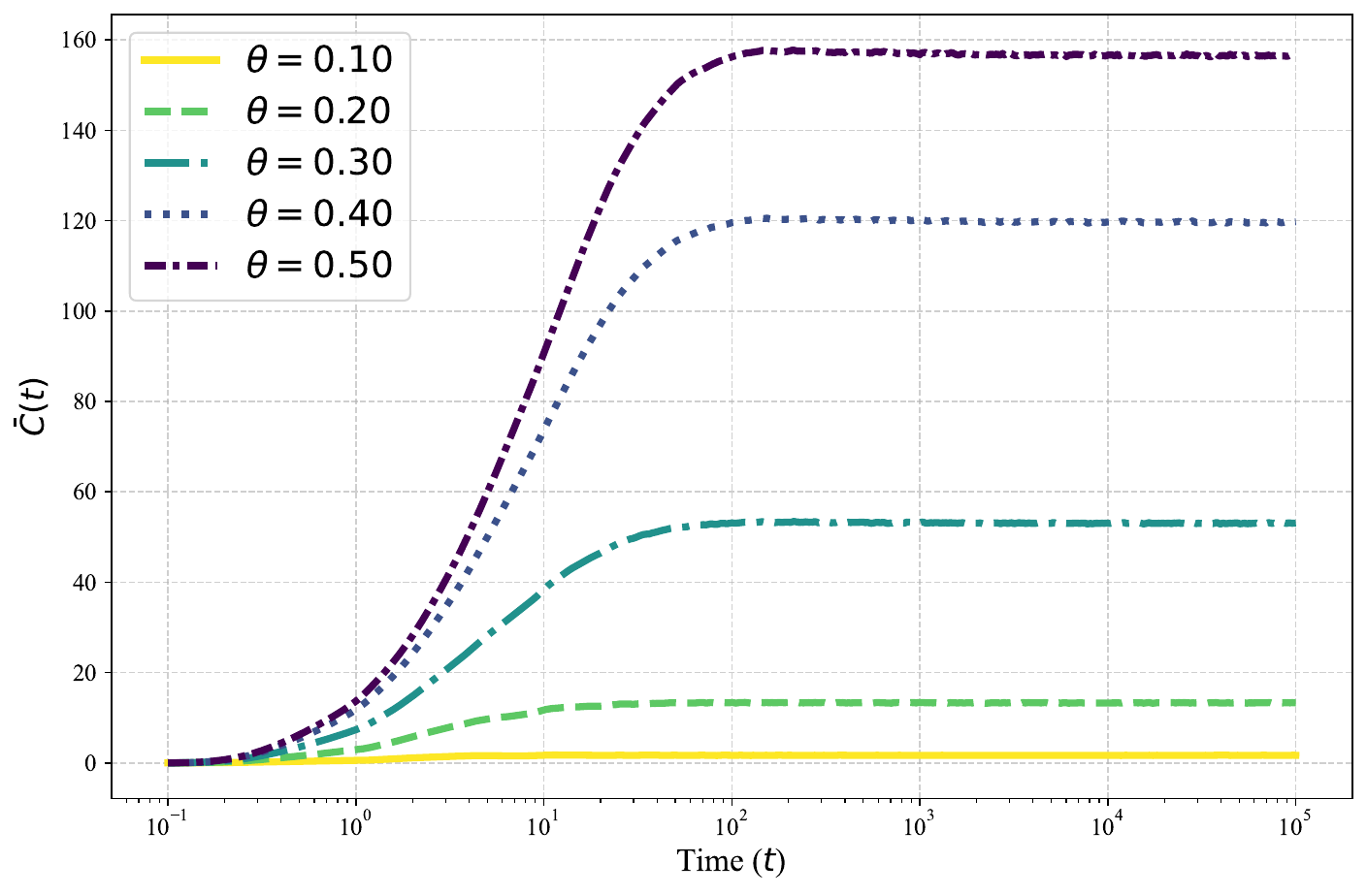}}\hfil
\subfloat[TNS $W=3.0$]{\includegraphics[width=7cm]{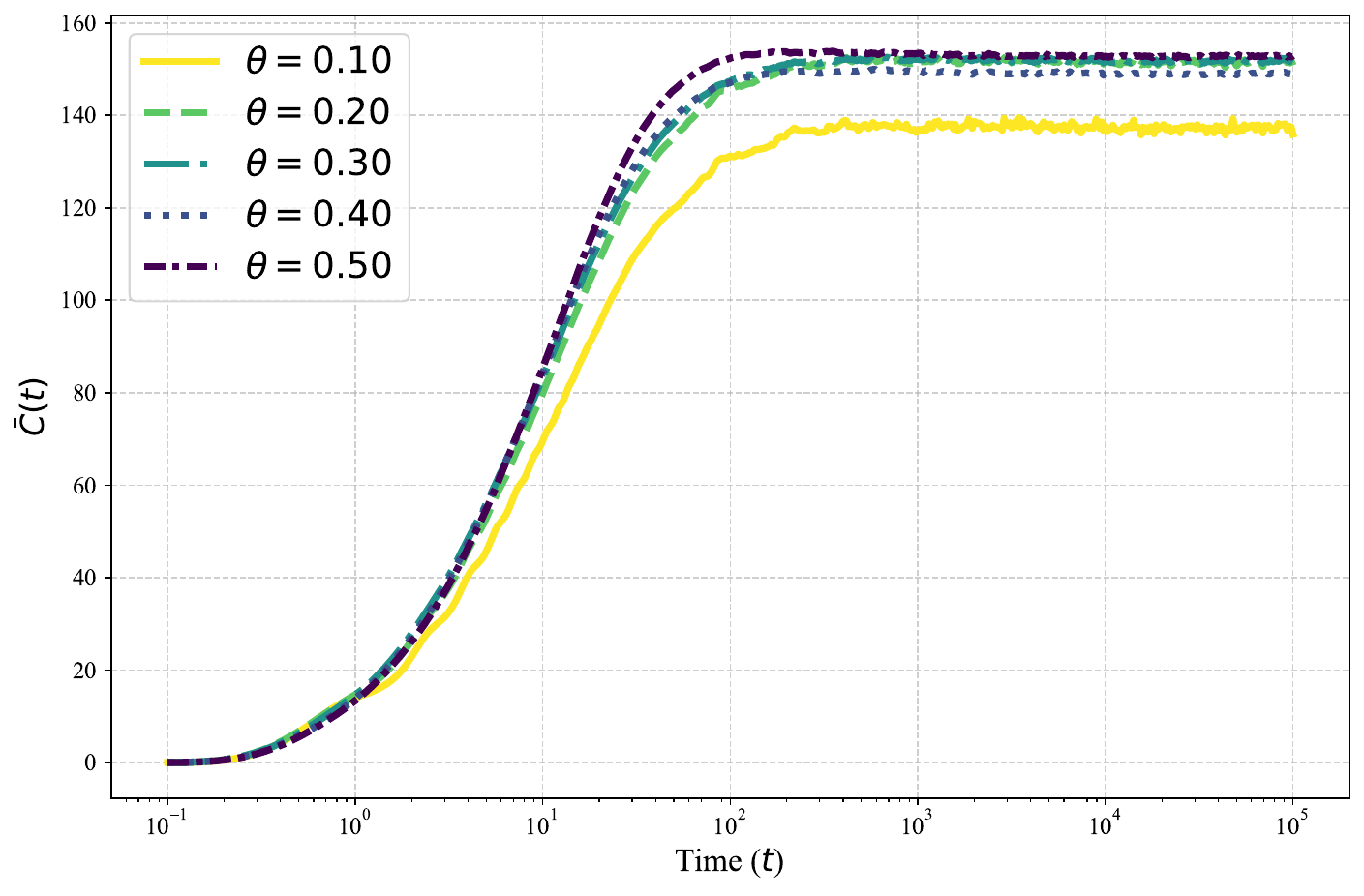}}\hfil 

\subfloat[TFS $W=3.3$]{\includegraphics[width=7cm]{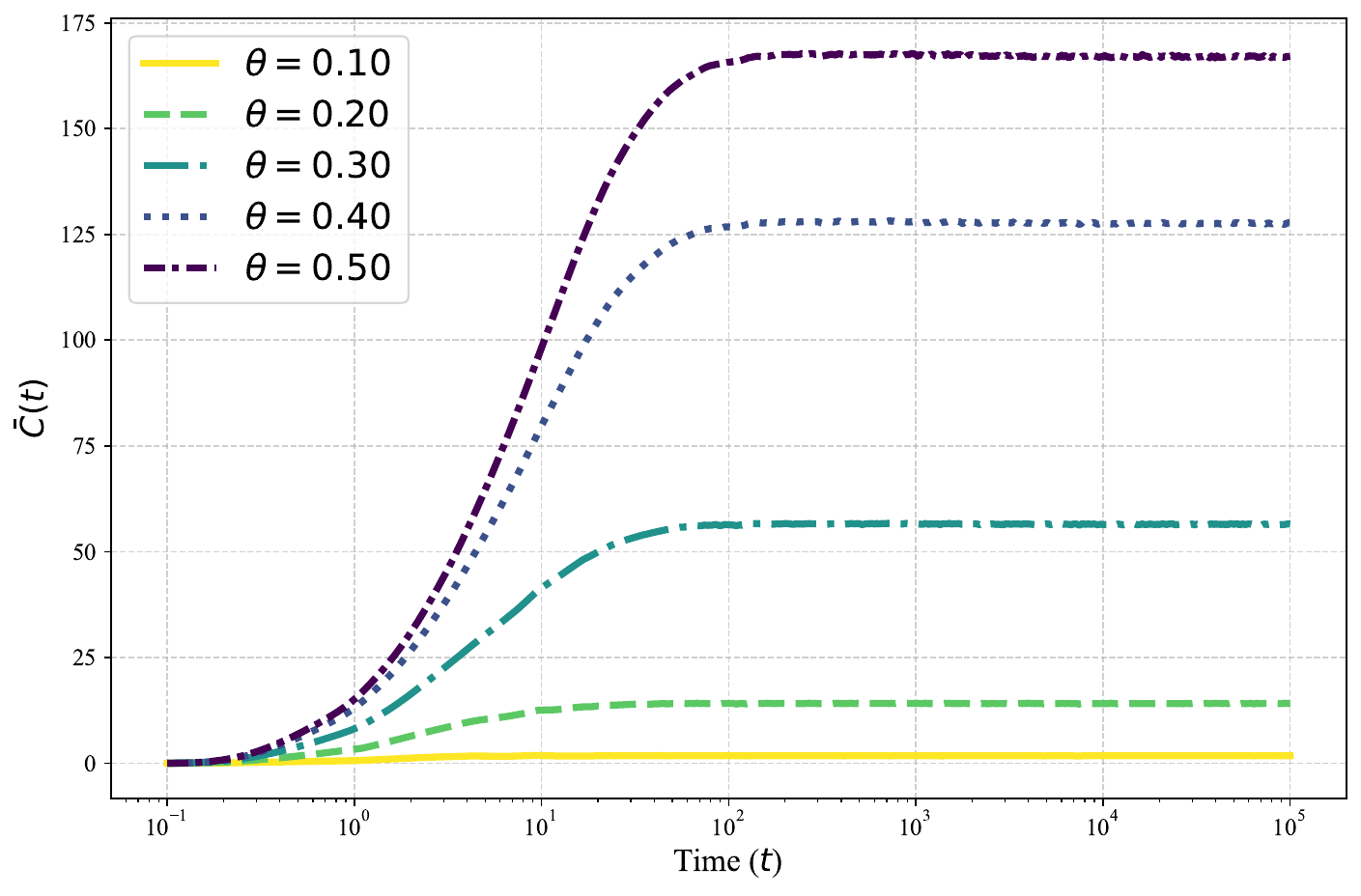}}\hfil
\subfloat[TNS $W=3.3$]{\includegraphics[width=7cm]{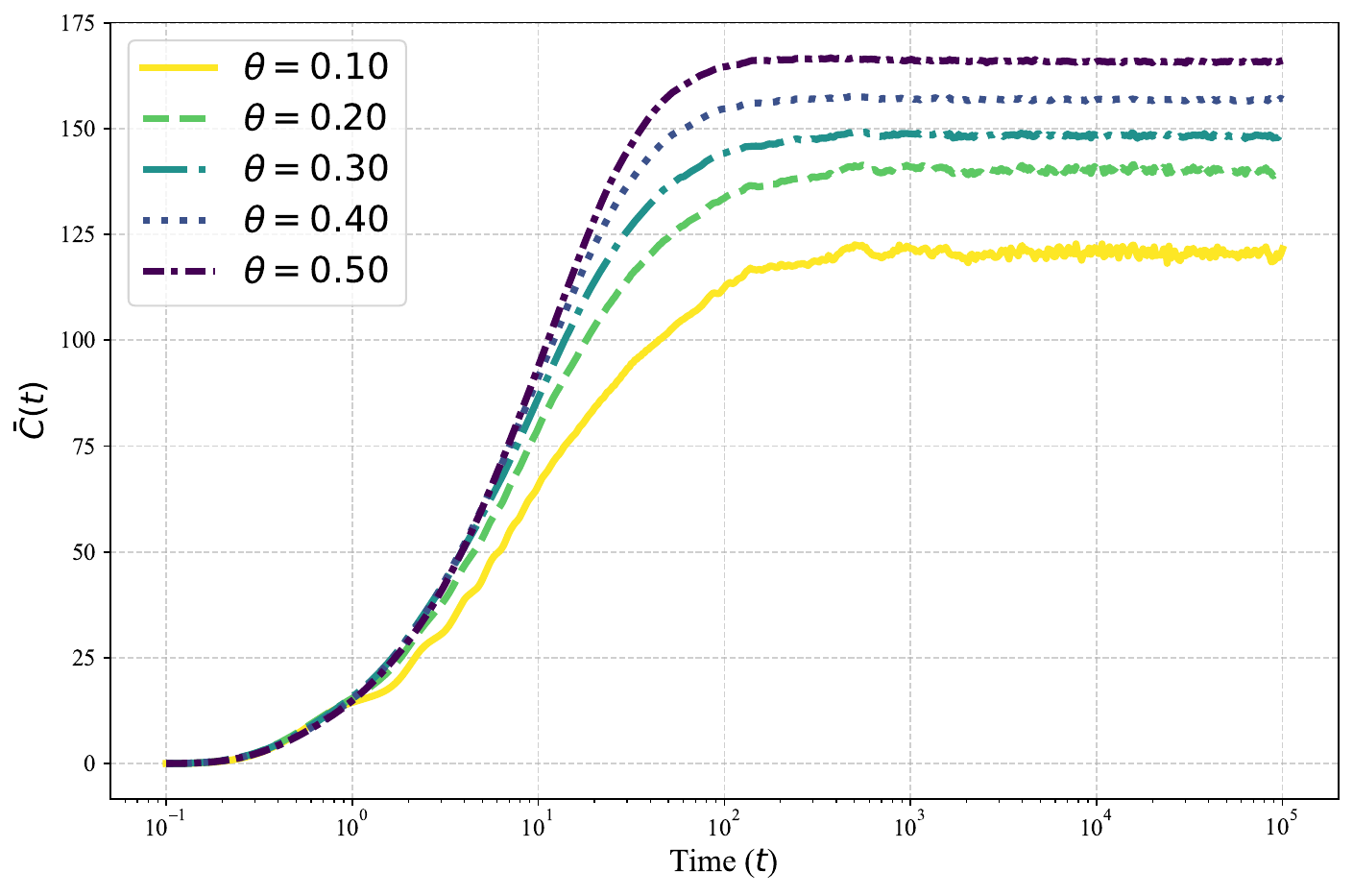}}\hfil 

\subfloat[TFS $W=3.6$]{\includegraphics[width=7cm]{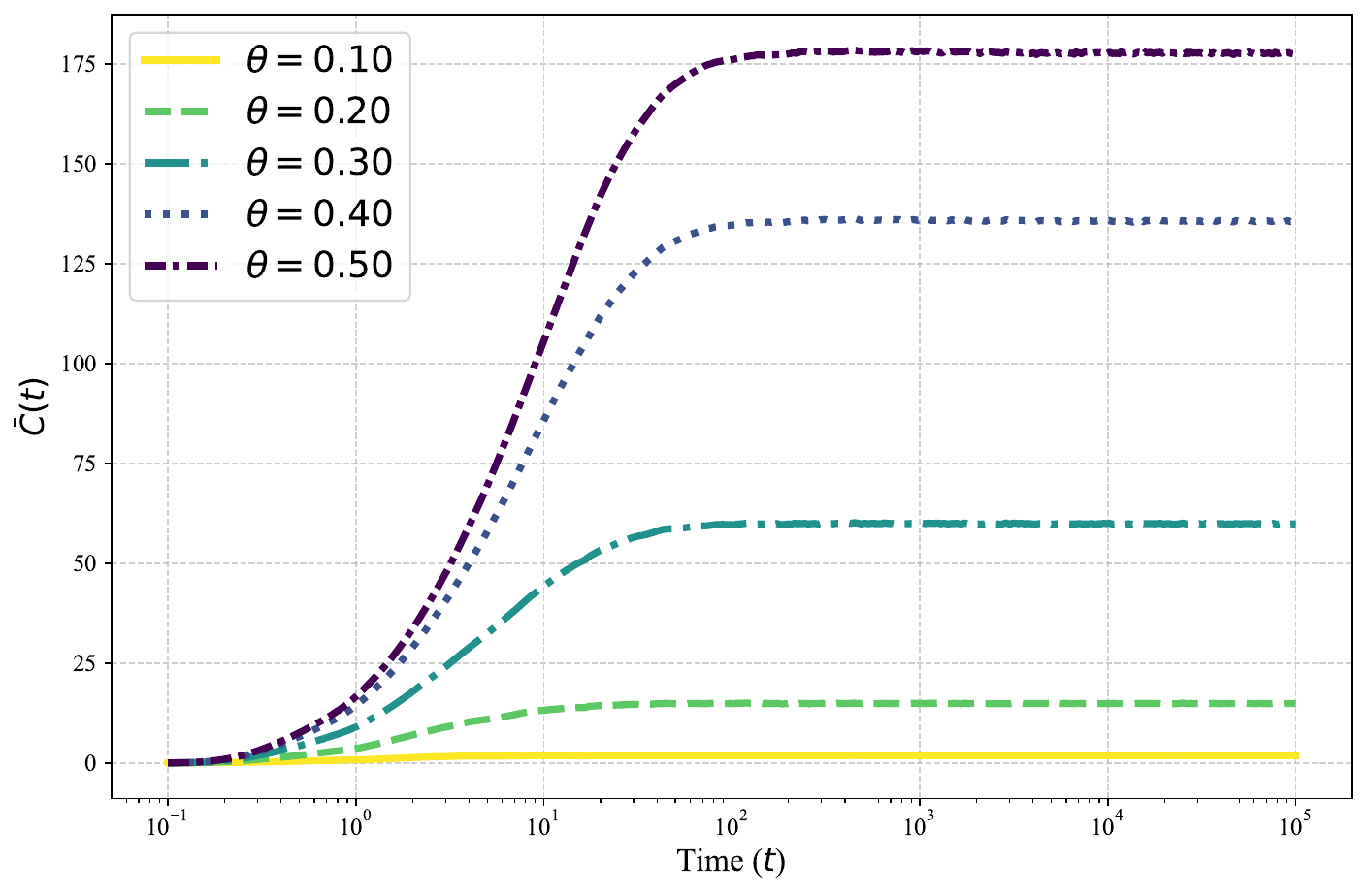}}\hfil
\subfloat[TNS $W=3.6$]{\includegraphics[width=7cm]{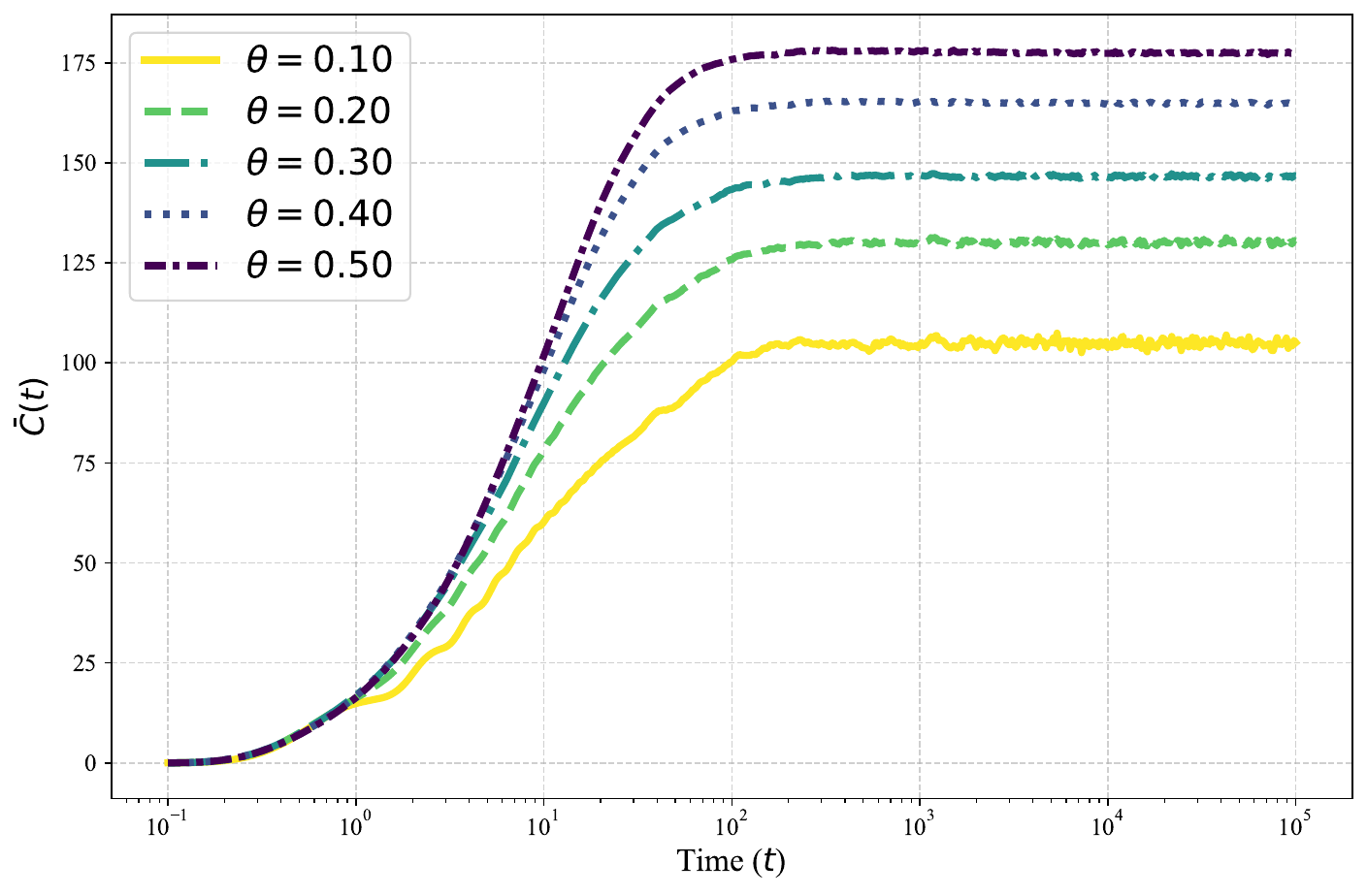}}\hfil 
\caption{Figures showing the probability-weighted SRKC ($\bar{C}(t)$) introduced in \cite{caputa2025block}. These plots should be compared to Fig. \ref{fig:diffusiveKC} in the main text.}
\label{fig:srkcqme}
\end{figure}

We show, in section \ref{sec:symmetricQME}, that the symmetric complexity is sensitive to the quantum Mpemba effect. Given that the Symmetry-Resolved Krylov Complexity (SRKC) shows similar features to that of symmetric complexity, albeit with nontrivial scaling in the features of the symmetric complexity, one might assume that the SRKC exhibits many of the same behaviors, such as crossing. If it were to do so, it would offer potential computational advantages over the symmetric complexity (at least in numerically intensive computations that require full reorthogonalization), due to the comparatively small sizes of the respective symmetry sector bases. The partner plots to Fig. \ref{fig:diffusiveKC} are presented in Fig. \ref{fig:srkcqme}.
Regarding Fig. \ref{fig:srkcqme}, one immediately notices that $\bar{C}(t)$ does not exhibit Mpemba-like crossings that are exhibited by $C_{K,S}(t)$. This is interesting for two reasons. First, both measures quantify the total symmetry sector diffusion, thus one might reasonably expect both to demonstrate similar crossing behavior. Second, there are large discrepancies between $\bar{C}(t)$ and $C_{K,S}(t)$, and understanding how these discrepancies might enhance/mask certain effects is a useful investigation in its own right.
\\ \\
Some features of the QME are still present; for the TNS states, the reordering of which tilts have higher saturation values does occur, albeit at lower values of the potential strength, while the TFS states are always ordered by tilt. Indeed, the ordering of saturation values of the tilted Néel states transitions from reverse-ordered (in terms of the tilt $\theta$) at $W=2.0$ to ordered by the tilt at $W=3.1$ (both not shown). Throughout the reordering transition, however, no clear Mpemba-like crossing is observed.

\subsection{Crossing Analysis}
\label{app:crossinganalysis}
\begin{figure}[!htb]
    \centering
    \includegraphics[width=0.95\linewidth]{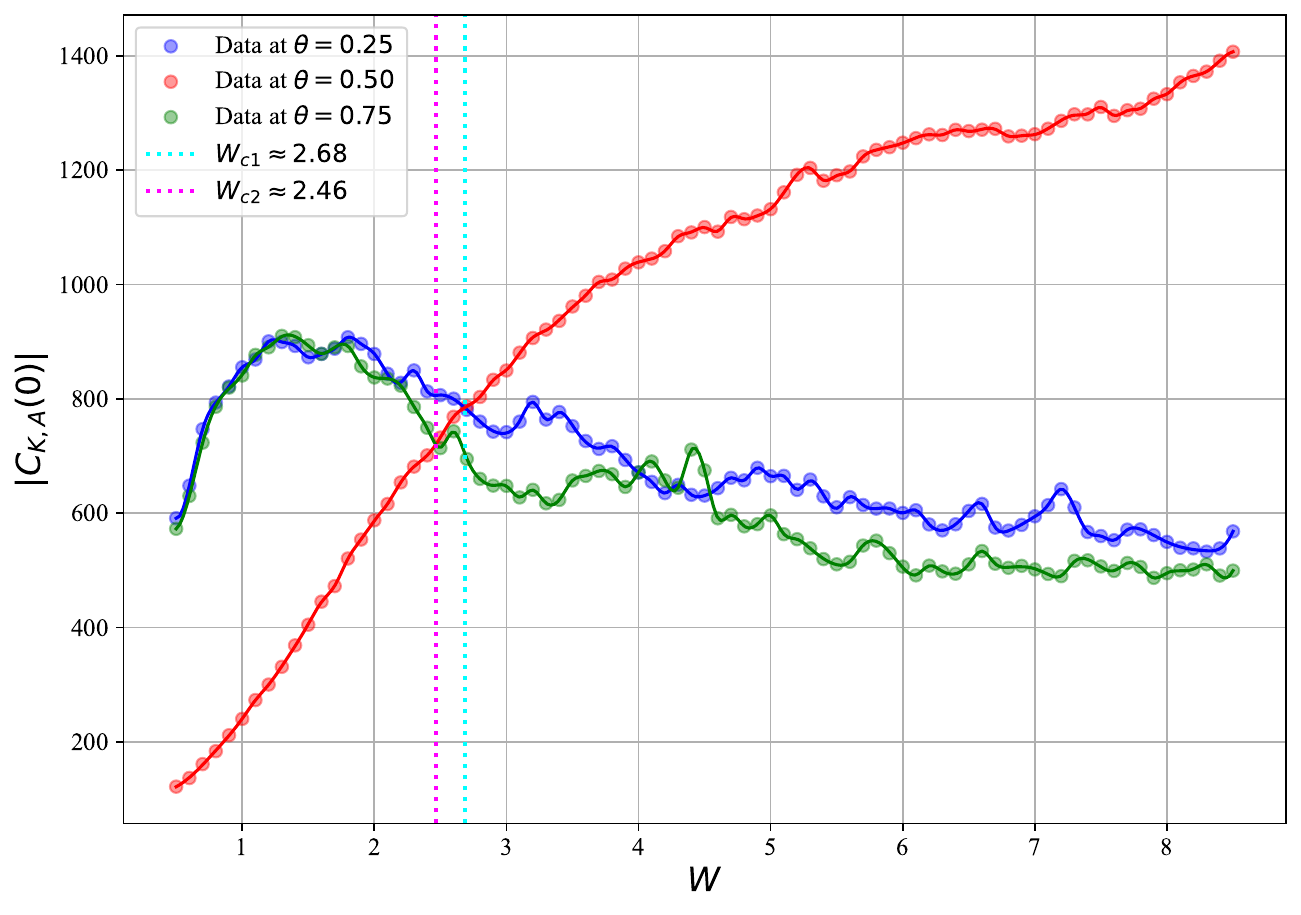}
    \caption{Peak analysis of the structural complexity, approximately identifying the values of $W$ for which the central (most asymmetric) peak at $\theta=0.5$ begins to dominate.}
    \label{fig:crossingstructure}
\end{figure}

As an approximation for where the maximal tilt dominates the structural complexity, we track the approximate location of the peaks of the structural complexity for the Néel state, shown in Fig. \ref{fig:crossingstructure}. Given that there are initially two peaks, we approximately track both of these peaks and indicate the values of W where the peak at $\theta=0.5$ becomes dominant. The figure clearly shows that the region where the $\theta=0.5$ peak dominates is at $W\approx2.5$. The data here is for a single realisation of the system, only, and is thus susceptible to realisation-specific fluctuations that would be suppressed if averaged over many realisations.

\subsection{Additional Comments on Symmetry Restoration versus Thermalization}
\label{app:thermomajorization}
As was discussed in the main text, one key challenge for any asymmetry quantifiers is that they equate asymmetry with the distance from a thermal/steady state. This is not an unnatural choice, since the vanishing of an asymmetry is a necessary (but insufficient) condition for thermal equilibrium. While the asymmetry is thus \textit{indicative} of the nearness to a thermal state, it does not account for how similar the initial state might be to the thermal state.
\\ \\
One way to quantify the similarity between a state and the thermal state is by using thermomajorisation curves \cite{Horodecki_2013,Van_Vu_2025}. Before proceeding, we emphasise that the thermomajorisation is typically considered for systems in contact with a bath at fixed temperature. In the ETH regime, therefore, thermomajorisation curves are meaningful as the tilted initial states will generally equilibrate; that is, we are generally able to predict the final state beforehand. In the MBL regime, however, nearness to a \textit{hypothetical} thermal Gibbs state is not necessarily indicative of the possibility of reaching that state, since it is expected that the state the system will relax to is \textit{not} the Gibbs state. Thus, the interpretation of thermomajorization breaks down in the localizing regime. While this is a fundamental limitation of the protocol, it is still interesting to consider how the tilt operator may create states that `appear' thermal-like in their probability distributions. A further consideration is that, given the conventional procedure of bath to set the temperature of the Gibbs state, thermomajorization curves are typically plotted for states that will evolve to the \textit{same} final state. The tilt operation induces different effective temperatures and, therefore, different probability distributions for the Gibbs state. Thus, the thermomajorization curves we plot will measure the difference between a given tilt's resultant probability distribution and its \textit{own} thermal state.
\\ \\
\subsubsection{Construction of Thermomajorization Curves}
\begin{figure}[!htbp]
\centering
\subfloat[TFS $W=0.1$]{\includegraphics[width=6.5cm]{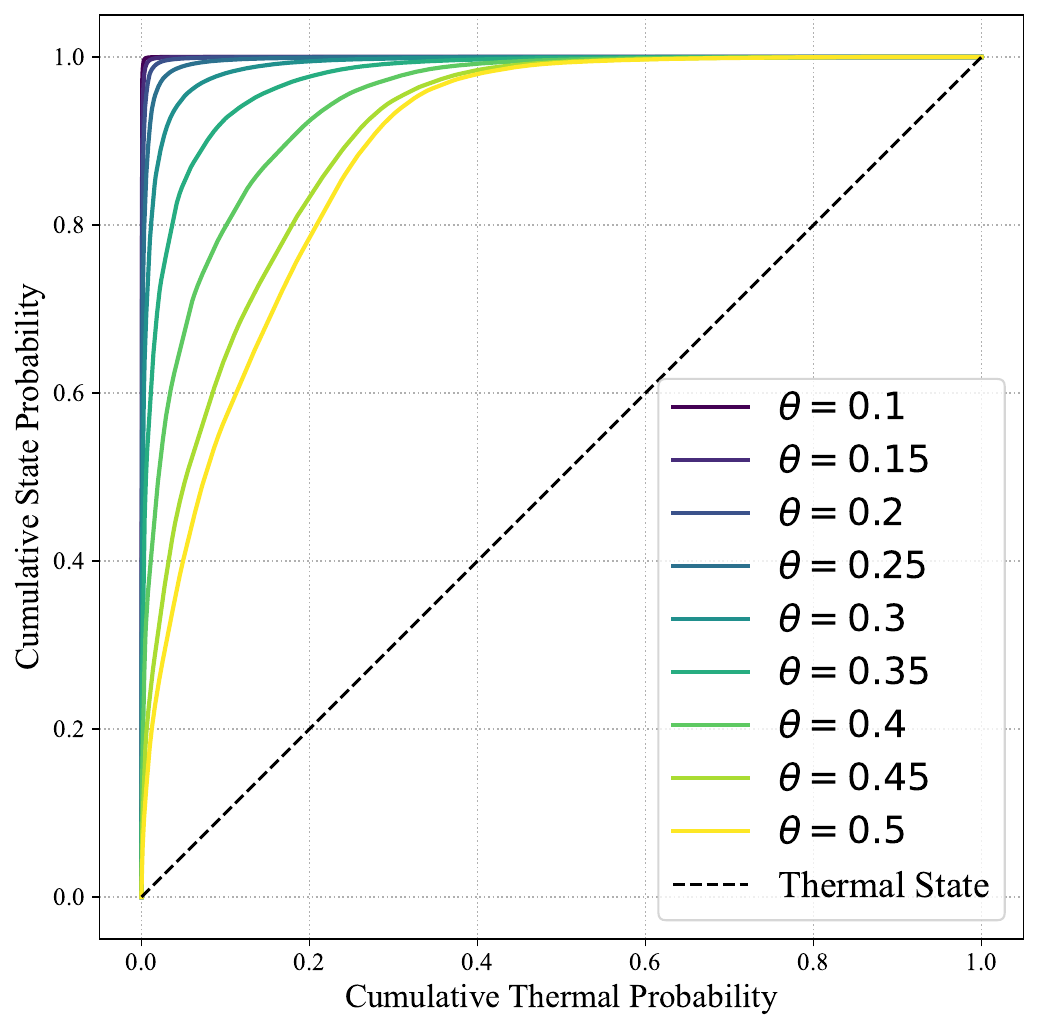}}\hfil
\subfloat[TNS $W=0.1$]{\includegraphics[width=6.5cm]{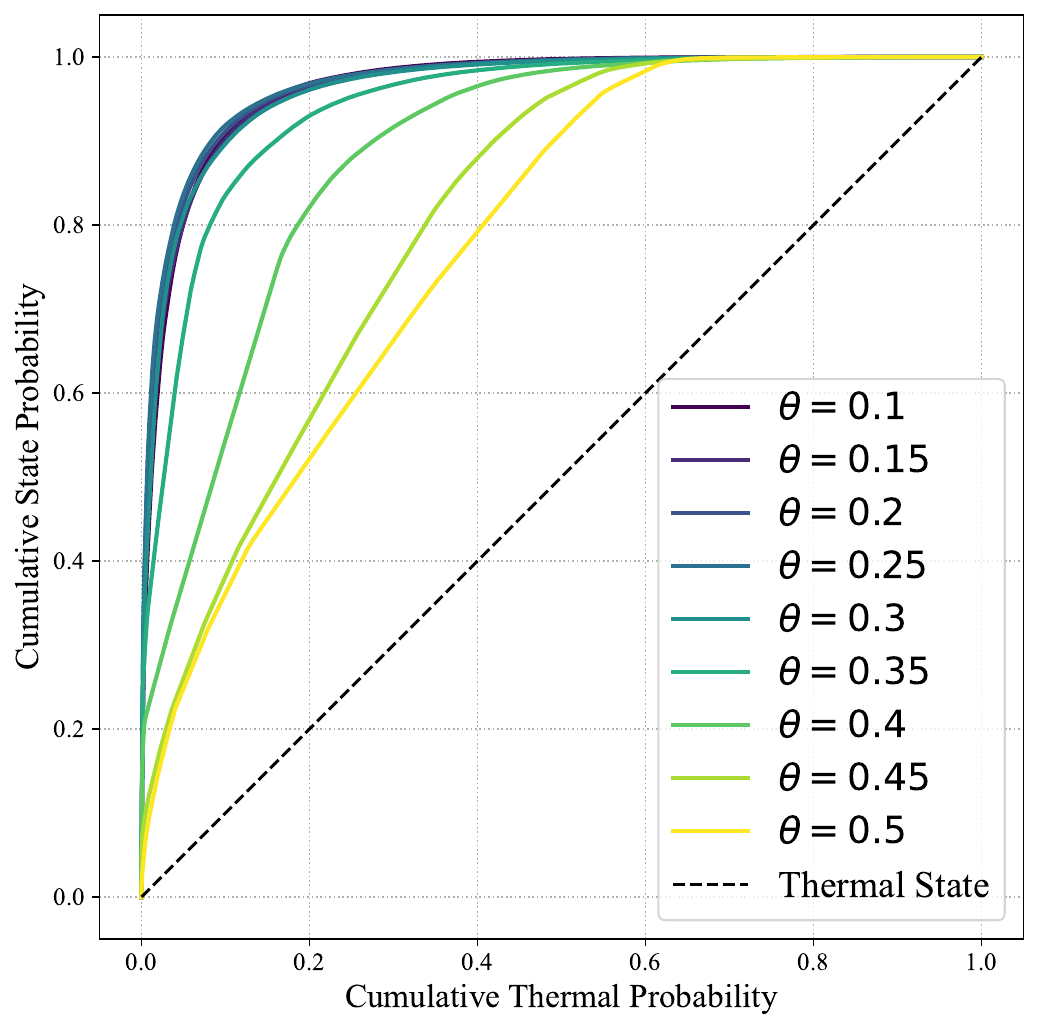}}\hfil 

\subfloat[TFS $W=3.3$]{\includegraphics[width=6.5cm]{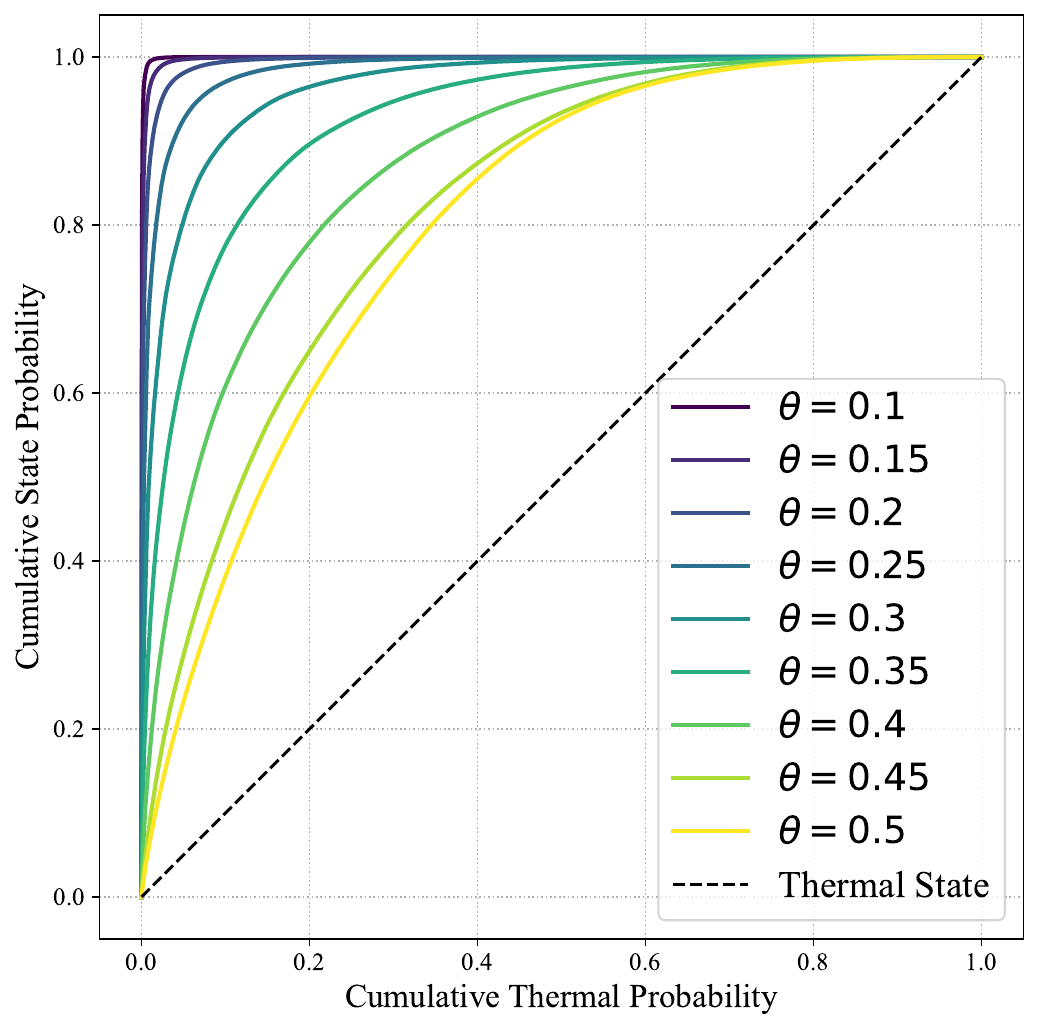}}\hfil
\subfloat[TNS $W=3.3$]{\includegraphics[width=6.5cm]{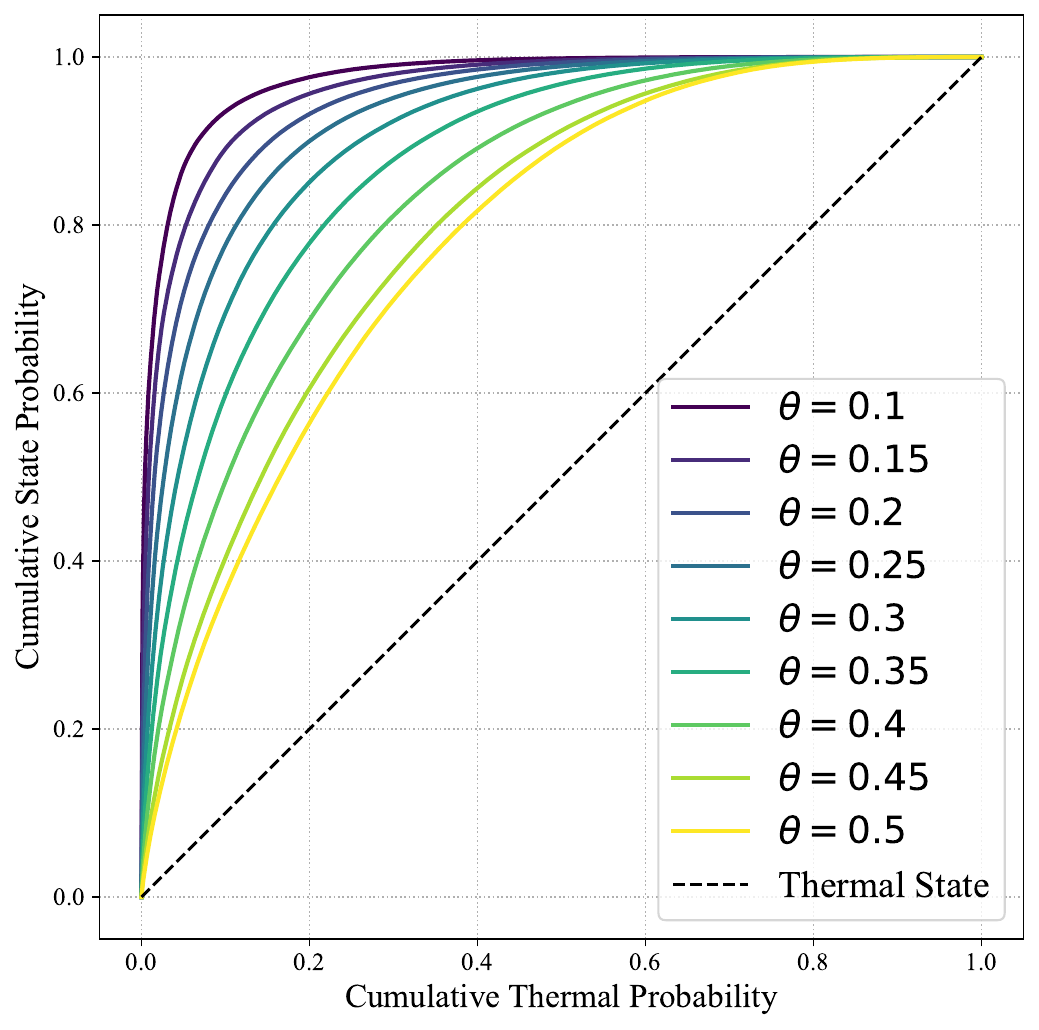}}\hfil 

\subfloat[TFS $W=5.0$]{\includegraphics[width=6.5cm]{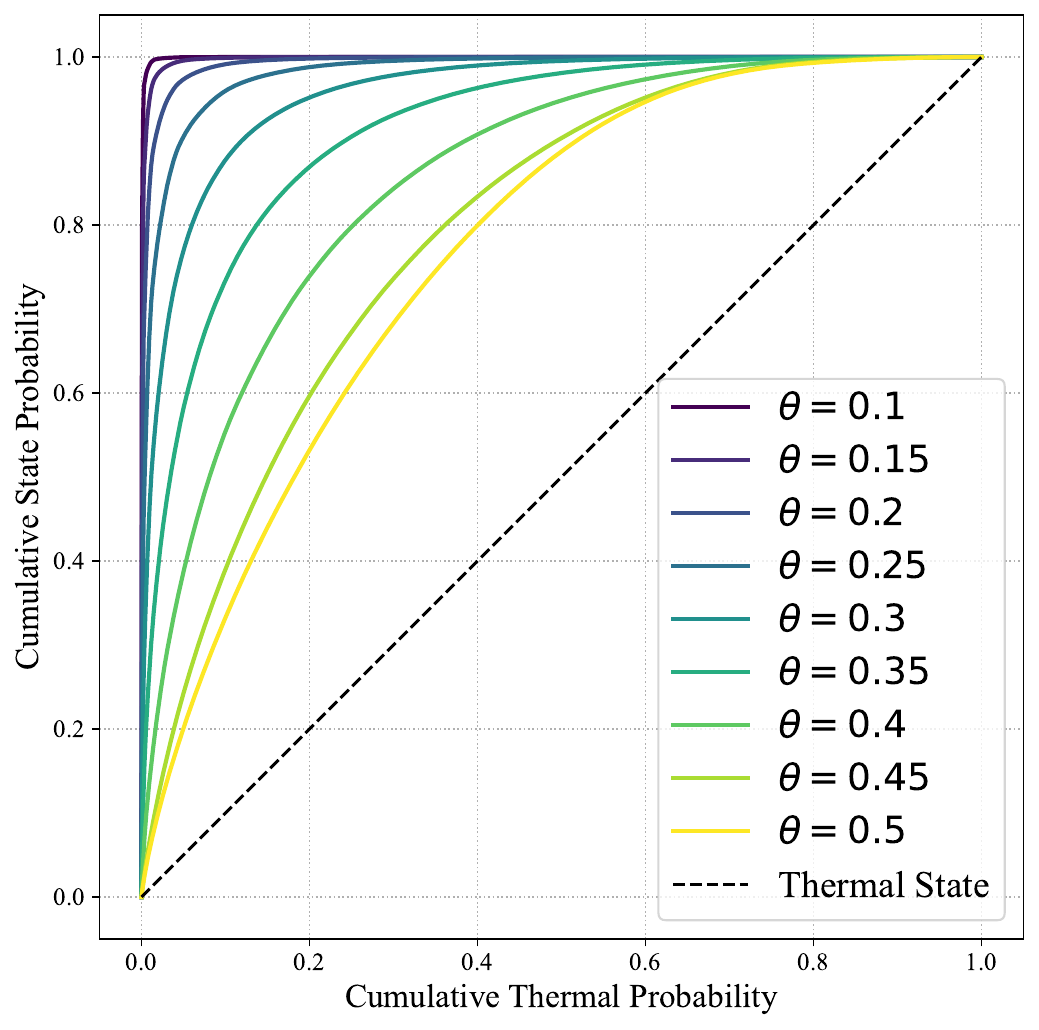}}\hfil
\subfloat[TNS $W=5.0$]{\includegraphics[width=6.5cm]{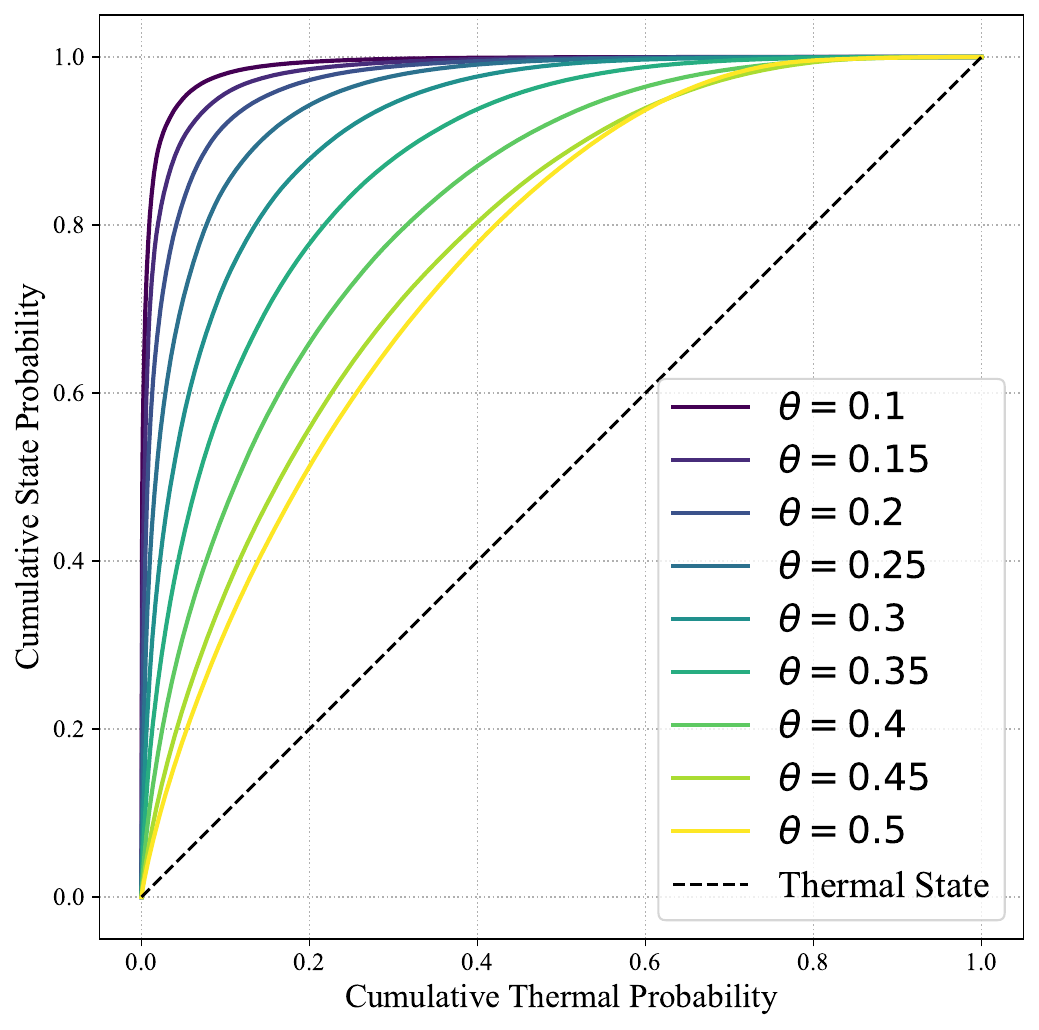}}\hfil 
\caption{Figures showing the thermomajorization curves of the initial tilted states at different potential strengths, averaged over 240 realisations of the AA model.}
\label{fig:thermomajorization}
\end{figure}

\begin{figure}[!htbp]
\centering
\subfloat[TFS]{\includegraphics[width=7cm]{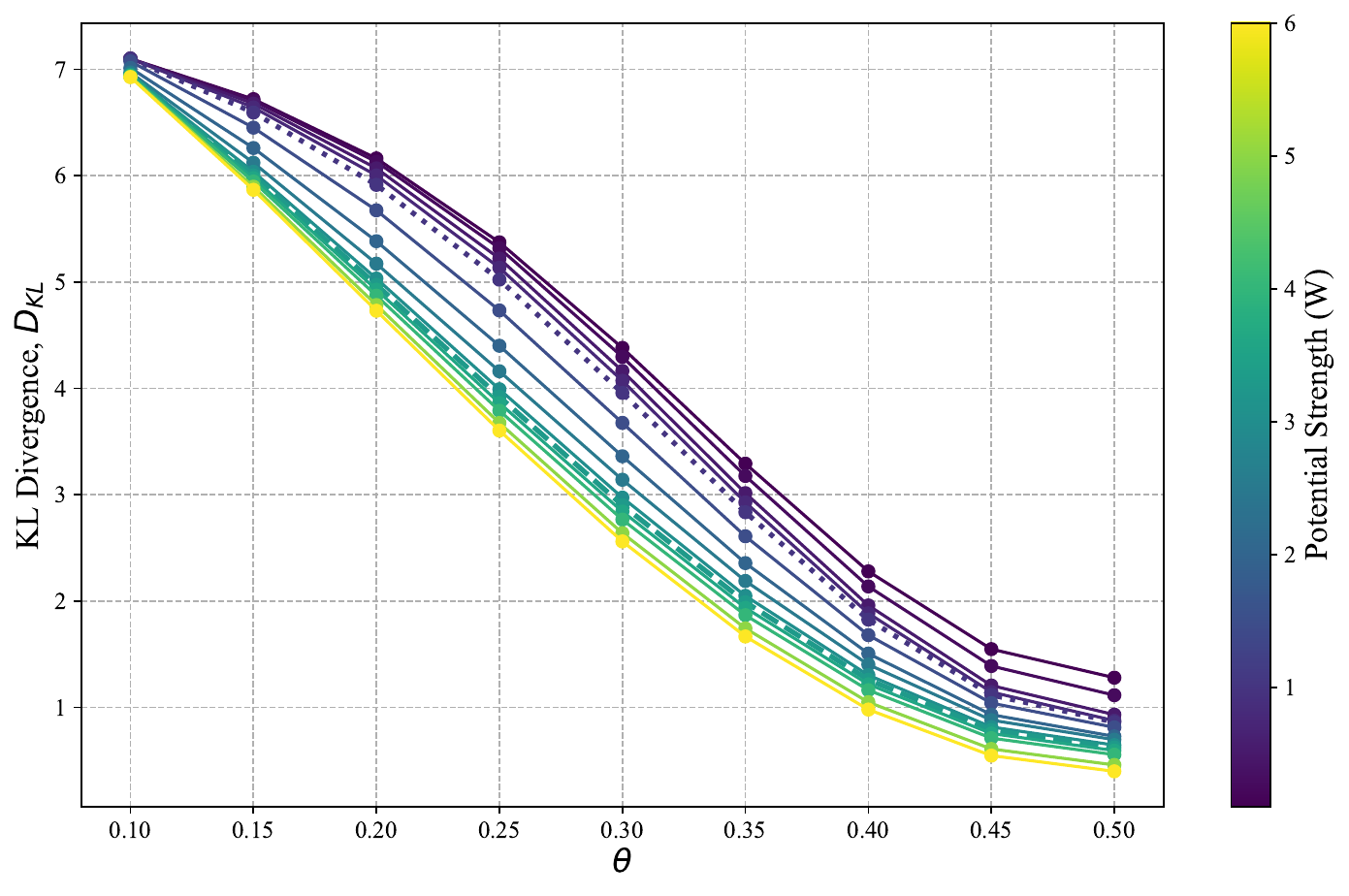}}\hfil
\subfloat[TNS]{\includegraphics[width=7cm]{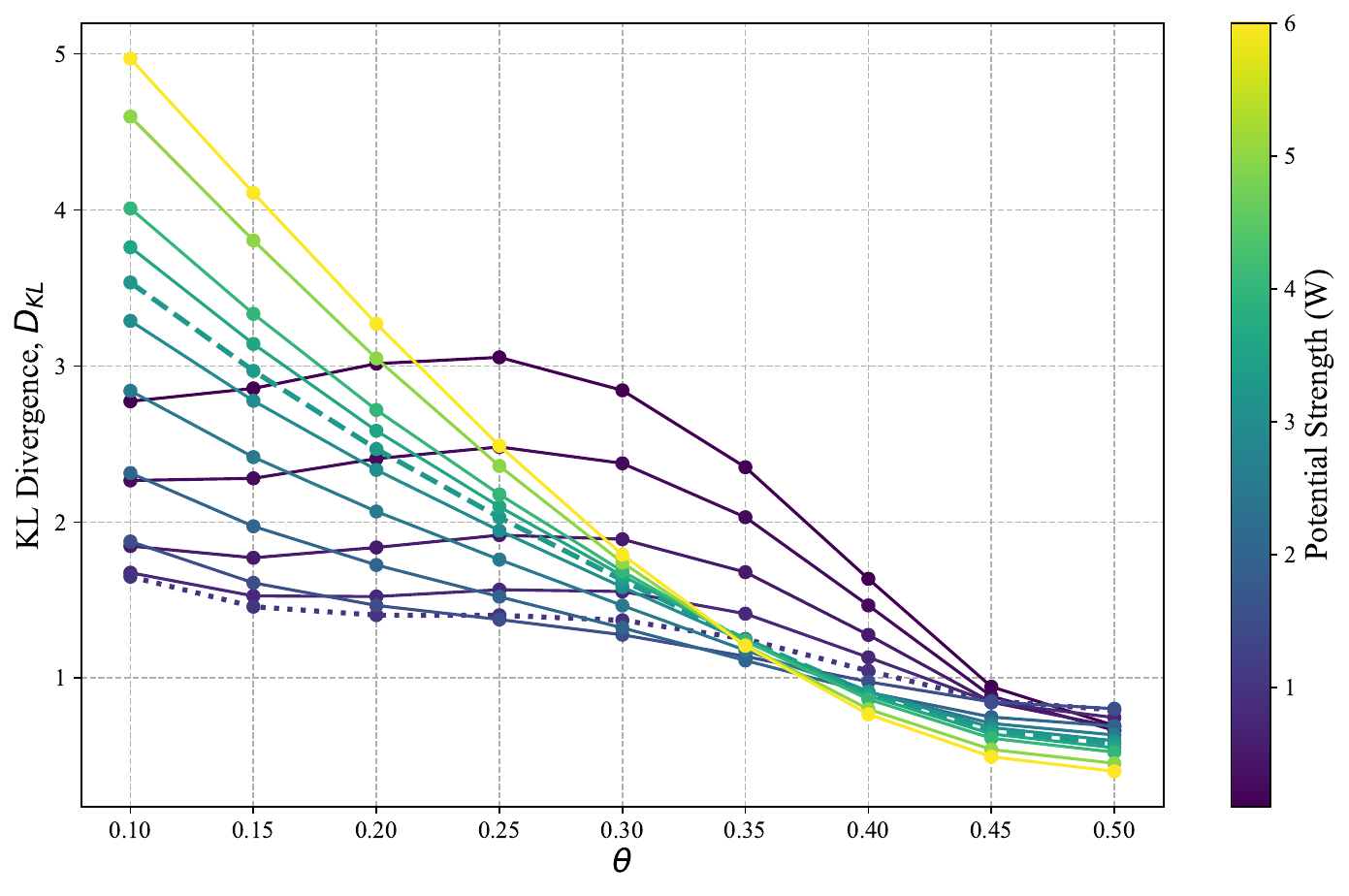}}\hfil 
\caption{Figures showing the KL divergence of the thermomajorization curves of Fig. \ref{fig:thermomajorization}. The dashed lines are used to highlight $W_c\approx3.3$, the critical point. The dotted line is used to indicate $W=1.0$, which corresponds to the lowest value of the potential strength for which the Néel state has a monotonically decreasing divergence.}
\label{fig:KLDivergence}
\end{figure}

The procedure for computing thermomajorization curves is as follows:
\begin{enumerate}
    \item Diagonalize the Hamiltonian to find the complete set of energy eigenstates $\{\ket{E_i}\}$ and eigenvalues $\{E_i\}$.
    \item For the tilted initial state, compute the state populations $p_i = |\braket{\psi(\theta)|E_i}|^2$.
    \item Numerically solve for the effective inverse temperature $\beta_{eff}$ of the system via 
    \begin{equation*}
        \frac{\sum_i E_i e^{-\beta_{eff}E_i}}{\sum_j e^{-\beta_{eff}E_j}}=E_{initial}= \bra{\psi(\theta)}H\ket{\psi(\theta)}
    \end{equation*}.
    \item Use $\beta_{eff}$ to construct the populations for the Gibbs state, $g_i=\frac{e^{-\beta_{eff}E_i}}{Z}$,\\ $Z=\sum_i e^{-\beta_{eff}E_i}$.
    \item Compute the population ratios $r_i = \frac{p_i}{g_i}$ and find the permutation $\pi$ that reorders these ratios in descending order $r_{\pi(1)}\geq r_{\pi(2)},...,r_{\pi(D)}$.
    \item Apply the permutation from the previous step to the populations of both the initial state ${p_{\pi(i)}}$ and the Gibbs state $g_{\pi(i)}$.
    \item Finally, use the cumulative sums of the reordered populations as coordinates for the thermomajorization curves:
    \begin{equation*}
        x_{k} = \sum_{i=1}^k g_{\pi(i)},\quad y_k=\sum_{i=1}^k p_{\pi(i)}.
    \end{equation*}
\end{enumerate}
The height of the resultant curve is a measure of how far the population induced by the tilt operation is from a thermal distribution (see Fig. \ref{fig:thermomajorization}). In the paradigm of quantum resource theory, the degree of \textit{athermality} of the distribution is a resource \cite{PhysRevLett.111.250404}; `hotter' states are ones with a higher KL divergence, while the `coldest' state (and the one with zero resource) is the thermal Gibbs state -- it is not the temperature difference since these plots have the same effective temperature. A thermal distribution on the thermomajorization curve would correspond to a diagonal line. The difference between the distributions can be computed using the Kullback–Leibler (KL) divergence
\begin{equation*}
    D_{KL}(p_{\pi}||g_{\pi})=\sum_{i}^Dp_i\log\Big( \frac{p_i}{g_i}\Big)
\end{equation*}
 (see Fig. \ref{fig:KLDivergence}), and accounts for the amount of the available. It is important to recognise that the states of different tilts are still going to converge to different steady states; the advantage of using thermomajorization curves is that they compare a tilted state to its \textit{own} (potentially hypothetical) thermal state.

\noindent
The weakness of this measure in the MBL regime ($W>3.3$) is highlighted when comparing the thermomajorization curves of Fig. \ref{fig:thermomajorization} to the survival probabilities shown in Fig. \ref{fig:ISCmpemba}. The observed behavior of the thermomajorization curves in the MBL regime (at $W=5.0$) is that larger tilts create states nearer to the (hypothetical) Gibbs state. On the other hand, the survival probabilities indicate that states with larger tilts reach steady states with which they have a lower overlap, and which, unsurprisingly, are decidedly athermal.
\\ \\
Conversely, the strength of this measure is that, in the ETH regime, it provides a measure of distance from the thermal state to which the system will generally converge. For the tilted ferromagnetic states, larger tilts (up to $\theta=0.5$) always correspond to a population distribution that is closer to that of the thermal one. The Néel state is different in that it exhibits non-monotonicity of the tilt operation for potential strengths $W<1.0$, whereafter it is also monotonically decreasing as a function of the tilt.
\\ \\
This data highlights the challenge of using entanglement asymmetry as a measure of distance from thermal (steady) state, especially in the ETH phase. While it is true that the tilt operation \textit{does} increase the asymmetry as measured by the entanglement asymmetry, it is not necessarily true that this asymmetry is aligned with the population of states being more distant from the thermal one.
\\ \\
What this analysis also highlights is an additional strength of the Krylov complexity as a probe of the QME in localizing systems; a region where thermomajorisation is not readily applicable as a meaningful measure, since the state will converge to a localized one, with a population distinct from the thermal one.

\subsection{Effects of averaging Lanczos Coefficients}
\label{app:avgLanczos}

In sec. \ref{sec:lanczos}, we present Fig. \ref{fig:b1s}, and study the behavior of the minimum and maximum of the $b_1$ values in the $(\theta, W)$ parameter space. In this appendix, we show how the behaviour of these minima and maxima are altered as more $b_n$ are included, corresponding to studying the average spread in Krylov space over increasing timescales. We will focus on the tilted Néel states, as the tilted ferromagnetic states display comparatively simple behavior, being approximately totally ordered by $n_{max}=10$.
\begin{figure}[!htbp]
\centering
\subfloat[TNS, $n_{max}=5$]{\includegraphics[width=7.0cm]{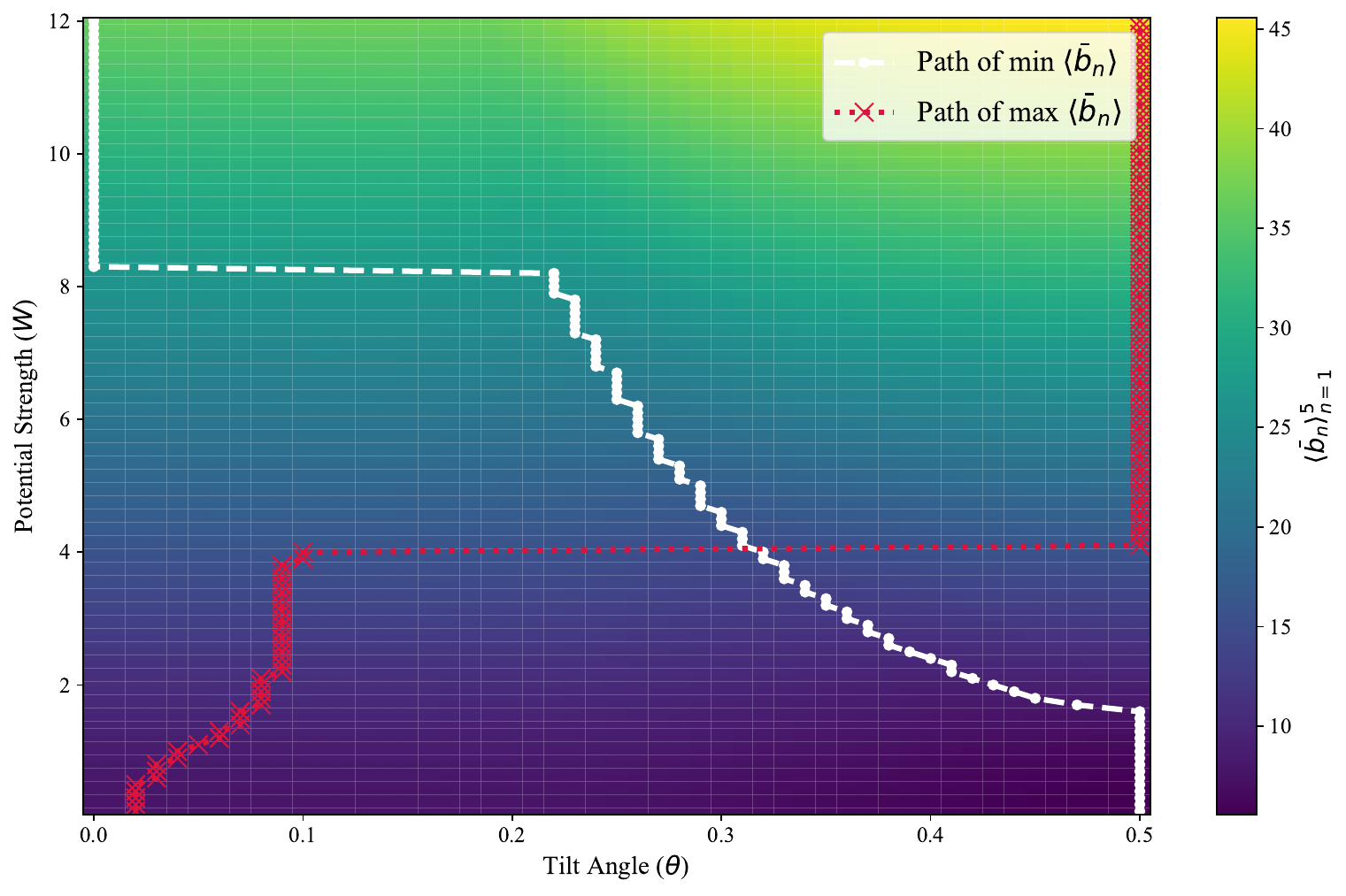}}\hfil
\subfloat[TNS, $n_{max}=10$]{\includegraphics[width=7.0cm]{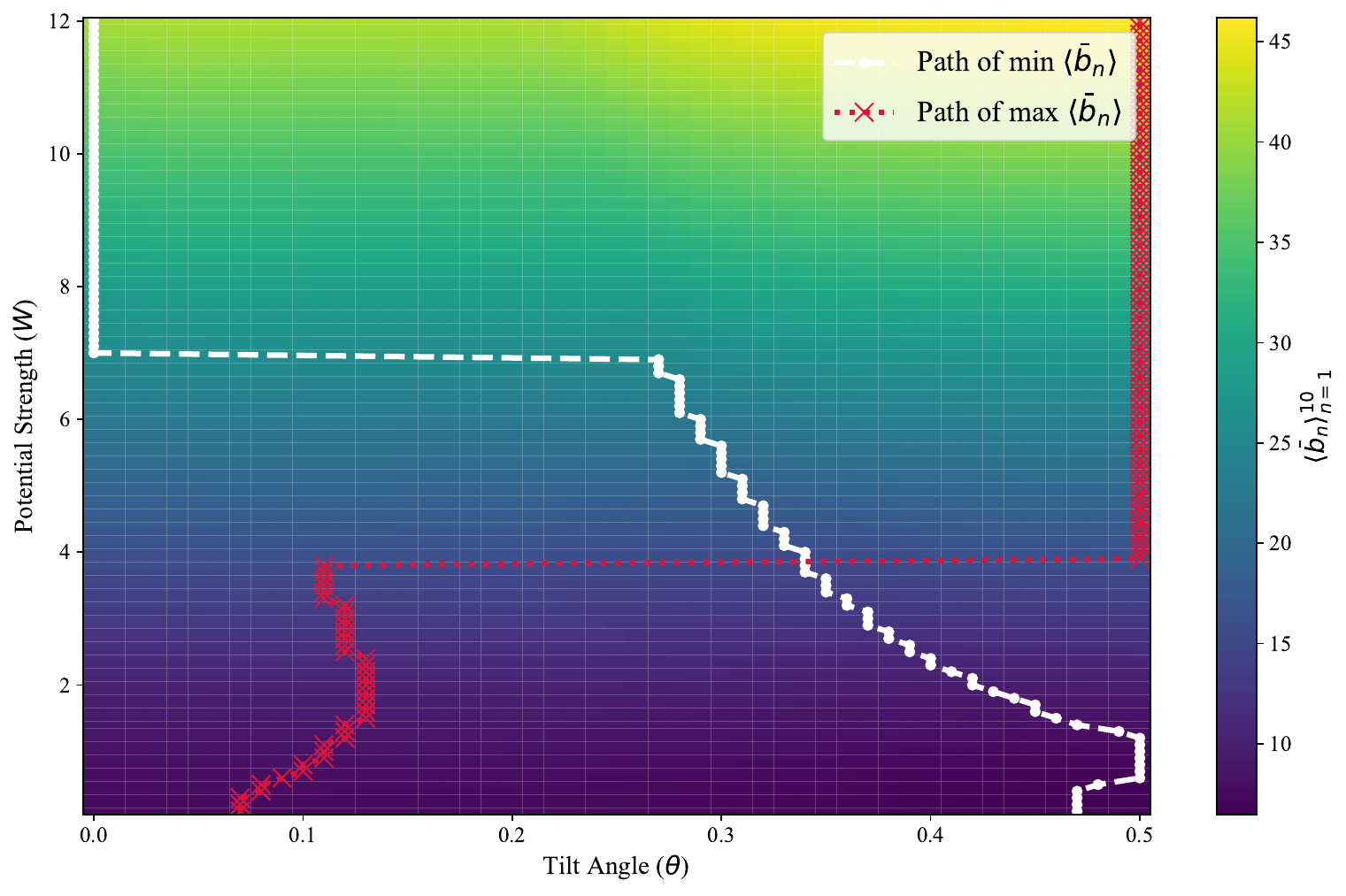}}\hfil 

\subfloat[TNS, $n_{max}=14$]{\includegraphics[width=7.0cm]{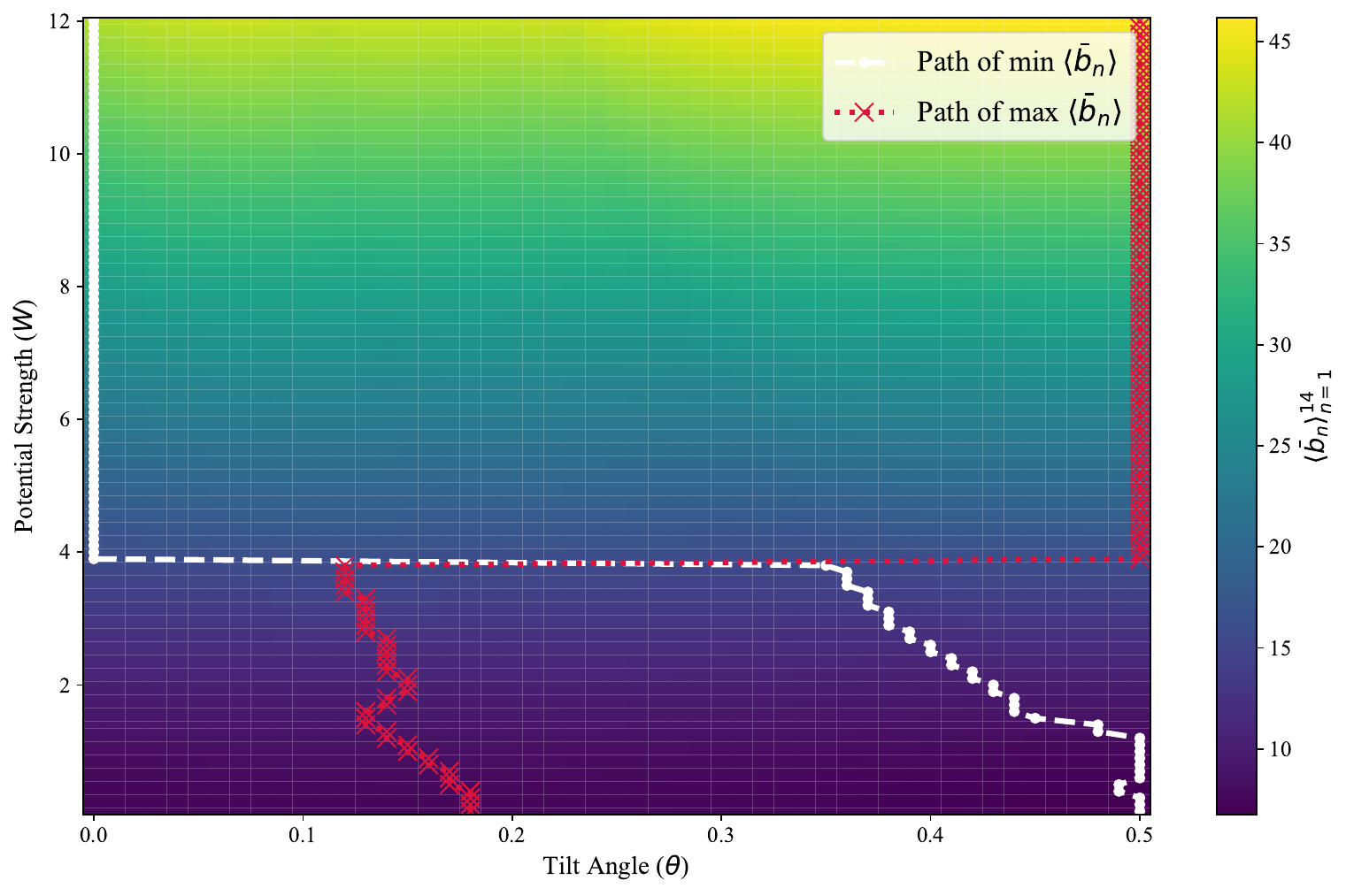}}\hfil
\subfloat[TNS, $n_{max}=28$]{\includegraphics[width=7.0cm]{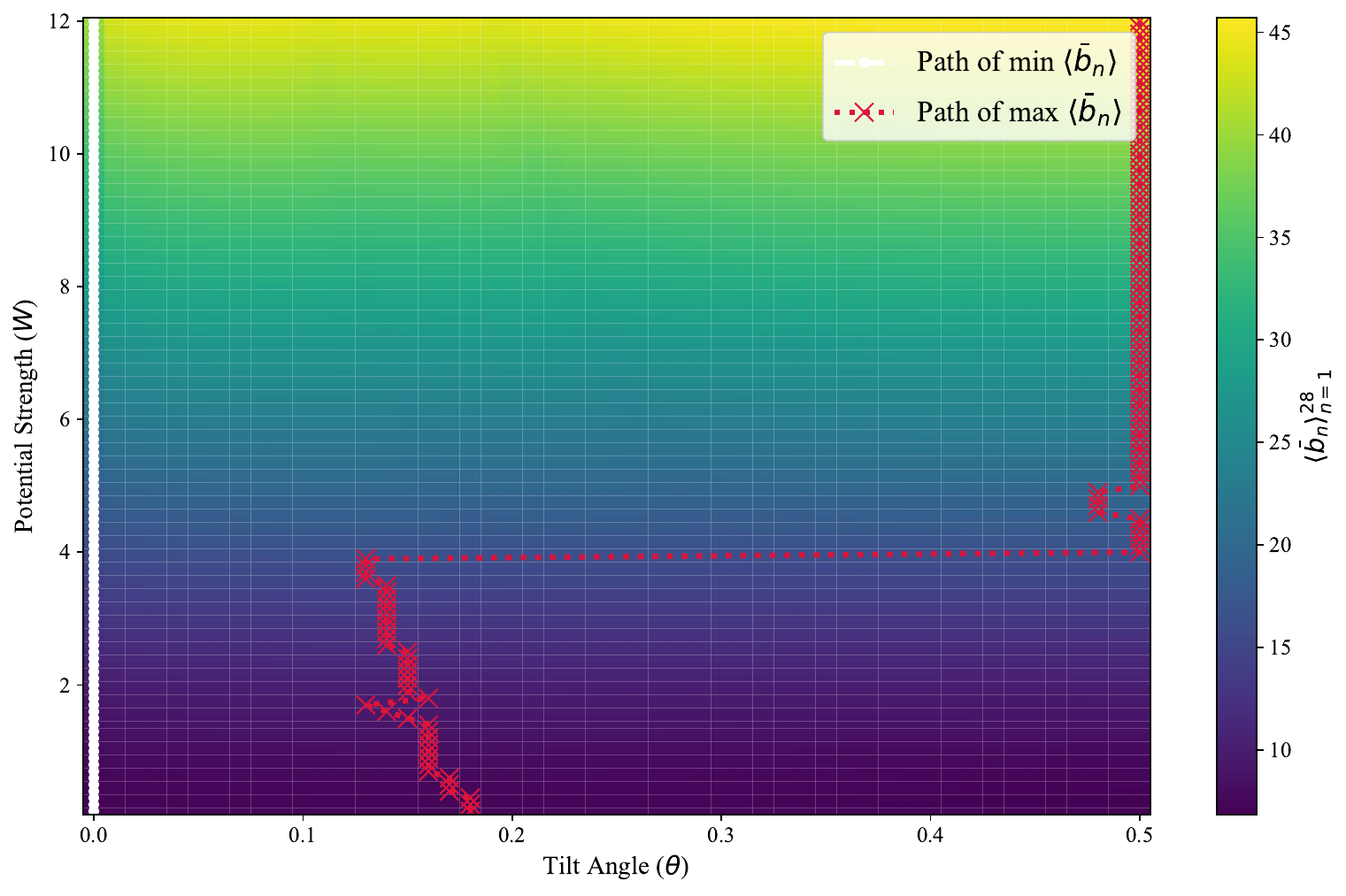}}\hfil 
\caption{Figures showing the heatmaps of the average of the first $n$ Lanczos coefficients in $(\theta,W)$ parameter space. Each $b_i$ is averaged over 240 realisations for an $N=12$ system, then the average of the first $n_{max}$ of these values is computed.}
\label{fig:bns}
\end{figure}
The effect of including additional Lanczos coefficients is clear -- it drives down the location of the minimum, eventually aligning with the minimum tilt for $28\leq n_{max}$. The stability of the location in the potential of the sudden maximum shift is interesting, lying between $W=3.6$ and $W=4.1$ for tested values of $n_{max}\in[1,50]$. It is interesting to note that the sudden shift of the minimum and the maximum overlaps at $n=14$, where the jump occurs going from $W=3.8$ to $W=3.9$.

\end{document}